%% file: rknew.tex
\documentstyle[11pt,amssymb,epsfig]{article}
\newcommand{\be}{\begin{equation}}
\newcommand{\ee}{\end{equation}}
\newcommand{\bea}{\begin{eqnarray}}
\newcommand{\eea}{\end{eqnarray}}
\newcommand{\ba}{\begin{array}}
\newcommand{\ea}{\end{array}}
\newcommand{\bmat}{\left(\ba}
\newcommand{\emat}{\ea\right)}
\def\3{\ss}

\def\d{\delta}
\def\ga{\gamma}
\def\Ga{\Gamma}

\def\g5{\gamma_5}
\def\mn{\mu\nu}
\def\rs{\rho\sigma}
\def\b{\beta}
\def\a{\alpha}
\def\ve{\varepsilon}
\def\r{\rho}
\def\si{\sigma}
\def\as2{\alpha^2_s}
\def\ha{{1\over 2}}

\def\Pon{P^{(0)n}}
\def\hPon{\hat P^{(0)n}}
\def\Q2{(Q^2_0)}
\def\zweib{\frac{2}{\beta_0}}
\def\vph{\varphi}
\def\nspm{NS\pm}
\def\gen{\gamma^{(1)n}}
\def\aspi{\frac{\a_s}{2\pi}}
\def\Pen{P^{(1)n}}
\def\hPen{\hat P^{(1)n}}    
\def\tolimit_#1{\mathrel{\mathop{\longrightarrow}\limits_{#1}}}
\def\tosim_#1{\mathrel{\mathop{\thicksim}\limits_{#1}}}
\begin{document}

%\title{Polarized Structure Functions and High Energy Polarized Scattering}
%\maketitle

\begin{flushright}
MPI-PhT/98-23 \\
DO-TH 98/02 \\
April 1998\\
\end{flushright}
%\vspace{70mm}
\begin{center}
\large {\bf Spin Physics and Polarized Structure Functions}\\ 
\mbox{ }\\
%\vfill
\normalsize
\vskip1cm
Bodo Lampe$^{\rm a,b}$ and Ewald Reya$^{\rm c}$
\vspace{0.5cm}\\
$^{\rm a}${\small\it
Sektion Physik der Universit\"at M\"unchen, 
Theresienstr. 37, D--80333 M\"unchen, FRG}\\
$^{\rm b}${\small\it
Max-Planck-Institut f\"ur Physik, 
F\"ohringer Ring 6, D-80805 M\"unchen, FRG}\\
$^{\rm c}${\small\it 
Institut f\"ur Physik, Universit\"at Dortmund, D-44221 Dortmund, FRG}
\vspace*{2cm}\\

%{\bf Bodo Lampe}
%\vskip0.3cm
%Max Planck Institut f\"ur Physik \\
%F\"ohringer Ring 6, D-80805 M\"unchen \\
%\vskip0.5cm
%{\bf Ewald Reya}
%\vskip0.3cm
%Institut f\"ur Physik \\ 
%Universit\"at Dortmund, D-44221 Dortmund \\ 
%\vspace{1cm}
\end{center}

\begin{abstract}
A review on the theoretical aspects and the experimental results of
polarized deep inelastic scattering and of other hard scattering 
processes is presented.
The longitudinally 
polarized structure functions are introduced and cross section
fromulae are given for the case of photon as well as $W^{\pm}$ 
and $Z^0$ exchange.
Results from the SLAC and CERN polarization experiments are
shown and compared with each other as well as their implications 
for the integrated $g_1(x,Q^2)$ are reviewed. 
More recent experiments presently underway (like HERMES at DESY) 
and future projects (like RHIC at BNL, HERA--$\vec N$ and a 
polarized HERA collider at DESY) are discussed too. The  
QCD interpretation and the LO and NLO $Q^2$--evolution
of $g_1$, i.e. of the longitudinally polarized parton densities,  
is discussed in great detail, in particular the role of the 
polarized gluon density, as well as the expectations for 
$x\rightarrow 0$.  
Particular emphasis is placed on the first moment of the polarized
structure function in various factorization schemes, 
which is related to the axial anomaly, and on its relevance 
for understanding the origin of the proton spin.
Sum rules (i.e. relations
between moments of the structure functions) are derived and
compared with recent 
experimental results. Various other phenomenological applications
are discussed as well, in particular the parametrizations
of polarized parton densities as obtained from recent data 
and their evolution in $Q^2$. Furthermore, jet, heavy quark 
and direct photon production
are reviewed as a sensitive probe of the polarized gluon density,  
and the physics prospects of the future polarized experiments 
at RHIC ($\vec p \vec p$) and a polarized HERA collider 
($\vec e \vec p$) are studied.
DIS semi--inclusive asymmetries and elastic neutrino--proton 
scattering are reviewed, which will help to disentangle the 
various polarized flavor densities in the nucleon.
The status of single and double spin asymmetries, and the 
observation of handedness in the final state, are discussed as well. 
Structure functions for higher spin hadrons and nuclei are defined
and possible nuclear effects on high energy spin physics are reviewed.
The theoretical concept of spin--dependent parton distributions 
and structure functions of the polarized photon is presented 
and possibilities for measuring them are briefly discussed. 
Various nonperturbative approaches to understand the origin of 
the proton spin are reviewed, such as the isosinglet $U_A(1)$ 
Goldberger--Treiman relation, lattice calculations and the chiral 
soliton model of the nucleon. 
The physical interpretation and model calculations 
of the transverse structure function $g_2$ are presented, as well 
as recent twist--3 measurements thereof,  
and the Burkhardt--Cottingham sum rule is revisited. Finally, the 
physics of chiral--odd 'transversity' distributions is described 
and experimental possibilities for delineating them are reviewed, 
which will be important for a complete understanding of the leading 
twist--2 sector of the nucleon's parton structure. 
In the Appendix the full 2--loop anomalous dimensions and Altarelli--Parisi
splitting functions governing the $Q^2$--evolution
of the structure function $g_1$
are given. 
\end{abstract}

\tableofcontents{}

\input{k1tex}

%\section{The Polarized Structure Functions}
\input{k2tex}

%experimental section 
\input{k3tex}

%g1
\input{k41tex}

\input{k42tex}
\input{k43tex}
\input{k44tex}

%first moment of g1
\input{k51tex}
\input{k52tex}

\input{k53tex}

\input{k54tex}

\input{k55tex}

%\end{multicols}

%phenomenology 
\input{k61tex}

\input{k62tex}

\input{k622tex}
\input{k63tex}

\input{k64tex}

\input{k65tex}
\input{k66tex}

\input{k67tex}

\input{k68tex}

\input{k69tex}

\input{k699tex}

\input{k6999tex}

%nonperturbative
\input{k71tex}

%transverse
\input{k81tex}
\input{k82tex}

%summary and appendix
%\input{kstex}
\input{katex}

\input{kfnewtex}
\end{document}

%% file: k1tex
\setcounter{equation}{0}
\section{Introduction}
%\subsection{Basics}
 
One of the most fundamental properties of elementary
particles is their spin because it determines their symmetry
behavior under space-time transformations. The spin degrees
of freedom may be used in high energy experiments to get
informations on the fundamental interactions which are more
precise than those obtained with unpolarized beams. For
example, the SLC experiment at SLAC is able to determine
$sin^2\theta_W$ with a higher precision by using polarized
$e^+e^-$ beams than current experiments at LEP with unpolarized
beams 
(for a recent review see, e.g., \cite{schaile}).
%\cite[Schaile, 1994]{schaile} 
 
Another aspect of polarization is the question how the spin of
non-pointlike objects like the nucleons is composed of the spins
of its constituents, the quarks and gluons . 
This question
can best be answered in high energy experiments because the
quarks and gluons behave as (almost) free particles at energy/momentum-scales
$Q>>\Lambda_{QCD}$. It is possible to attribute
numbers $\Delta\Sigma$ and $\Delta g$ to the quark and gluon
spin content of the nucleons which describe their total (integrated)
contribution
to the nucleon spin in the following sense \cite{jaffe,ji7,ji8}
\begin{equation}
{1\over 2} = {1\over 2} \Delta\Sigma + \Delta g + L_z \label{111}
\end{equation}
where on the left hand side we have the spin ($+{1\over 2}$)
of a polarized nucleon state and on the right hand side a
decomposition in terms of
$\Delta\Sigma (Q^2), \Delta g(Q^2)$
and the relative orbital angular momentum $L_z(Q^2)$ among all the
quarks \cite{sehgal} and
gluons. Furthermore,
$\Delta\Sigma = \Delta u + \Delta d + \Delta s + \Delta \bar{u} + \dots$
can be further decomposed into the contributions from the
various quark species which will be discussed in more detail in
Sections 4, 5 and 6.
 
\begin{figure} 
\begin{center}
\epsfig{file=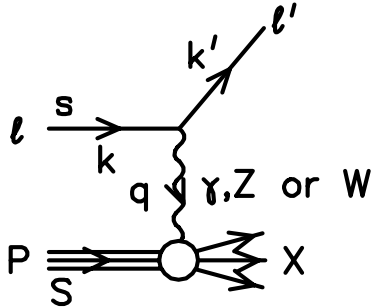,height=4cm}
\caption{The basic polarized deep inelastic scattering process}
\label{fig1}
\end{center}
\end{figure}

Unfortunately, the decomposition (\ref{111}) cannot directly be
measured in experiments. Instead various other combinations
of $\Delta\Sigma$ and $\Delta g$ appear in experimental
observables. The predominant role in the development of
understanding the spin structure of nucleons
is played by the deep inelastic
leptoproduction processes $\ell N\to \ell ' X$ (Fig. \ref{fig1})
because of their unique simplicity: The processes are
initiated by leptons and are
totally inclusive in the hadronic final state. More than 25 years ago
Bjorken \cite{bjorken} 
and others \cite{gourdin,ellis}  
have anticipated the
significance of these processes for the understanding of the
nucleon spin structure. But only recently, experiments have
become precise enough to test some of the theoretical ideas
developed so far. Two
earlier experiments at SLAC \cite{alguard,baum,baum1}   
and CERN \cite{papavassiliou,ashman,ashman1}   
were
followed by new results from both laboratories in the recent
years.
Whether or not there is agreement between the new CERN 
% \cite{adeva,adams,adams2} 
and SLAC 
% \cite{anthony,abe,abe1,abe2}   
data
will be discussed in detail and commented
in the course of this work. The Hermes experiment which 
takes place in the DESY--HERA tunnel 
\cite{coulter,dueren}
is expected to
have somewhat smaller errors and will hopefully lead to further
insights about the spin structure of the nucleons.

In Section 2 we summarize all relevant expressions for polarized 
DIS cross sections and structure functions for neutral and charged 
electroweak currents. Previous and recent results of longitudinally 
polarized DIS experiments for $g_1^N(x,Q^2)$ are presented and 
compared with each other in Section 3. The LO and NLO QCD renormalization 
group evolution of $g_1(x,Q^2)$ and of longitudinally polarized 
parton densities $\delta f(x,Q^2)$ are derived in Sect. 4, as well as 
their small--x behavior. Section 5 is devoted to the first moment 
(i.e. total helicities) of longitudinally polarized parton densities 
and of $g_1$ in various factorization schemes which is related 
to the axial anomaly and its relevance for understanding the origin 
of the proton spin. Here, the Bjorken and Drell--Hearn--Gerasimov 
sum rules are derived and compared with recent experimental results 
as well. 

Section 6 includes most of the phenomenological aspects relevant 
for longitudinally polarized processes. We start in Sect. 6.1 with 
a brief historical review of 'naive' parton model expectations 
for polarized parton densities; then we turn to recent developments 
for determining $\delta f(x,Q^2)$ in LO and NLO from recent data on 
$g_1^{p,n}(x,Q^2)$ and the implications for their first moments 
(total helicities). Here we also discuss briefly the present status 
of the orbital component $L_z=L_q+L_g$ in (\ref{111}), such as the 
$Q^2$--evolution equations for $L_{q,g}(Q^2)$ and how one might 
possibly relate them to measurable observables. 
In addition, hard processes initiated by doubly (singly) polarized
hadron--hadron collisions such as the production of heavy
quarks, of large-$p_T$ photons and jets, of Drell-Yan dimuons 
etc.\ will be also suitable to measure the polarized parton
distributions $\delta f(x,Q^2),f=q,\bar q,g$, in particular
the gluon distribution $\delta g(x,Q^2)$. 
Details will be discussed in the various subsections of Section 6. 
Furthermore, polarized $ep$
and $e^+e^-$ collisions can also shed light on the so far unmeasured
polarized parton desities of the photon which are theoretically 
formulated and discussed in Sect. 6.12. 

Various nonperturbative approaches to understand the origin of the 
proton spin are presented in Sect. 7, such as the isosinglet 
Goldberger--Treiman relation, lattice calculations and the chiral 
soliton model of the nucleon. Finally structure functions resulting 
from transverse polarizations are dealt with in Sect. 8. In Sect. 8.1 
the theoretical concepts and model calculations of the transverse 
structure function $g_2(x,Q^2)$ are presented as well as recent 
twist--3 measurements thereof, and the Burkhardt--Cottingham sum rule 
is revisited. The physics of the chiral--odd 'transversity' distributions 
is described in Sect. 8.2 and experimental possibilities 
for delineating them are reviewed. It should be remembered that 
a complete understanding of the leading twist--2 parton structure 
of the nucleon requires, besides the unpolarized and longitudinally 
polarized parton densities $f(x,Q^2)$ and $\delta f(x,Q^2)$, 
also the knowledge of the transversity densities $\delta_T q(x,Q^2)$ 
which are experimentally entirely unknown so far. 

The full 2--loop polarized Altarelli--Parisi splitting functions 
$\delta P_{ij}^{(1)}(x)$ and their Mellin n--moments (anomalous dimensions)
$\delta P_{ij}^{(1)n}$, governing the $Q^2$--evolution of 
$g_1(x,Q^2)$ and $\delta f(x,Q^2)$, are summarized in the Appendix.   

\subsection{Polarization of a Dirac Particle}

Let us start with a few basic facts about the polarization
of a relativistic spin ${1\over 2}$ particle. A free Dirac
particle of four-momentum $p$ and mass $m$ is described by a four
component spinor $u(p,s)$ which satisfies the equation
\begin{equation}
(p\llap{/} -m)u(p,s)=0
\label{121}\end{equation}
where $p\llap{/} =\gamma_\mu p^\mu$.
The polarization vector
$s$ is a pseudovector which fulfils $s^2\equiv{(s^0)}^2-(\vec s)^2=-1$
and $sp=0$.
The projection operator onto a state with polarization $s$
is known to be
\begin{equation}
P(s)= \ha (1+\g5 s\llap{/})
\label{122}\end{equation}
The transformation properties of $s$ are
given in Table 1
where we consider two Dirac particles which
move along the $z$-direction in the lab-frame, one of it
with transverse and the other one with longitudinal
polarization. The transverse polarization vector is
not changed when going from the rest frame to the lab--frame,
but the longitudinal is. The important point to notice is
that at high energies $E >> m$ the product $m s_L$ remains
finite and converges to $p$:
\begin{equation}
%ms_L @>>{E \rightarrow \infty}> p \/.
ms_L  \tosim_{E \rightarrow \infty} p \/.
\label{123}
\end{equation}
This fact will be used repeatedly in later applications. 
%\end{multicols}
\begin{table}[h]
\label{tab1}
\begin{center}
\begin{tabular}{|l|l|l|} 
\hline  
              & transverse polarization & longitudinal polarization  \\ 
\hline
 rest frame 
p=(m,0,0,0) & $s_T=(0,1,0,0)$ & $s_L=(0,0,0,1)$ \\ 
\hline
lab frame 
$p=(E,0,0,\sqrt{E^2-m^2})$ & $s_T=(0,1,0,0)$ & 
 $s_L={1\over m}(\sqrt{E^2-m^2},0,0,E)$ \\  
\hline
\end{tabular}
\bigskip
\end{center}
\caption{A transverse and a longitudinal polarized Dirac particle 
in their rest and laboratory frame}  
\end{table}

%% file: k2tex
%\begin{multicols}{2}
\setcounter{equation}{0}
\section{The Polarized Structure Functions}
\subsection{Basics of Pure Photon Exchange }
\begin{figure}
\begin{center}
\epsfig{file=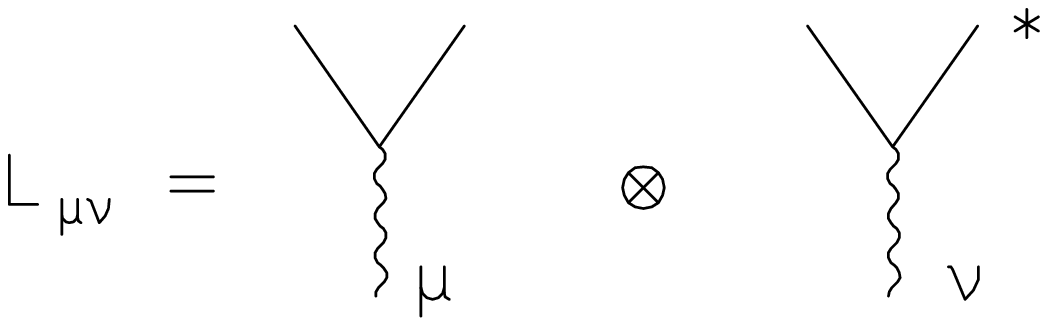,height=2cm}
\bigskip
\caption{The definition of the lepton tensor}
\label{fig2}
\end{center}
\end{figure}

Let us first consider Fig. \ref{fig1}
with photon exchange only. We assume that {\it both} the
incoming lepton {\it and} the incoming nucleon are polarized
(polarization vectors $s^{\mu}$ and $S^{\mu}$). We shall see below why
this is important. The procedure for polarized
particles is analogous to the case of unpolarized particles, i.e.\
the cross section is a product of a leptonic tensor
$L_{\mu\nu}$ which is known (cf.~Fig. \ref{fig2}) and a hadronic
tensor $W^{\mu\nu}$ which can be expanded into Lorentz
covariants whose coefficients define the structure functions
which are to be measured. In unpolarized $e(\mu)N$
scattering one has
the well known functions $F_1$ and $F_2$ whereas for polarized
particles two additional
functions $g_1$ and $g_2$ arise:
\begin{eqnarray} \nonumber 
W_{\mu\nu}&=&\int d^4x e^{iqx} < PS\vert J_\mu (x)J_\nu (0)\vert PS> 
\\ 
&=&W_{\mu\nu} (sym.) +i\frac{M}{Pq}\ve_{\mu\nu\rho\sigma}q^\rho 
\left[S^\sigma g_1(x,Q^2) +(S^\sigma-\frac{Sq}{Pq}P^\sigma)g_2(x,Q^2)
\right]
\label{211}
\end{eqnarray}
$\epsilon_{\mu\nu\rho\sigma}$ is the totally antisymmetric
tensor in 4 dimensions,
$\epsilon_{0123}=1,\epsilon_{1023}=-1$ etc. and the polarization 
vector of the proton is normalized to $S^2=-1$. Note that there
is no analogy $F_{1,2}\leftrightarrow g_{1,2}$ because the
contribution of $g_2$ to the cross section vanishes
in the limit of ultra-relativistic on-shell quarks
$(S^\sigma\sim P^\sigma)$ since there are not enough four-vectors
available anymore to form an antisymmetric combination in
Eq. (\ref{211}). Furthermore $Q^2 = -q^2$ and $x =
{Q^2\over 2Pq}$ are the usual (Bjorken) variables of the DIS process.
It is not entirely
trivial to see that (\ref{211}) is the most general form of the
antisymmetric hadron tensor. One has to make use of the
$\epsilon$-identity
\begin{equation}
g^{\alpha\beta}\ve^{\mu\nu\rho\sigma}=g^{\alpha\mu}
\ve^{\beta\nu\rho\sigma}+g^{\alpha\nu}\ve^{\mu\beta\rho\sigma}+
g^{\alpha\rho}\ve^{\mu\nu\beta\sigma}+g^{\alpha\sigma}
\ve^{\mu\nu\rho\beta}
\label{212}\end{equation}
to get rid of tensors like
\begin{equation}
\left[(p_\mu\ve_{\nu\alpha\beta\rho}-p_\nu\ve_{\mu\alpha\beta\rho})
q^\rho +p\cdot q\ve_{\mu\nu\alpha\beta}\right]S^\alpha p^\beta\/.
\label{213}\end{equation}
The incoming lepton (Fig. \ref{fig1})
is assumed to be polarized too.
Why is that necessary? To see that
let's have a look at the lepton tensor
\begin{equation}
L_{\mu\nu}=tr\left[(1+\g5 s\llap{/})(k\llap{/} +m_l)
\gamma_\mu (k\llap{/}'+m_l) \gamma_\nu\right] \/.
\label{214}\end{equation}
Obviously, $L_{\mu\nu}$ consists of a part independent of the 
lepton polarisation
$s^\beta$ and a part linear in $s^\beta$, the former being 
symmetric in $\mu$ and $\nu$, the latter antisymmetric:
\begin{equation}
L_{\mu\nu}=L_{\mu\nu}(sym.)+2im_l\ve_{\mu\nu\alpha\beta}
q^\alpha s^\beta\/.
\label{215}\end{equation}
The antisymmetry of the last term is due to the
$\gamma_5$ in (\ref{122}) and to the vector coupling of the photon
to fermions. With the symmetric part alone in Eq. (\ref{215}), $g_1$
and $g_2$ cannot be extracted from Eq. (\ref{211}).
One needs the antisymmetric part, i.e.
the lepton polarization.

From (\ref{215}) it seems that all polarization effects are
suppressed at high energy by a factor $m_l$.
However, in the case of longitudinal
polarization one has $m_l s^\beta\to k^\beta$
[according to (\ref{123})] and thus there is no suppression by factors
of $m_l$. In the following we shall always presume the
leptons to be longitudinally polarized.
 
\begin{figure}            
\begin{center}            
\epsfig{file=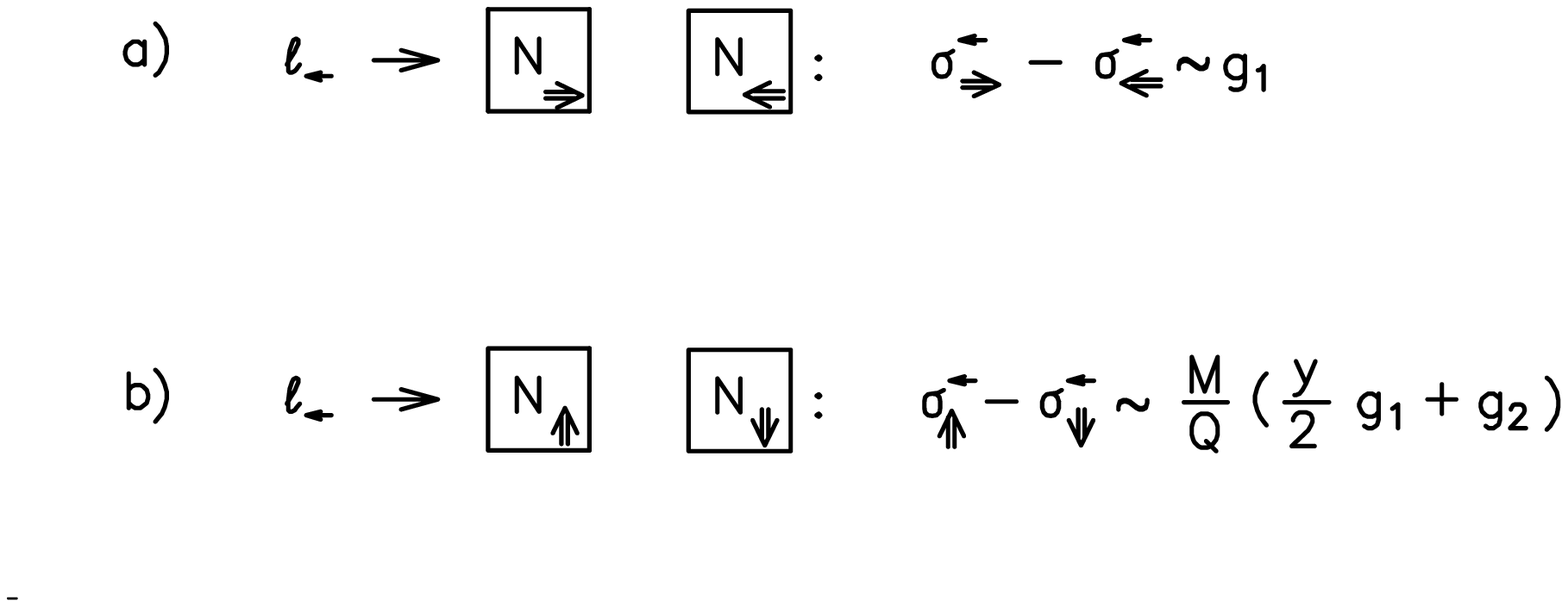,height=5cm}
\caption{The basic form of the polarized experiments}
\label{fig3}             
\end{center}              
\end{figure}

How to measure $g_1$ and $g_2$? The cross section $\sigma\sim
L_{\mu\nu} W^{\mu\nu}$ will be of the form
\be
L_{\mu\nu}W^{\mu\nu}=L_{\mu\nu}\hbox{ (sym.) }W^{\mu\nu}\hbox{ (sym.) } 
 +L_{\mu\nu}\hbox{ (antisym.) }W^{\mu\nu} \hbox{ (antisym.) } \/.
\label{216}\ee
One should try to get rid of the first term in (\ref{216})
because
it is the cross section for unpolarized scattering.
One
possibility is to consider differences of cross sections
with nucleons of opposite polarization
\cite{carlson,hey,jaffe1,anselmino1,roberts}  
as is depicted in
Fig. \ref{fig3}. 
In both parts of the figure one starts with a beam of
high energetic leptons with lefthanded helicity (=
longitudinally polarized with spin vector antiparallel to
the direction of motion). This beam is sent to two nucleon
probes with opposite longitudinal
polarization, i.e.\ with their spins along the direction
of the lepton beam and opposite to it ,
and to two probes with opposite
transverse polarization.
In the difference of the cross
sections [part a) of Fig.\ref{fig3}] the unpolarized structure
functions drop out and only $g_1$ survives (with respect to the
suppressed $(2 yx^2M^2/Q^2)g_2$ contribution, where $y=\frac{Pq}{Pk}$),
i.e. $g_1$ can
in principle be uniquely determined from measuring this
difference. Similarly, in the difference of cross sections
obtained from part b) of Fig.\ref{fig3}, the transverse polarization case
($kS={\vec k} \cdot {\vec S}=0$), the sum ${y\over 2} g_1+g_2$
appears. However, there is an overall
suppression factor $2xM/\sqrt{Q^2}$, where $M$ is the nucleon
mass so that $g_2$ can be obtained only from rather low
energy experiments. The appearance of this factor has, 
of course, to do with the transverse polarization.
We can take the fact that $g_2$ appears only in cross
sections with transverse polarized nucleons as a hint that
it is difficult to accomodate $g_2$ in the parton model.
There is no notion of transversality in the conventional parton
model.
We shall come back to this `transverse spin structure function'
in Section 8.

{\subsection{Quantitative Formulas for Pure Photon Exchange }
 
To be more specific let us write down the most general
cross section difference relevant for polarized deep inelastic fixed
target $\ell N$ scattering  
\cite{jaffe1,anselmino1} : 
\bea \nonumber 
 \frac{d^3[\sigma (\alpha)-\sigma (\alpha +\pi)]}{dxdy d\phi}=
 \frac{8\alpha^2}{Q^2}\Bigl\{\cos\alpha
\Bigl[(1-\frac{y}{2}-
\frac{y^2\gamma^2}{4})g_1(x,Q^2)
-\frac{y}{2}\gamma^2g_2(x,Q^2)\Bigr]  
\\
 -\sin\alpha\cos\phi\gamma\sqrt{1-y-\frac{y^2\gamma^2}{4}}
[\frac{y}{2}g_1(x,Q^2)+g_2(x,Q^2)]\Bigr\}\/.\label{221}   
\eea
This formula comprehends all information from the antisymmetric part
of the tensor Eq. (\ref{211}) where $\alpha$ is the angle between
the lepton beam momentum vector $\vec k$ and the nucleon-target
polarisation vector $\vec S$, $\phi$ is the angle between the $k-S$ plane
and the $k-k'$ lepton scattering plane (cf. Fig. \ref{fig4}), $\gamma=2Mx
/\sqrt{Q^2}$ and a scaling limit has not been taken. From Eq.
(\ref{221}) it is obvious that effects associated with $g_2$ are suppressed
(at least) by a factor ${2M \over \sqrt{Q^2}}$ with respect to the leading
terms.

\begin{figure}
\begin{center}
\epsfig{file=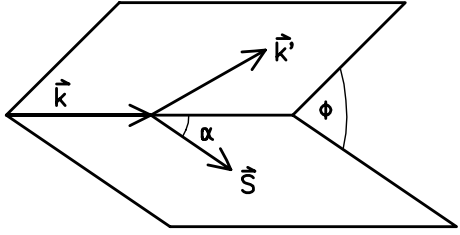,height=3cm}
\bigskip
\caption{The geometry of the polarized deep inelastic 
scattering process in the lab frame}
\label{fig4}
\end{center}
\end{figure}

More convenient than differences of cross sections are asymmetries
\begin{equation}         
A(\a)=\frac{\si(\a)-\si(\a +\pi)}{\si(\a)+\si(\a +\pi)}
\end{equation}           
as, for example, the longitudinal asymmetry
\begin{equation}         
A_L=\frac{\si^{\leftarrow}_{\Rightarrow}-\sigma^{\leftarrow}_{\Leftarrow}}
          {\si^{\leftarrow}_{\Rightarrow} +\si^{\leftarrow}_{\Leftarrow}}
\label{223}              
\end{equation}           
obtained for $\alpha =0$, and the transverse asymmetry $A_T$ obtained
for $\alpha=\pi/2$ and an asymmetric integration over $\phi$.

$A_L$ picks up the coefficient of $\cos\a$ in Eq. (\ref{221})
whereas $A_T$ picks up the coefficient of $\sin \, \a \, \cos \, \phi $.
Events with $\phi$ near $\pi/2$ or $3\pi/2$ (where the nucleon spin is
perpendicular to the scattering plane) can be obviously neglected in
the determination of $A_T$; they are not a good measure of 
$g_2+yg_1/2$.

The really interesting quantities are the virtual photon asymmetries
\begin{equation}         
A_1=\frac{\si_{1/2}-\si_{3/2}}{\si_{1/2}+\si_{3/2}}\label{224}
\end{equation}           
and                      
\begin{equation}         
A_2 = \frac{2\sigma_{TL}}{\si_{1/2}+\si_{3/2}}\label{225}
\end{equation}           
where                    
$\sigma_{1/2}$ and $\sigma_{3/2}$ are the virtual photoabsorbtion
cross sections when the projection of the total angular momentum of
the photon-nucleon system along the incident lepton direction is
1/2 and 3/2.             
Note that $\si_T=\ha (\si_{1/2}+\si_{3/2})$ and that
the term $\sigma_{TL}$ arises from the interference
between transverse and longitudinal amplitudes.
The significance of these quantities will be clarified in
subsection 2.3.          

$A_1$ and $A_2$          
can be related, via the optical theorem,
to the measured quantities $A_L$ and $A_T$ ,
or, equivalently, to the structure functions
by means of the following relations:
\begin{equation}
A_L = D(A_1+\eta A_2)  
\label{226}
\end{equation}   
\begin{equation} 
A_T = d(A_2-\xi A_1)
\label{226a}
\end{equation}           
with                     
\begin{equation}         
A_1 = (g_1-\gamma^2 g_2)/F_1 
\label{227}              
\end{equation}           
\begin{equation}
A_2 = \gamma (g_1+g_2)/F_1\/.
\label{227a}
\end{equation}           
The kinematic factors D,d and $\eta, \xi$ are defined by
%\begin{equation}         
%D = \frac{y(2-y)}{y^2+2(1-y)(1+R)} 
%\label{228}   
%\end{equation}           
%\begin{equation}
%\eta = 2\gamma (1-y)/(2-y) 
%\label{228a}              
%\end{equation}           
%\begin{equation}
%d = D\sqrt{\frac{2\epsilon}{1+\epsilon}}
%\label{228b}
%\end{equation}           
%\begin{equation}
%\xi = {\eta (1+\ve) \over 2\ve}
%\label{228c}
%\end{equation}           
\begin{eqnarray}
D &=& \frac{y(2-y)}{y^2+2(1-y)(1+R)}\nonumber\\
\eta &=& 2\gamma (1-y)/(2-y)\nonumber\\
d &=& D\sqrt{\frac{2\epsilon}{1+\epsilon}}\label{228}\\
\xi &=& \eta (1+\ve)/2\ve\nonumber
%\label{228}
\end{eqnarray}
with                     
\begin{equation}         
\ve ={(1-y) \over (1-y+y^2/2)}   
\label{229} 
\end{equation}           
being the degree of transverse polarization of the virtual
photon.D and d can be regarded as depolarization factors of the
virtual photon.          
Note that $\eta$ and $\xi$ are of order $M/\sqrt{Q^2}$. Finally,
$R$ is the ratio of cross sections for longitudinally and transversely
polarized virtual photons on an unpolarized target,defined by
\begin{equation}         
2xF_1(x,Q^2)=F_2(x,Q^2) \frac{1+\gamma^2}{1+R(x,Q^2)} \/.
\label{2210}
\end{equation}           
Note that in the limit $\gamma^2\equiv 4M^2 x^2/Q^2\ll 1$, 
\begin{equation} 
R={F_L \over 2xF_1}
\label{2211} 
\end{equation}
where $F_L\equiv F_2-2xF_1$. Furthermore, in the leading logarithmic
order of QCD one arrives asymptotically at the well known Callan-Gross
relation $2xF_1(x,Q^2)=F_2(x,Q^2)$.

%\setcounter{equation}{0}
%\vspace*{1cm}            
\noindent                
\subsection{A Look at the Forward Compton Scattering Amplitude}

Let us now briefly discuss the implications
of the optical theorem on polarized DIS.
The hadron tensor Eq. (\ref{211}) is the
absorptive part (imaginary part)
of the forward Compton scattering amplitude. This
amplitude in general has a decomposition into four independent
amplitudes               
which one may choose as  
\begin{eqnarray}         
\ha [T(1+\ha\to 1+\ha)+T (1-\ha\to 1-\ha)]\label{231}\\
     T (0+\ha\to 0+\ha) \label{232}\\
\ha [T (1+\ha\to 1+\ha)-T (1-\ha\to 1-\ha)]\label{233}\\
T (0+\ha\to 1-\ha) \/. \label{234}
\end{eqnarray}           
Their degrees of freedom 
correspond to the four structure functions $F_1,F_2,
g_1$ and $g_2$. More precisely, the combinations (\ref{231})-(\ref{234})
correspond 
\cite{hey,close,hey1,leader}  
via the optical theorem, to
$F_1$, $F_L$, $g_1-\gamma^2g_2$ and $\gamma(g_1+g_2)$,
respectively, with $R$ being defined in (\ref{2211}).

There are rigorous theoretical limits on the virtual photon
asymmetries $A_1$ and $A_2$ in Eqs.~(\ref{224}) and (\ref{225})
namely                   
\begin{equation}         
|A_1|\leq 1 
\label{235}
\end{equation}
\begin{equation}
|A_2|\leq\sqrt{R} \/. 
\label{236}
\end{equation}           
These inequalities follow from the hermiticity of the electromagnetic
current                  
\begin{equation}         
J_\mu =J_\mu^+           
\label{237}
\end{equation}           
which implies            
\begin{equation}         
a^*_\mu a_\nu W^{\mu\nu}\geq 0
\label{238}
\end{equation}           
for any complex vector $a_{\mu}$. Suitable choices of $a_{\mu}$
lead to the above inequalities.

\subsection{Effects of Weak Currents} 

Until now we have restricted ourselves to photon exchange
only, i.e. to the processes $eN\to eX, \mu N\to\mu X$, at
momenta $Q<<m_Z$, and have found two polarized structure
functions $g_1$ and $g_2$. If we take into account $W^{\pm}$ and
$Z$ exchange, three parity violating polarized structure functions
usually 
called $g_3, g_4$ and $g_5$ arise in addition. They will appear, for
example, in the scattering of neutrinos on polarized
nucleons $\nu N\to\ell 'X$
but become also important for the polarized extension of
HERA at large values of $Q^2$.
Explicitly, the hadron tensor has the form 
\cite{derman,anselmino,lampe,vogelsang3,mathews} 
\begin{eqnarray} \nonumber
 W_{\mu\nu}=W_{\mu\nu}(S=0)
+{M \over Pq}
 \biggl\{ i\ve_{\mu\nu\rs}q^\r S^\si g_1+i\ve_{\mu\nu\rs}q^\r
(S^\si -{Sq\over Pq}P^\si)g_2
 +qS\Bigl(-g_{\mu\nu}+{q_\mu q_\nu\over q^2}\Bigr) g_3 
\\
+{qS\over Pq}\Bigl(P_\mu-{Pq\over q^2}q_\mu\Bigr)
\Bigl(P_\nu - {Pq\over q^2}q_\nu\Bigr)g_4
 +\ha 
\biggl[
\Bigl(P_\mu -{Pq\over q^2}q_\mu\Bigr)
\Bigl(S_\nu -{qS\over q^2}q_\nu\Bigr)
 +
\Bigl(P_\nu -{Pq\over q^2}q_\nu \Bigr)\Bigl(S_\mu -
{qS\over q^2}q_\mu\Bigr)
\biggr]
 g_5 \biggr\} \/.
\label{241}
\end{eqnarray}
In this expression $g_1,g_3$ and $g_4+g_5$ are the "longitudinal"
structure functions which survive in the high energy limit, 
and $g_2$ and $g_4-g_5$ are the transverse ones.
The appearance of symmetric tensors which are
linear in $S^{\mu}$ is due to the axial vector component of the
$W^{\pm}$ and Z couplings to fermions.
Notice that terms proportional to $q^{\mu}$ or $q^{\nu}$
can be dropped in the definition of $W_{\mn}$ because they give no
contribution in the limit $ {m_l \over E} \rightarrow 0$
when contracted with the appropriate lepton tensors $L_{\mn}$.

If one considers longitudinally
polarized nucleons ($S^{\mu} \sim P^{\mu}$),
 the structure functions $g_2$
and $g_4-g_5$ are clearly not of interest. The cross
section difference $\sigma_{\Rightarrow}^{\leftarrow}
- \sigma_{\Leftarrow}^{\leftarrow}$
of part a) of Fig. \ref{fig3} will measure a linear combination 
\cite{anselmino1,derman,anselmino} 
of $g_1, g_3$
and $g_4 + g_5$ with coefficients depending on the vector
and axial vector couplings of the $W/Z$ to the lepton.

Let us first
consider {\it neutrino} nucleon scattering.
Here the lepton tensor is the same for the polarized
and the unpolarized case
\be 
L_{\mu\nu}\hbox{(long.pol.)} =L_{\mu\nu}\hbox{(unpol.)}
=2(k'_\mu k_\nu +k'_\nu k_{\mu} -g_{\mu\nu} kk'+
i\ve_{\mu\nu\a\b}k^\a k^{'\b})
\label{242}
\ee
because the $ \nu W^+$--interaction is
$V-A$ so that the neutrino is forced to be lefthanded. As a
result
\be 
 L_{\mu\nu}\hbox{(long.pol.)} W^{\mu\nu}\hbox{(long.pol.)}=
L_{\mu\nu}\hbox{(unpol.)}W^{\mu\nu}\hbox{(unpol.)}
\Big|_{
{\scriptstyle F_3\rightarrow g_1,}
{\scriptstyle F_1\rightarrow g_3,}
{\scriptstyle F_2\rightarrow g_4+g_5}}
\label{243}
\ee
i.e. one can formally get the polarization cross section
$( \sigma_{\Rightarrow}^{\leftarrow}
- \sigma_{\Leftarrow}^{\leftarrow} )^{\nu N}$
by taking the well known $\nu N \rightarrow l^- X$
cross section for unpolarized
beams and replace the unpolarized structure functions $F_i$ by
the polarized ones (here $g_i \equiv g_i^{\nu N}$ and $F_i \equiv 
F_i^{\nu N}$
are the structure functions specific to $\nu N$-scattering):  
\begin{eqnarray} \nonumber
 {d^2(\sigma_{\Rightarrow}^{\leftarrow}
-\sigma_{\Leftarrow}^{\leftarrow})^{\nu N} \over dxdy} =
{G_F^2s \over 2 \pi}{1 \over (1+{Q^2 \over m_W^2})^2}
\bigl\{
y(1-{y \over 2}-{xyM \over 2E})xg_1^{\nu N}-
{x^2yM \over E}g_2^{\nu N}
+y^2x(1+{xM \over E})
g_3^{\nu N}
\\  
+(1-y-{xyM \over 2E})
[(1+{xM \over E})g_4^{\nu N}+g_5^{\nu N}]
\bigr\}
\label{244}
\end{eqnarray}
where $s=2ME$ in the nucleon's rest frame.
In contrast, if the nucleon is transversely polarized one
finds
\begin{equation}
d (\sigma_{\Uparrow}^{\leftarrow}
-\sigma_{\Downarrow}^{\leftarrow})^{\nu N} \sim
 L_{\mu\nu}\hbox{(long.pol.)}W^{\mu\nu}
 \hbox{(transv. pol.)}
\label{245}
\end{equation} 
i.e.
\begin{eqnarray} \nonumber 
 {d^2(\sigma_{\Uparrow}^{\leftarrow}
-\sigma_{\Downarrow}^{\leftarrow})^{\nu N} \over dxdy} =
{MG_F^2 \over 16 \pi^2}{1 \over (1+{Q^2 \over m_W^2})^2}
 \sqrt{xyM \left[ 2(1-y)E-xyM \right] }
\bigr\{-2xy({y\over 2}g_1^{\nu N}+g_2^{\nu N})
\\  
+xy^2g_3^{\nu N}
+ (1-y-{xyM \over 2E})g_4^{\nu N}
-{y\over 2}g_5^{\nu N}\bigr\} \/.
\label{246}
\end{eqnarray}   
To obtain this result one should make use of the relations
$kS = 0$ (transverse polarization),
$PS=0\/,\/ kP={Pq\over y}\/,\/ 2kq=q^2\/,\/ P^2=k^2=0
   \hbox{ and } x={Q^2\over 2Pq}$.
In Eq. (\ref{246}) we recognize the term $\sim {y\over 2} g_1+g_2$
which was mentioned already earlier for the
case of pure $\gamma$-exchange.
Note that the transverse cross section is suppressed by
${M \over Q}$ with respect to the longitudinal cross section 
in (\ref{244}).

The corresponding cross section for antineutrinos,
$\bar{\nu} N \rightarrow l^+X $ ($W^-$ exchange) can be
obtained by reversing the sign of $g_1$ and $g_2$ in Eqs.  
(\ref{244}) and (\ref{246}).
Note that, to get a nonvanishing effect, one must have $kS = 0$ but
$qS\ne 0$.

In Sections 4 and 6 a physical (parton model) interpretation will be
given for the structure functions $g_1,g_3$ and $g_4+g_5$.
It should be stressed that until today no satisfactory
physical model exists for the 'transverse' structure
functions $g_2$ and $g_4-g_5$.

Let us now come to neutral and charged current interactions
initiated by charged {\it leptons}, since 
neutrino induced reactions on polarized targets are not
very realistic, because large nucleon targets are difficult
to polarize. For sufficiently large $Q^2$, charged (and neutral)
current exchange occurs in $l^{\pm}$-induced processes as well.
As will become clear below, such an experiment would yield some very
important informations on the polarized parton densities and will
therefore hopefully be carried out in the next century.

One has separate cross sections for charged and neutral current exchange.
For the charged current processes, $lN \rightarrow \nu N$
one can take over the results from above, Eqs.  
(\ref{244}) and (\ref{246}). For the neutral
current the cross section consists of 3 terms,
for $\gamma$-, for Z-exchange and for $\gamma$-Z-interference,
\begin{equation}
d \sigma \sim
 \eta^{\gamma \gamma} L_{\mu\nu}^{ \gamma \gamma}W^{\mu \nu \gamma \gamma}
+\eta^{\gamma Z}L_{\mu\nu}^{ \gamma Z}W^{\mu\nu\gamma Z}
+\eta^{ZZ}L_{\mu\nu}^{ Z Z}W^{\mu\nu Z Z}
\label{247}
\end{equation}
where
\begin{equation}
L_{\mu\nu}^{ \gamma \gamma}
=2(k'_\mu k_\nu +k'_\nu k_{\mu} -g_{\mu\nu} kk'
-i\lambda\ve_{\mu\nu\a\b}k^\a k^{'\b})
\label{248}
\end{equation}
\begin{equation}
L_{\mu\nu}^{ \gamma Z}=(v_l- \lambda a_l) L_{\mu\nu}^{ \gamma \gamma}
\label{249}
\end{equation}
\begin{equation}
L_{\mu\nu}^{ Z Z}=(v_l- \lambda a_l)^2 L_{\mu\nu}^{ \gamma \gamma}
\label{2410}
\end{equation}
$
v_l=-{1 \over 2} +2s_W^2
$ and
$
a_l=-{1 \over 2}
$ are the Z--$l^-$--
couplings and $\lambda=  \pm 1$ is the helicity of
the incoming lepton.
Furthermore, we have defined
\begin{equation}
\eta ^{ \gamma \gamma}=1
\label{2411}
\end{equation}
\begin{equation}
\eta ^{ \gamma Z}=
{G_Fm_Z^2 \over 2 \sqrt{2} \pi \alpha}
{1 \over (1+{Q^2 \over m_Z^2})^2}
\label{2412}
\end{equation}
\begin{equation}
\eta ^{ Z Z}=\Bigl(
{G_Fm_Z^2 \over 2 \sqrt{2} \pi \alpha}
{1 \over (1+{Q^2 \over m_Z^2})^2}\Bigr)^2 \/. 
\label{2413}
\end{equation}
Note that 
${G_Fm_Z^2 \over 2 \sqrt{2} \pi \alpha}={1 \over 4s_W^2c_W^2} \approx 1.4$ 
and 2ME=s. 
All 3 hadron tensors $W^{\gamma \gamma}$,
$W^{\gamma Z}$ and $W^{ZZ}$ have an expansion of the form of Eq. 
(\ref{241}) ,
but of course for $W^{\gamma \gamma}$ one has $g_3^{\gamma \gamma}=
g_4^{\gamma \gamma}=g_5^{\gamma \gamma}=0$ due to parity
conservation. All in all, there are 12 free independent  
polarized structure
functions, among them 7 ($g_1^{\gamma \gamma}$,$g_1^{\gamma Z}$,
$g_1^{ZZ}$,$g_3^{\gamma Z}$,$g_3^{ZZ}$,
$g_{4+5}^{\gamma Z}$,$g_{4+5}^{ZZ}$)
with
and 5 without a parton
model interpretation.

The neutral current cross section for longitudinally polarized leptons
is given by \cite{anselmino} 
\begin{eqnarray} \nonumber  
 {d^2\sigma^{\ell N}_{NC} \over dx \, dy}(\lambda, S=S_L)&=&
4 \pi M Ey \, {\alpha^2\over Q^4} 
\sum_{i=\gamma \gamma, ZZ, \gamma Z} \eta^i C^i  
 \biggl\{ 2xy\,F_1^i + 
{2\over y} \left( 1-y-{xyM\over 2E}\right) (F_2^i + g_5^i) 
\\ \nonumber 
& & -2\lambda x\left( 1-{y\over2} \right)F_3^i
-2\lambda x\left(2-y-{xyM\over E}\right)g_1^i
-{2\over y}\left( 1+{xM\over E}\right)\left(1-y-{xyM\over 2E}
\right)g_4^i 
\\
& & +4\lambda {x^2M\over E}\, g_2^i
+2xy\left(1+{xM\over E}\right) g_3^i \biggl\} 
\label{2414}
\end{eqnarray}
where, for negatively charged leptons,
\begin{equation}
C^{\gamma \gamma}=1, C^{\gamma Z}=v_l- \lambda a_l,
C^{ZZ}=(v_l- \lambda a_l)^2
\label{2415}
\end{equation}
and for positively charged leptons one simply replaces $a_l$
by $-a_l$. Notice that when the lepton flips its helicity,
$\lambda$ changes sign, and when the nucleon flips its spin,
all terms containing a polarized structure function $g_{1,2,3,4,5}$
also change sign. 
Upon averaging over $\lambda$ and S one
obtains the unpolarized cross section
\be 
 {d^2\sigma^{\ell N}_{NC} \over dx \, dy}({\rm unp.}) =
{1\over 4} \sum_{\lambda, S}
{d^2\sigma^{\ell N}_{NC} \over dx \, dy}(\lambda, S) 
= 4 \pi M Ey \, {\alpha^2\over Q^4} 
 \sum_i \eta^i C^i
\left\{ 2xy \, F_1^i +{2\over y} \left(1-y-{xyM\over 2E}\right)
F_2^i\right\}
\label{2416}
\ee
where we have again used s=2ME appropriate for a fixed nucleon 
target in its rest frame. Alternatively, ${d^2 \sigma \over dx dQ^2}$ 
can be simply obtained from ${d^2 \sigma \over dx dQ^2}= 
{1 \over sx} {d^2 \sigma \over dx dy}$.

In the case of nucleons with transverse polarization, i.e.
with a spin orthogonal to the lepton direction (z-axis)
at an angle $\alpha$ to the x-axis, one has
\begin{eqnarray} \nonumber
{d^3\sigma^{\ell N}_{NC} \over dx \, dy \, d\phi}(\lambda, S=S_T)&=&
2M Ey \, {\alpha^2\over Q^4} \sum_i \eta^i C^i 
 \biggl\{ 2xy\,F_1^i +   
{2\over y} \left( 1-y-{xyM\over 2E}\right)F_2^i
- 2\lambda x\left( 1-{y\over2} \right)F_3^i 
\\ \nonumber 
& & +{\sqrt{xyM[2(1-y)E-xyM]}\over E} \, \cos(\alpha-\phi) 
 \Bigl[ -2\lambda xg_1^i 
-4\lambda {x\over y} \, g_2^i
+{1\over y} \, g_5^i 
\\
& & +{2\over y^2} \left( 1-y-{xyM\over 2E}\right) g_4^i
-2xg_3^i \Bigr] \biggl\}
\label{2417}
\end{eqnarray}

To experimentally unravel the whole set of independent structure
functions one should make use of leptons of opposite charges
and/or polarizations, in which cases the structure
functions enter with different weights. Furthermore, use can be
made of the propagator structure ${1 \over (1+{Q^2 \over m_Z^2})^{0,1,2}}$,
so that one can separate the $\gamma \gamma$, $\gamma Z$ and $ZZ$
components
by a measurements at different $Q^2$-values.
We shall come back to these processes in Sect. 6.7 where 
the parton model interpretation and some phenomenological 
applications will be given. 

%% file: k3tex
\setcounter{equation}{0}
\section{Polarized Deep Inelastic Scattering (PDIS) Experiments}
\subsection{Results from old SLAC Experiments}

After the famous Stern-Gerlach discovery it became possible to
produce and use polarized atomic beams. However,experiments
with polarized $lepton$ beams are a fairly recent
development because before 1972 it was not possible to produce
a polarized electron beam.
In 1972 physicists from the Yale
university succeeded to polarize elctrons by photoionization
from polarized alkali atoms.
This was used afterwards for
elastic scattering experiments between polarized electrons
and polarized atoms.     
As time went by, beam energies increased from
the eV level to the level where deep inelastic experiments
can be performed.        

This is -- in short -- the prehistory of the SLAC experiments.
The CERN experiments followed another route. They used polarized muons
from high energetic pions. These muons are automatically
polarized because of the weak V-A nature of the decay.

But let's start with SLAC.
High energetic electrons from a polarized alkali
source were used in the SLAC experiment E80
which took place in 1976 \cite{alguard,baum}. The electrons
in this experiment with energies between 6 and 13 GeV, 
scattered off a polarized butanol target, were
detected by the 8 GeV spectrometer built for all DIS
experiments at SLAC. Average beam and target polarizations
of experiment E80 were rather high (50 resp.\ 60 \%).
In the butanol           
target only the hydrogen 
atoms contribute to the polarized scattering because
carbon and oxygen are spin-0 nuclei. Therefore the effective
target polarization is reduced by a "dilution factor"
$f={10 \over 74}$ (=ratio of number of hydrogen nucleons over
total number of nucleons). This dilution effect together
with other nuclear uncertainties is a problem of all present
experiments. The HERMES experiment \cite{coulter} 
at DESY
is going to use a hydrogen gas target in which this
problem is absent.       

\begin{figure}
\begin{center}
\epsfig{file=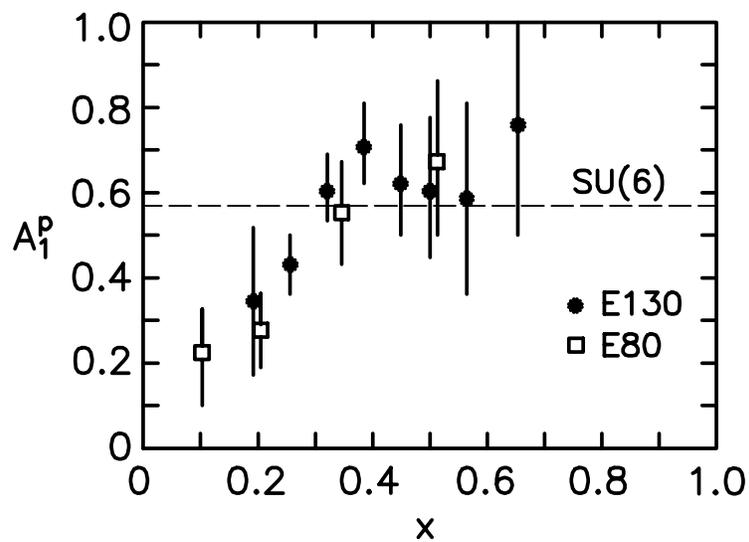,height=7cm}
\vskip 0.5cm
\caption{The spin asymmetry of the proton from the "old" 
SLAC data 
from 1976 and 1983 \protect\cite{hughes1}.  
In the naive SU(6) model one has $A_1^p={5 \over 9}$ and $A_1^n=0$ 
\protect\cite{close}.}
\label{fig5}
\end{center}
\end{figure}

The main deficiency of the E80 experiment was that its
polarized source was rather limited in beam
current. Nevertheless, it was possible to select about
two million scattered electron events and to determine
the asymmetry $A_1$, Eq.~(\ref{224}),
for the proton for several $x$ values
between 0.1 and 0.5, and rather low values of $Q^2$ (about 2 GeV$^2$).
The asymmetry turned out to be rather large,
in rough agreement with expectations based on the quark-parton model
\cite{alguard,baum}. 

The desire to reduce higher twist effects motivated a second
SLAC experiment in 1983 (E130) \cite{baum1}. 
This experiment was run at an
electron beam            
energy of about 23 GeV.  
The beam polarization was increased to about 80\%
and a new M\"oller polarimeter was built which allowed
for continuous beam polarization measurements during the
experiment.              
The detector was improved as well
so that the kinematic coverage
extended in $x$ from 0.2 to 0.65 and in $Q^2$ from 3.5 to 10~GeV$^2$.
The experiment concentrated on measurements of rather high
$x$ values and consequently collected only about one million
events.                  

In Fig. \ref{fig5} the results for the proton asymmetry $A_1^p$ are
shown for E80 and E130 together. Their average of 
\cite{baum1}
\begin{equation}         
\Gamma^p_1(<Q^2>\simeq 4\; GeV^2) = 0.17 \pm 0.05\qquad({\rm SLAC~E80,
                 E130)}\label{311}
\end{equation}           
is in good               
agreement with the value (${5 \over 9}$) of the static SU(6) quark model
\cite{hey1,close,leader}
and the Ellis-Jaffe 'sum rule' \cite{gourdin,ellis} 
to be discussed in Section 5.
Here we have defined the 'first moment' of $g_1$ by
\begin{equation}         
\Gamma^{p,n}_1(Q^2)\equiv\int^1_0 g_1^{p,n} (x,Q^2)dx \/.
\label{312}
\end{equation}           
However, the result (\ref{311}) is plagued by a large
error whose main source comes from the extrapolation into
the unmeasured $x$ ranges, in particular as $x\to 0$.
We shall see that a discrepancy
with the CERN data exists which 
originates from data 
taken at small x (between 0.01 and 0.1).

\subsection{The CERN Experiments}

The CERN PDIS experiments were started as an addendum 
\cite{papavassiliou,ashman,ashman1} to the
unpolarized EMC deep inelastic muon-nucleon experiments. A polarized
beam source of muons with energies $E_{\mu}=100-200$~GeV
was available to hit a polarized ammonia ($NH_3$) target.
Due to the high muon energy, $x$-values as low as 0.01 could be
reached.                 

The results from this experiment which was meant to supplement
the SLAC data at small-x and to confirm the Ellis-Jaffe 'sum
rule' 
\cite{gourdin,ellis}
came as a major surprise.
Actually, they confirmed 
the SLAC measurements in the common large-$x$ region but found too
small asymmetries in the small-$x$ region, in disagreement
with the expectations of Gourdin, Ellis and Jaffe 
and the naive quark parton model.

The high energetic muon beam is a low-luminosity beam because
it is a secondary beam from the semileptonic decay of pions which are
produced in proton collisions. Its main advantage is its
high energy which not only allows to enter the small-$x$ region
but also                 
guarantees higher $Q^2$-values
in the region of intermediate-$x$ and a corresponding suppression
of possible higher twist effects.
In the rest system of the pions the emitted muons are 100 \%
left-handed, due to the V-A nature of the decay. In the laboratory
frame the muons have a degree of polarization which depends
on the ratio $E_{\mu}/E_{\pi}$. For the EMC experiment the
beam was selected such that the polarization of the muon beam
was about 80 \%.         
The muon beam polarization was determined from the muon
event distribution via a Monte Carlo study of muon
production -- an indirect and not really satisfactory method.

The ammonia target was quite large, with a length of about
2 meters, and            
separated into two halves with opposite
polarization. On the average only a fraction $f$, the dilution factor,
of the target protons were polarized.
The polarization of the target
could be determined as a function of its length (by NMR
coils placed along the target).

The scattered muons were detected
by a well established muon tracking spectrometer.
It was possible to reconstruct the muon scattering vertex
to determine that half of the target
in which the scattering took place.
The target spins were reversed (once per week)
eleven times in the experiment.
Changes in the spectrometer acceptance from one spin reversal to
another was the main systematic uncertainty in the experiment.

\begin{figure}
\begin{center}
\epsfig{file=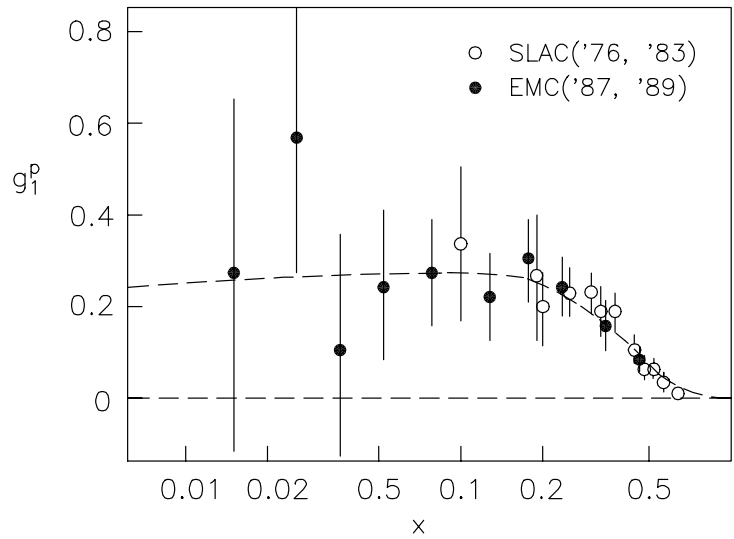,height=7.0cm}
\vskip 0.5cm
\caption{Combined results of SLAC and EMC on $g_1^p$ 
\protect\cite{ashman1}. 
The dashed curve is an EMC 
parametrization 
of the data.}
\label{fig6} 
\end{center}
\end{figure}

In Fig. \ref{fig6} the combined results from SLAC and CERN on the
structure function $g_1^p$ are presented. The low values of
the asymmetries at small-$x$ translate into low values for
the proton structure function and a low value for its
first moment 
\cite{papavassiliou,ashman,ashman1}
\begin{equation}
\Gamma^p_1(<Q^2>)=0.126\pm 0.010\pm 0.015\qquad({\rm EMC,~SLAC})
\label{321}\end{equation}
where the first is the statistical and the second is the
systematic error which includes, as always, uncertainties from Regge
extrapolations to $x\to 0$; the average values of $Q^2$ are
$<Q^2>=10.7\; GeV^2$ (EMC) and $<Q^2>=4\; GeV^2$ (SLAC). It should be
emphasized that almost all of $\int^1_0g^p_1dx$ stems from the region
$x\geq 0.01$ which gives 
\cite{papavassiliou,ashman,ashman1}, instead of (\ref{321}),
\begin{equation}
\int\limits^1_{0.01} g^p_1(x,<Q^2>)dx\simeq 0.123\/.
\label{322}\end{equation}
If, on the contrary, one does not combine the EMC and SLAC
measurements, the EMC asymmetry measurements alone imply
similar results, namely 
\cite{papavassiliou,ashman,ashman1}
\begin{equation}
\Gamma^p_1(<Q^2>=10.7 \; GeV^2)=0.123\pm 0.013\pm 0.019\qquad {\rm (EMC)}\/.
\label{323}\end{equation}

\begin{figure}
\begin{center}
\epsfig{file=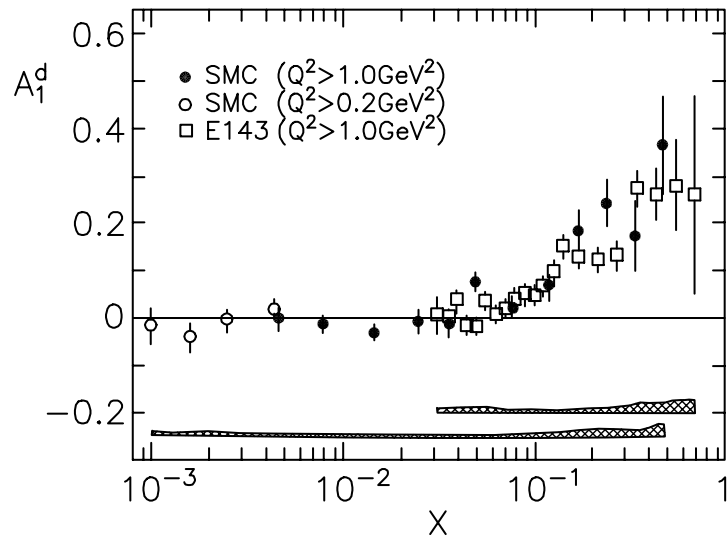,height=7cm}
\vskip 1.cm
\caption{The virtual photon--deuteron cross section    
asymmetry as measured 
by SMC \protect\cite{adams6} and compared to the SLAC E143 data 
\protect\cite{abe1}. 
Only statistical errors 
are shown. 
The size 
of 
the systematic errors is indicated by the 
shaded area.}
\label{fig7a}
\end{center}
\end{figure}

The surprising EMC result triggered a second CERN experiment
with polarization, the so-called SMC experiment 
\cite{adeva,adams6}.
Instead of ammonia a polarized butanol target has been used.
In butanol the only polarized nucleons are the protons
($\simeq$~12 \%) and deuterons ($\simeq$ 19~\%).
The goal of this experiment was to
infer information about the neutron from the
difference $g_1^d-g_1^p$.
Butanol allows for a     
much more rapid spin reversal than ammonia, this way reducing
the large systematic uncertainty from the varying detector
acceptance. With the alcohol targets, spin reversals took one hour
and were implemented every 8 hours.

The main improvement of the SMC experiment was in the
measurement of the beam polarisation.
It was obtained from the energy spectrum of positrons from
muon decay. The positron energy spectrum is rather sensitive
to the muon beam polarization and provides a direct
measurement.

\begin{figure}
\begin{center}
\epsfig{file=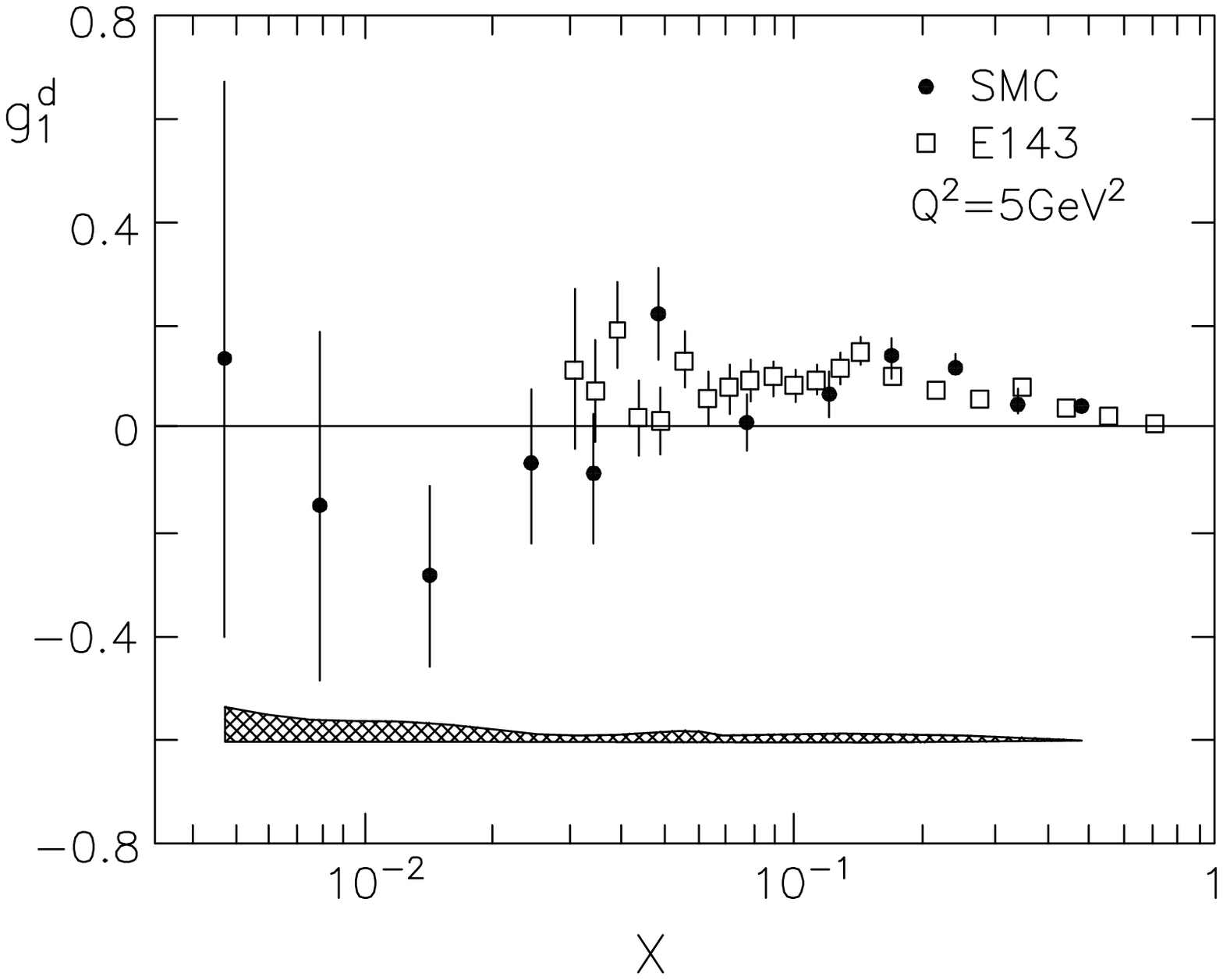,height=7cm} 
\vskip 1.cm
\caption{$g_1^d$ deduced from $A_1^d$ in Fig. \ref{fig7a} 
evolved to a common $Q^2=5$ GeV$^2$ \protect\cite{adams6}. 
Only statistical 
errors are shown.
The size of the systematic errors 
is 
indicated by the
shaded area.}
\label{fig7b}
\end{center}
\end{figure}

\begin{figure}
\begin{center}
\epsfig{file=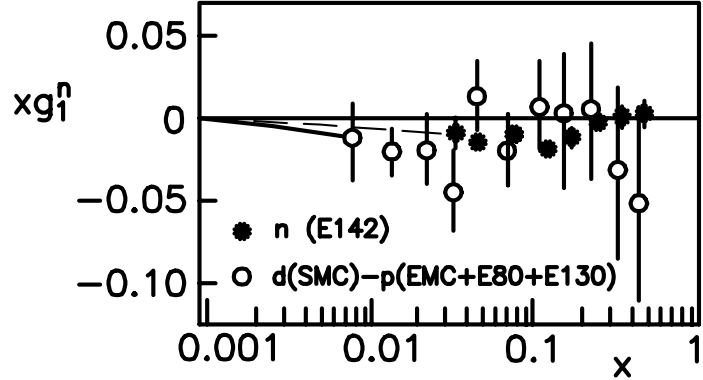,height=5cm}
\vskip 0.5cm
\caption{$g_1^n$ as a function of x at $Q^2$= 5 $GeV^2$. The full 
circles 
are the data
from SLAC E142 
experiment. The dashed and solid curves 
show the extrapolations to low x using the E142 data and 
using the combined data, respectively \protect\cite{adeva1}. }
\label{fig8}
\end{center}
\end{figure}

In Figs. \ref{fig7a} and \ref{fig7b} 
the SMC results for $A_1^d$ and $g_1^d$ are shown which
are plotted as a function of $\ln x$ to make the small-$x$
results more transparent. For $x$ less than 0.1 the results are
compatible with zero although with rather limited statistics.
The theoretically more interesting longitudinally polarized
neutron structure function $g_1^n(x,Q^2)$ can be obtained via the
relation                 
\begin{equation}         
g^d_1(x,Q^2)=\ha [g^p_1(x,Q^2)+g^n_1(x,Q^2)](1-{3\over 2}w_D)
\label{324}              
\end{equation}           
where $w_D(\simeq 0.058)$ accounts for the $D$-state
admixture in the deuteron wave function \cite{adeva}. 
The resulting
\cite{adeva1} 
$g^n_1 (x,Q^2= 5\; GeV^2)$ is shown in Fig. \ref{fig8}
where the
'old' EMC 
\cite{papavassiliou,ashman,ashman1} 
and SLAC \cite{alguard,baum,baum1} 
data have been used for $g^p_1$,
all reevaluated at $Q^2=5\; GeV^2$ under the assumption that the asymmetries
$A_1^{d,p}$ are independent of $Q^2$. (This latter
assumption is theoretically questionable, at least in the small-$x$
and larger-$Q^2$ region where no data exist so far, as will be discussed
in Section 6). Also shown in Fig. \ref{fig8} 
are the SLAC (E142) 
data \cite{anthony}, 
to be discussed next, which agree with the SMC results
in the $x$ region of overlap. The SMC results of Fig. \ref{fig8} imply 
\cite{adeva}
\begin{equation}         
\Gamma^n_1(Q^2=5\; GeV^2)=-0.08\pm 0.04\pm 0.04\qquad (SMC)
\label{325}              
\end{equation}           
which still deviates from the Ellis-Jaffe expectation $-0.002\pm 0.005$.

\begin{figure}
\begin{center}
\epsfig{file=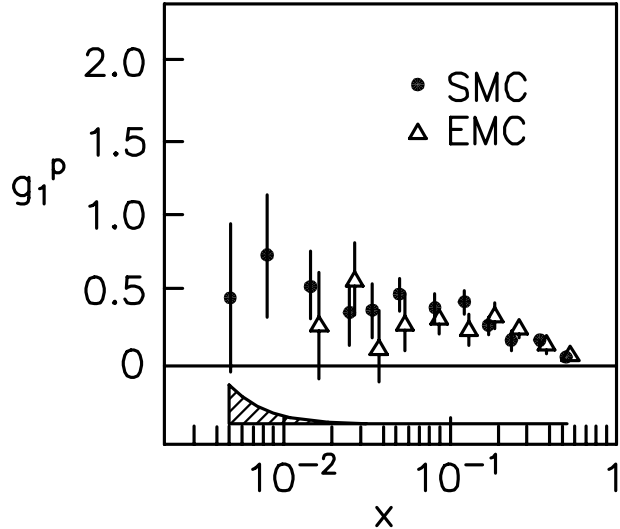,height=7.0cm}
\vskip 0.8cm 
\caption{SMC/EMC results for $g_1^p$ at the average $Q^2$ for 
each x bin \protect\cite{adams,adeva3}. 
Only statistical errors 
are shown. 
The size of the 
systematic 
errors is 
indicated by the shaded area.}
\label{fig9a}
\end{center}
\end{figure}

\begin{figure}
\begin{center}
\epsfig{file=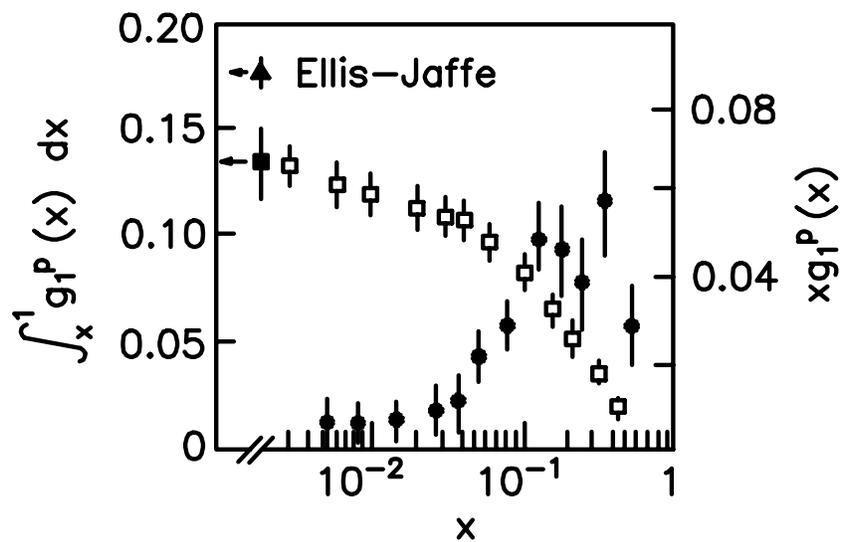,height=7cm}
\vskip 0.8cm 
\caption{The solid circles show the SMC results for 
$g_1^p$ as a 
function of x, at $Q_0^2=10\;GeV^2$. 
The open boxes show  
the integral from x to 1 
(left hand axis). Only 
statistical errors are shown.  
The solid square shows  
the result for the first moment  
(integral from 0 to 1), with 
statistical and systematic 
errors combined in quadrature 
\protect\cite{adams}. }
\label{fig9b}
\end{center}
\end{figure}

More recently, SMC has also measured 
$g^p_1(x,Q^2)$
\cite{adams,adams7} 
shown in Figs. \ref{fig9a} and \ref{fig9b}, which results in
\begin{equation}         
\Gamma^p_1(Q^2=10\; GeV^2)=0.136\pm 0.013\pm 0.011 \/. \qquad  (SMC)
\label{326}              
\end{equation}           
This result increases to \cite{adams,adams7} 
\be           
\Gamma^p_1(Q^2=5\; GeV^2)=0.141 \pm 0.011 
 \qquad (SMC, EMC, SLAC)
\label{327}              
\ee           
if the 'old' EMC and SLAC measurements are included which led to
(\ref{321}). Equation (\ref{327}) represents, for the time being,
probably the best estimate of the full first moment $(0\leq x\leq 1)$
of $g^p_1 (x,Q^2)$ at $Q^2=5$ to $10\; GeV^2$. All the above
results are still significantly below the Gourdin-Ellis-Jaffe 
\cite{gourdin,ellis}
expectation (assuming a vanishing total polarization of strange sea 
quarks) of about $0.18\pm 0.01$
which will be discussed in more detail in Sections 5 and 6.1.

SMC has also presented semi--inclusive $\pi^{\pm}$ measurements  
\cite{adeva2} which give direct access to the polarized valence 
densities $\delta u_v (x,Q^2)$ and $\delta d_v (x,Q^2)$. 
More details will be discussed in Sect. 6.5.  

Due to the use of (longitudinally) polarized high current electron
beams, measurements of $g_1^{p,n}(x,Q^2)$ with much higher statistics
are obtained from the latest round of SLAC experiments (although at
significantly lower values of $Q^2$ due
to the lower beam energy) to which
we turn now.             

\subsection{The New Generation of SLAC Experiments }

The SLAC E142 experiment \cite{anthony} 
used a different electron source
than the old SLAC experiments 
\cite{alguard,baum,baum1}
relying on developments on
solid state GaAs cathodes. They were able to produce
high current electron beam pulses with energies 19.4, 22.7
and 25.5 GeV and through that a rather high statistics experiment.
Altogether, 300 million events could be collected.
In addition, reversal of beam spin direction could be implemented
randomly.                

The target was polarized $^3 He$. From the measurements on $^3He$
the neutron structure function can be directly inferred
because the polarization effects from the two protons
inside the Helium compensate each other, due to the Pauli
principle. This is true to the extent that the Helium nucleus
is in its S-state. The probability for this is about 90 $\%$.
The remaining 10 $\%$ probability with which the two
proton spins are parallel can be corrected for.

There is another nuclear uncertainty at low $Q^2$ ($<Q^2>\simeq
2\; GeV^2)$, namely the possible
exchange of mesons ($\rho$'s  and $\pi$'s) which is a bit more dangerous
for $^3He$ than for the deuteron because $^3He$ has a larger binding
energy per nucleon than d. Furthermore, its magnetic moment
is a worse approximation to the free neutron than $\mu_d$ is
to $\mu_p+\mu_n$.        

The $^3He$ target is polarized by optical pumping. Circularly
polarized near infrared laser light illuminates the target cell with
$^3He$ and rubidium vapor. The outer shell electrons in the
rubidium become polarized and transfer their polarization to
the $^3He$ nucleus via spin exchange collisions.Once achieved,
the polarization of the $^3He$ is rather stable. The entire
target chamber was placed in a constant magnetic field
which holds the $^3He$ spins in a fixed orientation.
Target spin reversal was achieved several times per day
and used to reduce the systematic error.

%\begin{figure}
%\begin{center}
%\epsfig{file=kfig10-57.eps,height=6cm}
%\vskip 0.8cm
%\caption{Combined results from CERN (SMC) and SLAC (E142) 
% for $A_1^n$ at the average $Q^2$ of the respective experiments.}
%\label{fig10a}
%\end{center}
%\end{figure}

\begin{figure}
\begin{center}
\epsfig{file=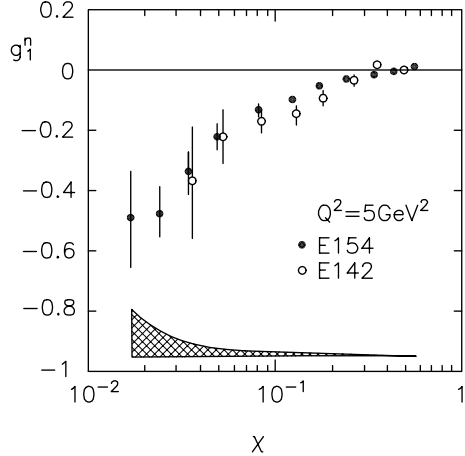,height=6cm}
\vskip 0.8cm
\caption{Result for $g_1^n$ at 
$Q^2=5$ GeV$^2$ from SLAC (E142) \protect\cite{anthony}. The 
most recent SLAC (E154) results are also shown 
for comparison \protect\cite{abe5}. }
\label{fig10b}
\end{center}
\end{figure}

\begin{figure}
\begin{center}
\epsfig{file=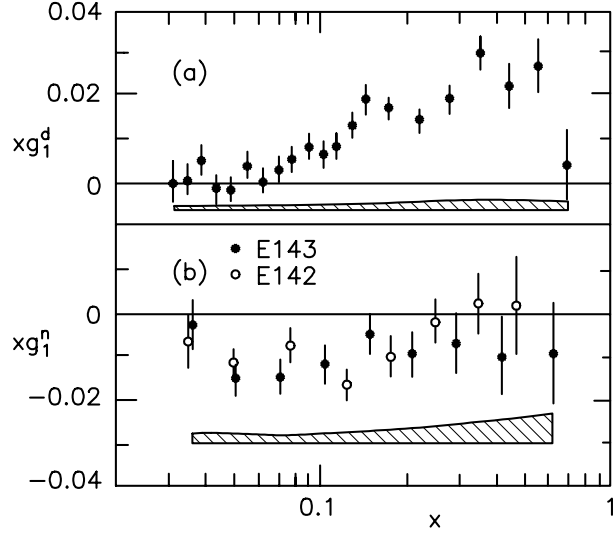,height=7cm}
\vskip 0.8cm
\caption{$xg_1$ for (a) the deuteron at $Q^2$=3 $GeV^2$ and 
(b) the neutron at $Q^2$=2 $GeV^2$ 
as measured by E142 and E143 \protect\cite{abe1}. 
Systematic errors are indicated by the shaded bands.}
\label{fig11}
\end{center}
\end{figure}

\begin{figure}
\begin{center}
\epsfig{file=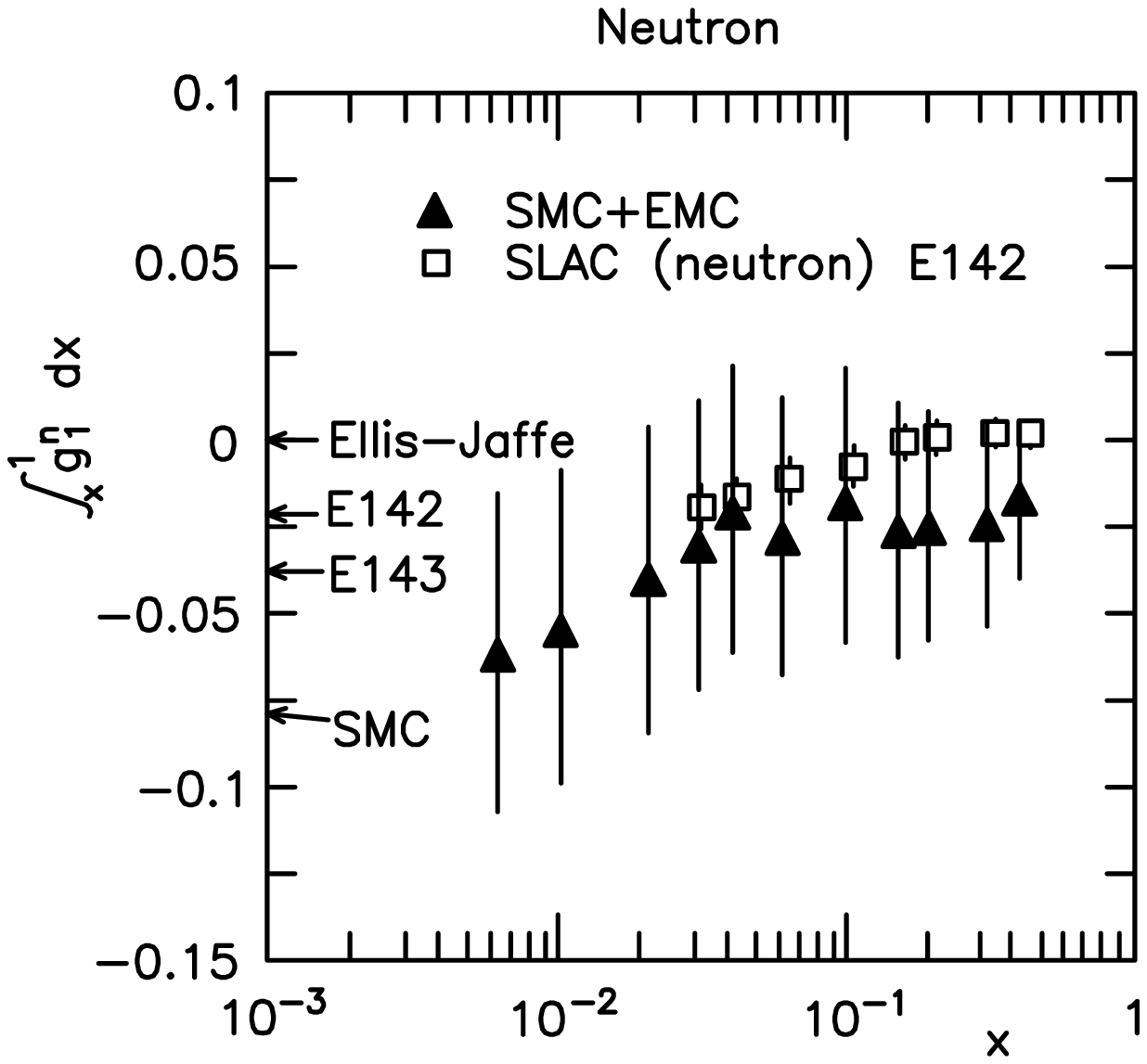,height=7cm}
\vskip 0.8cm
\caption{Comparison between SLAC and CERN of the 
  integral of $g_1^n$ at the average $Q^2$ of the 
 respective experiments. }
\label{fig13}  
\end{center}
\end{figure}

The results for $g_1^n$ are shown in Fig.  
\ref{fig10b} 
and can be compared with the more recent SLAC E143 experiment 
\cite{abe1}
in Fig. \ref{fig11}. 
These experiments span the $Q^2$-range $1<Q^2<10\; GeV^2$,
corresponding to $0.029< x<0.8$ at their respective energies, where
the E143 polarized electron beam energies refer to 9.7, 16.2 and
29.1~GeV. The E143 data for $g^n_1$ in Fig. \ref{fig11} imply 
\cite{abe1}
\begin{equation}         
\Gamma^n_1(<Q^2>=2\; GeV^2)=-0.037\pm 0.008\pm 0.011\quad ({\rm E143})
\label{328}              
\end{equation}
compared with $-0.022\pm 0.007\pm 0.009$ from the E142 data 
\cite{anthony}.
These results are consistent with the less accurate SMC measurement
in (\ref{325}). The E143 experiment 
\cite{abe1,abe2} 
uses a deuterated ammonia $(ND_3)$ target, polarized by
dynamic nuclear polarization in a 4.8~T magnetic field, in order to
check the SMC experiment at lower $Q^2$. The comparison of their
measured asymmetries $A^d_1(x,Q^2)$ in Fig. \ref{fig7a} demonstrates that
these two experiments are consistent whith each other although
it should be kept in mind that the $Q^2$ -value of SMC is about
twice as large, for each specific $x$-bin, as of E143; the average
$Q^2$ of the latter experiment varies from 1.3~GeV$^2$ (at low $x$) to
9~GeV$^2$ (at high $x$). A comparison of the integral of
$g^n_1,\int^1_x g^n_1(x',Q^2) dx'$, for each $x$-bin
as lower integration limit is shown in Fig. \ref{fig13} which eventually (as
$x\to 0$) leads to the full 'first moment' of $g^n_1$ as stated in
Eqs.~(\ref{325}) and (\ref{328}). The results shown in Fig. \ref{fig13} 
are particularly interesting because they clearly demonstrate
that the difference between SLAC and SMC/EMC comes mainly from the
small-$x$ region, $x<0.03$, where SLAC has no data points and
thus has to fully rely on assumptions about the behavior of $g_1$ as
$x\to 0$.                

Furthermore the E143 experiment used also an ammonia target
\cite{abe} in order to check, at lower $Q^2$,
the SMC/EMC proton-measurements
for $g^p_1$. Their results are again compatible with the ones of SMC
in the $x$-region of overlap and give \cite{abe}
\begin{equation}         
\Gamma^p_1(Q^2=3\; GeV^2)=0.127\pm 0.004\pm 0.010 \qquad (E143)
\label{329}              
\end{equation}           
to be compared with Eqs.~(\ref{321})-(\ref{323}) and (\ref{326}).
Since the latter SMC result/estimate (\ref{326}) holds at
$Q^2=10\; GeV^2$, it is more appropriate to calculate the integral at
$Q^2=3\; GeV^2$, assuming $g^p_1/F^p_1\simeq A^p_1$ to be
independent of $Q^2$, which gives $0.122\pm 0.011\pm 0.011$, instead of
Eq.~(\ref{326}), and compares better with the E143 result (\ref{329}).
These various consistent measurements imply that by now, after 20 years
of having performed polarized deep inelastic experiments, we have
available a rather reliable and sufficiently precise result for
$\int^1_0 g^p_1(x,Q^2)dx$ which is confidently more than two standard
deviations {\it below} the Ellis-Jaffe-Gourdin sum rule expectation
[no polarized strange sea 
\cite{gourdin,ellis}]  
of $\Ga^p_1(Q^2=3\; GeV^2)_{EJ}=0.160\pm
0.006$ \cite{abe}. 
We shall come back to this point in Section 5.

%\begin{figure}
%\begin{center}
%\epsfig{file=kfig12-63.eps,height=7cm}
%\vskip 0.8cm
%\caption{Comparison of the results for $A_1^d$ from E143  
% and SMC \cite{abe1}.  
%Note that 
%for the SLAC 
%experiment $Q^2$  
%varies in the range 1.3 to 9 GeV$^2$, 
%whereas at SMC $Q^2$ is 
%about twice as large.}
%\label{fig12}
%\end{center}
%\end{figure}

The integrated quantities $\Ga^{p,n}_1(Q^2)$ which resulted from all
polarization experiments discussed and performed thus far 
\cite{voss}
are finally summarized in Table 2. 

There are several new experiments at SLAC [see e.g. 
\cite{hughes}], the experiments E154 and E155 with a $^3He$ 
target \cite{hughes1} and an ammonia target 
\cite{arnold}, respectively, which  
run at 50 GeV beam energy and which are the follow-up experiments 
to E142 and E143. Clearly, with SLAC statistics, these  
measurements will provide us with a powerful test of the larger 
$Q^2$ and small x ($\gtrsim 0.01$) dependence of the spin structure functions 
$g_1^{p,n}$, as can be seen from the comparison 
with the first results of E154 with E142 in Fig. 
\ref{fig10b}.  

%\end{multicols}
\begin{table} 
\label{tab2}  
\begin{center}
\begin{tabular}{|l|l|l|}
\hline
Experiment & $\Gamma_1(Q^2/GeV^2)$ & Reference  \\
\hline
EMC,SLAC & $\Gamma^p_1(10.7)=0.126\pm0.010\pm0.015$ & \cite{ashman,ashman1} 
                                                   \\ 
SMC & $\Gamma^p_1(10)=0.136\pm0.013\pm0.011$ & \cite{adams,adams7} \\ 
E143 & $\Gamma^p_1(3)=0.127\pm0.004\pm0.010$ & \cite{abe} \\ 
SMC & $\Gamma^n_1(10)=-0.063\pm0.024\pm0.013$ &  \cite{adams2} \\ 
SMC & $\Gamma^n_1(5)=-0.08\pm0.04\pm0.04$ & \cite{adeva} \\     
SMC & $\Gamma^n_1(5)=-0.048\pm0.022$ & \cite{adams7} \\
E142 & $\Gamma^n_1(2)=-0.022\pm0.007\pm0.009$ & \cite{anthony} \\ 
E143 & $\Gamma^n_1(2)=-0.037\pm0.008\pm0.011$ & \cite{abe1}  \\
E154 & $\Gamma^n_1(5)=-0.041\pm0.004\pm0.006$ & \cite{abe5} \\ 
\hline
\end{tabular}
\bigskip
\caption{Experimental results for the $g_1$-integrated quantity
$\Gamma_1(Q^2)$ in Eq.~(\ref{312})} 
\end{center}
\end{table}
%\begin{multicols}{2}

%\setcounter{equation}{0}
\subsection{Future Polarization Experiments}

The HERMES experiment 
\cite{coulter,barber}, being a fixed target experiment,
takes place in the HERA tunnel with a longitudinally 
polarized electron beam of about 30~GeV incident on a polarized
$H, D$ or $^3He$ gas jet target. Although this experiment
is performed at a similarly low energy as present SLAC
measurements, the novel technique of polarized gas-jet
target technology is expected to allow for high precision measurements
since the target atoms are present as pure atomic species and
hence almost no dilution of the asymmetry occurs in the scattering 
from
unpolarized target material. 
The main advantages: HERMES will use a pure proton gas target
in a thin wall storage cell,
so that the dilution factor will be close to one
and there is almost no background from windows effects. The thin targets
will still provide very good statistics because the beam
current is enormous.     
Even a measurement of the structure function $g_2$ is conceivable.
Finally, the spectrometer allows for multiparticle identification
and thereby measurements of semiinclusive cross sections
from which additional information on valence,sea and
strange quark polarization can be obtained.
In addition, polarized internal gas
targets allow for a rapid reversal of the target spin 
\cite{coulter}
which will be crucial for understanding and minimizing systematic
errors. This rapid spin reversal will be also crucial for measuring
{\it directly neutron} asymmetries by using a  $^3He$ gas
target where spins of the two protons are practically in opposite
directions and thus the $^3He$ target acts as an effective neutron
target. Actual data taking has started in 1995/96 and the
accessible kinematic region 
($0.02 \leq x \leq 0.8, \; 1 \leq Q^2 \leq 10 \; GeV^2$)  
will be similar to the one of present SLAC experiments. It is intended to 
measure $g_1^{p,n,d}, g_2^{p,n}$, etc., and in particular 
semi-inclusive asymmetries $A_{p,n}^{\pi}$ from $\vec e \vec p 
(\vec n) \rightarrow e \pi X$ which allows to extract 
separately the polarised valence quark densities $\d d_v(x,Q^2)$ 
and $\d u_v(x,Q^2)$. More details will be discussed in Section 6. 

Very recently, the SMC group has submitted a proposal for measuring 
semi-inclusive (D--mesons, etc.) reactions in deep inelastic 
$\vec \mu \vec p (\vec d)$ scattering at 100-200 GeV $\vec \mu$--beam 
energies \cite{nappi,baum2}. This will be a 
dedicated experiment for measuring, among other things, the 
polarized gluon distribution $\d g(x,Q^2)$  
which plays a predominant role in understanding the 
nucleon spin structure and which so far is experimentally entirely 
unknown. It will be deduced from the production of heavy quarks 
(like charm) via the fusion process $\vec \gamma^{\ast}  
\vec g \rightarrow c \bar{c}$ responsible for open charm production 
which is one of the most promising and cleanest processes for 
extracting $\d g(x,Q^2)$ since it occurs already in the leading 
order (LO) of QCD with no light 
quark contributions present (see Section 6.2). 
It is expected to achieve a sensitivity for ${\d g \over g}$ 
of about $15 \%$ and the data taking could start in 1998. 
Note however, that this cross section is {\it not} sensitive to the  
'anomalous' (=first moment) part of the polarized gluon density, because the 
first moment contribution of polarized gluons to heavy quark 
production vanishes. 
%The 'anomalous' part is the contribution 
%due to the ABJ anomaly. Many theorists tend to believe that 
%the anomalous part of the polarized gluon density is  
%potentially large, and gives a large contribution 
%to the first moment of the 
%structure function $g_1$. It turns out, however, that the 
%first moment  
%contribution to the production of heavy quarks via 
%$\vec \gamma^{\ast}  
%\vec g \rightarrow c \bar{c}$  is small. 
These issues will be 
discussed in detail in sections 5 and 6.  

So far we have concentrated on deep inelastic $\vec l \vec N$ reactions. 
Purely hadronic reactions are presently being studied at the 
Fermilab Spin Physics Facility which consists of a 200 GeV longitudinally 
polarized proton or antiproton beam  
incident on a fixed polarized proton target 
\cite{adams1,yokosawa1}. Apart
from the non-uniquely polarized pentanol target, the beam
energy is rather low for a purely hadronic reaction
($\sqrt{s}$=19.4 GeV). Nevertheless, first measurements
resulted in a small, almost vanishing longitudinal
spin asymmetry $A_{LL}^{\pi^0}$ for inclusive $\pi^0$-production at
small $p_T^{\pi^0}$ (between 1 and 4 GeV).
This measurement is naively more consistent with a
small polarized gluon component in the proton than with a
large one, but has a very large error. An updated result
is being prepared by E-704 as well as results on
the totally inclusive polarized cross section and
polarized hyperon, 
direct photon and $J/ \psi$ production \cite{adams3,adams4}.
All of these have limited statistics and thus will be of very limited
use for perturbative QCD interpretations.
Using the 200 GeV proton beam, E-704 have also measured transverse  
single spin asymmetries in inclusive pion production, 
$p^{\uparrow} + p \rightarrow \pi^{0\pm}+X$ and found 
appreciable asymmetries \cite{adams8}. More details will be 
discussed in Sect. 6.8.

On the other hand there is the upcoming very promising
experimental spin program of the Relativistiv Heavy Ion Collider
(RHIC) Spin Collaboration (RSC) 
\cite{bunce,yokosawa} 
at the Brookhaven National
Laboratory, to which many of the E-704 physicists have switched.
At RHIC {\it both} proton beams, with an average energy of 250 GeV each,
will be polarized, using the 'Siberian snake' concept 
\cite{derbenev,derbenev1,krisch,roser}, 
to an expected polarization of about 70 $\%$.
Due to the high luminosity of the order of $\sim 10^{32} cm^{-2} s^{-2}$
(corresponding to an integrated luminosity of about 800 $pb^{-1}$)
the polarized RHIC pp collisions will play a decisive role
for measuring the polarized gluon density. This can be achieved
either via heavy quark production ($\vec{g}
\vec{g} \rightarrow Q \bar{Q}$, etc.) or via direct photon
production ($\vec{g} \vec{q} \rightarrow
\gamma q$, etc.). The latter process is particularly promising,
and a sensitivity of about 5 $\%$
is expected for ${\d g \over g} $ .
The theory of these processes will be discussed in Section 6.
Of course, polarization asymmetries for semi-inclusive
pion production and jet production will be also measured, in
particular for $W^\pm$ and Z production which give access to
to the various polarized quark densities
$\d u$,$\d \bar{u}$,$\d d$ and $\d \bar{d}$ separately
(cf. Section 6). Furthermore, transversity distributions (see Sect. 8.2)
will also be accessible.
Some other theoretical aspects of RHIC are discussed in \cite{bourrely1}. 

\begin{figure}           
\begin{center}
\epsfig{file=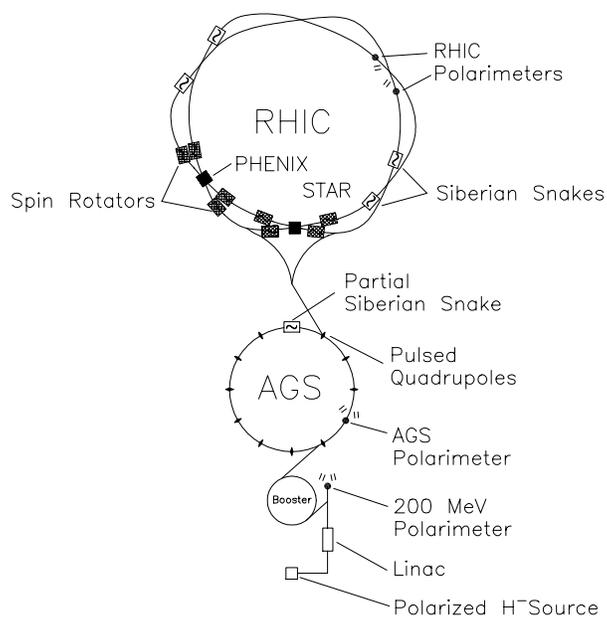,height=8cm,angle=0}
\caption{The built-up of the RHIC collider}
\label{figrhic}
\end{center}
\end{figure}

The build-up of the RSC experiment is depicted in Fig. \ref{figrhic}. 
The protons are taken from a polarized $H^{-}$ source and
are succesively accelerated by a LINAC, a Booster and the
Alternating Gradient Synchrotron (AGS)
to an energy of 24.6 GeV. In the AGS a 'partial snake'
is built in to maintain the polarization. A successful
partial snake test has been carried out in 1994
by the E880 collaboration \cite{blinov,huang,goodwin}.
Within the RHIC tunnel two full Siberian snakes will be
installed , more precisely four split Siberian snakes, two
per ring, $180^o$ apart.
This is expected to be complete in 1999, at
which time the PHENIX \cite{gregory} and STAR \cite{beddo} 
detectors should also
be completed. First data taking is scheduled for January 2000.

\begin{figure}             
\begin{center}
\epsfig{file=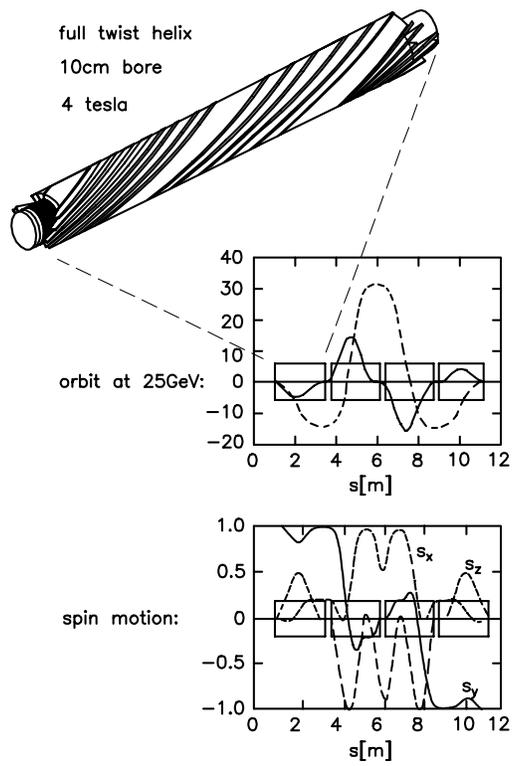,height=10cm,angle=0}
\caption{The layout of a Siberian snake}
\label{figsnake}
\end{center}
\end{figure}

For the nonexperts we include here a qualitative description
of how a 'Siberian snake' works.
Siberian snakes \cite{derbenev,derbenev1} 
are localized spin rotators
distributed around the ring to overcome the effects
of depolarizing resonances.
For spin manipulations
on protons at high energy, one should use transverse
magnetic fields, because the impact of longitudinal fields
disappears at high energy. In a transverse B-field the spin
of the high energy protons rotates $\kappa_p \gamma$ times faster
than motion ($\gamma ={E \over m}$ and $\kappa_p =1.79$ is the 
anomalous magnetic moment of the proton).
During acceleration, a depolarizing resonance is crossed
if this product is equal to an integer or equals the
frequency with which spin-perturbing magnetic fields are
encountered.
A Siberian snake turns the spin locally by an angle $\d$
($\d =180^0$ for a full snake, $\d \neq 0$ for a partial snake)
so that the would-be resonance condition is violated.
For the two full Siberian snakes in the RHIC collider the number
of $360^o$ spin rotations per turn is ${1 \over 2}$, i.e. the
resonance condition can never be met. The evolution of the spin and
orbit motion in the snake area are shown in Fig. \ref{figsnake}.

Coming back to the future of polarized experiments, we
note that the possibility of polarized protons is
also being considered at HERA.
The concept for HERA is in principle very similar 
to RHIC, the role of the AGS being taken by DESYIII
and PETRA which successively accelerate the protons to
40 GeV \cite{bluemlein}. Recently, a "Workshop on Future 
Physics at HERA" has taken place in Hamburg, where the 
option of polarized high energy protons at HERA has 
been discussed. We refer to the proceedings of this 
workshop for extensive discussions of this topic \cite{ingelman}. 
Within the framework of this workshop, 
another experiment ('HERA--$\vec{N}$')
utilising an internal polarized fixed nucleon target in the 820 GeV
HERA proton beam has also been examined, see also \cite{korotkov} 
and references therein. 
Conceivably, this would be the only place where to study high energy
nucleon--nucleon spin physics besides the dedicated RHIC spin program. 
An internal polarized nucleon target offering unique features such as
polarization above 80\% and no or small dilution, can be safely
operated
in a proton ring at high densities up to $10^{14}$ atoms/cm$^2$
\cite{steffens}.
As long as the polarized target is used in conjunction with an
unpolarized proton
beam, the physics scope of  HERA--$\vec{N}$ would be focussed to
'Phase I', i.e. measurements of single spin asymmetries (to be 
discussed in Sect. 6.8).
Once later polarized protons should become available, the same set--up
would 
be readily available to measure a variety of double spin asymmetries.
These 'Phase II' measurements would constitute an alternative
fixed target approach to similar physics which will be
accessible to
the collider experiments  STAR and  PHENIX at the low end of
the RHIC energy scale ($\sqrt{s} \simeq 50$ GeV).

Furthermore, there are several polarized low energy ($E_{beam}^{e^-} 
\lesssim 5 \; GeV$) facilities, such as AmPS-NIKHEF, MIT-Bates, CEBAF-Newport 
News, ELSA-Bonn and MAMI-Mainz, some of which are already operating 
and will provide us with low-energy precision measurements of $\vec e \vec
N$ reactions. The interested reader is referred to the respective 
review articles in \cite{faessler}. 

%% file: k41tex
\setcounter{equation}{0}
\section{The Structure Function $g_1$ and Polarized Parton Distributions}
\subsection{The Quark Parton Model to Leading Order of QCD}
 
The parton model is a very useful tool for the
understanding of hadronic high energy reactions. This is due
to its simplicity and comprehensiveness as well as to its
universality, i.e. applicability to any hadronic process. In
the form of the  QCD improved parton model it has had
tremendous successes in the understanding of unpolarized
scattering. One is therefore tempted to apply it to
processes with polarized particles as well.
 
In the following we will assume that the nucleon is
longitudinally polarized. The notion of transversality is
difficult to adopt in the parton model, only at the price of
loosing many of its virtues. The main aim of this section is to
represent the polarized structure functions in terms of
``polarized'' parton densities, in a similar fashion as the
spin averaged structure functions can be represented in
terms of spin averaged parton densities. For definiteness,
let us consider the process $\mu p\to \mu X$ at scales
$Q^2 \ll m_{W,Z}^2$ (only photon exchange needed), with a
proton of positive longitudinal polarization (= right handed
helicity). Within this reference proton there are (massless)
partons with positive and negative helicity which carry a
fraction $x$ of the proton momentum and to whom one can
associate quark densities $q_+ (x,Q^2)$ and $q_-(x, Q^2)$.
The difference
\begin{equation}
\delta q(x,Q^2) = q_+(x,Q^2) - q_-(x,Q^2)
\label{411}
\end{equation}
measures how much the parton of
flavor $q$ ``remembers'' of its parent proton polarization.
Similarly,one may define
\begin{equation}
\delta \bar q(x,Q^2)=\bar q_+(x,Q^2) -\bar q_-(x,Q^2)
\label{412}\end{equation}
for antiquarks.
Note that the ordinary, spin averaged parton densities are
given by
\begin{equation}
q(x,q^2)=q_+(x,Q^2)+q_-(x,Q^2)
\label{413}\end{equation}
and
\begin{equation}
\bar q(x,Q^2)=\bar q_+(x,Q^2)+\bar q_-(x,Q^2)\/.
\label{414}
\end{equation}
In the quark parton model, to leading order (LO) in QCD,
$g_1$ can be written as a linear
combination of $\delta q$ and $\delta \bar{q}$ 
\cite{feynman,altarelli,altarelli1},
\begin{equation}
g_1(x,Q^2)=\ha\sum_q e^2_q [\delta q(x,Q^2)+\delta \bar q(x,Q^2) ]
\label{415} 
\end{equation}
where $e_q$ are the electric charges of the (light) quark-flavors
$q=u,d,s$.
Notice that in the case of spin averaged structure functions
$F_{1,2}$ the negative helicity densities $q_-,\bar q_-$
enter with an opposite
sign, e.g.\ $F_1(x,Q^2)=\ha \sum_q e^2_q(q+\bar q)$. 
This has to do with the opposite charge conjugation 
property of $\gamma_\mu$ and $\gamma_\mu\gamma_5$.
In the case of the polarized $\nu N$ structure 
functions $g_3$ etc. the situation
is reversed.There the $\delta \bar q(x,Q^2)$ enter with a
negative sign, $g_3 \sim \delta q - \delta \bar q$, 
cf. Section 6; and for 
the unpolarized $\nu N$ structure function  
$F_3$ one has $F_3 \sim q - \bar q$. 
 
Furthermore, Eq.\ (\ref{415}) can be decomposed into a flavor
nonsinglet (NS) and singlet (S) component
\begin{equation}
g_1(x,Q^2)=g_{1,NS}(x,Q^2)+g_{1,S}(x,Q^2)
\label{416}\end{equation}
where 
\begin{equation}
g_{1,NS}=\ha\sum\limits_q(e^2_q-<e^2>)(\delta q+\delta\bar q)
\label{417}\end{equation}
with $<e^2> = { 1 \over f} \sum_q e^2_q$ (e.g.\ for $f=3$ light
$u,d,s$ flavors $<e^2>={2 \over 9}$) and
\begin{equation}
g_{1,S} =\ha <e^2>\sum\limits_q (\delta q+\delta\bar q)
        \equiv \ha <e^2>\delta\Sigma\/. 
\label{418}
\end{equation}
In the last equation we have defined the singlet combination (sum of)
quarks and antiquarks by
\begin{equation}
\delta\Sigma (x,Q^2)=\sum_q [\delta q(x,Q^2)+\delta\bar{q}(x,Q^2)]
\label{419}
\end{equation}
where the sum usually runs over the light quark-flavors
$q=u,d,s$, since the heavy quark contributions $(c,b,\ldots)$
have preferrably to be calculated perturbatively from the
intrinsic light quark ($u,d,s$) and gluon $(g)$ partonic-constituents
of the nucleon which will be discussed in Sect. 6.2.
 
\begin{figure}
\begin{center}
\epsfig{file=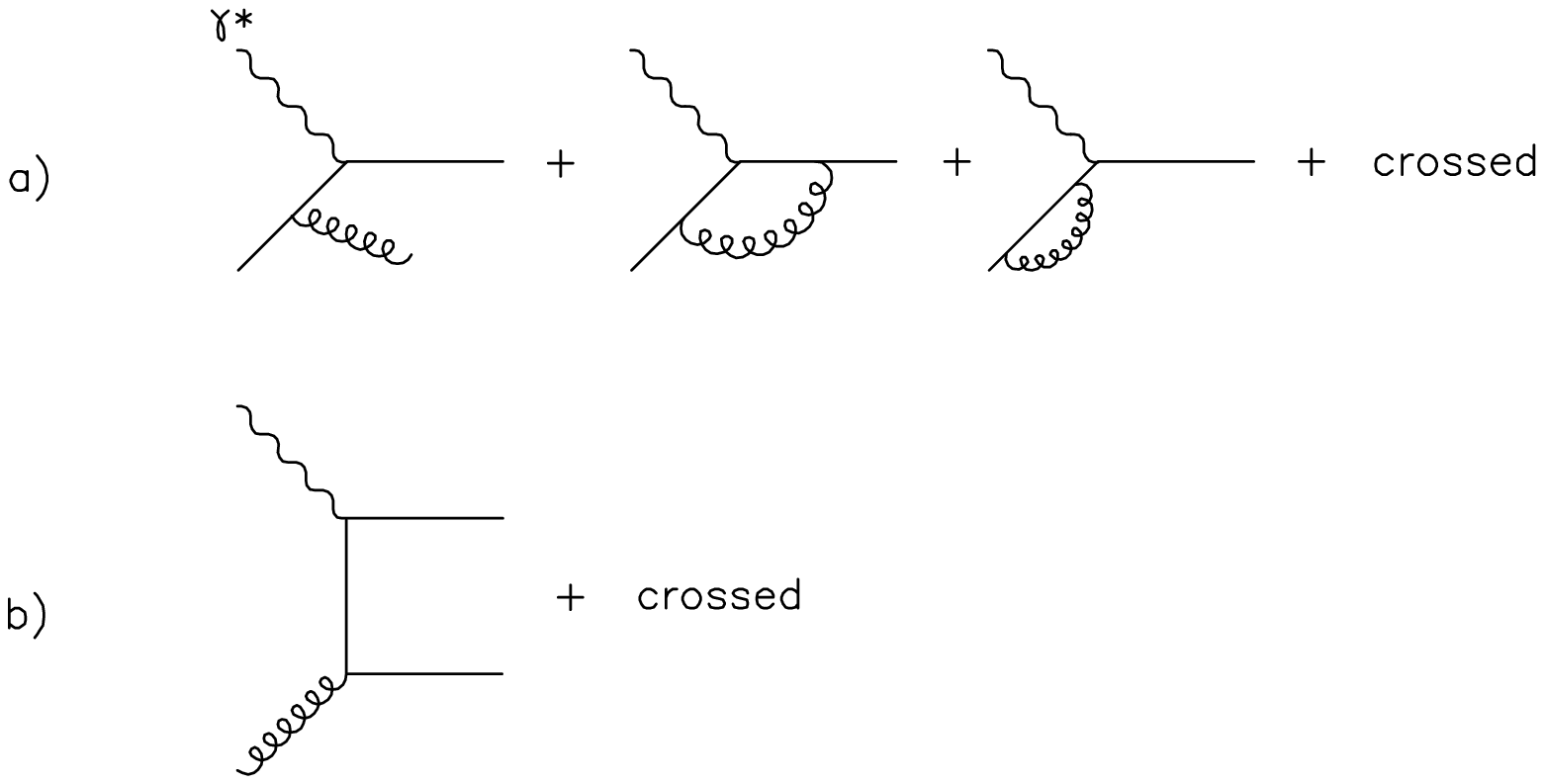,height=5cm}
\vskip 0.5cm
\caption{The parton subprocesses $\gamma^{\ast}q\to gq$ and $\gamma^{\ast}
g\to q\bar q$}
\label{fig14}  
\end{center}
\end{figure}

The QCD scale-violating $Q^2$-dependence of the above
structure functions
and parton distributions is inherently introduced dynamically due to
gluon radiation $(q\to qg)$ and gluon-initiated $(g\to q\bar q)$
subprocesses depicted, to LO, in Fig. \ref{fig14}.
To the leading logarithmic order
(LO), these $Q^2$-corrections have been calculated in
\cite{altarelli,ahmed,ahmed1,sasaki}.
The next-to-leading order NLO
(two-loop) results (Wilson coefficients and in particular splitting
functions) have recently been calculated 
\cite{mertig,vogelsang2} and will be
discussed in detail in Sec. 4.2. One of the main ingredients from
QCD (or, more generally, from any strongly interacting quantum field
theory) is the appearance of gluon distributions in the nucleon in the
form $\delta g(x,Q^2)$ which is the longitudinally polarized gluon
density, probed at a scale $Q^2$, and is defined as follows. Assume that
in our reference proton of positive helicity the gluons have momenta of
the form $p_g=E(1,0,0,1)$ [in the Breit-frame $P=(\sqrt{P^2+M^2},0,0,P)$,
$q=(0,0,0,-Q)$ in which $E=xP$]. The two possible polarization vectors
of the gluon are 
$\ve^\mu_g ={1 \over \sqrt{2}} (0,1,\pm i,0)$ which correspond to
positive and negative circular polarization. To each state of polarization
one can attribute a gluon density, $g_+(x,Q^2)$ and $g_-(x,Q^2)$.
The ordinary, spin averaged gluon density is given by $g(x,Q^2)=
g_++g_-$, whereas $\delta g$ is defined as
\begin{equation}
\delta g(x,Q^2)= g_+(x,Q^2)- g_-(x,Q^2)\/.
\label{4110}
\end{equation}
Since the individual parton distributions with definite helicity
$f_{\pm}(x,Q^2)$, $f=q,\bar q, g$, in Eqs.~(\ref{411})-(\ref{414})
and (\ref{4110}) are by definition positive definite, their
difference $\delta f$ have to satisfy the general positivity constraints
\begin{equation}
|\delta f(x,Q^2)|\leq f(x,Q^2)\/.
\label{4111}
\end{equation}
 
In LO the gluon distribution does {\it not directly} contribute to
the structure function $g_1(x,Q^2)$ in (\ref{415}),
but only indirectly via the
$Q^2$--evolution equations. Furthermore it is a pure
flavor-{\it singlet},
as is $\delta\Sigma (x,Q^2)$ in (\ref{419}), because each massless
quark flavor $u,d,s$ is produced by gluons at the same rate.
 
\begin{figure}
\begin{center}
\epsfig{file=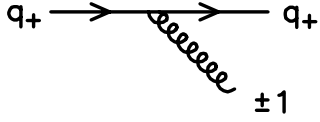,height=1.5cm}
\caption{The leading order splitting process $q \rightarrow qg$}
\label{fig15}  
\end{center}
\end{figure}

The LO $Q^2$-evolution (or renormalization group) equations are
as follows. Only flavor-nonsinglet (valence)
combination $\delta q_{NS}$
[$=\delta u-\delta\bar u,\delta d-\delta \bar d$,
$(\delta u+\delta \bar u)-
(\delta d+\delta \bar d)$, $(\delta u+\delta\bar u)+
(\delta d+\delta \bar d)-
2(\delta s+\delta \bar s)$, etc.], i.e.\
where sea- and gluon-contributions cancel,
evolve in the {\it same} way in LO:
\begin{equation}
\frac{d}{dt} \delta q_{NS}(x,Q^2)=\frac{\alpha_s(Q^2)}{2\pi}\delta
P^{(0)}_{NS}\otimes\delta q_{NS}
\label{4112}
\end{equation}
where $t=\ln\frac{Q^2}{Q^2_0}$, with $Q_0$ being the appropriately chosen
reference scale at which $\delta q_{NS}$ is determined (mainly from
experiment), and
\begin{equation}
\frac{\alpha_s(Q^2)}{4\pi}\simeq \frac{1}{\beta_0\ln 
{Q^2 \over \Lambda^2_{LO}}}
\label{4113}
\end{equation}
with $\beta_0=11-{2 \over 3}f$ 
and $f$ is the number of active ('light') flavors.
The convolution ($\otimes$) is given by
\begin{equation}
(P\otimes q)(x,Q^2)=\int\limits^1_x \frac{dy}{y} P(\frac{x}{y}) q(y,Q^2)
\label{4114}
\end{equation}
which goes over into a simple ordinary product if one considers Mellin
$n$-moments to be discussed later. The $LO~NS$ splitting function,
\begin{equation}
\delta P^{(0)}_{NS}(x)=\delta P^{(0)}_{qq} (x)\equiv
P^{(0)}_{q_+q_+}-P^{(0)}_{q_-q_+} \/,
\label{4115}\end{equation}
where $P_{q_\pm q_+}$ corresponds to transitions from
a quark $q_+$ with positive helicity to a quark $q_\pm$ with
positive/negative helicity, 
is given by \cite{altarelli}
\begin{equation}
\delta P^{(0)}_{qq}(x)=P^{(0)}_{qq}(x)= C_F\left(\frac{1+x^2}{1-x}\right)_+
\label{4116} \end{equation}
with $C_F=4/3$. The fact that 
$\delta P^{(0)}_{qq}$ turns out to be equal to
the unpolarized splitting function $P^{(0)}_{qq}$, i.e.\
\begin{equation}
P^{(0)}_{q_-q_+}(x)=0
\label{4117}\end{equation}
in (\ref{4115}), is a consequence of helicity conservation, i.e.\ the
fact that no transition between quarks of opposite helicity are
allowed in massless perturbative QCD -- at least to leading order.
In suitable (chirality respecting) regularization schemes this
statement, i.e.\ Eq.~(\ref{4117}), can be generalized to higher
orders as we shall see later. The LO diagram is shown in Fig. \ref{fig15}:
The conservation of the quark helicity is a consequence of the
vector-like coupling between quarks and gluons. Furthermore, since
the gluon has spin $+1$ or $-1$, a finite angular momentum
between the quark and gluon is always produced in such a process.
Finally, the convolution (\ref{4114}) with the $(~~)_+$ distribution
\cite{altarelli}
in (\ref{4116}) can be easily calculated using
\be 
\int\limits^1_x\frac{dy}{y} f(\frac{x}{y})_+ g(y)= 
\int\limits^1_x\frac{dy}{y} f(\frac{x}{y}) \left[ g(y)-\frac{x}{y}
g(x)\right]  
-g(x)\int\limits^x_0 dy f(y)\/.
\label{4118}
\ee

In contrast to (\ref{4112}),   
the LO $Q^2$--evolution equations in the flavor-singlet section are 
{\it coupled} integro-differential equations,
\begin{equation}
\frac{d}{dt}{\delta\Sigma(x,Q^2) \choose \delta g(x,Q^2)}
=\frac{\alpha_s (Q^2)}{2\pi}{\delta P^{(0)}_{qq} \quad 2f\delta P^{(0)}_{qg}
                      \choose\delta P^{(0)}_{gq} \quad \delta P^{(0)}_{gg}}
\otimes {\delta\Sigma\choose\delta g}
\label{4119}
\end{equation}
with $\delta P^{(0)}_{qq}(x)$ given by Eq.~(\ref{4116}). The remaining
longitudinally polarized splitting functions are defined, in analogy to
(\ref{4115}),
\begin{equation}
\delta P^{(0)}_{AB}(x)\equiv P^{(0)}_{A_+B_+}-P^{(0)}_{A_-B_+}
\label{4120}
\end{equation}
with $A,B=q,g$, which fulfil $P^{(0)}_{A_{\pm}B_+}=P^{(0)}_{A_{\mp}B_-}$
due to parity invariance of the strong interactions (QCD).
The spin averaged splitting functions are given by the sum
$P^{(0)}_{AB}(x)=P^{(0)}_{A_+B_+}+P^{(0)}_{A_-B_+}$. Note that
$\delta \Sigma$ refers to the {\it sum} of all quark flavors and
antiflavors and therefore the factor $2f$ in front of $\delta P^{(0)}_{qg}$
in (\ref{4119}). Besides $\delta P^{(0)}_{qq}$ in (\ref{4116}), the
remaining polarized LO splitting function in (\ref{4119}) are given
by \cite{altarelli}
%\begin{equation}
%\delta P^{(0)}_{qg}(x)=T_R[x^2-(1-x)^2]=T_R(2x-1) 
%\label{4121a}
%\end{equation}
%\begin{equation}
%\delta P^{(0)}_{gq}(x)=C_F\frac{1-(1-x)^2}{x}=C_F(2-x) 
%\label{4121b}
%\end{equation}
%\begin{multline}
%\delta P^{(0)}_{gg}(x)=C_A\bigl\{(1+x^4)(\frac{1}{x}+\frac{1}{(1-x)_+}) \\ 
%                        - \frac{(1-x)^3}{x} 
%                       +(\frac{11}{6}-\frac{f}{9})\delta (1-x)\bigr\} \\
%\label{4121c}
%\end{multline}
\begin{eqnarray} \nonumber      
\delta P^{(0)}_{qg}(x)&=&T_R[x^2-(1-x)^2]=T_R(2x-1) \\  \nonumber 
\delta P^{(0)}_{gq}(x)&=&C_F\frac{1-(1-x)^2}{x}=C_F(2-x) \\  
\delta P^{(0)}_{gg}(x)&=&C_A\bigl\{(1+x^4)(\frac{1}{x}+\frac{1}{(1-x)_+}) 
  - \frac{(1-x)^3}{x} 
                       +(\frac{11}{6}-\frac{f}{9})\delta (1-x)\bigr\} 
\label{4121}
\end{eqnarray} 
with $T_R=\ha$ and $C_A=N_c=3$. The advantage of introducing the
singlet-set of differences of parton distributions $(\delta\Sigma,\delta g)$
in (\ref{4119}), instead of using the individual positive and negative
helicity densities $q_\pm$ and $g_\pm$ in (\ref{411}) and (\ref{4110}),
is that they evolve {\it in}dependently in $Q^2$ of the set of the conventional
unpolarized densities $(\Sigma ,g)$ where
$\Sigma\equiv\sum_f(q+\bar q)$. This is in contrast to the individual
densities $q_\pm$ and $g_\pm$ of definite helicity. The evolution
equations (\ref{4112}) and (\ref{4119}) can be solved numerically
by iteration directly in Bjorken-$x$ space. In many cases it is, however,
more convenient and physically more transparent to work in Mellin
$n$-moment space where the LO as well as the NLO evolution equations
can be solved analytically to a given (consistent) perturbative
order in $\alpha_s$. This is due to the fact that for
moments, defined by
\begin{equation}
f^n(Q^2)\equiv\int^1_0 x^{n-1} f(x,Q^2)dx\/,
\label{4122}
\end{equation}
the convolution (\ref{4114}) appearing in (\ref{4112}) and (\ref{4119})
factorizes into simple ordinary products:
\be 
\int^1_0 dx~x^{n-1}f\otimes g \equiv\int^1_0 dx~x^{n-1} 
\int^1_x\frac{dy}{y}f(y)g(\frac{x}{y})      
=\int^1_0dx~x^{n-1}\int^1_0 dydz\delta (x-zy)f(y)g(z)=f^ng^n\/.
\label{4123}
\ee
In moment space the LO nonsinglet and singlet evolution equations
(\ref{4112}) and (\ref{4119}) are thus simply given by
\begin{equation}
\frac{d}{dt}\delta q^n_{NS}(Q^2)=\frac{\alpha_s(Q^2)}{2\pi}
\delta P^{(0)n}_{qq}\delta q^n_{NS}(Q^2)
\label{4124}
\end{equation}
\begin{equation}
\frac{d}{dt}{\delta\Sigma^n(Q^2) \choose \delta g^n(Q^2) }=
\frac{\alpha_s (Q^2)}{2\pi}
{\delta P^{(0)n}_{qq} \quad 2f\delta P^{(0)n}_{qg}\choose
 \delta P^{(0)n}_{gq} \quad \delta P^{(0)n}_{gg}}
{\delta\Sigma^n(Q^2) \choose \delta g^n(Q^2) }
\label{4125}
\end{equation}
where 
the $\delta P^{(0)n}_{ij}$ are simply the $n$-th moment of
Eqs.~(\ref{4116}) and (\ref{4121}) :
\begin{eqnarray} \nonumber
\delta P^{(0)n}_{qq}&=&\frac{4}{3}\left[\frac{3}{2} +\frac{1}{n(n+1)}-
2 S_1(n)\right]      \\  \nonumber 
\delta P^{(0)n}_{qg} &=&\ha\frac{n-1}{n(n+1)}     \\  \nonumber 
\delta P^{(0)n}_{gq} &=&\frac{4}{3}\frac{n+2}{n(n+1)}   \\ 
\delta P^{(0)n}_{gg} &=&3\left[\frac{11}{6}+\frac{4}{n(n+1)}-2 S_1(n)\right]
-\frac{2}{3}\frac{f}{2} \/.
\label{4126}
\end{eqnarray}
%\delta P^{(0)n}_{qq}&=\frac{4}{3}\left[\frac{3}{2} +\frac{1}{n(n+1)}-  
%2 S_1(n)\right]      \\ 
%\delta P^{(0)n}_{qg} &=\ha\frac{n-1}{n(n+1)}     \\
%\delta P^{(0)n}_{gq} &=\frac{4}{3}\frac{n+2}{n(n+1)}   \\
%\delta P^{(0)n}_{gg} &=3\left[\frac{11}{6}+\frac{4}{n(n+1)}-2 S_1(n)\right]  
%-\frac{2}{3}\frac{f}{2} \/.
Here $S_1(n)\equiv\sum\limits^n_{j=1} \frac{1}{j}=\psi (n+1)+\gamma_E$,
$\psi(n)\equiv{\Gamma'(n) \over \Gamma(n)}$ and $\gamma_E=0.577216$.
 
The evolution equations (\ref{4124}) and (\ref{4125}) 
in $n$-moment space are
usually referred to as LO renormalization group (RG) equations which
were originally derived 
\cite{christ,callan,kogut,gross,gross1,georgi,curci,floratos,furmanski}
from the operator product (light-cone) expansion for unpolarized structure
functions [for reviews see, 
for example, \cite{altarelli1,politzer,reya}].
The moments $\delta P^{(0)n}_{ij}$ are called (or, more precisely,
related 
\footnote{the anomalous dimensions $\delta\gamma^n$, being
usually defined as an expansion in ${\alpha_s \over 4\pi}$, 
$\delta\gamma^n=
{\alpha_s \over 4\pi}\delta\gamma^{(0)n}+({\alpha_s \over 4\pi})^2\delta
\gamma^{(1)n}+\cdots$, in terms of the LO (1-loop)
$\delta\gamma^{(0)n}$ and NLO (2-loop) $\delta\gamma^{(1)n}$
expressions, are related to the $\delta P^n$ via
$\delta P^{(0)n}_{ij}=-{1 \over 4}\delta \gamma^{(0)n}_{ij}$, 
$\delta P^{(1)n}_{ij}=-{1 \over 8}\delta \gamma^{(1)n}_{ij}$,
etc. where the 2-loop splitting functions $\delta P^{(1)}_{ij}$
will become important for the NLO evolutions to be discussed in
Section 4.2.\label{fn1}}
%++++++++++++ende der fussnote+++++++++++++++++++++++++++++
to the) 'anomalous dimensions' because they determine the logarithmic
$Q^2$-dependence of the moments of parton distributions and thus
of $g_1$ as we shall see below.
 
The solution of the simple $NS$ equation (\ref{4124}) is straightforward:
\begin{equation}
\delta q^n_{NS}(Q^2)=L^{-\frac{2}{\beta_0}\delta P^{(0)n}_{qq}} 
\delta q^n_{NS}(Q^2_0)
\label{4127}
\end{equation}
with $L(Q^2)\equiv\alpha_s(Q^2)/\alpha_s(Q^2_0)$, 
$\beta_0$ being defined after (\ref{4113}) and $\delta q^n_{NS}(Q^2_0)$
is the appropriate NS-combination of polarized parton densities
fixed (mainly) from experiment at a chosen input scale $Q^2_0$.
The solution of the coupled singlet evolution equations
(\ref{4125}) is formally similar to (\ref{4127}):
\begin{equation}
{\delta\Sigma^n (Q^2)\choose\delta g^n(Q^2)}=L^{-\frac{2}{\beta_0}
\delta\hat P^{(0)n}}{\delta\Sigma^n (Q^2_0)\choose \delta g^n(Q^2_0)}
\label{4128}
\end{equation}
where $\delta \hat P^{(0)n}$ denotes the $2\times 2$ singlet matrix of
splitting functions in Eq.~(\ref{4125}).
The treatment of this exponentiated matrix follows the standard
diagonalization technique 
, see e.g. \cite{gross,gross1},  
where one projects onto the larger and smaller eigenvalues
$\lambda^n_{\pm}$ of $\delta\hat P^{(0)n}$ with the help of the
$2\times 2$ projection matrices $\hat P_{\pm}$ given by
\begin{equation}
\hat P_\pm \equiv\pm{\delta\hat P^{(0)n}-\lambda^n_\mp 1 \over
 \lambda^n_+-\lambda_-^n}  
\label{4129}
\end{equation}
with
\be  
\lambda^n_{\pm} = \ha\big[\delta P^{(0)n}_{qq}+\delta P^{(0)n}_{gg}
\pm
\sqrt{(\d\Pon_{qq}-\d\Pon_{gg})^2+8f\d\Pon_{qg}\d\Pon_{gq}}\Big]\/.
\label{4130}
\ee
The projection matrices $\hat P_{\pm}$ have the usual properties 
$\hat P^2_\pm = \hat P_\pm,\quad \hat P_+\hat P_-=\hat P_-\hat P_+=0$ 
and $\hat P_++\hat P_- =1$. 
Since $\d \hat P^{(0)n} =\lambda^n_+\hat P_++\lambda^n_-\hat P_-$, 
the matrix expression in (\ref{4128}) can be explicitly calculated using
\begin{equation}
f(\d\hPon)=f(\lambda^n_+)\hat P_++f(\lambda^n_-)\hat P_-\/,
\label{4131}
\end{equation}
i.e.
\begin{equation}
L^{-\frac{2}{\beta_0}\d\hPon}=L^{-\frac{2}{\beta_0}\lambda^n_+}
\hat P_+ + L^{-\frac{2}{\beta_0}\lambda^n_-}\hat P_-\/. 
\label{4132}
\end{equation}
Thus the solution of (\ref{4128}) is explicitly given by
\be 
\delta\Sigma^n(Q^2)=[\alpha_n\delta\Sigma^n\Q2 +\beta_n\d g^n\Q2]
L^{-{\zweib}\lambda^n_-}  
 +[(1-\alpha_n)\delta\Sigma^n\Q2 -\beta_n\delta g^n\Q2]L^{-{\zweib}
\lambda^n_+}
\label{4133}
\ee
\be 
\delta g^n(Q^2) = 
[(1-\alpha_n)\delta g^n\Q2 +\frac{\alpha_n(1-\alpha_n)}
{\beta_n} \delta\Sigma^n\Q2]L^{-{\zweib}\lambda^n_-}
+[\alpha_n\delta g^n\Q2 -\frac{\alpha_n(1-\alpha_n)}{\beta_n}
\delta\Sigma^n\Q2]L^{-{\zweib}\lambda^n_+}
\label{4134}
\ee
with
%\begin{equation}
%\begin{split}
%\alpha_n &={\delta \Pon_{qq}-\lambda^n_+ \over \lambda^n_--\lambda^n_+} \\ 
%\beta_n &={2f\delta\Pon_{qg} \over \lambda^n_--\lambda^n_+} \/.
%\label{4135}
%\end{split}
%\end{equation}
\begin{equation}
\alpha_n ={\delta \Pon_{qq}-\lambda^n_+ \over \lambda^n_--\lambda^n_+} 
\label{4135a}  
\end{equation}
\begin{equation}
\beta_n ={2f\delta\Pon_{qg} \over \lambda^n_--\lambda^n_+} \/.
\label{4135b}
\end{equation}
Once the parton distributions are fixed at a specific input scale
$Q^2=Q^2_0$, mainly by experiment and/or theoretical model constraints,  
their evolution to any $Q^2>Q^2_0$ is uniquely predicted by the
QCD-dynamics due to the uniquely calculable 'anomalous dimensions'
$\delta P^{(0)n}_{ij}$ in LO of $\alpha_s$. Of course a LO calculation
by itself is in general not sufficient since neither the $\Lambda$-parameter
in $\alpha_s(Q^2)$ can be unambiguously defined nor can one prove
the perturbative reliability of the results which requires at least a NLO
analysis, to which we will turn in the next subsections.
 
To obtain the $x$-dependence of structure functions and parton
distributions, usually required for practical purposes, from the
above $n$-dependent exact analytical solutions in Mellin-moment
space, one has to perform 
a numerical integral 
in order to invert the Mellin-transformation in
(\ref{4122}) according to
\begin{equation}
f(x,Q^2)=\frac{1}{\pi}\int\limits^\infty_0 dz Im\left[
e^{i\vph} x^{-c-ze^{i\vph}}f^{n=c+ze^{i\vph}}(Q^2)\right]
\label{4136}
\end{equation}
where the contour of integration, and thus the value of $c$, has to
lie to the right of all singularities of $f^n(Q^2)$ in the complex
$n$-plane, i.e., $c>0$ since according to Eq.~(\ref{4126})  
the dominant pole of {\it all} $\delta P^n_{ij}$ is located at $n=0$. 
For all practical purposes one may choose $c \approx 1$, $\vph=135^o$
and an upper limit of integration, for any $Q^2$, of about $5+10/\ln x^{-1}$,
instead of $\infty$, which suffices to guarantee accurate and stable
numerical results \cite{glueck,glueck1}. 
Note that it is advantageous to use $\vph > {\pi \over 2}$
in (\ref{4136}) because then the factor $x^{-z exp(i\vph)}$
dampens the integrand for increasing values of $z$ which allows for a
reduced upper limit in the numerical integration; this is in contrast
to the standard mathematical choice $\vph ={\pi \over 2}$ which corresponds
to a contour parallel to the imaginary axis.

%% file: k42tex
\subsection{Higher Order Corrections to $g_1$}
The LO results discussed so far originated from calculating
the logarithmic $O(\alpha_s)$ contributions of the parton subprocesses
$\gamma^{\ast}q\to gq$ and $\gamma^{\ast} g\to q\bar q$ (Fig. \ref{fig14}) 
to the zeroth order
'bare' term $\gamma^*q\to q$ of $g_1$ \cite{altarelli1,altarelli}:
\begin{eqnarray} \nonumber  
g_1(x,Q^2)=\ha\sum_{q,\bar q} e^2_q
\biggl\{
\delta q_0 (x)
+\frac{\a_s (Q^2)}{2\pi}
\int\limits^1_x\frac{dy}{y}\delta q_0(y)
\Big[
t\delta P^{(0)}_{qq}(\frac{x}{y})+
\delta f_q(\frac{x}{y})\Big] \biggr\}
\\  
+\ha\left(\sum\limits_{q,\bar q}e^2_q\right)\frac{\a_s(Q^2)}{2\pi}
\int\limits^1_x\frac{dy}{y}\delta g_0(y)
\Big[t\delta P^{(0)}_{qg}
(\frac{x}{y})+\delta f_g(\frac{x}{y})\Big] 
\label{421}
\end{eqnarray}
where $\delta q_0, \delta g_0$ denote the unphysical and unrenormalized
(i.e. scale independent) bare parton distributions and the function
$\delta f_q$ and $\delta f_g$ are sometimes called
'constant terms' or 'coefficient functions' because they are related to the
$\ln Q^2$-independent terms and to the Wilson coefficients usually
introduced
within the framework of the operator product expansion. In the leading
logarithmic order (LO) one assumes the dominance of the universal 
$t=\ln {Q^2 \over Q^2_0}$ terms which are independent of the regularization
scheme adopted, in contrast to $\delta f_{q,g}$. 

Eq.~(\ref{421}) as it stands is physically meaningless since $Q^2_0$ is
an entirely arbitrary scale which, among other things, serves
as an effective cutoff for the emitted parton-$k_T$ (or angular)
phase space integration in order to avoid collinear (mass)
singularities when the emitted partons in Fig. \ref{fig14} are along the 
intitial quark/gluon direction $(k_T=0$ or $\theta=0)$.
Therefore only the {\it variation} with $Q^2$ can be uniquely
predicted for the physical and renormalized (i.e. scale dependent)
'dressed' quark distribution, 
\be 
\delta q(x,Q^2)  \equiv  \delta q_o(x)+\frac{\a_s(Q^2)}{2\pi}
t(\delta q_0\otimes\delta P^{(0)}_{qq}+\delta g_0\otimes
\delta P^{(0)}_{qg})
\label{422}  
\ee
and similarly for a suitably defined 'dressed' polarized
gluon density $\delta g(x,Q^2)$. These redefinitions result in the RG 
evolution equations (\ref{4112}) and (\ref{4119})
\cite{altarelli}.

In NLO, i.e.\ if one goes beyond the leading logarithmic order,
the 'finite' terms $\delta f_{q,g}$ in (\ref{421}) have to be
included, as well as the contributions of the 2-loop splitting 
functions $\delta P^{(1)}_{ij} (x)$. These additional quantities
have the unpleasant feature that they depend on the
regularization scheme adopted. For calculational convenience
one often chooses dimensional regularization and the 't Hooft-Veltman
prescription for $\gamma_5$ \cite{hooft,breitenlohner,chanowitz}.
In $D=4-2\ve$ dimensions one obtains for the diagrams in
Fig. \ref{fig14} 
\cite{mertig,ratcliffe,vogelsang1,bodwin,manohar,ellwanger}  
\begin{eqnarray} \nonumber  
g_1 (x,Q^2) = \ha\sum\limits_{q,\bar q} e^2_q\biggl\{\delta
q_0(x)+\frac{\a_s(Q^2)}{2\pi}
\int\limits^1_x \frac{dy}{y}
\delta q_0(x)\Big[(\ln {Q^2 \over \mu^2}-\frac{1}{\ve}+\gamma_E-\ln 4\pi)
\delta P_{qq}^{(0)}
(\frac{x}{y}) +\delta C_q(\frac{x}{y})\Big] \biggr\}
\\
+\ha\left(\sum_{q,\bar q} e^2_q\right)
\frac{\a_s(Q^2)}{2\pi}
\int\limits^1_x\frac{dy}{y}\delta g_0(y)
\Big[(\ln {Q^2 \over \mu^2}-\frac{1}{\ve}
+\gamma_E-
\ln 4\pi)
\delta P_{qg}^{(0)}(\frac{x}{y})+
\delta C_g(\frac{x}{y})\Big] 
\label{423}
\end{eqnarray}
where the dimensional regularization mass parameter $\mu$ is 
usually chosen to equal $Q$, and 
\be 
\delta C_q(z)=\frac{4}{3}\biggl[(1+z^2)\left(\frac{\ln(1-z)}{1-z}\right)_+
-\frac{3}{2}\frac{1}{(1-z)_+} 
-{1+z^2 \over 1-z}\ln z+2+z
-\left(\frac{9}{2}+\frac{\pi^2}{3}\right)\delta (1-z)\biggr] 
\label{424q}
\ee
\begin{equation}
\delta C_g(z)=\frac{1}{2}[(2z-1)(\ln {1-z \over z}-1)+2(1-z)] \/.
\label{424}
\end{equation}
The $(~~)_+$ distribution is, as usual, defined by
\begin{equation}
\int\limits^1_0 dx\frac{f(x)}{(1-x)_+}\equiv\int\limits^1_0 dx
\frac{f(x)-f(1)}{1-x}
\label{425}
\end{equation}
and Eq. (\ref{4118}) is again useful for calculating the 
convolutions. 

The unphysical 'bare' quark and gluon densities $\delta q_0$ and
$\delta g_0$ in (\ref{423}) have to be redefined in order to get
rid of the singularities present (for $\ve =0$). The renormalized
'dressed' quark distribution is defined by
\be  
\delta q (x,Q^2)  \equiv  \delta q_0(x)
+\frac{\a_s(Q^2)}{2\pi}
(\ln {Q^2 \over \mu^2}-\frac{1}{\ve} +\gamma_E-\ln 4\pi)
(\delta q_0\otimes\delta
P_{qq}^{(0)}+\delta g_0\otimes\delta P_{qg}^{(0)})
\label{426}
\ee
and there is a similar equation for the redefined gluon density. 
Eq. (\ref{426}) refers to the 
'modified minimal subtraction' $(\overline{MS})$
factorization scheme since also the combination $\gamma_E-\ln 4\pi$, being
just a mathematical artifact of the phase space in $4-2\ve$ dimensions,
has been absorbed, together with ${-1 \over \ve}$, into the definition of
$\delta q (x,Q^2)$. Besides dimensional regularization one has of course
the possibility to use masses of quarks and/or gluons to regulate the
divergences. However, in those schemes 
NLO calculations become usually far more cumbersome. 
Furthermore, one obtains results for $\delta C_q$ and $\delta C_g$ which
differ from the ones in (\ref{424q}) and (\ref{424}). This, however, is not a
fundamental problem. It can be traced to a difference in the
definition of parton densities and NLO
splitting functions $\delta P^{(1)}_{ij}$ in the various schemes.
We shall come back to this point at the end of Section 4.3 and 
in Section 5. 

It should be mentioned that the straightforward $\overline{MS}$ result
for $\delta C_g$ in (\ref{424}), which gives rise to a vanishing
'first moment'
\begin{equation}
\Delta C_g\equiv\int\limits^1_0 dz\delta C_g(z)
=\ha[-1+1]=0\/,
\label{427}
\end{equation}
has been a matter of dispute during the past years 
\cite{altarelli2,efremov,carlitz,fritzsch,efremov1,altarelli3,jaffe,
berger1,veneziano, 
altarelli4,altarelli6,altarelli7,bodwin,fritzsch1,fritzsch2,
manohar,ellwanger,shore,
bass1,cheng,reya1,reya2,ioffe}.
This vanishing is caused by the second term $+2(1-z)$ in
square brackets of $\delta C_g(z)$ in (\ref{424}) which derives from 
the soft non-perturbative collinear region where 
$k^2_T\sim m^2_q<<\Lambda^2$
\cite{mankiewicz,vogelsang1,mankiewicz1};
therefore it appears to be reasonable that this term
should be absorbed, besides the $\delta P_{qg}^{(0)}$
piece in (\ref{423}), into the definition of the 
light (non-perturbative) quark distribution 
$\delta q(x,Q^2=Q^2_0)$ in (\ref{426}). 
This implies that, instead of $\delta C_g(z)$ in 
({\ref{424}), we have
\begin{equation}
\delta\tilde C_g(z)=\ha (2z-1)(\ln\frac{1-z}{z}-1)
\label{428}
\end{equation}
which has a {\sl non}-vanishing first moment $\Delta \tilde C_g=-\ha$,
in contrast to (\ref{427}). 
It should be mentioned that a first moment of $-\ha$ for the 
Wilson coefficient of the gluon can be obtained {\it without any 
subtractions} in the so called off--shell scheme, to be discussed 
at the end of Section 4.3 and in Section 5. 
This scheme therefore {\it directly} reproduces the ABJ anomaly 
contribution to the first moment of $g_1$. For more 
details see Section 5. However, from a technical point of view 
the off--shell scheme is 
somewhat impractical, because 2-loop splitting functions are very 
difficult to calculate in this scheme, and  
only the 
$\overline{MS}$ results (\ref{424q}) and (\ref{424}) comply 
with the NLO ($\overline{MS}$) result 
\cite{mertig,vogelsang2}
for the 2-loop splitting functions 
$\delta P^{(1)}_{ij} (x)$ to be discussed next.

Within the $\overline{MS}$ factorization scheme the NLO 
contributions to $g_1(x,Q^2)$ are finally given by
\be 
g_1(x,Q^2)= \ha\sum\limits_q e^2_q\bigl\{\delta q(x,Q^2)+\delta \bar q(x,Q^2)
+\frac{\a_s(Q^2)}{2\pi}[\delta C_q\otimes (\delta q+\delta\bar q)+
2\delta C_g\otimes\delta g]\bigr\} \/.
\label{429}
\ee
The NLO parton densities $\delta f(x,Q^2), f=q,\bar q, g$, 
evolve according to the NLO evolution equations where the 
(scheme dependent) 2-loop splitting functions $\delta P^{(1)}_{ij}(x)$ 
enter and to which we now turn.

\bigskip 
\bigskip 
\begin{figure}
\begin{center}
\epsfig{file=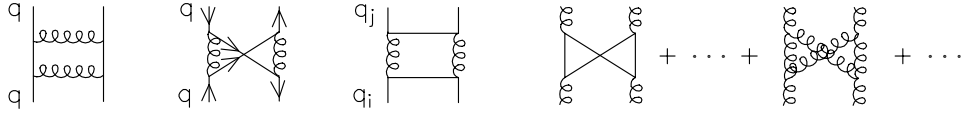,height=12cm,angle=270}
\bigskip  
\caption{Diagrams relevant for 2-loop splitting functions}
\label{fig16}
\end{center}
\end{figure}     

In NLO the evolution equations (\ref{4112}) and (\ref{4119}) have 
to be generalized since, in contrast to the LO in $\alpha_s$, the 
$O(\a_s^2)$ 2-loop splitting functions $\delta P^{(1)}_{ij}$ 
allow for transitions between quarks and antiquarks and among the 
different quark flavors as illustrated in Fig. \ref{fig16}. The situation 
is completely analogous to the unpolarized, i.e.\ spin-averaged 
case \cite{altarelli1,curci,floratos,furmanski}. 
The NLO flavor non-singlet renormalization group equations
read
\begin{equation}
\frac{d}{dt}\delta q_{NS\pm}(x,Q^2)=\delta P_{NS\pm}
\otimes \delta q_{NS\pm}
\label{4210}
\end{equation}
where
\be  
\delta P_{NS\pm}= \frac{\a_s(Q^2)}{2\pi}
\delta P^{(0)}_{qq}(x) 
+ \left(\frac{\a_s(Q^2)}{2\pi}\right)^2
\delta P^{(1)}_{NS\pm} (x)
\label{4211}
\ee
and
\begin{equation}
\frac{\a_s(Q^2)}{4\pi}\simeq\frac{1}{\beta_0\ln Q^2/\Lambda^2}
-\frac{\b_1}{\b_0^3}\frac{\ln\ln Q^2/\Lambda^2}{(\ln Q^2/\Lambda^2)^2}
\label{4212}
\end{equation}
with $\b_1=102-{38\over 3} f$ (e.g. $\beta_0=9$, $\b_1=64$ 
for the $f=3$ light active flavors). 
There are two different, independent NS evolution equations 
in NLO because of the additional transitions between different,
non-diagonal flavors ($u\to d$, $u\to\bar s$, etc.) and 
$q\bar q$-mixing $(u\to\bar u$, etc.) which start at 
2-loop ($\a^2_s$) order as illustrated in Fig. \ref{fig16}. 
Thus, opposite to the situation of unpolarized (spin-averaged) parton
distributions 
\cite{curci,glueck}, 
$\d q_{NS+}$ corresponds to the NS combinations $\d u -\d \bar u \equiv 
\d u_v$ and $\d d -\d \bar d \equiv \d d_v$, while 
$\d q_{NS-}$ corresponds to the 
combinations $\d q+\d\bar q$ appearing in the $NS$ expressions
$(\d u+\d\bar u)-(\d d+\d\bar d)$ and $(\d u+\d\bar u)+
(\d d+\d\bar d)-2(\d s+\d\bar s)$. The required 2-loop
splitting functions $\d P^{(1)}_{NS\pm}(x)$ 
in (\ref{4211}) are the same \cite{kodaira1,zijlstra}
as the unpolarized ones 
\footnote{More
explicitly, $P^{(1)}_{NS\pm}(x)$ is given, in the notation of 
\cite{curci}
by $P^{(1)}_{NS\pm}=P^{(1)}_{(\pm)}=P^{(1)}_{qq}\pm P^{(1)}_{q\bar q}$
according to Eqs.~(4.8), (4.35) and (4.50)-(4.55) of 
\cite{curci}, which are summarized in the Appendix. 
\label{fn2}}
%++++++++++ ende der fussnote +++++++++++++++++++
, $\d P^{(1)}_{NS\pm}=P^{(1)}_{NS\pm}$
in the (chirality conserving) $\overline{MS}$ regularization
scheme \cite{curci,floratos} --
similarly to the LO
splitting function in Eq.~(\ref{4116}). The NLO flavor-{\it singlet}
$Q^2$-evolution equations are similar
to the LO ones in (\ref{4119}):
\begin{equation}
\frac{d}{dt} {\d\Sigma (x,Q^2)\choose \d g(x,Q^2)}=
\d\hat P\otimes {\d\Sigma\choose \d g}
\label{4213}
\end{equation}
with $\d\Sigma$ being defined in (\ref{419}), and
\begin{equation}
\d\hat P=\frac{\a_s(Q^2)}{2\pi}\d \hat P^{(0)} (x)
+\left(\frac{\a_s(Q^2)}{2\pi}\right)^2\d\hat P^{(1)}(x)
\label{4214}
\end{equation}
where the LO $2\times 2$ matrix $\d\hat P^{(0)}$ of singlet
splitting functions is as in Eq.~(\ref{4119}) and the NLO 2-loop
singlet matrix is given by
\begin{equation}
\d\hat P^{(1)}(x)=
\left(\begin{array}{cc}\d P^{(1)}_{qq} &2f\d P^{(1)}_{qg}\\
                       \d P^{(1)}_{gq} &\d P^{(1)}_{gg}\end{array}
\right) \/.
\label{4215}
\end{equation}
The calculation of all these NLO singlet splitting functions
$\d P^{(1)}_{ij}(x), \quad i,j=q,g$, has been recently completed in the
$\overline{MS}$ factorization scheme 
\cite{mertig,vogelsang2}  
\footnote{These
are given by Eqs.~(3.65)-(3.68) of 
\cite{mertig} where
$\d P^{(1)}_{qq}=\d P^{(1)}_{NS-}+\d P^{(1)}_{PS,qq}$ with
$\d P^{(1)}_{PS,qq}$ being given by Eq.~(3.65) of 
\cite{mertig}. Note, however, that the $\d P^{(1)}_{ij}$ have been 
defined relative to $({\alpha_s \over 4\pi})^2$ in \cite{mertig},
cf. footnote \ref{fn1}, and that the factor 2f in (\ref{4215}) has been
absorbed into the definition of $\d P^{(1)}_{qg}$ in 
\cite{mertig}.
\label{fn3}}
%+++++++++++ ende der fussnote +++++++++++++               
and is summarized in the Appendix. 
As in the LO case,the independent $NS$ evolution equations   
(\ref{4210}) and the coupled singlet equations (\ref{4213}) can
be solved numerically by iteration directly in Bjorken-$x$ 
space in order to calculate $g_1(x,Q^2)$ in (\ref{429})    
to NLO. It will, however, be more convenient and physically
more transparent to work in Mellin $n$-moment space where the
evolution equations can be solved analytically.            

In moment space the NLO evolution equation (\ref{4210}) and
(\ref{4213}) are simply given by
\be  
\frac{d}{dt}\d q^n_{\nspm}(Q^2)= \Big[\aspi\d P^{(0)n}_{NS}+
\left(\aspi\right)^2
\d P^{(1)n}_{\nspm}\Big]
\d q^n_{\nspm}(Q^2)
\label{4216}
\ee
\be 
\frac{d}{dt}{\d\Sigma^n(Q^2)\choose\d g^n(Q^2)}= 
\Big[\aspi \d\hPon
+\left(\aspi\right)^2
\d\hPen\Big]{\d\Sigma^n(Q^2) \choose \d g^n(Q^2)}
\label{4217}
\ee
with the moments of the LO splitting functions given in
(\ref{4126}) and the 
moments of the NLO $(\overline{MS})$ flavor-NS
splitting function $\d\Pen_{\nspm}$ are again equal to the
unpolarized ones
\footnote{In the notation of 
\cite{floratos},
the 'anomalous dimensions' (see footnote \ref{fn1}) are given by
$\d P^{(1)n}_{\nspm} =-\gen_{NS}(\eta=\pm 1)/8$ with
$\gen_{NS}(\eta)$ given by Eq. (B.18) of 
\cite{floratos} which are summarized in the Appendix.
\label{fn4}}
%+++++++++++ ende der fussnote ++++++++++++++
. The moments of the NLO ($\overline{MS}$) flavor-singlet splitting
functions $\d\Pen_{ij}$ appearing in $\d\hPen$, as        
defined in (\ref{4214}), have also been presented in 
\cite{mertig};
however, in a form which is not adequate for analytic continuation
in $n$ as required for a (numerical) Mellin inversion to Bjorken-$x$
space. The appropriate NLO ($\overline{MS}$) anomalous dimensions can
be found in 
\cite{glueck2} 
\footnote{In the notation of
\cite{mertig,glueck2}, $\d\Pen_{ij}=-\d\gen_{ij}/8$
with $\d\gen_{ij}$ being given by Eqs.~(A.2)-(A.6) of
\cite{glueck2}.
Moreover, $2f\d\Pen_{qg}=-\d\gen_{qg}/8$ since the factor $2f$ has been
absorbed into the definition of $\d\gen_{qg}$ in
\cite{mertig,glueck2}. \label{fn5}}
%++++++++++++ende der fussnote ++++++++++++
and are summarized in the Appendix. 
The more complicated matrix equation (\ref{4217}) can
be easily solved in a compact form by introducing
\cite{furmanski}
an evolution matrix (obvious $n$-dependencies are suppressed)
\begin{equation}
\hat E(Q^2)=\left(1+\frac{\a_s(Q^2)}{2\pi}\hat U\right)
L^{-\zweib \d\hPon}\/,
\label{4218}
\end{equation}
defined as the solution of the equation [cf. (\ref{4213})]
\begin{equation}
\frac{d}{dt}\hat E=\d\hat P^n\hat E
\label{4219}
\end{equation}
with $L(Q^2)\equiv{\a_s(Q^2) \over \a_s\Q2}$, the NLO $\a_s$ being
given in (\ref{4212}), and where $\hat U$ accounts for the
2-loop contributions as an extension of the LO expression
(\ref{4128}). It satisfies \cite{furmanski}
\begin{equation}
[\hat U,\d \hat P^{(0)n}]=\frac{\b_0}{2}\hat U+\hat R
\label{4220}
\end{equation}
with $\hat R\equiv\d\hPen -{\b_1\over 2\b_0}\d\hPon$, which yields
\be 
\hat U=-\zweib (\hat P_+\hat R\hat P_+ +\hat P_- \hat R\hat P_-)
+\frac{\hat P_-\hat R\hat P_+}{\lambda^n_+-\lambda^n_- -\ha\b_0}+
\frac{\hat P_+\hat R\hat P_-}{\lambda^n_- -\lambda^n_+ -\ha\b_0}
\label{4221}
\ee
where the projectors $\hat P_\pm$ and $\lambda^n_\pm$ are given in
(\ref{4129}) and (\ref{4130}). The singlet solutions are then
given by
\be 
{\d\Sigma^n(Q^2)\choose\d g^n(Q^2)}=\Biggl\{
L^{-\frac{2}{\b_0}\d\hPon}
+\frac{\a_s(Q^2)}{2\pi}\hat U
L^{-\zweib\d\hPon}
-\frac{\a_s\Q2}{2\pi} L^{-\zweib\d\hPon}
\hat U\Biggr\}
{\d\Sigma^n\Q2\choose\d g^n\Q2}+O(\as2)
\label{4222}
\ee
where the remaining matrix expressions can be explicitly
calculated using Eq.~(\ref{4131}).

For the flavor nonsinglet evolution equations (\ref{4216}), which
do not involve any matrices, Eq.~(\ref{4220}) simply reduces to
$U_{NS}=-{2 \over \b_0}R_{NS}$ and (\ref{4222}) obviously reduces to
\be 
\d q^n_{NS\pm}(Q^2)= \Biggl\{1+
\frac{\a_s(Q^2) -\a_s\Q2}{2\pi}
(-\zweib)
\Bigl(\d\Pen_{\nspm}-\frac{\b_1}{2\b_0}\d\Pon_{qq}\Bigr)\Biggr\}
L^{-\zweib\d\Pon_{qq}}\d q^n_{\nspm}\Q2 +O(\as2) \/.
\label{4223}
\ee

These analytic solutions (\ref{4222}) and (\ref{4223}) in $n$-moment
space can now be inverted to Bjorken$-x$ space by numerically
performing the integral in (\ref{4136}). The final predictions
for $g_1(x,Q^2)$ can then be calculated according to (\ref{429}).
Alternatively, one can use directly the moment solutions
(\ref{4222}) and (\ref{4223})
and insert them into the n-th moment of (\ref{429}),
\be  
g^n_1(Q^2)= \ha\sum\limits_q e^2_q\bigl\{(1+\aspi\d C^n_q)
[\d q^n(Q^2)+\d\bar q^n(Q^2)]
+\aspi 2\d C^n_g\d g^n (Q^2)\bigr\}
\label{4224}
\ee
where \cite{kodaira2}
\be  
\d C^n_q= \frac{4}{3}\biggl[-S_2(n)+(S_1(n))^2
+\left(\frac{3}{2}-
\frac{1}{n(n+1)}\right)
S_1(n)
+\frac{1}{n^2}
+\frac{1}{2n}+\frac{1}{n+1}-\frac{9}{2} \biggr]
\label{4225}
\ee
is the $n$-th moment of $\delta C_q(z)$ in Eq.~(\ref{424q})
with $S_1(n)$ given after
(\ref{4126}) and  
$S_2(n)\equiv\sum^n_{j=1}\frac{1}{j^2}={\pi^2 \over 6}-\psi '(n+1)$
where $\psi '(n)={d^2\ln\Gamma (n) \over dn^2}$. The $n$-th moment of
$\delta C_g(z)$ in Eq.~(\ref{424}) is
\begin{equation}
\delta C^n_g=\ha\biggl[-\frac{n-1}{n(n+1)}(S_1(n)+1)
-\frac{1}{n^2}+
\frac{2}{n(n+1)}\biggr]
\label{4226}
\end{equation}
%\begin{multline}
%\delta C^n_g=\ha\biggl[-\frac{n-1}{n(n+1)}(S_1(n)+1)
%\\
%-\frac{1}{n^2}+
%\frac{2}{n(n+1)}\biggr]
%\label{4226}
%\end{multline}
which gives rise to the vanishing first $(n=1)$ moment, 
$\delta C^1_g=0$. On the other hand, in the factorization 
scheme of Eq. (\ref{428}) 
we have
\begin{equation}
\delta\tilde C_g^n=\delta C^n_g -\ha \frac{2}{n(n+1)}
\label{4227}
\end{equation}
i.e.\ $\delta\tilde C^1_g=-\ha$. With these expressions at hand
one can now obtain $g_1(x,Q^2)$ from Eq.~(\ref{4224}) by performing
a {\it single} numerical integration according to (\ref{4136}).
It should be remembered that a theoretically consistent NLO
analysis can, for the time being, performed only within the $\overline{MS}$
factorization scheme where all $\delta P^{(1)}_{ij}$ are known
\cite{mertig,vogelsang2}, 
using $\delta C_{q,g}$ of Eqs.~(\ref{4225}) and
(\ref{4226}). This is particularly relevant for the parton
distributions which have to satisfy the fundamental positivity
constraints (\ref{4111}) at any value of $x$ and scale $Q^2$, as
calculated by the unpolarized and polarized evolution equations,
within the {\it same} factorization scheme.

From the above solutions (\ref{4222}-\ref{4224}) it becomes
apparent that, as usual, a NLO analysis requires 2-loop O$(\as2)$
splitting functions $\delta P^{(1)}_{ij}$ and 1-loop O($\a_s$)
Wilson coefficient, i.e.\ partonic cross sections, $\delta C_i$.
Furthermore, in any realistic analysis beyond the LO, the
coefficient and splitting functions are not uniquely
defined to the extent that it is a mere matter of a theorist's
convention of how much of the NLO corrections are attributed
to $\delta C_i$ and how much to $\delta P^{(1)}_{ij}$ [see,
for example, the discussion which led to Eq.~(\ref{428})].
This is usually referred to as 'renormalization/factorization
scheme convention'. What is, however, important is that, to a given
perturbative order in $\a_s$, any physically directly measurable
quantity must be {\it in}dependent of the convention chosen
('scheme independence') and that the convention dependent terms appear
only beyond this order in $\alpha_s$ considered which are perturbatively
(hopefully) small. The requirements of convention independence of
our NLO analysis can be easily derived 
\cite{furmanski,glueck3}
from Eqs.~(\ref{4222}-\ref{4224}): Choosing a different
factorization scheme in the NS sector $(\delta C_{NS}=\delta C_q)$
according to
\begin{equation}
\delta C^n_{NS}\to\d C^{'n}_{NS}=\d C^n_{NS}+\Delta_{NS}^n\/,
\label{4228}
\end{equation}
this change has to be compensated, to $O(\as2)$, by an
appropriate change of $\d P^{(1)}_{NS}$,
\begin{equation}
\d\Pen_{NS}\to\d P^{(1)'n}_{NS}
=\d\Pen_{NS}
+\frac{\b_0}{2}\Delta_{NS}^n \/.
\label{4229}
\end{equation}
Similarly in the singlet sector, where we have to deal with
$2\times 2$ matrices, a change of the factorization scheme
\begin{equation}
\delta\hat C^n\to\delta\hat C^{'n}=\delta\hat C^n+\hat\Delta^n
\label{4230}
\end{equation}
implies \cite{furmanski,glueck3} 
\begin{equation}
\d\hPen\to\d \hat P^{(1)'n}
=\d\hPen +\frac{\b_0}{2}
\hat\Delta^n-[\hat\Delta^n,\d \hPon]
\label{4231}
\end{equation}
in order to guarantee convention (scheme) independence to
order $\as2$. 
The
upper row of $\delta\hat C$ corresponds to our fermion (quark)
$\delta C_{qq}\equiv\d C_q$ and gluon $\delta C_{qg}\equiv \d C_g$
Wilson coefficients in an obvious notation. To keep the treatment
as symmetric as possible we introduced in addition hypothetical
Wilson coefficients $\d C_{gq}$ and $\d C_{gg}$ in the lower row
of $\d\hat C$ which do {\it not\/} directly contribute to deep
inelastic lepton-hadron scattering and indeed drop out from the
final results relevant for $g_1$. In general, the transformation
of splitting functions in (\ref{4231}) and of the gluon density is
{\it not} fixed by the change of the physical first row coefficient
functions alone since the lower row of $\hat\Delta^n$ remains
undetermined. This is partly in contrast to the unpolarized situation
\cite{furmanski,glueck3,glueck} where the energy-momentum
conservation constraint (for $n=2$) may be used together with the
assumption of its analyticity in $n$. 

From these results it is clear that a consistent,
i.e.\ factorization scheme independent NLO analysis of
$g_1(x,Q^2)$ in (\ref{4224}) requires the knowledge of
{\it all} polarized 2-loop splitting functions
$\d\Pen_{ij}$ [or $\d P^{(1)}_{ij}(x)$], besides the coefficient
functions $\d C^n_{q,g}$ [or $\d C_{q,g}(x)$]. It is clear
that, for the time being, such an analysis can be performed only
in the $\overline{MS}$ factorization scheme where all
$\d\Pen_{ij}$ are known 
\cite{mertig,vogelsang2} -- a situation very similar
to the unpolarized case.

One could of course choose to work within a different factorization
scheme, in particular one which leads to (\ref{428}) and (\ref{4227})
or any other specific scheme. In this case, however, one has for
consistency reasons to calculate {\it all} polarized NLO quantities
($\d C_i,\d\Pen_{ij}$, etc.), and not just their first $(n=1)$
moments, in these specific schemes {\it as well as} also NLO subprocesses
of purely hadronic reactions to which the NLO (polarized/unpolarized)
parton distributions are applied to. So far, complete NLO calculations
have only been performed in the $\overline{MS}$ scheme.

%% file: k43tex
\subsection{Operator Product Expansion for $g_1$}

The phenomena discussed so far in the QCD improved parton model 
can also be understood in the framework of the operator product 
expansion (OPE). In contrast to the QCD parton model, which is universally 
valid, the OPE is designed exclusively to the understanding of 
deep inelasic lepton nucleon scattering. One starts with an 
expansion of the Fourier transform of the
time ordered product of two currents, i.e., the virtual 
Compton amplitude 
\begin{equation}
T_{\mn}=i\int d^4xe^{iqx}T(J_\mu (x)J_\nu (0))
\label{431}
\end{equation}
near the light-cone $x^2 \sim \frac{1}{Q^2} \approx 0$,
i.e.\ in powers of ${1\over Q^2}$, and 
is led to a description of the moments of structure functions
in terms of anomalous dimensions, Wilson coefficients and 
matrix elements. The OPE has been 
used very successfully to derive results for
unpolarized scattering \cite{gross,gross1,georgi,politzer,reya},
and results for polarized scattering exist as well 
\cite{ahmed,ahmed1,sasaki,kodaira1,mertig}. 
The particular feature of {\it polarized} DIS is the
appearance of antisymmetric ($\mu \leftrightarrow \nu$) 
terms in the expansion of $T_{\mn}$.
We shall start with a discussion of the structure
function $g_1$ 
for the case of photon exchange. 
The analysis of
$g_2$ is complicated by the appearance of transverse effects
and will be discussed in Section 8.
We may then assume 
longitudinal polarisation $P_{\si} =MS_{\si}$ of the proton. 
One has
\be 
T_{\mn}\mbox{(antisymm.,em. current)}=i\ve_{\mn\lambda\si}
q^\lambda\sum\limits_{n=1,3,5,\ldots}({2\over Q^2})^n
q_{\mu_1}\ldots q_{\mu_{n-1}}
\sum\limits^{9}_{i=0}R_i^{\si\mu_1\ldots\mu_{n-1}}
E^n_i(Q^2/\mu^2,\alpha_s)  \label{432}
\ee
where 
\begin{equation} 
R^{\si\mu_1\ldots\mu_{n-1}}_i =i^{n-1}\bar\psi\g5
\gamma^{\{\si}D^{\mu_1}\ldots D^{\mu_{n-1}\}}
\lambda_i\psi  \label{433}
\end{equation}
with 
$ i=0,\ldots,8$;
$n=1,3,5,\ldots$ and
\begin{equation}
R^{\si\mu_1\ldots\mu_{n-1}}_{9} =i^{n-1}\ve^{\a\b\gamma\{\si}
G_{\b\gamma}D^{\mu_1}\ldots D^{\mu_{n-2}\}}G^{\mu_{n-1}}_\a
\label{434}
\end{equation}
with $n=3,5,7,\ldots$ 
are the relevant operators, $\lambda_i$ are the Gell-Mann matrices
($\lambda_0$= unit matrix) and  
$\{\}$ denotes symmetrization
on the indices $\si\mu_1\ldots\mu_{n-1}$. In order to obtain 
operators of definite twist and spin, the appropriate subtraction 
of trace--terms is always implied. 
$R_{1-8}$ are the nonsinglet and $R_0$ and
$R_{9}$ the singlet contributions.
For each n=3,5,7,... there are 8 NS operators and two 
singlet operators. For n=1 (the first moment case) there are 
8 NS operators $\bar\psi\g5\gamma_{\si}\lambda_{1-8}\psi$ 
but only one singlet operator $\bar\psi\gamma_5\gamma_{\si}\psi$,
the axial vector singlet current.This operator is of particular 
interest because it carries the triangle anomaly.Through 
this anomaly the gluon enters the scene , 
via $\d \tilde C_g$ in Eqs. (\ref{428}) and (\ref{4227})
even though formally 
no gluon operator $R_{9}$ exists for n=1. 
The case of the first moment will be discussed in great detail 
in Section 5. 

We shall denote the matrix elements of the operators $R_i$
by $M_i$ ($S_{\si}=P_{\si}/M$):                          
\begin{equation}
\langle PS\vert R_i^{\si\mu_1\ldots\mu_{n-1}}\vert PS\rangle =
-M^n_iS^{\{\si}P^{\mu_1}\ldots P^{\mu_{n-1}\}}\/.
\label{435}
\end{equation}
Furthermore, the imaginary part of
$\langle PS \vert T_{\mn} \vert PS \rangle$ 
is related to the hadronic tensor $W_{\mn}$ in Eq.~(\ref{211})  
which determines the cross section and contains the structure
functions \cite{christ,gross,gross1,georgi,politzer}.
Therefore, through the optical
theorem, $g_1$ can be related to the $E^n_i$ and
the matrix elements $M_i^n$:
\begin{equation}
\int\limits^1_0 dx x^{n-1}g_1(x,Q^2)=\phantom{-}\ha\sum\limits_i M_i^n 
E^n_i (Q^2/\mu^2,\alpha_s) 
\label{436}
\end{equation}
for $n=1,3,5,\ldots$. 
Notice that 
the knowledge of the moments for $n=1,3,5,\ldots$ together with 
the fact that $g_1$ is even in $x \leftrightarrow -x$ completely 
determines $g_1$ by analytic continuation. 
The functions $E^n_i$ , 
the "Wilson coefficents", have a $Q^2$--dependence 
which is determined by the anomalous dimensions $\delta\gamma^n$ 
defined in Section 4.2 namely 
\be
E^n_i (Q^2/\mu^2,\alpha_s) = \sum\limits_j E^n_j (1,\alpha_s(Q^2))
  T \exp
\biggl \lgroup -\int\limits^{\alpha_s(Q^2)}_{\alpha_s(\mu^2)}
\frac{\d \gamma^n (\a_s)}{2\beta (\a_s)}d\a_s\biggr \rgroup _{ij}
\label{437}
\ee
where $\mu$ is the renormalization point . 
T indicates "time" ordering, 
\be
T \exp \int\limits^b_a f(x)dx = 1+ \int\limits^b_a f(x)dx 
  + \int\limits^b_a  dx \int\limits^b_x dy f(x)f(y) +  \ldots
\label{438}
\ee

When one expands Eq. (\ref{437}) in powers of $\alpha_s$, it turns out 
that the two-loop $\beta$-function and anomalous dimensions
enter the first order correction.
This will be shown explicitly in Eq. (\ref{4311}). 
Those two-loop quantities
as well as the first order Wilson coefficients are 
in general scheme and convention dependent.By scheme 
dependence we mean a dependence on the regularization 
procedure as well as on the renormalization prescription.
But this dependence
cancels in the combination which enters Eq. (\ref{4313}) below,
i.e. it cancels 
in the prediction for the physical quantities.
This is complementary to what we said about different 
definitions of quark densities in Section 4.2 and 
is 
exactly analogous to what happens in unpolarized scattering 
\cite{furmanski,reya,glueck3}.

Comparing (\ref{436}) and (\ref{4224}) there is a one-to-one
correspondence between the OPE and the parton model description. 
The matrix elements $M_i^n$ correspond to the moments of suitable 
combinations of parton densities $\d q$,$\d \bar{q}$ and $\d g$.
For example, 
\begin{equation}
M_8^n =\frac{1}{18} \int\limits_0^1 dx x^{n-1} \left[\d u + \d \bar{u}
+ \d d + \d \bar{d} -2 ( \d s + \d \bar{s} )\right] \/.
\label{439}
\end{equation}
The anomalous dimensions of the OPE govern the $Q^2$-evolution 
of the parton densities and the Wilson coefficients arise 
as the 'constant terms' discussed in Eq. (\ref{423}).
This correspondence holds for general n. For n=1 there is the
peculiar situation that only one singlet operator exists 
whereas in the parton model there are two degrees of freedom, 
the first moment of $\d \Sigma $ and the 
first moment of $\d g$. Therefore the translation between the
two schemes is somewhat more subtle for n=1 than for general n 
and will be discussed in detail in Section 5.

For each $n=3,5,...$ the anomalous dimension matrix 
$\delta \gamma^n$ decomposes into
a $2\times 2$ singlet and a $8\times 8$ nonsinglet block.
Let us first discuss the nonsinglet part in some detail.
In the nonsinglet block the operators are multiplicatively 
renormalizable with effectively one anomalous dimension 
$\delta \gamma_{NS}^n$, proportional to 
the moment of the evolution kernel 
$\delta P_{NS}$ introduced in the last section (see footnote \ref{fn1}).
It can be expanded in powers of $\alpha_s$ 
\begin{equation}
\delta \gamma_{NS}^n=\frac{\a_s}{4\pi}
\delta \gamma_{NS}^{(0)n}+
\left({\a_s\over 4\pi}\right)^2\delta \gamma_{NS}^{(1)n} +
O(\a_s^3)
\label{4310}\end{equation}
where $\d \gamma_{NS}^{(0,1)n}$ are given by the moments
of the quantities $\d P_{qq}^{(0)}$ and $\d P_{NS-}^{(1)}$ 
introduced in Eq. (\ref{4211}).
They are identical to the analogous quantities in  
unpolarized scattering.
More precisely,one has 
$\d \gamma_{NS}^{(0)n} = -4 \d P_{qq}^{(0)n} $ and
$\d \gamma_{NS}^{(1)n} = -8 \d P_{NS-}^{(1)n} $
(see footnotes \ref{fn1} and \ref{fn4})
because of the use of $4 \pi$ instead of $2 \pi$ in the 
expansion parameter.
$\d P_{NS+}^{(1)}$ plays no role for $g_1$ but is only important
for $g_3$ and $g_4+g_5$ in (\ref{241}) (cf. Sect. 6.7). 
Since the OPE method 
provides only the values of the even or odd moments depending on 
the crossing parity of the particular structure functions, only 
specific NS splitting functions ($\d P_{NS-}^{(1)}$ {\it or} 
$\d P_{NS+}^{(1)}$) are allowed in the evolution kernel. This is 
in contrast to the more general parton model method discussed before, 
which makes no restriction on the value of n. In other words, 
the parton model formulae, besides reproducing the OPE results, 
provide also the analytic continuation of the OPE results to those 
values of n which are artificially forbidden in the OPE. 

Inserting Eq. (\ref{4310}) in (\ref{437}) one obtains
\be 
T \exp \biggl \lgroup  -\int\limits^{\alpha_s(Q^2)}_{\alpha_s(\mu^2)}
\frac{\d \gamma_{NS}^n (\a_s)}{2\beta (\a_s)} d\alpha_s \biggr \rgroup
=
\biggl \lgroup\frac{\alpha_s(Q^2)} {\alpha_s(\mu^2)} \biggr \rgroup
^{\delta \gamma_{NS}^{(0)n} / 2 \beta_0} 
\biggl \lgroup  1+ 
\frac{\a_s(Q^2)-\a_s(\mu^2)}{4\pi}\left(
\frac{\delta\gamma_{NS}^{(1)n}}{2\b_0}-
\frac{\b_1\delta\gamma_{NS}^{(0)n}}{2\b^2_0}\right)  
+O(\a_s^2) \biggr \rgroup
\label{4311}
\ee
where we have used
\begin{equation}
{\b(\a_s) \over \a_s}=-\b_0\frac{\a_s}{4\pi}-\b_1
\left(\frac{\a_s}{4\pi}\right)^2
\label{43be}
\end{equation}
with $\b_0=11-{2\over 3}f$ and $\b_1=102-{38\over 3}f$ 
as in the previous section.
This makes explicit that the two-loop anomalous dimensions
and $\beta$-function enter the NLO one-loop analysis of cross
sections, i.e. Wilson coefficicents. 

The Wilson coefficient has the form 
\begin{equation}
E^n(1,\alpha_s(Q^2))=1+
\frac{\a_s}{2\pi} \d C_{NS}^n+O(\alpha_s^2)
\label{4312}
\end{equation}
with $\d C_{NS}^n$ given by Eq. (\ref{4225})
in the $\overline{MS}$--scheme. 

One can combine the above expansions to obtain 
\begin{eqnarray} \nonumber  
\int\limits^1_0 dx x^{n-1}g_{1,NS}(x,Q^2)=
\left(\frac{\a_s(Q^2)}{\a_s(\mu^2)}\right)
^{\delta \gamma_{NS}^{(0)n} / 2 \beta_0} 
\Biggl\{ 1+\frac{\a_s(Q^2)}{2\pi}\d C_{NS}^n +
\frac{\a_s(Q^2)-\a_s(\mu^2)}{4\pi} 
\\   
\times 
\left(\frac{\delta\gamma^{(1)n}_{NS}}{2\b_0}
-\frac{\b_1\delta\gamma_{NS}^{(0)n}}{2\b_0^2}\right) \Biggr\} 
\ha \sum_i M_i^n(\mu^2)
\label{4313}
\end{eqnarray}
which should be compared with Eqs. (\ref{4223}) and (\ref{4224}).
This equation shows how the $O(\alpha_s)$ Wilson coefficient 
$\d C_{NS}^n$ and the anomalous dimensions combine with the 
matrix element to form the n-th moment of $g_1$. 
As discussed earlier, the scheme dependence of the 
quantities $\d C_{NS}^n$ and $\d \gamma_{NS}^{(1)n}$ must 
cancel in the combination in Eq. (\ref{4313}) because it gives a physical
observable. 

Similar features arise in the singlet sector and in higher 
orders although the formalism becomes more complicated.In the 
singlet sector the anomalous dimensions form a nondiagonal 
2x2 matrix 
and there are two coefficients,one for the quark type 
operator $R_0$ and the other one for the gluon operator 
$R_{9}$.To first order $\alpha_s$ the coefficient for the 
quark type operator is the same as for the nonsinglet 
operators, and is of the form 
\begin{equation}
E_q^n(1,\a_s(Q^2))=1+\frac{\a_s}{2\pi}\d C_{q}^{n}+O(\a_s^2)
\label{4314}
\end{equation}
with $\delta C^n_q$ given again by Eq. (\ref{4225}).
The coefficient for the gluon 
operator is of the form 
\begin{equation}
E_g^n(1,\alpha_s(Q^2))=\frac{\a_s}{2\pi} \d C_{g}^n+O(\a_s^2)
\label{4315}
\end{equation}
with $\d C_g^n$ given in the $\overline{MS}$ scheme by
Eq. (\ref{4226}).
The singlet anomalous dimension matrix has an expansion
\begin{equation}
\delta \gamma_{S}^n=\frac{\a_s}{4\pi}
\delta \gamma_{S}^{(0)n}+
\left(\frac{\a_s}{4\pi}\right)^2\delta \gamma_{S}^{(1)n} +
O(\a_s^3)
\label{4316}
\end{equation}
where $\delta \gamma_{S}^n$, $\delta \gamma_{S}^{(0)n}$
and $\delta \gamma_{S}^{(1)n}$ are $2\times 2$ matrices with
indices $ij=qq,qg,gq,gg$.
With these expansions one can now calculate the singlet part 
of $\int\limits^1_0 dx x^{n-1}g_1(x,Q^2)$, 
using the one-to-one correspondence between the parton model 
densities and the OPE matrix elements,
with the same methods 
as described after Eq. (\ref{4217}). The result is exactly
analogous to Eqs. (\ref{4222}) and (\ref{4224}). An alternative to
present this somewhat cumbersome matrix result is 
the following.
If one goes to a basis of the quark and gluon matrix elements, 
in which $\d \gamma _S^{(0)n}$ is a diagonal matrix, one can
calculate the exponential matrix 
\begin{equation}
T \exp \biggl \lgroup -\int\limits^{\alpha_s(Q^2)}_{\alpha_s(\mu^2)}
\frac{\d \gamma_S^n (\a_s)}{2\b (\a_s)} d\alpha_s \biggr \rgroup 
= 
\left( \begin{array}{ll}\d E_{++} &\d E_{+-}\\
         \d E_{-+} &\d E_{--}\end{array}\right)
\label{4317}
\end{equation}
with 
%\begin{multline}
%\d E_{\pm \pm}=
%\biggl \lgroup  { \alpha_s(Q^2) \over \alpha_s(\mu^2) } \biggr \rgroup
%^{\delta \gamma_{\pm}^{(0)n} / 2 \beta_0}
%\biggl \lgroup  1+
%\\
%{\delta \gamma_{\pm \pm}^{n(1)} - \beta_1 \delta \gamma_{\pm}^{n(0)}
%\over 8 \pi \beta_0} (\alpha_s(\mu^2) - \alpha_s(Q^2))
%+ ... \biggr \rgroup
%\label{4318}
%\end{multline}
%\begin{multline} 
%\d E_{\pm \mp}={\d \gamma_{\pm \mp}^{n(1)} \over \d \gamma_{\pm}^{n(0)}
%- \d \gamma_{\mp}^{n(0)} +\beta_0}
%\\
%\biggl \lgroup   
%\biggl \lgroup  { \alpha_s(\mu^2) \over \alpha_s(Q^2) } \biggr \rgroup
%^{\delta \gamma_{\pm}^{n(0)} / 2 \beta_0}
%{\alpha_s(\mu^2) \over 2 \pi} -
%\biggl \lgroup  { \alpha_s(\mu^2) \over \alpha_s(Q^2) } \biggr \rgroup
%^{\delta \gamma_{\mp}^{n(0)} / 2 \beta_0}
%{\alpha_s(Q^2) \over 2 \pi}
% \biggr \rgroup
%\label{4319}
%\end{multline}
\be 
\d E_{\pm \pm}= 
\left(\frac{\a_s(Q^2)}{\a_s(\mu^2)}\right)
^{\d \gamma_{\pm}^{(0)n}/2\beta_0}
\Biggl\{ 
1+\frac{\a_s(Q^2)-\a_s(\mu^2)}{4\pi}
\left(\frac{\d\gamma_{\pm\pm}^{(1)n}}{2\beta_0}-
\frac{\b_1\d\gamma_{\pm}^{(0)n}}{2\b_0^2}\right)  
\Biggr\} 
\label{4318}
\ee

\be 
\d E_{\pm\mp}= \frac{\d\gamma_{\pm \mp}^{(1)n}}
     {\d\gamma_{\pm}^{(0)n}-\d\gamma_{\mp}^{(0)n} -2\beta_0}
\Biggl\{
\frac{\alpha_s(\mu^2)}{4\pi}
\biggl \lgroup \frac{\a_s(Q^2)}{\a_s(\mu^2)} \biggr \rgroup
^{\delta\gamma_{\pm}^{(0)n}/2\beta_0}
- \frac{\a_s(Q^2)}{4\pi} 
\biggl \lgroup
\frac{\a_s(Q^2)}{\a_s(\mu^2)} \biggr \rgroup
^{\delta \gamma_{\mp}^{(0)n}/2\beta_0}
\Biggr\}
\label{4319}
\ee
where $\d \gamma_{\pm}^{(0)n}=-4\lambda_{\pm}^n$ 
are the eigenvalues of $\d \gamma _S^{(0)n}$ [cf. Eq. (\ref{4130})], 
and $\d \gamma _S^{(1)n}$ is assumed to be of the form 
\begin{equation}
\d \gamma _S^{(1)n}=
\left(\begin{array}{ll}
\d \gamma_{++}^{(1)n} &\d \gamma_{+-}^{(1)n}\\
         \d \gamma_{-+}^{(1)n} &\d \gamma_{--}^{(1)n}\end{array}
     \right)
\label{4320}
\end{equation}
in the new basis.

The first order Wilson coefficients 
and the two-loop anomalous dimensions are scheme dependent.
However, for consistent schemes the scheme dependence
 must be such that the predictions for the physical 
quantities, the moments of $g_1$ , are scheme independent.
This implies that the combinations of coefficients and 
 anomalous dimensions , which appear in the 
representation of $g_1$, are scheme independent as discussed
at the end of the previous section.
For example, from Eq. (\ref{4313}) we see that
$2\d C _{NS}^{n} + {\gamma_{NS}^{(1)n} \over 2\beta_0}$
must be a scheme independent combination. 
A similar scheme independent combination exists in the 
singlet sector.   
In the following we want to compare some results for 
$\d C_{NS,q,g}^{n}$ and $\d \gamma _{NS,S}^{(1)n}$
in various schemes.
Note that the first order anomalous dimensions $\d \gamma ^{(0)n}$
[given in Eq.~(\ref{4126})]  
as well as the $\beta$-function coefficients $\beta_0$
and $\beta_1$ 
are scheme independent.
In all schemes singular collinear pole contributions arise,  
which are to be factorized into the quark distributions.
The various schemes are: 
 
\begin{itemize}
\item The $\overline{MS}$ scheme in dimensional regularization
using the reading point method \cite{koerner}
for the treatment of $\gamma_5$ ($\gamma_5$ is appearing due to the projector
on the quark's helicity).  
In this scheme the factorized singular terms in the 
coefficients are of the form $\d \gamma^{(0)n} (-\frac{1}{\epsilon}
-\ln 4\pi+\gamma_E)$. The reading point method is the more
systematic generalization of the CFH \cite{chanowitz}
$\gamma_5$-presription, i.e.\ a totally anticommuting $\gamma_5$,
in which the $\gamma^{\ast} q$-vertex is defined 
to be the starting point, from which the Dirac trace is read.
From the calculational point of view this scheme is the most tractable 
one, because no extra mass parameters or counter terms have to be
introduced. Indeed, it is this scheme, for which all 
coefficients $\d C^n_i$ are known and all anomalous dimensions
$\d \gamma^{(1)n}_{ij}$ have been recently calculated for the first
time \cite{mertig}.
The coefficients are given in Eqs.\ (\ref{4225}) and (\ref{4226}) and the
2-loop anomalous dimensions are given in 
\cite{mertig,glueck2}
as described in footnotes \ref{fn3} and \ref{fn5}. 
It is interesting to note, that
for the first moment $(n=1)$
\begin{equation}
\d C_{NS}^1=\d C_q^{1} = -{3 \over 2}C_F =-2, 
%\qquad 
\label{4321a}  
\end{equation}
\begin{equation}
\d C^1_g=0
\label{4321b}
\end{equation}
and 
\begin{equation}
\d \gamma_{NS-}^{(1)1} = 0\/,
%\quad
\label{4322a}  
\end{equation}
\begin{equation}
\d\gamma_{qq}^{(1)1}=24 C_F {f\over 2}
\label{4322b}
\end{equation}
where $\d \gamma^{(1)}_{NS-}\equiv\gamma^{(1)}_{NS} (\eta=-1)$  
(see footnote \ref{fn4}).
The relation (\ref{4321a}) implies that
the correction to the Bjorken sum rule is saturated by 
$ \d C_{NS}^{1}$, i.e. $\d \gamma_{NS-}^{(1)1}=0$ in Eq. (\ref{4313}) 
due to (\ref{4322a}).
The relation (\ref{4322b}) implies that
in this scheme 
the quark contribution to the first moment of $g_1$ is
not conserved ($Q^2$-dependent). It will be shown 
in Section 5 that this implies that in this scheme there is 
no gluon contribution to the first moment of $g_1$, i.e.\
$\delta C^1_g=0$ as made explicit in Eq.~(\ref{4321b}).
The {\it same} results for $\delta C^n_i$ and 
$\delta\gamma_{ij}^{(1)n}$
are obtained \cite{mertig} 
if one adopts the HVBM \cite{hooft,breitenlohner} method for
treating the $\gamma_5$ matrix in $D \not= 4$ dimensions,
{\it provided} an additional renormalization constant (counter term)
\cite{larin} is introduced in order to guarantee the
conservation of the NS axial vector operators $R_{1-8}$ in
(\ref{433}). It is mandatory to keep these NS axial vector
currents conserved \cite{kodaira1} due to the absence of
gluon-initiated triangle $\gamma_5$-anomalies in the flavor
non-singlet sector which {\it dictates} the vanishing of
$\delta\gamma^{(1)1}_{NS-}$.

It should be noted, however, when using 'naively' the original HVBM
prescription for the treatment of
$\gamma_5$ one obtains 
\cite{weber,vogelsang} a result for
$\d C^n_{NS}=\d C^n_q$ which implies for the first moment
$\delta C^1_q=-{7\over 2}C_F$ which is different than the one in
Eq.~(\ref{4321a}). This corresponds, however, to a non-zero
value for $\delta\gamma^{(1)1}_{NS-}$ [on account of
the scheme invariance of 
$2\d C^1_{NS}+{\delta\gamma_{NS-}^{(1)1}\over 2\b_0}=-3 C_F$,
using Eqs. (\ref{4321a}) and (\ref{4322a})] in contradiction to the
conservation of the NS axial vector current.

\item 
The regularization with (on--shell) massless quarks and off--shell 
gluons.
From the calculational point of view this scheme is 
difficult, because the gluon off--shellness is difficult to handle
in two-loops, and consequently the $\d \gamma^{(1)n}_{ij}$ are not
known for general $n$. However, this scheme is rather meaningful 
physically for polarized DIS, because it corresponds best to the
notion of constituent quarks 
\cite{altarelli3,altarelli6,altarelli7,reya2}
as will be discussed in detail
in Section 5. Furthermore, it avoids the fundamental difficulties 
with $\gamma_5$ present in dimensional regularization. The 
Wilson coefficients in this scheme are given by 
\cite{chiappetta,kodaira}
\be 
\delta C^n_q=C_F\Big[-\frac{9}{4}-\frac{3}{2n}+\frac{3}{n+1}+\frac{2}{n^2}
-\frac{1}{(n+1)^2} +
(\frac{3}{2}-\frac{1}{n(n+1)}) S_1(n) 
+(S_1(n))^2-3S_2(n)\Big]   
\label{4323}                      
\ee                    
\begin{equation} 
\d C^n_g=2 T_R\Big[\frac{1}{n}-\frac{2}{n+1}-\frac{1}{n^2}+
\frac{2}{(n+1)^2}\Big]
\label{4324}
\end{equation}
where, as previously, $S_k(n)\equiv\sum\limits^n_{j=1}\frac{1}{j^k}$.
In particular one has
\begin{equation}
\delta C^1_{NS}=\delta C^1_q=-2\/,
%\quad 
\label{4325a}  
\end{equation}
\begin{equation}
\d C^1_g=-\ha
\label{4325b}
\end{equation}
and \cite{altarelli6}
\begin{equation}
\d\gamma^{(1)1}_{NS-}=\delta\gamma_{qq}^{(1)1}=0\/.
\label{4326}
\end{equation}
These results correspond to a nonvanishing gluon  
contribution to the first moment of $g_1$ and to 
conserved, i.e.\ $Q^2$-independent, first moments of the polarized 
quark densities. 
 
A theoretically consistent NLO analysis of polarized
structure functions and parton distributions cannot be performed
for the time being, apart from their first $(n=1)$ moments since it would
require the knowledge of all $\d\gamma_{ij}^{(1)n}$, or equivalently of
$\delta\gamma_{ij}(x)$, in this particular regularization/factorization
scheme.

%Based on the suggestion in \cite{glueck3} for unpolarized 
%scattering there may be the possibility of a transformation
%of the $\d \gamma_{ij}^{(1)n}$ from the $\overline{MS}$ scheme 
%as given by \cite{mertig} 
%to the off--shell scheme. Namely,one should use the     
%transformation formula Eq. (\ref{4230}) for the coefficients 
%and try to calculate the matrix
%$\hat \Delta $ appearing in that equation. $\hat \Delta $ contains,
%in addition to the ordinary Wilson coefficients 
%$\delta C_q(\overline{MS})-\delta C_q(off)$ and 
%$\delta C_g(\overline{MS})-\delta C_g(off)$ 
%two hypothetical coefficients 
%$\delta C_{gq}(\overline{MS})-\delta C_{gq}(off)$ and
%$\delta C_{gg}(\overline{MS})-\delta C_{gg}(off)$.                                
%In the order of $\alpha_s$, in which we are working,
%these do not get contributions from DIS,because the
%gluon does not couple directly to the photon. There is
%only a contribution from gluon operator matrixelements (cf. Figs. 
%\ref{figopegq} and \ref{figopegg}). 
%To obtain     
%$\hat \Delta_{gq}$ and $\hat \Delta_{gg}$, 
%these one-loop matrixelements  
%have to be calculated both in the $\overline{MS}$ scheme and in  
%the offshell scheme. 
%Once the matrix $\hat \Delta$ is known, it can be put into 
%Eq. (\ref{4231}) to obtain the $\d \hat P^{(1)n}(off)$ from 
%the known $\d \hat P^{(1)n}(\overline{MS})$.  
%We are presently checking the possibility  
%whether this is a tractable approach.   

Just as the $\overline{MS}$ scheme 
the off--shell scheme naturally fulfills 
$\delta C_q^1=-2$. This implies that the coefficient alone 
saturates the 
Bjorken sum rule and 
reproduces the standard correction factor $1-{\alpha_s \over \pi}$ 
for the first moment of the singlet contribution. 

\item 
The regularization with massive quarks. 
Some of the collinear singularities in the coefficient functions 
can be regularized by introducing a quark mass. 
The remaining singularities are again regularized by a nonzero 
offshellness/mass of the gluon. 
For example,for the Wilson coefficient of the quark field 
one obtains to one-loop order 
\cite{stratmann}
\be  
\delta C_q^n=C_F\Big[-\frac{5}{2}-\frac{5}{2n}+
\frac{2}{n+1}+\frac{1}{n^2}
-\frac{2}{(n+1)^2} 
+(\frac{7}{2}+\frac{1}{n(n+1)}) S_1(n)-3 S_2(n)
-(S_1(n))^2   \Big]
\label{4327}
\ee
For the first moment this yields $\delta C_q^1=-\frac{7}{2}C_F$, 
i.e. not directly the expected correction to the Bjorken sum 
rule (cf. Section 5 for more explanations on the scheme 
dependence of the first moment). 

Although
this scheme is not very tractable in a two-loop calculation,
and for the light quarks is usually not considered very 
physical, its results are sometimes interesting for comparative 
reasons. 
For example, in \cite{altarelli6} it was shown that the difference 
$\delta \gamma_{qq}^{(1)1}$(massive quarks)--
$\delta \gamma_{qq}^{(1)1}$(massless quarks) is of such a form that 
it cancels the corresponding change in $\delta C^1_g$ so that 
a scheme independent result arises 
(for more details see Section 5). 

In the case of heavy quarks ($m_q^2 \gg \Lambda_{QCD}^2$)
a calculation with quark masses has to be done, but then the 
quark mass is not just a regulator ($O(m)$-terms to be neglected), 
but has a real physical meaning \cite{watson,glueck8}. 
The heavy quark contribution to PDIS will 
be discussed in detail in Section 6. 

\item 
Keeping the
incoming parton (quark or gluon) off--shell is
another possibility to renormalize; 
$\delta C_q$ is then given by \cite{kodaira1,stratmann}
\be 
\delta C_q^n=C_F\Big[-\frac{3}{2n}+\frac{2}{n+1}+\frac{2}{n^2}
-\frac{2}{(n+1)^2} 
+\frac{3}{2} S_1(n)-4 S_2(n)\Big]
\label{4328}
\ee
and $\delta C^n_g$ is as in Eq. (\ref{4324}). 
The first moments are again as in Eqs. (\ref{4325a}) and 
(\ref{4325b}), and the ones for the quark anomalous 
dimensions are given by Eq. (\ref{4326}).  
Again the anomalous
dimensions for arbitrary $n$ are not known.

\end{itemize}

%% file: k44tex
\subsection{The Behavior of $g_1(x,Q^2)$ at Small $x$}

The behavior of $g_1$ at small $x$ is an important issue because the
experimental determination of the first moment of $g_1(x,Q^2)$, as defined
in Eq. (\ref{312}), depends on it rather strongly. Therefore it has an
influence on tests of the fundamental Bjorken
sum rule \cite{bjorken} and of the Gourdin-Ellis-Jaffe
expectations for $\Gamma_1^{p,n}$, which will be discussed in detail
in Section 5, and also on the question how large the contribution
from the various parton species to the proton spin is, cf. Eq. (\ref{111}).
Due to the very different polarized and unpolarized splitting functions
(except for $\d\Pon_{qq}=\Pon_{qq}, \d\Pen_{\nspm}=\Pen_{\nspm})$
one expects different $Q^2$ evolutions of $g_1(x,Q^2)$ and $F_{1,2}(x,Q^2)$,
respectively, and thus different small--$x$ predictions for these two
cases especially in the medium-- to small--$x$ region $(x<0.2)$
dominated by polarized flavor-singlet contributions $\delta\Sigma$ and
$\d g$. According to our results in the two previous subsections, e.g.\
Eqs.~(\ref{4128}) and (\ref{4222}), these quantum field theoretic
renormalization group predictions depend on the input
densities $\d\Sigma (x,Q^2_0)$ and $\d g(x,Q^2_0)$ to be fixed mainly
by experiment. Unfortunately, the present polarization experiments
\cite{ashman,ashman1,adeva,adams,adams2,anthony,abe,abe1,abe2}  
with their scarce statistics constrain 
these
singlet input densities rather weakly, 
see, e.g.\ \cite{altarelli5,gehrmann,glueck4,glueck2}, 
in particular $\d g(x,Q^2_0)$ remains almost
entirely arbitrary, which is in constrast to unpolarized structure
functions (see, e.g., \cite{glueck5,glueck9,glueck6}). 
One therefore has, for
the time being, to rely on theoretical prejudices and guesses.

There is a suggestion from Regge theory for the small-x behavior 
of $g_1$. Under the assumption that there is no spin dependent 
diffractive scattering, one may assume that the 
$a_1(1260)$ trajectory dominates the Regge asymptotics 
\cite{brodsky,ellis1} and obtains  
\begin{equation}
g_1(x,Q_0^2) \tosim_{x\to 0} x^{-\alpha_{a_1}(0)}  , \qquad 
-0.5 \lesssim \alpha_{a_1}(0) \lesssim 0 
\label{441}
\end{equation}
where $\a_{a_1}(0)$ is the
intercept of the degenerate $a_1(1260)$, $f_1(1285)$ and
$f_1(1420)$ Regge trajectories. 
The scale $Q^2_0$ 
where this asymptotic expectation is supposed to hold is entirely
unrestricted by Regge arguments. 
A naive guess would be that
$g_1(x,Q^2_0)$ should behave more or less like a constant as $x\to 0$ 
($g_1\sim x^0$), to be compared with a fit \cite{brodsky,ellis1} 
to the EMC data \cite{ashman,ashman1}, which has been done 
for the region $x<0.2$ and yields $g^p_1(x,Q^2)\sim
x^{0.07^{+0.32}_{-0.42}}$. The main uncertainty lies of course in the
value of $Q^2=Q^2_0$ where this Regge behavior is implemented. If,
for example, we implement Eq. (\ref{441}) at a scale $Q^2_0\simeq 1$~GeV$^2$
then, according to the QCD evolution, $g_1^p(x,Q^2)$ as well as
$\delta q (x,Q^2)$, $\delta \bar q (x,Q^2)$ and $\delta g(x,Q^2)$
derived from it will be steeper as $x\to 0$ than in Eq.~(\ref{441})
at $Q^2 > 1~GeV^2$, e.g.\ at $Q^2\simeq 10$~GeV$^2$ relevant for
some recent experiments 
\cite{ashman,ashman1,adeva,adams,adams2}.
The only somewhat reliable conclusion we can draw from this is that
$g_1$ 
and thus $\d q_v$, $\d\bar q$ and $\d g$ 
will {\it not} diverge
at the same strength as $x\to 0$ as the unpolarized structure
functions $F_1$ and $F_2/x$ since $q_v\sim x^{-\a_{\rho}(0)}
\sim x^{-1/2}$
and $\bar q, g\sim x^{-\a_P(0)}\sim x^{-1}$. The 
divergence of the unpolarized 
structure functions for $x\to 0$ is 
driven by the $\frac{1}{x}$ singularity of the
vacuum (Pomeron) exchange which is {\it not} present 
in $\d q_v(x,Q^2)$, $\d\bar q(x,Q^2)$
and $\d g(x,Q^2)$.

Accepting a particular input behavior at $Q^2=Q^2_0$, the perturbative 
QCD prediction at $Q^2>Q^2_0$ for the singlet sector follows from
Eqs.~(\ref{4128}--\ref{4135b}) and (\ref{4222}). Even in the small--$x$
limit one has to resort to the full solutions of the evolution
equations since the 'leading pole' or 'asymptotic $\frac{1}{x}$'
approximation \cite{einhorn,berera,close1,ball} to the
polarized splitting functions is quantitatively (and partly even
qualitatively) {\it not} appropriate \cite{glueck4,gehrmann1}, at
least for $x$--values of experimental relevance, $x\gtrsim 10^{-3}$.
For qualitative purposes it is nevertheless instructive to recall
the leading pole (in $n$) or asymptotic $\frac{1}{x}$
approximation, presented here 
in LO for simplicity. In polarized scattering {\it all}
$\d P^{(0)}_{ij}$ in Eqs. (\ref{4121}) 
and (\ref{4126})    
are less singular as $x\to 0$ than in unpolarized scattering:
$\d \Pon_{ij}\sim \frac{1}{n}$ [$\d P^{(0)}_{ij}(x)\sim$ const.]
as compared to $\Pon_{gi}\sim\frac{1}{n-1}$ [$P^{(0)}_{gi}(x)\sim
\frac{1}{x}$] and $\Pon_{qi}\sim\frac{1}{n}$ 
[$P^{(0)}_{qi}(x)\sim$ const.]. This complicates the situation to
some extent because polarized quark and gluon densities contribute at the
same level. Thus the $2\times 2$ matrix of splitting functions in
Eq. (\ref{4125}), required for the LO solution (\ref{4128}), reduces to
\begin{equation}
\d \hPon
\tosim_{x\to 0}
\frac{1}{n}\left( \begin{array}{cc} C_F\/ \/ &-2fT_R\\
                               2C_F\/ \/&4C_A\end{array}\right)
\label{442}
\end{equation}
with $C_F={4\over 3}, T_R=\ha, C_A=3$. We first diagonalize this matrix 
and obtain for the eigenvalues [cf. (\ref{4130})]  
\begin{equation}
\lambda^n_\pm = \frac{1}{n}\bar\lambda_\pm   \,  ,
\qquad 
\bar\lambda_\pm =
{1 \over 2}
\left(\frac{40}{3}\pm\frac{32}{3}\sqrt{1-\frac{3f}{32}}\right)\/.
\label{443}
\end{equation}
For $f=3$ active flavors we have $\bar\lambda_+
=11.188\gg\bar\lambda_-=2.145$. This means that 
the second term in Eq. (\ref{4131}) which involves the $\lambda^n_-$
renormalization group exponent is subleading 
as $x \rightarrow 0$ and we obtain approximately
\be  
\left( \begin{array}{ll}\d\Sigma^n (Q^2)  \\ \d g^n(Q^2)\end{array}\right)
 \simeq  \frac{1}{\bar\lambda_+-\bar\lambda_-}
\left( \begin{array}{cc} C_F-\bar\lambda_-  &-2fT_R\\
       2C_F\/         &4C_A-\bar\lambda_-\end{array}\right)
\left(\begin{array}{ll}\d\Sigma^n(Q^2_0) \\ \d g^n(Q^2_0)\end{array}\right)
e^{\zweib\frac{\bar\lambda_+}{n}\xi}
\label{444}
\ee
where 
$\xi=\xi (Q^2)\equiv\ln L^{-1}=\ln[\alpha_s (Q_0^2)/ \alpha_s (Q^2)]$.
Approximately the same result can be obtained, if one uses just  
the dominant $\d P^{(0)}_{gg}$ contribution $4 C_A$ and neglects 
the remaining entries in (\ref{442}): In that case 
one has 
$\bar\lambda_+=4 C_A=12$ and 
$\bar\lambda_-=0$, i.e.\ a sufficiently accurate
approximation \cite{close1}.
Assuming the input densities to be flat in $x$ as $x\to 0$ [cf. 
Eq. (\ref{441})], $\d \Sigma (x,Q^2_0)\sim$ const. and
$\d g(x,Q^2_0)\sim$ const., i.e.\ $\d\Sigma^n(Q^2_0)\sim\frac{1}{n}$
and $\d g^n(Q^2_0)\sim\frac{1}{n}$, Eq.\ (\ref{444}) can be easily
Mellin-inverted, cf.\ (\ref{4136}), on account of 
\cite{glueck7,martin,degrand}
\be 
\d f(x,Q^2) = \frac{1}{2\pi i}\int\limits^{c+i\infty}_{c-i\infty} dn 
     x^{-n}\d f^n (Q^2)
\sim \frac{1}{2\pi i}\int\limits^{c+i\infty}_{c-i\infty} dn
x^{-n} {1 \over n} e^{\frac{a}{n}}=I_0(2\sqrt{a\ln\frac{1}{x}}).
\label{445}
\ee
Using the asymptotic expression for the modified Bessel function
$I_0(z)\sim {e^z \over \sqrt{2\pi z}}-O(\frac{1}{z})$ for large $z$,
we arrive at the 'double leading log' (DLL) formula 
\begin{equation}
\d \Sigma (x,Q^2),\d g(x,Q^2) 
\tosim_{x\to 0} 
  C_{\Sigma, g}
exp\left[ 2\sqrt{\zweib\bar\lambda_+\xi (Q^2)\ln\frac{1}{x}}\right]
\label{446}
\end{equation}
where all (partly unknown) constants are lumped into
$C_{\Sigma, g}$. This gives the dominant $x\to 0$ behavior of the singlet
component $g_{1,S}$ of $g_1=g_{1,NS}+g_{1,S}$ which
dominates $g_1$ because the nonsinglet part (\ref{4127}) leads, analogously
to the above derivation using $\d \Pon_{qq}\sim {C_F \over n}$, to
\begin{equation}
\d q_{NS}(x,Q^2)
\tosim_{x\to 0}
C_{NS} exp\left[ 2\sqrt{\zweib C_F
\xi (Q^2)\ln\frac{1}{x}}\right]
\label{447}
\end{equation}
which is again subleading since $\bar\lambda_+\gg C_F$. 
These results illustrate that for example a positive $\delta g$ 
input ($C_g > 0$) drives $\delta \Sigma (x,Q^2)$ negative as 
$x\rightarrow 0$ and $Q^2$ increases, due to the negative 
large matrix element $-2fT_R$ in (\ref{444}). Therefore 
$g_1^p$ is eventually driven negative as well for $x \lesssim 10^{-3}$ 
and $Q^2 > 1$ GeV$^2$ \cite{ball,altarelli10}. Thus, 
estimates of small--x contributions to the first moment 
$\Gamma_1(Q^2)$ of $g_1(x,Q^2)$ using Regge extrapolations 
alone will be unreliable, underestimating their size, 
for a positive $\delta g$,  
and sometimes even giving them the wrong sign 
\cite{altarelli10,ball}.

Equation (\ref{446})
may be compared with the asymptotic $\frac{1}{x}$ behavior of the
unpolarized structure function $F_{1,2}(x,Q^2)$: Assuming that the
gluon with $\Pon_{gg}\sim {2 C_A \over n-1}$ dominates the evolution of the
singlet quark combination $\Sigma = q+\bar q$, the quantity which couples
directly to $F_{1,2}$, one obtains the standard DLL result 
\cite{glueck7,martin,degrand}
\begin{equation}
xg(x,Q^2)
\tosim_{x\to 0} exp\Big[2\sqrt{\zweib 2C_A\xi (Q^2)\ln\frac{1}{x}}
\Big]
\label{448}
\end{equation}
which is in principle similar to the result in Eq. (\ref{446}) but has
an additional factor $\frac{1}{x}$ in front due to the dominant vacuum
(Pomeron) exchange allowed for spin-averaged structure functions as
discussed at the beginning of this subsection.

This relatively simple exercise demonstrates explicitly that $A_1(x,Q^2)$
in Eq. (\ref{227}) will {\it not} be independent of $Q^2$ in the
small--$x$ region, as commonly assumed 
\cite{papavassiliou,ashman,ashman1,adeva,adams,adams2,anthony,abe,abe1}
for extracting $g_1(x,Q^2)$ 
[because $\bar\lambda_+\simeq 4 C_A > 2C_A$
according to Eqs. (\ref{446}) and (\ref{448})], i.e.\
\be 
A_1(x,Q^2)  \simeq  \frac{g_1(x,Q^2)}{F_1(x,Q^2)}\simeq 2x
\frac{g_1(x,Q^2)}{F_2(x,Q^2)}\tosim_{x\to 0} 
x~exp\Bigl\{2(\sqrt{2}-1)\sqrt{\zweib 2C_A\xi(Q^2)\ln\frac{1}{x}}\Bigr\}.
\label{449}
\ee
Note that $A_1(x,Q^2)\sim const.$ as $x\to 1$ due to $\delta\Pon_{qq}=
\Pon_{qq}$ and $\d\Pen_{NS\pm}=\Pen_{NS\pm}$ which dominate in the
large--$x$ region. It should, however, be emphasized that the above
asymptotic results in (\ref{446}) and
(\ref{447}) are {\it not} even qualitatively sufficient for $x$
as low as $10^{-3}$ when compared with the exact LO/NLO
results for $A_1$ \cite{glueck4,gehrmann1,glueck2}. They might 
become relevant 
for $x\lesssim 10^{-4}$, depending on the input densities 
\cite{glueck4,gehrmann1}. For practical
purposes, however, the very small--$x$ region appears to be not very relevant
since, according to Eq.~(\ref{449}), $A_1(x,Q^2)\simeq {2x g_1 \over F_2} 
\to 0$ as
$x\to 0$ is already unmeasurably small (of the order $10^{-3}$) for
$x\lesssim 10^{-3}$. Thus the small--$x$ region is unlikely to be
accessible experimentally for $g_1(x,Q^2)$, in contrast to the situation
for the unpolarized $F_{1,2}(x,Q^2)$.

In addition, the evolution of parton distributions $f(x,Q^2)$
involve in general convolutions $P\otimes f$, see e.g.~Eqs.(\ref{4112}),
(\ref{4114}) and (\ref{4119}), which are sensitive to the shape of these
distributions in the large--$x$ region, even for 
$(P \otimes f)(x\to 0)$.
This can be easily envisaged by considering according to Eq.~(\ref{441}),
for example,
\begin{equation}
f(x)=x^{-\alpha}(1-x)^{a} 
=x^{-\alpha} \sum_{n=0}^a\left({a\atop n}\right) x^n
\end{equation}
which implies for the convolution, using $P(x)={1\over x^p}$
with $p\geq 0$,
\be 
(P \otimes f)(x\to 0)  \simeq  \frac{1}{x^p}\sum_n\left({a\atop n}\right)
\frac{1}{n+p-\alpha}-\frac{x^{-\a}}{p-\a}
+O(x^{p-\a})
\ee
Thus, because of the remaining $a$-dependence, the behavior of
$P \otimes f$ in the small--x limit is closely correlated with the
distributions in the {\it large}-$x$ region! A more detailed
quantitative analysis can be found in \cite{gehrmann1}. 
Therefore the mere knowledge of $P(x)$ and $f(x,Q^2)$ as $x\to 0$ is
insufficient to predict $g_1(x,Q^2)$ in the small--$x$ limit. It is
thus apparent that reliable results can
only be obtained by performing {\it full} LO and/or NLO
analyses as described in the previous sections 4.1 and 4.2 
\cite{glueck4,glueck2}. 
Anyway, the scale-violating $Q^2$-dependence
of $A_1(x,Q^2)$ is a general and specific feature of perturbative QCD
as soon as gluon and sea densities become relevant. This is due to the
very different polarized and unpolarized splitting functions
$\d P^{(0,1)}_{ij}$ and $P^{(0,1)}_{ij}$, respectively (except for
$x\to 1$ as discussed above, after Eq. (\ref{449})).

Finally, let us turn to the more speculative non-perturbative 
approach to the 
small--$x$ behavior of $g_1$. The result in Eq. (\ref{441}) is obtained
if there is no spin dependence in diffractive scattering. The
possibility of spin dependent diffractive scattering has been examined
in \cite{close1,bass} where it has been shown that a
logarithmic rise of $g_1$ at small $x$ could in principle be induced
\begin{equation}
g_1(x,Q^2)\tosim_{x\to 0} \ln\frac{1}{x}
\label{4410}
\end{equation}
with the scale $Q^2$ being entirely unspecified. An explicit calculation
\cite{bass}, being based on the exchange of two non-perturbative
gluons, has manifested this $\ln x$ behavior, i.e.\ $g_1\sim 2\ln
\frac{1}{x}-1$. The behavior in Eq.~(\ref{4410}) is by no means
compelling but it shows that there is a considerable amount of theoretical
uncertainty, at least as far as the non-perturbative input distributions
are concerned. A numerical analysis shows that this leads to an
uncertainty in the determination of the first moment, defined in
Eq.~(\ref{312}), typically about 10\% \cite{close1}. This can be seen
by taking the present experimental results with cut values $x\gtrsim 0.01$,
cf.\ Eq.\ (\ref{322}), and fitting Eqs.\ (\ref{441}) and (\ref{4410})
in the small--$x$ region. 

Even extreme double-logarithmic contributions
$g_1\sim\ln^2 {1\over x}$ 
are conceivable and, in fact, claimed to be present 
at $x\ll 1$ \cite{close1,ermolaev,bartels,bartels1}, 
which are not included in the usual RG evolution equation. 
More precisely, in a simple ladder approach   
one has for each (except for the first) power of 
$\alpha_s$ terms of the form $(\alpha_s \ln^2 {1\over x})^2$. 
Summing up to arbitrary order in $\alpha_s$ one obtains 
$g_1\sim ({1\over x})^{c_1\sqrt{\alpha_s}}$. This is to be 
contrasted to the unpolarized structure function $F_1$ for which 
one gets in this approach $F_1\sim ({1\over x})^{c_2\alpha_s}$. 
The latter result is obtained by resumming single logarithms of 
${1\over x}$ where double logarithms are not present in 
the unpolarized case. 
Numerically, the exponent $c_1\sqrt{\alpha_s}$ turns out to be larger than 1 
for the singlet contribution to $g_1$. 
Therefore in this approach the first moment of $g_1$ does 
not exist. This probably signals a breakdown of the  
approximation used 
and suggests that other, yet unknown, terms in addition to the 
logarithms of x have to be summed as well. 
 
A recent more detailed and consistent analysis 
\cite{bluemlein1,bluemlein5} 
has demonstrated, however, that such more singular terms are not present  
in the flavor nonsinglet contribution to $g_1$ because potentially 
large $\ln^n x$ contributions get cancelled by similar singular terms 
in the Wilson coefficients. Such a conclusion can not be reached 
for the singlet contribution to $g_1$ since less singular terms of 
the NNLO (3--loop) singlet splitting functions have not yet 
been calculated \cite{bluemlein2}.

%% file: k51tex
\setcounter{equation}{0}
\section{The First Moment of $g_1$}
\subsection{The First Moment and the Gluon Contribution}
Quantities of particular significance are the first moments of polarized
parton distributions $\delta f(x,Q^2)$,
%Gleichung 5.1
\begin{equation}
\Delta f(Q^2) \equiv \int^1_0 dx\, \delta f(x,Q^2)\,\,\, , \quad
  f=q, \bar{q}, g
\label{51}
\end{equation}
since they enter the fundamental spin relation Eq.~(\ref{111}). According to
Eqs.~(\ref{411}) and (\ref{412}), 
$\Delta (q+\bar{q})$ is the net number of righthanded
quarks of flavor $q$ inside a righthanded proton and thus $\frac{1}{2}
\Delta\Sigma(Q^2)$, cf. Eq.~(\ref{419}), 
is a measure of how much all quark flavors
contribute to the spin of the proton.  Similarly, $\Delta g(Q^2)$ in 
(\ref{111}) 
represents the total gluonic contribution to the spin of the nucleon.
Later we shall try to answer the important question of how much these
first moments contribute numerically to the spin of the proton.  The
present experimental results on the first moment $\Gamma _1(Q^2) \equiv
\int^1_0 dx\, g_1(x,Q^2)$, defined in 
Eq.~(\ref{312}) and reviewed in Sect. 3,
have a better statistics than $g_1(x,Q^2)$ itself, just because 
$\Gamma_1(Q^2)$ is
an average.  It should, however, be kept in mind that the determination
of $\int^1_0 dx\, g_1(x,Q^2)$ relies on theoretical assumptions concerning
the extrapolations for $x\to 0$ and $x\to 1$ since the actual $(x,Q^2)$
dependent data extend only over a limited range of $x$.

Let us first study the scale $(Q^2)$ dependence of $\Delta\Sigma (Q^2)$ 
and $\Delta g(Q^2)$ in LO.  The $n=1$ moment of the singlet evolution
equations in (\ref{4125}) is given by
\begin{eqnarray}
 \frac{d}{dt} \left( \begin{array}{c}
    \Delta\Sigma (Q^2) \\ \Delta g(Q^2) \end{array} \right)
& = & \frac{\alpha _s(Q^2)}{2\pi} \left( \begin{array}{cc}
      \Delta P^{(0)}_{qq}, & 2f\Delta P^{(0)}_{qg} \\
      \Delta P^{(0)}_{gq}, & \Delta P^{(0)}_{gg} \end{array} \right)
%\nonumber \\ & &
    \left( \begin{array}{c}
         \Delta\Sigma (Q^2) \\ \Delta g(Q^2) \end{array} \right)
     + {\cal{O}}(\alpha _s^2) \nonumber \\
& = &  \frac{\alpha _s(Q^2)}{2\pi} \left( \begin{array}{cc}
       0, & 0 \\ 2, & {\beta_0 \over 2} \end{array} \right)
        \left( \begin{array}{c}
         \Delta\Sigma (Q^2) \\ \Delta g(Q^2) \end{array} \right)
%\nonumber \\ & & 
     + {\cal{O}}(\alpha _s^2)
\label{52}
\end{eqnarray}
%\be
% \frac{d}{dt} \left( \begin{array}{c}
%    \Delta\Sigma (Q^2) \\ \Delta g(Q^2) \end{array} \right)
%=  \frac{\alpha _s(Q^2)}{2\pi} \left( \begin{array}{cc}
%      \Delta P^{(0)}_{qq}, & 2f\Delta P^{(0)}_{qg} \\
%      \Delta P^{(0)}_{gq}, & \Delta P^{(0)}_{gg} \end{array} \right)
%   \left( \begin{array}{c}
%         \Delta\Sigma (Q^2) \\ \Delta g(Q^2) \end{array} \right)
%     + {\cal{O}}(\alpha _s^2) 
% =   \frac{\alpha _s(Q^2)}{2\pi} \left( \begin{array}{cc}
%       0, & 0 \\ 2, & {\beta_0 \over 2} \end{array} \right)
%        \left( \begin{array}{c}
%         \Delta\Sigma (Q^2) \\ \Delta g(Q^2) \end{array} \right)
%     + {\cal{O}}(\alpha _s^2) 
%\label{52}
%\ee
according to Eq.~(\ref{4126}) , and $\beta_0 =\frac{1}{3}(11N_c-2f)$ is 
the
coefficient appearing in the renormalization group equation for 
$\alpha_ s, \frac{d\alpha _s}{dt} = -\beta_0\frac{\alpha_s^2}{4\pi} +
{\cal{O}}(\alpha_s^3)$, with $N_c=3$ and $f=3$ for the relevant 'light' 
$u,d,s$ flavors.  The $Q^2$--independence of $\Delta\Sigma$ is trivial due 
to $\Delta P_{qq}^{(0)} = \Delta P_{qg}^{(0)} = 0$ which holds also for each
$NS$ combination in (4.24) due to $\frac{d}{dt}\Delta q_{NS}(Q^2) =
0+{\cal{O}}(\alpha _s^2)$.  Thus the total polarization of each (anti)quark 
flavor 
is {\it conserved}, i.e.\ $Q^2$-independent in LO:  
$\Delta\!\!\!\!\stackrel{(-)}{q}\!\!(Q^2) = {\rm{const}}$. The vanishing of 
$\Delta P_{qg}^{(0)}$ can be understood as
follows:  assume that a gluon of positive helicity $+1$ splits into a
$q\bar{q}$ pair.  Then the helicity of the quark will always be
$+\frac{1}{2}$ and that of the antiquark 
$-\frac{1}{2}$, and therefore $\Delta P_{qg}^{(0)}=0$. (Note that in 
this process angular momentum is produced.)  $\Delta P_{qq}=0$
is a consequence of $\delta P_{qq}^{(0)}(x) = P_{qq}^{(0)}(x)$
together with the so-called Adler sum rule \cite{adler} which says that the
number of quarks of a certain flavor inside the proton is $Q^2$--independent.
In chirality preserving regularization schemes the $Q^2$--independence of
$\Delta\!\!\!\!\stackrel{(-)}{q}\!\!$ 
%  $(\delta P_{qq}(x) = P_{qq}(x))$ 
holds true even
beyond the leading order.  Furthermore, Eqs.~(\ref{52}) imply for 
$\alpha_s\Delta g$
\begin{equation}
\frac{d}{dt}[\alpha_s(Q^2)\Delta g(Q^2)] = 0+{\cal{O}}(\alpha_s^2)
\label{53}
\end{equation}
and thus $\alpha_s(Q^2)\Delta g(Q^2)\simeq {\rm{const.}}$, i.e.\
$Q^2$-independent in LO \cite{lam,glueck10,altarelli2}, 
but not in higher orders.
This is a rather peculiar property of the $n=1$ moment of 
$\delta g(x,Q^2)$ and derives formally from the appearance of
$\beta_0$ in $\Delta P_{gg}^{(0)}$.  Therefore the product 
$\alpha _s\Delta g$ behaves more like an object of order ${\cal{O}}(1)$, 
although strictly speaking it refers to a combination which enters
only in NLO, and any contribution $\sim \alpha_s\Delta g$ to 
$\Gamma _1(Q^2)$ could be in principle potentially large, irrespective
of the value of $Q^2$.  From the theoretical point of view it is 
important to note that $\alpha _s(Q^2)\Delta g (Q^2)$ becomes $Q^2$-dependent
beyond the LO.  However, for practical purposes the $Q^2$-dependence
is too small to distinguish $\Delta \Sigma $ and $\alpha _s(Q^2) 
\Delta g(Q^2)$ by examining this $Q^2$-dependence.

In the LO-QCD parton model one has, using the $SU(3)$ flavor decomposition
Eq.~(\ref{416}) , 
\begin{equation}
\Gamma _1^{p,n}(Q^2) = \frac{1}{2} \left( \pm\, \frac{1}{6}A_3 + 
    \frac{1}{18}A_8 + \frac{2}{9} A_0 \right)
\label{54}
\end{equation}
where
%equations 5.5, 5.6, 5.7
\begin{eqnarray}
A_3 & = & \Delta u +\Delta \bar{u} -\Delta d -\Delta \bar{d} 
\label{55} \\
A_8 & = & \Delta u + \Delta \bar{u} + \Delta d + \Delta \bar{d} -
              2(\Delta s + \Delta \bar{s}) 
\label{56} \\
A_0 & \equiv & \Delta\Sigma = \sum_q(\Delta q + \Delta \bar{q}) =
              A_8 + 3(\Delta s + \Delta \bar{s})
\label{57}
\end{eqnarray}
with $A_i$ being related to $M_i^{n=1}$ in 
Eq.~(\ref{436})  in an obvious way.
As discussed above, the two flavor nonsinglet combinations ($A_{3,8}$) and
the singlet $A_0\equiv\Delta\Sigma$ are independent of $Q^2$ in LO QCD.
As discussed already in Sect. 4.3 (in particular 
after Eq.~(\ref{4322b})  ), the
NS combinations have to remain $Q^2$-independent, i.e. conserved to any
order in $\alpha _s$ due to the absence of the gluon-initiated 
$\gamma _5$-anomaly in
the NS sector.  This is in contrast to the singlet component $A_0$
which can and will become $Q^2$-dependent beyond the LO, depending of
course on the specific factorization scheme chosen (e.g.\ due to the
nonvanishing $\Delta\gamma_{qq}^{(1)}$ in (\ref{4322b})  in the 
$\overline{MS}$ scheme).
The fundamental conserved $NS$ quantities $A_3$ and $A_8$ can be fixed
by the Gamov-Teller part of the (flavor changing) octet hyperon $\beta$
-decays ($F,\, D$ values) \cite{sehgal,anselmino1}:
\begin{eqnarray}
A_3 & = & F+D = {g_A \over g_V} = 1.2573\pm 0.0028 
\label{58} \\
A_8 & = & 3F-D = 0.579\pm 0.025
\label{59}
\end{eqnarray}
with the $F,\, D$ values taken from \cite{montanet,close3}.  
It should be noted that the
result for $A_3$ relies only on the fundamental $SU(2)$ isospin symmetry,
i.e.\ is obtained just from the neutron $\beta$-decay. In order to obtain
$A_8$ one has, however, to extend the phenomenological analysis of the
${g_A \over g_V}$ ratios to the full $SU(3)$ baryon octet, the spin $\frac{1}{2}$
hyperons $p,\, n,\,\Lambda,\,\Sigma^{0,\pm}$ and $\Xi^{0,-}$.  The
$\beta$-decay of some of these baryons or, more precisely, the transition
of strange into nonstrange components, can be used to get information about
$A_8$.  Thus the main assumptions used are $SU(3)_f$ symmetry and the 
approximation of massless quarks -- both are not very precise but are to
some extent reasonable.  Then the hyperon transition matrix elements of the
octet of the axial vector currents are of the general form
\begin{equation}
\langle H_j\, PS|J_{5\mu}^i|H_k\, PS \rangle = \, S_{\mu}
  (-if_{ijk} F + d_{ijk} D)
\label{510}
\end{equation}
where the hyperons are denoted by $H_i,\, i=1, \ldots ,\, 8$ ($H_1 =
\Sigma^0$, etc.) and $f_{ijk}$ and $d_{ijk}$ are 
the totally antisymmetric and symmetric $SU(3)$ 
group constants, respectively.  The expression (\ref{510}) is completely
fixed by the two constants $F$ and $D$, with a recent experimental
fit resulting in \cite{montanet,close3} 
\begin{equation}
F = 0.459 \pm 0.008\, ,\quad \quad D = 0.798 \pm 0.008\, ,
\label{511}
\end{equation}
i.e.\ ${F \over D} = 0.575 \pm 0.016$, which gives $A_8$ in 
Eq.~(\ref{59}). 
Moreover, Eqs.~(\ref{58}) and (\ref{59}), together with 
(\ref{55}) and (\ref{56}), allow
to express $\Delta (u+\bar{u})$ and $\Delta (d+\bar{d})$ in terms of
$\Delta (s+\bar{s})$:
\begin{equation}
\Delta u + \Delta \bar{u}  =  \frac{1}{2} (A_8 + A_3)
   + \Delta s + \Delta \bar{s} = 0.92 + \Delta s + \Delta\bar{s}\, ,
\label{512} 
\end{equation} 
\begin{equation}
\Delta d + \Delta \bar{d}  =  \frac{1}{2} (A_8 - A_3)
   + \Delta s + \Delta \bar{s} = -0.34 + \Delta s + \Delta\bar{s}\ 
\label{513}.
\end{equation}
The crude assumption $\Delta s = \Delta \bar{s} = 0$ essentially 
corresponds to the so called Ellis-Jaffe sum rule to be discussed below.
A finite $\Delta s = \Delta \bar{s} \neq 0$ (as likely to be the case
experimentally) obviously implies significant changes of the total 
polarizations carried by $u$ und $d$ quarks in 
Eqs.~(\ref{512}) and (\ref{513}).  
Apart from contributing differently to the nucleon's spin, such changes
of $\Delta (u+\bar{u})$ and $\Delta(d+\bar{d})$ from their 'canonical'
values $\frac{1}{2}(A_8 \pm A_3)$ in 
(\ref{512}) and (\ref{513}) may be of
astrophysical relevance 
\cite{ellis2,ellis3,ellis7,ellis4,griest} as well as of substantial consequences
for, e.g., laboratory searches of supersymmetric dark matter candidates,
and for the flux of neutrinos from supersymmetric dark matter annihilation
in the sun (which will be reduced due to the reduced photino/neutralino
trapping rate in the sun).

The constraint equations (\ref{58}) and (\ref{59}) 
are the ones used in most analyses
performed so far and we shall refer to them as the $SU(3)_f$ symmetric
"standard" scenario.  While the validity of (5.8) is unquestioned since
it depends merely on the fundamental $SU(2)_f$ isospin rotation 
($u \leftrightarrow d$) between charged and neutral axial currents, the
constraint Eq.~(\ref{59}) depends critically on the assumed $SU(3)_f$ flavor
symmetry between hyperon decay matrix elements of the flavor changing
charged weak axial currents and the neutral ones relevant for 
$\Delta f(Q^2)$.  There are some serious objections to 
\cite{sehgal,kaplan,anselmino2,lipkin}, as well
as some arguments in favor of \cite{song} 
this latter full $SU(3)_f$ symmetry,
i.e.\ to the constraint Eq.~(\ref{59}).  
We shall come back to these $SU(3)_f$
symmetry breaking effects later.

Inserting the constraints Eqs.~(\ref{58}) and (\ref{59}) 
into Eq.~(\ref{54}) gives, using
(\ref{57}),
\be
\Gamma_1^{p,n}  =  \pm\,\, \frac{1}{12}(F+D) + \frac{5}{36}\,\, (3F-D) + 
  \frac{1}{3} (\Delta s + \Delta \bar{s}) \nonumber \\
\rule{0mm}{10mm}
 =  \left\{ \begin{array}{r@{\quad \pm \quad}l}
                 0.185 & 0.004 \\ 
                -0.024 & 0.004 \end{array} \right\} 
            + \frac{1}{3} (\Delta s + \Delta \bar{s})
\label{514}
\ee
where the contribution from the strange sea remains unknown since the flavor
changing $(NS)$ hyperon $\beta$-decay data cannot constrain the singlet
quantity $A_0\equiv\Delta\Sigma$ in Eq.~(\ref{57}).  {\it Assuming}
naively $\Delta s=\Delta\bar{s}=0$, Eq.~(\ref{514}) gives
\begin{equation}
\Gamma _{1,EJ}^p \simeq 0.185
\label{515}
\end{equation}
which is the so-called Ellis-Jaffe "sum rule" originally derived in 
\cite{gourdin,ellis}.
This "naive" theoretical expectation lies, however, significantly 
{\it above} present measurements (cf.\  Table 2).  This fact is
usually referred to as the "spin crisis" or more appropriately "spin
surprise" since a sizeable negative polarization of the strange sea is
required \cite{glueck10,ellis2,ellis3} 
in Eq.~(\ref{514}) ($\Delta s=\Delta\bar{s}\simeq -0.05$ to
$-0.1$) in order to {\it reduce} $\Gamma _{1,EJ}^p$, i.e.\ to
reduce significantly the singlet contribution $\Delta\Sigma$ in 
(\ref{57}) 
relative to $A_8$; thus $\frac{1}{2}\Delta\Sigma\ll\frac{1}{2}$, i.e.\
the quark flavors seem to contribute marginally to the spin of the proton,
Eq.~(\ref{111}), which is surprising indeed!  Alternatively, a negative 
polarization of the light sea quarks ($\Delta\bar{u} \simeq\Delta\bar{d}<0$,
keeping $\Delta s = \Delta\bar{s} = 0$) could also account for a suppression
of $\Delta\Sigma$ in Eq.~(\ref{57}) 
when $SU(3)_f$ symmetry breaking effects are 
taken into account \cite{lipkin1,lipkin2,lichtenstadt}, i.e.\ when the 
constraint Eq.~(\ref{59}) does not hold
anymore; or a negative $\alpha _s\Delta g$ contribution to $\Delta\Sigma$
could equally account for the required reduction 
\cite{altarelli2,efremov} of $A_0$ in Eq.~(\ref{57}).
The latter scenario of a large gluon contribution will be discussed in
detail below.  More detailed quantitative analyses and results will be
discussed in Sect. 6.  Here it suffices to remark that one does not 
expect intuitively a large total polarization of strange sea quarks due
to the fact that it is easier for a gluon to create a nonstrange `light'
pair ($u\,\bar{u},\, d\,\bar{d}$) than a heavier strange pair -- a 
situation very similar to the unpolarized broken $SU(3)$ sea 
\cite{glueck5,glueck9,glueck6} 
as observed by neutrino-nucleon scattering experiments 
\cite{abramowicz,foudas,rabinowitz}.

It is perhaps also interesting to compare the above results with the
expectations of a (nonperturbative) "constituent" quark model 
\cite{hey1,close,leader}.
The constituent models generally fulfil $\Delta\Sigma =1$, i.e.\ the
entire nucleon spin is saturated by valence quark spins.  Among the
constituent models, the $SU(6)$ model 
\cite{hey1,close,leader} is the most favorable, because
it is able to explain some of the static properties of nucleons.  
For example, the $SU(6)$ model explains the measured ratio $\frac{\mu_n}
{\mu_p}$ to be about $-{2 \over 3}$ 
(experimental value $= -0.685$).  
\footnote{It is instructive 
to remind the reader that
constituent quark models may be used to represent the nucleon
magnetic moments $\mu_N$ in terms of quark magnetic moments
$\mu_q=\frac{e}{2\hat{m}_q}$ where $\hat{m}_q$ is the constituent quark
mass, e.g.\ $\hat{m}_u = 336$ MeV \cite{bartelski,karl,cheng1,hogaasen}. 
This fact can be used to derive the successful relation 
$\frac{\mu _n}{\mu _p}=-\frac{2}{3}$.  Writing 
\begin{displaymath}
\mu _p = \mu_u(u_+-u_-) +\mu_d(d_+ -d_-) + \mu_s(s_+ - s_-)
\end{displaymath}
it is even possible to include strange quark contributions 
\cite{bartelski,karl,hogaasen}.  However,
a real understanding of $\Gamma_1^p$ within constituent models will
never be possible.  The reason for this is that in magnetic moments 
antiquarks count with opposite sign than in $\Gamma_1^p$.  Constituent
quark models are able to account for antiquarks but informations from
magnetic moment measurements cannot be used to get informations on 
$\Gamma_1^p$ because in magnetic moments combinations
\begin{displaymath}
\Delta q - \Delta\bar{q} = q_+ - q_- - \bar{q}_+ + \bar{q}_-
\end{displaymath}
enter, whereas $\Gamma _1^p$ is determined by combinations
\begin{displaymath}
\Delta q + \Delta\bar{q} = q_+ - q_- + \bar{q}_+ - \bar{q}_- \,\, .
\end{displaymath} 
\label{fn6}}

However, it fails to predict $\frac{g_A}{g_V}\simeq 1.26$ correctly but
gives $\frac{g_A}{g_V}=\frac{5}{3}$.  Since $\frac{g_A}{g_V}$ is the
triplet ingredient of $\int_0^1g_1(x,Q^2)dx$ one suspects that $SU(6)$
will fail for $\int_0^1g_1(x,Q^2)dx$ as well.

In a static picture of the nucleon, in which the nucleon consists 
purely of valence quarks, its wave function can be found by counting all
possible antisymmetric combinations of three quark states.  If there is
negligible $L_z$ in the system, one quark spin is always antiparallel to
the other two.  The proton can then be described by a wave function which
is a member of a 56-plet of $SU(6)$, and the probabilities to find 
$u_+,\, u_-,\, d_+$ and $d_-$ in the proton turn out to be 5/3, 1/3, 1/3
and 2/3, respectively.  For the neutron the role of $u$ and $d$ are to
be interchanged.  Thus $\Delta u = 4/3,\,\Delta d = -1/3$ and 
$\Delta\bar{u} =\Delta\bar{d}=\Delta s =\Delta\bar{s}=0$, i.e.
\begin{equation}
A_3^{SU(6)}=\frac{5}{3}\,\, , \quad A_8^{SU(6)}=A_0^{SU(6)} = 1
\label{516}
\end{equation}
so that
\begin{equation}
\Gamma _1^{p,SU(6)}=\frac{5}{54}\,\, , \quad \Gamma _1^{n,SU(6)} = 0\, .
\label{517}
\end{equation}
These numbers deviate only by about 30\% in the $NS$ sector from the 
experimental results Eqs.~(\ref{58}) and (\ref{59}), 
but the difference is huge in the
singlet sector $A_0\ll A_0^{SU(6)} = 1$ -- again a rather surprising
result.  It should be kept in mind, however, that constituent quarks are
strictly nonperturbative objects which cannot be reached by perturbative
evolutions in contrast to partonic quarks which should rather be identified
with the current quark content of hadrons.

The NLO corrections to Eq. (\ref{54}) 
in the ${\overline{MS}}$ scheme can be directly
read off Eq. (\ref{4224}) :
\begin{equation}
\Gamma _1^{p,n}(Q^2)=\left( 1-\frac{\alpha_s(Q^2)}{\pi}\right)
 \left[ \pm\ \, \frac{1}{12}A_3+\frac{1}{36}A_8+\frac{1}{9}\Delta\Sigma(Q^2)\right]
\label{518}
\end{equation}
where we used Eqs.~(\ref{4321a}) and (\ref{4321b}); 
the $NS$ combinations $A_{3,8}$ remain
$Q^2$--independent due to the vanishing of $\Delta\gamma_{NS-}^{(1)}$
in (\ref{4322a}) in contrast to the singlet combination 
$A_0(Q^2)\equiv\Delta\Sigma(Q^2)$ since 
$\Delta\gamma_{qq}^{(1)}\neq 0$ in (\ref{4322b}).  The total
polarizations of the parton distributions themselves evolve according to
Eqs.~(\ref{4216}) and (\ref{4217}) with $n=1$ 
and the LO $\Delta P_{ij}^{(0)}$ given
in Eq. (\ref{52}) and the NLO $\Delta P_{ij}^{(1)}$ given by 
\cite{mertig} (see footnotes \ref{fn4} and \ref{fn5})
\begin{eqnarray}
\Delta P_{NS-}^{(1)}=0 \, , \qquad \Delta P_{NS+}^{(1)} \approx -0.3197 
%=-\frac{5}{9} 
\nonumber \\ 
\Delta P_{qq}^{(1)}=-2f \, , \qquad \Delta P_{qg}^{(1)}=0 \, ,  \nonumber \\ 
  \Delta P_{gq}^{(1)}=\frac{1}{3} (59-{2f \over3}) \, , \qquad  
    \Delta P_{gg}^{(1)} = {\beta_1 \over 4}   
\label{519}
\end{eqnarray}
where again $\beta_1 = 102-{38f \over 3}$. Note that $\Delta g(Q^2)$
in Eq. (\ref{4224}) 
does {\it not} contribute to $\Gamma _1$ in (\ref{518}) in
the ${\overline{MS}}$ factorization since here the first moment decouples
due to $\Delta C_g\equiv \delta C_g^{n=1}=0$, 
cf.\ (\ref{4321b}).  It should be
emphasized, however, that an actual analysis of present data which are
available only in a limited $x$-range
%, $0<x<1$, 
requires {\it all}
moments in (\ref{4224}) where $\delta g^n(Q^2)$, or equivalently 
$\delta g(x,Q^2)$, represent an important contribution to $g_1(x,Q^2)$
as we shall see in Sect. 6.  Furthermore such an analysis is, for the
time being, possible only in the ${\overline{MS}}$ scheme where 
{\it all} 2-loop $\delta P_{ij}^{(1)}$ are known 
\cite{mertig,vogelsang2}, unless
one allows for transformations 
(\ref{4231}) to other factorization
schemes.  Moreover, since the polarized and unpolarized parton densities
originate from the {\it same} densities of definite positive 
and negative helicities, $\delta f=f_+ - f_-$ and $f=f_+ +f_-$ according
to Eqs.~(\ref{411})--(\ref{414}) and (\ref{4110}), 
it is also desirable to remain within the
factorization scheme commonly used in all present
NLO analyses of unpolarized deep inelastic/hard processes --  
which is the 
${\overline{MS}}$ scheme.  
This is of
particular importance for a consistent implementation of the fundamental
positivity constraints (\ref{4111}).

For completeness it should be mentioned that the NLO(${\overline{MS}}$)
result in Eq. (\ref{518}) has been extended to even higher orders 
\cite{larin1}:
\begin{eqnarray} \nonumber 
\Gamma_1^{p,n}(Q^2)  =  \left( \pm\, \frac{1}{12} A_3 +
        \frac{1}{36} A_8 \right)
 \biggl \lgroup 1-\frac{\alpha _s(Q^2)}{\pi}
        -3.5833\,\,
        \left( \frac{\alpha _s(Q^2)}{\pi}\right) ^2 \biggr \rgroup
\\ 
 +\frac{1}{9}\Delta\Sigma (Q^2) \biggl \lgroup 1-\frac{\alpha _s(Q^2)}{\pi}
  -1.0959\,\, \left( \frac{\alpha _s(Q^2)}{\pi}\right) ^2 \biggr \rgroup
\label{520}
\end{eqnarray}
for $f=3$ light flavors. The evolution of $\Delta\Sigma(Q^2)$
involves now, in contrast to the NLO result (\ref{518}), two-- 
{\it and} three--loop anomalous dimensions of the singlet
axial current \cite{larin1}.  The $Q^2$ corrections to the $NS$ contributions
$A_{3,8}$ have been calculated 
\cite{larin2} even to ${\cal{O}}(\alpha _s^3)$. 
This is relevant for the (nonsinglet) Bjorken sum rule to be
discussed in Sect. 5.4.

Although a complete NLO analysis for all moments can, for the time 
being, only be performed in the ${\overline{MS}}$ scheme, there is a 
factorization scheme for the {\it first} moment, which
is closer to the constituent quark picture Eq. (\ref{516}); it allows, 
at least in principle, for larger contributions of $\Delta\Sigma$ to 
the spin sum rule (\ref{111}). 
This is the scheme corresponding to Eq. (\ref{428}),
or the off--shell scheme resulting in 
Eqs.~(\ref{4325a}) and (\ref{4325b}), where Eq.~(\ref{4224})
yields
\begin{eqnarray} \nonumber 
\Gamma _1^{p,n}(Q^2) & = & \left( 1-\frac{\alpha _s(Q^2)}{\pi} \right)
 \biggl \lgroup \pm\, \frac{1}{12} A_3+\frac{1}{36}A_8
 + \frac{1}{9}
       \Delta\Sigma_{\rm{off}} \biggr \rgroup
  -\frac{1}{9}f
         \frac{\alpha _s(Q^2)}{2\pi} \Delta g(Q^2)
\\  
 & = & \left( 1-\frac{\alpha _s(Q^2)}{\pi} \right)
        \biggl \lgroup \pm\, \frac{1}{12}A_3 + \frac{1}{36}A_8
  + \frac{1}{9}
          A_0(Q^2) \biggr \rgroup + {\cal{O}}(\alpha _s^2)
\label{521}
\end{eqnarray}
for $f=3$ light quark flavors and where the singlet contribution is now
given by
\begin{equation}
A_0(Q^2) = \Delta\Sigma_{\rm{off}} - f\frac{\alpha _s(Q^2)}{2\pi} \Delta g(Q^2)
\label{522}
\end{equation}
in contrast to the identification (\ref{57}) which appears also in 
(\ref{518}).
Here, $\Delta\Sigma_{\rm{off}}$ is different from the quantity $\Delta\Sigma
(Q^2)$ appearing in (\ref{518}) and is $Q^2$ {\it in}dependent due to
the vanishing of $\Delta\gamma _{qq}^{(1)}$ in (\ref{4326}).  Whether this
conserved $\Delta\Sigma_{\rm{off}}$ is used to define the actual polarized
quark densities or alternatively the "renormalized" non-conserved quantity
$\Delta\Sigma(Q^2)\equiv A_0(Q^2)$, is a matter of theoretical convention
(choice of factorization scheme) or 'intuitive' physical interpretation
\cite{altarelli4,altarelli7,reya1,reya2} 
as shall be discussed below.  It should be remembered that only
the {\it products}  $\Delta C_i A_i$ and $\Delta C_g\Delta g$
in Eq.~(\ref{521}) are renormalization convention {\it in}dependent
quantities to the ${\cal{O}}(\alpha _s^2)$ considered, which can therefore
be meaningfully related to physical, i.e.\ experimental measurements.
In this sense the second line in Eq.~(\ref{521}) is identical to the first one
since the NNLO terms ${\cal{O}}(\alpha _s^2)$ are disregarded.  It is 
this latter expression which has been frequently used for first moment
analyses in the past, for example by \cite{altarelli3,altarelli5,gehrmann}.  
It should be again mentioned
that in these (non-${\overline{MS}}$) factorization schemes only
the $n=1$ moments of splitting functions (anomalous dimensions) have 
been calculated so far to which we will now turn.

Originally, NLO contributions to the first moment $\Gamma _1(Q^2)$
have been calculated by Kodaira \cite{kodaira} with the help of the operator
product expansion near the light-cone using an off--shell factorization
scheme with all external lines kept off--shell $(p^2<0)$ and assuming
massless quarks and gluons.  They have been confirmed and substantially
reinterpreted in the framework of the QCD based  parton model in recent
years \cite{altarelli2,efremov,carlitz,altarelli6}.  
The upshot of the result in the off--shell scheme is 
that Eq.~(\ref{54}) goes over to
\be
\Gamma_1^{p,n}(Q^2) = \left( \pm\, \frac{1}{12}A_3+\frac{1}{36}A_8 \right)
\left( 1-\frac{\alpha_s(Q^2)}{\pi} \right)
+ \frac{1}{9} A_0
     \left[ 1-\frac{\alpha_s(Q^2)}{\pi}\,\,
        \left( 1-\frac{2f}{\beta_0} \right) \right]
\label{523}
\ee
where $A_0 \equiv A_0(Q^2)\left[ 1-\frac{\alpha_s(Q^2)}{\pi}\,
\frac{2f}{\beta_0} \right]$ and $A_0(Q^2)$ is given by 
(\ref{522}) for $f=3$
light quark flavors.  This result fixes the effect of the quarks and 
gluons and incorporates all scaling violations up to next-to-leading order.
Remember that all the $\Delta\!\!\!\!\stackrel{(-)}{q}$ in $\Delta\Sigma$
are $Q^2$ independent quantities due to Eq.~(\ref{4326}).  

Several remarks are
in order:  the decomposition of $A_0(Q^2)$ into a quark and a gluon
contribution, Eq.~(\ref{522}), was not known to Kodaira 
\cite{kodaira} because he
worked in the framework of the OPE where the basic objects, and in particular
the singlet part $A_0(Q^2)$, are matrix elements.  The decomposition becomes
only apparent in the QCD parton model where the gluon term is induced by
the photon-gluon fusion process in Fig.  \ref{fig14}.  
The coefficient $f=3$ of
$\Delta g$ in Eq.~(\ref{522}) is the number of light quarks because heavy 
quarks (with masses much larger than $\Lambda_{\rm{QCD}}$) can be shown to
yield a vanishing first moment contribution \cite{watson,lampe1,glueck8}.  
This is a consequence
of the matrix element calculation which will be discussed in Sect. 5.3.

\begin{figure}
\begin{center}
\epsfig{file=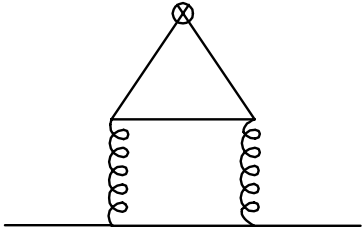,height=3cm}
\vskip 1.cm
\caption{The 2--loop triangle diagram which contributes 
to the singlet $\Delta \gamma_{qq}^{(1)}$.}
\label{fig22}
\end{center}
\end{figure}

The factors $1-\frac{\alpha_s(Q)}{\pi}$ in Eq.~(\ref{523}) arise from the 
Wilson coefficient of the quarks, both for the nonsinglet and singlet
contribution.  The fact that $\Delta\Sigma$ and $\Delta g$ appear in a
factorized form in Eq.~(\ref{522}) has to do with the fact that for the 
singlet there is only one operator.  Thus the expansion of the RG exponent
is formally the same as for the nonsinglet case in 
(\ref{4311}) or (\ref{4313}) which
contributes $\frac{\alpha_s}{4\pi}\,{\Delta \gamma_{qq}^{(1)} 
\over 2\beta_0} =
%\frac{\alpha_s}{4\pi}\, 16f/2\beta_0 = 
\frac{\alpha_s}{\pi}\, {2f \over \beta_0}$
in the OPE scheme used by Kodaira \cite{kodaira}, 
using $\Delta\gamma_{qq}^{(0)}=0$.
Since this 2-loop term $\sim {2f \over \beta_0}$ in 
Eq.~(\ref{523}) has been explicitly
factored out from the (fermionic) singlet matrix element, according to
Kodaira's original calculation \cite{kodaira}, the $A_0$ in 
Eq.~(\ref{523}) [but not 
$A_0(Q^2)$] is $Q^2$-{\it in}dependent in this specific way of
writing.  The $A_0$ in Eq.~(\ref{523}) 
should in fact depend on the renormalization
(input) scale $\mu$.  This dependence has been suppressed here but will be 
made explicit later [cf.\ Eq.~(\ref{553})].  
Note that the term $\sim {2f \over \beta_0}$
does not arise in the nonsinglet part in (\ref{523}), but is inherently
a singlet contribution and a two-loop effect of the triangle anomaly
(Fig.  \ref{fig22}).

It is evident from the above results and discussions that the representation
Eq.~(\ref{522}) of $A_0(Q^2)$ in 
terms of $\Delta\Sigma$ and $\Delta g$ depends on
the regularization scheme, despite the fact that the remaining 
$Q^2$-dependent corrections in Eq.~(\ref{523}) are scheme independent.  
The appearance
of the gluon in (\ref{522}), 
with the $-{\alpha_s \over 2\pi}$ coefficient as calculated
directly from the photon--gluon fusion diagram in Fig.  \ref{fig14} 
using an off--shell regularization,
could account for the reduction of the singlet contribution to
$\Gamma_1(Q^2)$ 
\cite{altarelli2,efremov,carlitz} 
required by experiment as discussed in particular
after Eq.~(\ref{515}).  For example, the recent experimental average result
for $\Gamma _1^p$ in (\ref{327}) at $Q^2=5$ to $10$ GeV$^2$ implies, 
using Eq.~(\ref{521})
together with (\ref{58}) and (\ref{59}) for $A_{3,8}$,
\begin{equation}
A_0(Q^2)\equiv \Delta\Sigma_{\rm{off}} -3\frac{\alpha_s(Q^2)}{2\pi}\Delta g
  (Q^2) = 0.27 \pm 0.13
\label{524}
\end{equation}
at $Q^2=5$ to $10$ GeV$^2\,$.
\footnote{For consistency 
note that (\ref{524}) implies,
via Eq.~(\ref{521}), for the neutron $\Gamma_1^n(Q^2)=-0.055\pm 0.014$ in
agreement with the neutron measurements in Table 2.\label{fn7}}   
Making the extreme
and equally naive assumption that $\Delta s=\Delta\bar{s}=0$ and
disregarding any effects due to flavor $SU(3)_f$ breaking, one obtains
a large $\Delta\Sigma = A_8+3(\Delta s +\Delta\bar{s}) = A_8 =
0.579 \pm 0.025$ according to Eqs.~(\ref{57}) and (\ref{59}).  
If one wants to keep a large and conserved 
$\Delta\Sigma \simeq 0.6$ and at the same time reduce the 
Ellis-Jaffe prediction (\ref{515}) to the experimental measurement 
in Eq.~(\ref{327}), a value 
\begin{equation}
\Delta g(Q^2=10\,\,{\rm{GeV}}^2) = 3.4\pm 1.5
\label{525}
\end{equation}
is required by Eq.~(\ref{524}). This number is obtained 
with the NLO $\alpha_s(10 GeV^2) / \pi = 0.061$ for 
$\Lambda_{{\overline{MS}}}=0.2$ GeV. 
Realistically, however, there is no reason for a vanishing
total helicity of strange quarks and one expects a combination of 
{\it both} effects ($\Delta g \neq 0$ and $\Delta s = \Delta\bar{s}
\neq 0$), probably with additional flavor $SU(3)_f$ breaking 
effects, to account for the required singlet suppression in 
Eq.~(\ref{524}).
As we shall see in the next chapter, present data are far too scarce
for distinguishing or confirming the various scenarios.  Before turning
to a discussion of more detailed $(x,Q^2)$ dependent analyses of present
experiments, let us concentrate on a few theoretical aspects concerning
the total helicities (first moments) of quarks and gluons for the remainder
of this chapter.

As already mentioned above, it is a matter of theoretical convention to
interpret $\Delta\Sigma$ as the actual total helicity of the singlet
quark densities or effectively the entire singlet expression 
(\ref{522})
\begin{equation}
\Delta\Sigma (Q^2)\equiv\Delta\Sigma_{\rm{off}}-3\frac{\alpha_s(Q^2)}{2\pi}
    \Delta g(Q^2)
\label{526}
\end{equation}
which, in contrast to $\Delta\Sigma_{\rm{off}}$, 
is $Q^2$-dependent and effectively has
to be interpreted as the singlet contribution entering the ${\overline{MS}}$
result (\ref{518}).  The real question then arises is which $\Delta\Sigma$
enters, for example, Eq.~(\ref{111}).  
Intuitive arguments have been forwarded
\cite{altarelli4,altarelli7,reya1,altarelli6} in favor of 
the {\it conserved} $Q^2$-independent
$\Delta\Sigma_{\rm{off}}$ in 
Eq.~(\ref{526}): in this case the flavor singlet quark 
contribution
can be kept sizeable, $\Delta\Sigma_{\rm{off}}=A_8\simeq 0.6$ (still assuming
$\Delta s=\Delta\bar{s}=0$), as compared to the (constituent) $SU(6)$
prediction in (\ref{516}),
\begin{equation}
\Delta\Sigma_{\rm{off}}\simeq 0.6 \stackrel{<}{\sim} \Delta\Sigma^{SU(6)} = 1
\label{527}
\end{equation}
in complete analogy to the flavor nonsinglet sector where, using 
Eqs. (\ref{58}), (\ref{59}) and (\ref{512}),
\begin{equation}
A_3\simeq 1.26 \stackrel{<}{\sim} A_3^{SU(6)} =\frac{5}{3} , \,\,  \quad
  A_8\simeq 0.58 \stackrel{<}{\sim} A_8^{SU(6)}=1\, . 
\label{528}
\end{equation}
In {\it both} cases there are only 30-40\% reductions from the
$SU(6)$ expectations which are attributed to helicity non-conservation
induced at low energy scales by finite quark mass effects which break
chirality and thus the symmetry in the non-perturbative region which
creates a difference in the initial values for the perturbative QCD
evolution \cite{altarelli4,altarelli7}.  
Note that only for {\it conserved} quantities
one expects, in the absence of chirality breaking effects, constituent
and parton results to coincide.  
Furthermore the choice of a $Q^2$-independent 
$\Delta\Sigma_{\rm{off}}$
fulfills the requirement of an extended Adler sum rule.  The usual Adler
sum rule \cite{adler} 
says that the number of quarks of a certain flavor inside
the nucleon is conserved, i.e.\ $Q^2$-independent, to any order in 
$\alpha _s$.  The extended Adler sum rule extends this to the conservation
of the number of quarks of a certain helicity.  Unfortunately, our present
ignorance of all moments of 2-loop splitting functions 
$\delta P_{ij}^{(1)n}
$, or equivalently of $\delta P_{ij}^{(1)}(x)$,
calculated in the off--shell scheme prevents us
from a detailed $(x,Q^2)$ dependent analysis of present data, in contrast
to the ${\overline{MS}}$ regularization/factorization scheme 
where such an analysis has been performed. 
In the ${\overline{MS}}$ scheme one is confronted with a less
plausible large difference between the singlet sector
\begin{equation}
\Delta \Sigma (Q^2)\equiv A_0(Q^2)\simeq 0.3\ll A_0^{SU(6)} = 1\,\, ,
\label{529}
\end{equation}
at $Q^2=5$ to $10$ GeV$^2$ according to 
(\ref{524}), and the nonsinglet quantities
in (\ref{528}).  

%% file: k52tex
\subsection{The First Moment and the Anomaly }

\begin{figure}
\begin{center}
\epsfig{file=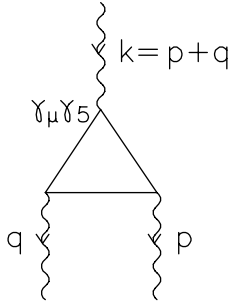,height=4cm}
\vskip 1.cm
\caption{Triangle anomaly diagram giving
rise to Eqs.~(\ref{534}) 
and (\ref{540}).}
\label{fig23}
\end{center}
\end{figure}

In the OPE approach
the possible decomposition Eq.~(\ref{522}) into quark and gluon
was not realized because
the fundamental quantities of the OPE are not parton desities but
matrix elements of certain operators between proton states. 
$\langle PS\vert\bar q\gamma_\mu\gamma_5 q\vert PS\rangle$
are the matrix elements relevant for the total helicities, i.e. for 
the first moment of $g_1$ \cite{sehgal}. 
More precisely, in the framework of the operator
product expansion one obtains the result Eq.~(\ref{523}) with
$A_0, A_3$ and $A_8$ defined by
\begin{equation}
A_3 S_\mu =\langle PS\vert\bar\psi\ga_\mu\g5\lambda_3\psi\vert PS\rangle
\label{530}
\end{equation}
\begin{equation}
A_8 S_\mu =\sqrt{3}\langle PS\vert\bar\psi\ga_\mu\g5\lambda_8\psi\vert
PS\rangle
\label{531}
\end{equation}
\begin{equation}
A_0 S_\mu =\langle PS\vert\bar\psi\ga_\mu\g5\psi\vert PS\rangle
\label{532}
\end{equation}
where $\psi =(u,d,s)^T$ and $\lambda_a$ are the
Gell-Mann matrices, $a=1,\ldots,8$.  
Longitudinal polarization $S^{\sigma} \sim P^{\sigma}$ of the proton 
is assumed throughout this section.  
In the naive quark parton model
one has the identity
\begin{equation}
(\Delta q +\Delta \bar{q} ) S_\mu  
\stackrel{naive}{=} \langle PS\vert\bar q\ga_\mu\g5 q\vert PS\rangle
\label{533}
\end{equation}
between the first moments $\Delta q +\Delta \bar{q} $ (q=u,d,s) 
and the matrix elements. 
However, one should note that $j^5_\mu =\bar\psi\ga_\mu\g5\psi$
which appears in Eq.~(\ref{532}) is the axial vector singlet current
which is not conserved, $\partial^\mu j^5_\mu \not= 0$, but
is 'anomalous' in the sense of Adler \cite{adler1} and Bell/Jackiw
\cite{bell} 
\begin{equation}
\partial^\mu j^5_\mu ={f\over 2}{\a_s\over 2\pi}
\ve^{\mn\a\b}G_{\a\b}^aG_{\mn}^a \/.
\label{534}
\end{equation}
The appearance of the gluon field strength $G_{a\mn}$ in this
equation is the reason why Eqs.~(\ref{522}) and (\ref{532}) are
not in contradiction and instead of Eq.~(\ref{533}) one has 
(in the particular factorization scheme called the off--shell scheme) 
\begin{equation}
(\Delta q +\Delta \bar{q} -{\a_s\over 2\pi}\Delta g)S_\mu 
=\langle PS\vert\bar q\ga_\mu\g5 q\vert
PS\rangle  \/.
\label{535}
\end{equation}
The anomaly enters in one-loop order through the diagram Fig. \ref{fig23},	
which, in the gluon off--shell scheme, 
determines the (Wilson) coefficient $-{1 \over 2}$ of 
$2{\a_s\over 2\pi}\Delta g$,and in two-loop 
order through the diagram Fig. \ref{fig22} from which Kodaira 
\cite{kodaira} has 
obtained his anomalous 
dimension. 

$\Delta q +\Delta \bar{q}$ and $-{\a_s\over 2\pi}\Delta g$ 
appear with the same 
coefficient in Eq.~(\ref{535}) because there is only one operator 
for the first moment singlet contribution. For the gluon there 
is no direct operator,but only a representation of $\Delta g$ 
\begin{equation}
\langle PS \vert K^{\mu} \vert PS \rangle =-S^{\mu} \Delta g
\label{536}
\end{equation}
in terms
of the gauge dependent current 
\begin{equation}
K_{\mu}={1 \over 2} \epsilon^{\mu \nu \rho \sigma} A_{\nu}^a 
(G_{\rho \sigma}^a -{g_s \over 3} f_{abc} A_{\rho}^b A_{\sigma}^c ) 
\label{537}              
\end{equation}
whose divergence is 
\begin{equation}         
\partial^\mu K_{\mu}={1 \over 2} \epsilon^{\mu \nu \alpha \beta} 
G_{\alpha \beta}^a G_{\mu \nu}^a 
\label{538}              
\end{equation}
(so that $j_{\mu}^5-f{\a_s\over 2\pi}K_{\mu}$ is conserved). 

An explicit computation by Forte \cite{forte} of 
$\langle PS \vert {\alpha_s \over 2\pi} K^{\mu} \vert PS \rangle$ 
in (\ref{536}) has revealed that 
${\alpha_s \over 2\pi} \Delta g ={\alpha_s \over 2\pi} \Delta g' 
+ \Omega $ where $\Delta g'$ corresponds to the gauge invariant 
hard piece of the anomaly which may be identified with the 
usual perturbative partonic definition used thus far; the remaining 
gauge variant Chern--Simons (instantonic) piece $\Omega$ is the 
soft non--perturbative contribution to the triangle anomaly 
which does not correspond to a hard (large--$k_T$) subprocess 
and thus should be absorbed into the definition of the conserved 
total quark polarisation, i.e. 
$\Delta \Sigma_{off} -f{\alpha_s \over 2\pi}\Delta g =
(\Delta \Sigma_{off} -f\Omega )-f{\alpha_s \over 2\pi }\Delta g' 
\equiv \Delta \Sigma '-f{\alpha_s \over 2\pi}\Delta g'$. This 
leaves us formally with our original expression (\ref{526}) 
with both terms being now separately gauge invariant. The term 
$-f\Omega$ may be interpreted as an additional sea polarization 
induced by instantonic effects. A first attempt \cite{mandula1} 
to evaluate 
$\langle PS \vert {\alpha_s \over 2\pi} K^{\mu} \vert PS \rangle$
on a (small) lattice yielded the bound 
$3 \vert {\alpha_s \over 2\pi}\Delta g'+ \Omega \vert < 0.05$. 
This surprisingly small result can obviously not be used to 
extract some information about $\Delta g'$.  

It should be noted that Eq.~(\ref{535}) is really a representation 
of (chiral invariant) massless QCD. If fermion masses are introduced,  
the anomaly Eq.~(\ref{534}) is modified according to 
\begin{equation}
\partial^\mu j^5_\mu = 2i \bar\psi \g5 M \psi + 
{f\over 2}{\a_s\over 2\pi}
\ve^{\mn\a\b}G_{\a\b}^aG_{\mn}^a
\label{539}
\end{equation}
where $M$=diag$(m_u,m_d,m_s)$ is the fermion mass matrix.  
The mass term in this equation is able to conceal the effect 
of the anomalous term under certain circumstances. 
To examine this effect, we write down the anomaly equation 
Eq.~(\ref{539}) 
in momentum space for the case of one fermion 
with mass m 
\begin{equation}
(p+q)^{\a} \Ga_{\a}^{\mn}(p,q) = 2m \Ga^{\mn}(p,q)-{\a_s\over \pi} 
\ve^{\mn\a\b}p_{\a}q_{\b} \, .
\label{540}
\end{equation}
Here $\Ga_{\a}^{\mn}(p,q)$ is the triangle diagram 
(c.f. Fig. \ref{fig23}) and 
%\begin{eqnarray} \nonumber & &
%\Ga^{\mn}(p,q)=m{\a_s\over \pi}\ve^{\mn\a\b}p_{\a}q_{\b}
%\\ & &
%\times \int^1_0 dx \int^{1-x}_0 {dy \over m^2-(p+q)^2xy} \/.
%\label{541}
%\end{eqnarray}
\begin{equation} 
\Ga^{\mn}(p,q)=m{\a_s\over \pi}\ve^{\mn\a\b}p_{\a}q_{\b}
 \int^1_0 dx \int^{1-x}_0 {dy \over m^2-(p+q)^2xy} \/.
\label{541}
\end{equation}   
the contribution from the mass term. If $(p+q)^2 \leq m^2$,  
the two terms on the right hand side of Eq.~(\ref{540}) 
(i.e. the anomalous contribution
and the mass contribution) 
cancel each other, up to terms of order ${(p+q)^2 \over m^2}$, 
so that the anomaly is concealed.   

A similar effect occurs in Eq.~(\ref{535}) if one includes the 
quark mass. Namely,the anomalous gluon term $\sim \Delta g$ 
disappears and one formally recovers the naive Eq.~(\ref{533}). 
Thus the contribution to $g_1$ from the anomaly           
is hidden in the massive theory (as well as in the 
$\overline{MS}$ subtraction scheme).                           
This result will be derived in detail in the next section.  

The anomaly induces a pointlike interaction between the 
axial-vector current and the gluons because the amplitude 
$\Ga^{\a \mn}(p,q)$ does not depend on the momentum 
transfer $p-q$ when $m=0$. Furthermore it is known,that 
the anomaly is not affected by higher order 
corrections \cite{adler2} and it has been argued \cite{hooft1} that 
this is true even beyond perturbation theory. 
This leads us to believe
that the identification Eq.~(\ref{535}) remains true 
in higher orders, at least in suitable chirality preserving 
schemes like the gluon off--shell scheme.

%% file: k53tex
\subsection{Detailed Derivation of the Gluon Contribution }

The most straightforward way to derive the gluon contribution is the
QCD parton model. In the following we shall only consider the 
flavor singlet
piece of $g_1$ 
because the gluon contributes only to the singlet. In the QCD
parton model the evolution of the first moments of the polarized
quark and gluon density is given by
\be 
{d\over dt}\pmatrix{\Delta \Sigma\cr \Delta \Gamma\cr} 
 =  \left({\a_s(t)\over 2\pi}
\right)^2\pmatrix{\Delta P_{qq}^{(1)} &\Delta P^{(2)}_{qg}\cr
                  \Delta P^{(0)}_{gq} &\Delta P^{(1)}_{gg}-{\b_1 \over 4} \cr}
\pmatrix{\Delta \Sigma\cr \Delta \Gamma} 
+O(\a_s^3) \/.
\label{542}
\ee
This is an extension of Eq.~(\ref{52}) to second order 
in $\alpha_s$, in which
the 2-loop contribution $\b_1$ to the $\b$ function arises, 
cf. Eq.~(\ref{4212}), 
because instead of $\Delta g$ we have introduced the product
$\Delta \Ga ={\a_s (t)\over 2\pi}\Delta g$ (see the discussion in Sect. 5.1). 
In Eq.~(\ref{542}) we have used quantities 
$\Delta P_{ij}^{(k)}$ which are the first moments of 
$\delta P_{ij}^{(k)}$ defined in Eq.~(\ref{4214})  
\be 
\Delta P_{ij}=\int_0^1 dx \delta P_{ij}(x)=
\Delta P^{(0)}_{ij} +{\a_s\over 2\pi}\Delta P^{(1)}_{ij}
+
\left({\a_s\over 2\pi}\right)^2\Delta P^{(2)}_{ij}+O(\a_s^3) \, .
\label{543}
\ee
They are related to the corresponding anomalous dimensions 
by the usual factors of $-{1 \over 4}$ and $-{1 \over 8}$ in 
the 1- and 2-loop case 
(see footnote \ref{fn1}). 

As is well known, for a complete discussion of first order
effects the knowledge of second order ($k=1$)  
anomalous dimensions is mandatory.
These second order (2-loop) anomalous dimensions depend on the calculational
(i.e.\ regularization) scheme. For example, it is not true in
general that $\Delta P^{(1)}_{qq}=\Delta P^{(2)}_{qg}=0$.
However, some specific (chirality preserving) 
schemes have this property, so that
$\Delta\Sigma$ is constant in $Q^2$ and the generalized Adler sum rule
holds. Note furthermore that in deriving the evolution equation 
Eq.~(\ref{542}) we have used $\Delta P_{qg}^{(1)}=0$:  
$\Delta P_{qg}^{(1)}$ has to vanish on general grounds 
in any scheme 
\cite{altarelli2,kaplan}, since the gauge invariant axial  
current $j_{\mu}^5$ is multiplicatively renormalizable and therefore 
cannot mix with the gauge dependent current $K_{\mu}$ in 
Eq.~(\ref{537}).  

In the QCD parton model it is well known how to calculate the
contribution of $\Delta g$ to $\Gamma_{1,S}= \int^1_0 dx g_{1,S}$.
Namely, one just has to calculate the diagrams in Fig.\ref{fig14}b 
and take the
first moment. We call the amplitude squared 
corresponding to these diagrams $\Omega_{\mn\rs}$. It has four
indices, $\mu,\nu$ for the photon , and $\r,\si$ for the gluon.
To get the contribution proportional to $\Delta g$ in 
Eqs.~(\ref{522}) and (\ref{523})  
one has to
contract it with $\ve^{\mn \alpha \beta } p_{\alpha} q_{\beta } \times 
\ve^{\rs \gamma \delta} p_{\gamma} q_{\delta }$ because 
$\ve^{\rs \gamma \delta} p_{\gamma} q_{\delta }$
is proportional to the difference of products
of gluon polarization tensors $\ve_{\r_+}\ve_{\si_+}^*-\ve_{\r_-}
\ve_{\si_-}^*$ of gluons with positive and negative helicity.
This one can see, for example, in the Breit frame where
$p=E_g (1,0,0,1)$ is the gluon momentum and $q=(0,0,0,-Q)$.
In this frame the
polarization tensors are $\ve_{\pm}={1\over\sqrt{2}}(0,1,\pm i,0)$.

To calculate the diagrams or, more precisely, the coefficient of
the $\Delta \Ga$ contribution to $\Ga_{1,S}$ it is more
appropriate to go to the c.m.s. of the photon and gluon in which
\begin{eqnarray} 
\nonumber 
q &=& (q_0,0,0,-p_1) \\ 
\nonumber
p\phantom{'}&=&(p_0,0,0,p_1) \\ 
\nonumber
k &=& (k_o,0,k_1\sin\theta,k_1\cos\theta) \\ 
p' &=& (p'_0,0,-k_1\sin\theta,-k_1\cos\theta)
\label{544}
\end{eqnarray} 
where
%\begin{eqnarray} 
%\nonumber 
%q_0&=&\sqrt{p^2_1-Q^2} \\ 
%\nonumber
%p_0&=&\sqrt{p^2_1-P^2} \\ 
%\nonumber
%p^2_1&=&
%\sqrt{(pq)^2-P^2Q^2\over s} \\ 
%k_1&=&p'_0=k_0={\sqrt{s} \over 2} \, .
%\label{545}
%\end{eqnarray} 
\begin{equation}
q_0=\sqrt{p^2_1-Q^2}, \qquad p_0=\sqrt{p^2_1-P^2}, \qquad 
p^2_1=\sqrt{(pq)^2-P^2Q^2\over s}, \qquad 
k_1=p'_0=k_0={\sqrt{s} \over 2} \, .
\label{545}
\end{equation} 
Note that $s=(p+q)^2$ and the fraction of
momentum of the gluon which is carried by the intermediate quark 
in Fig. \ref{fig14}b 
is given by $z={Q^2\over s+Q^2}$. This is the quantity with respect to
which the first moment has to be taken. A phase space integration over
the production angle $\theta$ is also necessary.

In Eq.~(\ref{545}) a gluon off--shellness $P^2=-p^2$ has been introduced.
It is needed for regularization purposes because although the
final result turns out to be finite, singular expressions arise
in intermediate steps of the calculation (from the collinear
region $\theta\to 0)$. In the off--shell sheme $(P^2\not= 0)$ the
coefficient of $\Delta \Ga$ in Eq.~(\ref{522}) turns out to be
\be
f\int\limits^1_0 dz (1-2z)
\int\limits^1_{-1}d \cos\theta{1\over 4}
 \biggl \lgroup 2-{1\over u}-{1\over t}
+P^2{z^2\over Q^2}({1\over u^2}+
{1\over t^2})\biggr \rgroup
=f\int\limits^1_0 dz (2z-1)\ln{Q^2(1-z)\over P^2 z}
=-f+0(P^2)  
\label{546}
\ee
where
\begin{eqnarray}
\nonumber
u = -(p-p')^2{z\over Q^2}\approx \ha (1+\cos\theta)+P^2{z^2\over Q^2}
\\
t = -(k-p)^2{z\over Q^2}\approx\ha (1-\cos\theta)+P^2{z^2\over Q^2}
\label{547}
\end{eqnarray} 
and $f=3$ is the number of light quarks. In Eq.~(\ref{546})
terms like ${P^2\over u}$ or ${P^2\over t}$ which do not
contribute in the limit $P^2\to 0$ have been left out. However, the
terms $\sim {P^2 \over u^2}$ and $\sim {P^2 \over t^2}$ are important. After
integration over $\theta$ a term
$\sim \Delta P_{qg} \ln P^2$ arises which drops out
because the first moment of $\delta P_{qg}^{(0)}=
\ha (2z-1)$ vanishes.
There is an overall factor of $f$ because each quark
flavor can be produced.

Instead of $P^2$ a quark mass $m$ may be introduced to
regulate the collinear singularity (`on--shell scheme').
This is the more appropriate procedure for heavy 
quarks (c and b) but less useful for the light 
quarks (u,d and s).
In that scheme one has 
instead of Eq.~(\ref{545}) 
%\begin{eqnarray} \nonumber  
%q_0 &=&\sqrt{p^2_1-Q^2} 
%\\ \nonumber
%p_0^2&=&p_1^2={(pq)^2\over s}
%\\ \nonumber
%k_0 &=&p'_0={\sqrt{s}\over 2} 
%\\
%k_1&=&{\sqrt{s}\over 2}
%\sqrt{1-4m^2/s} \, . 
%\label{548}
%\end{eqnarray}
\begin{equation}
q_0=\sqrt{p^2_1-Q^2}, \qquad p_0^2=p_1^2={(pq)^2\over s}, \qquad 
k_0=p'_0={\sqrt{s}\over 2}, \qquad k_1={\sqrt{s}\over 2} 
\sqrt{1-4m^2/s} \, . 
\label{548} 
\end{equation}  
The coefficient of $\Delta \Gamma$ is now
\begin{eqnarray} \nonumber 
 \int\limits^1_0 dz\int\limits^1_{-1}d\cos\theta
\biggl  \lgroup \ha (2z-1)\Bigl({1\over 1-\b\cos\theta}
    +
{1\over 1+\b\cos\theta}
-1\Bigr)   
\\
+{2m^2(1-z)\over s}
 \left({1\over (1-\b\cos\theta)^2}
+
{1\over (1+\beta\cos\theta)^2}\right)\biggr \rgroup
=O(m^2)
\label{549}
\end{eqnarray}
where $\beta=\sqrt{1-{4m^2\over s}}$. This time the effect of the
double pole term in Eq.~(\ref{549}) $\sim {2m^2 \over s}$ 
is such that there is no
gluon contribution (for small quark masses $m\to 0$).

If used for the light quarks, Eq.~(\ref{549}) 
seems to be in contradiction with Eq.~(\ref{518}) because they seem to
imply, at some conveniently chosen input scale $Q^2=\mu^2$, 
two different relations between the axial vector
singlet current matrix element and $\Delta \Sigma$ and $\Delta \Ga$
namely
\begin{equation}
A_0(\mu^2)=\cases{\Delta \Sigma_{on}(\mu^2)     & on--shell scheme
\cr
\Delta \Sigma_{off}-f\Delta\Ga(\mu^2)     &off--shell scheme\cr}
\, .
\label{550}
\end{equation}
$\Delta \Gamma$ is scale independent in first order but picks
up a scale dependence in higher orders.
It is well known,
for example, from the calculation
of QCD corrections to unpolarized structure functions , 
that different regularization schemes
can lead to different results. Usually
this is interpreted in such a way that the use of
different regularization schemes corresponds to different
definitions of the quark density. In our case this means
that one has to deal with two different $\Delta \Sigma$'s
which are denoted by $\Delta \Sigma_{on}$ and $\Delta \Sigma_{off}$
in Eq.~(\ref{550}).
Going from the second to the first relation in Eq.~(\ref{550}) 
means that one absorbs
the gluon contribution into a redefinition of $\Delta \Sigma_{off}$. 
The obtained result, $\Delta \Sigma_{on}$, corresponds to the one in the 
conventional $\overline{MS}$ scheme in Eq.~(\ref{518}) where 
$\Delta \Gamma (\mu^2)$ does not explicitly occur due to 
$\Delta C_g=0$ in Eq.~(\ref{4321b}). Although in principle it is 
a mere matter of convention to choose a particular factorization 
scheme the use of the off--shell scheme (where a potentially large gluon
contribution exists) has the advantage that $\Delta \Sigma_{off}$ 
is conserved, i.e. scale independent, as will be shown below. 
This way the Adler sum rule \cite{adler} 
%Usually, this would be no problem because usually the terms by
%which the quark density is redefined are small contaminations because
%they are higher order and eventually vanish at large Q.
%In the case at hand
%this is not true. We have seen in section 5.1 that $\d\Ga$ is not
%a priori small. In this case a {\it physical\/} definition of the
%quark contribution is needed. We propose to use the definition
%of the off--shell scheme (where a potentially large gluon contribution
%exists) because it can be shown that $\d\Sigma_{off}$ is
is extended to higher
orders of polarized scattering. Furthermore,
for the case of the light quarks $(u,d,s$ with $m<\Lambda_{\rm QCD})$
the massless scheme might be more physical because it could be related 
to the notion of constituent quarks \cite{altarelli4,reya2}. 
To really understand
why the two definitions in Eq.~(\ref{550}) do not contradict 
each other, one must 
write down the full result for the first moment of the
singlet component of $g_1$ in Eq.~(\ref{416}), $\int^1_0 dxg_{1,S}$, in both
schemes and compare it with Kodaira's original result, cf. Eq.~(\ref{523}) 
\cite{kodaira}. In order to do this it should be noted that  
$\int^1_0 dxg_{1,S}$ is a physical observable and therefore the
prediction for it must be scheme independent. 
The physical prediction is always a product of coefficient functions,
anomalous dimensions and the parton densities/matrix elements, i.e. 
\begin{eqnarray}
\int\limits^1_0 dx g_{1,S}(x,Q^2) = {1\over 9} C_{off} (Q) E_{off}
(Q,\mu)\pmatrix{\Delta\Sigma_{off}\cr\Delta\Ga (\mu)\cr}
\label{551}
\\ 
={1\over 9} C_{on} (Q)E_{on} (Q,\mu)
  \pmatrix{\Delta\Sigma_{on} (\mu)\cr \Delta\Ga (\mu)\cr}
\label{552}
\\
\, ={1\over 9}C_{\rm Kodaira} (Q)E_{\rm Kodaira} (Q,\mu)A_0(\mu)
\label{553}
\end{eqnarray}
in the three schemes to be compared, with  
\begin{equation}
C_{\rm Kodaira}=1+{\a_s (Q)\over 2\pi}(-{3\over 2} C_F)
\label{554}
\end{equation} 
and
\begin{equation}
E_{\rm Kodaira} =1+{\a_s (\mu)-\a_s (Q)\over \pi\b_0}
(-{3\over 2} C_F f)
\label{555}
\end{equation}
are taken from 
the work of Kodaira \cite{kodaira} 
which was done with massless quarks
and off--shell gluons. For simplicity we work below the charm threshold 
so that $f=3$. 
Furthermore \cite{altarelli6}, 
\begin{equation}
C_{off} =(1,-f)+{\a_s (Q)\over 2\pi}(-{3\over 2} C_F,
        c_{\Gamma_{off}})
\label{556}
\end{equation}
\begin{equation}
E_{off} = 1+{\a_s (\mu)-\a_s (Q)\over \pi\b_0}
\pmatrix{0                  &0\cr
        {3\over 2}C_F &\Delta P^{(1)}_{gg,off}-{\b_1 \over 4} \cr} \, .
\label{557}
\end{equation}
Here
$c_{\Gamma_{off}}$ is a second order correction to the first order 
gluon
photon--fusion process (Fig. \ref{fig14}b). Eq.~(\ref{550}) 
and the
entry $-{3\over 2}C_F$ in Eq.~(\ref{556})  give rise to the
QCD correction factor $1-{\alpha_s (Q)\over\pi}$ in the 
coefficient of $A_0$ in Eq.~(\ref{523}).
Both $c_{\Gamma_{off}}$ and 
$\Delta P^{(1)}_{gg,off}$ can be
determined by comparison with Kodaira's result, 
Eqs.~(\ref{554}) and (\ref{555}),  
i.e.\ by the
consistency requirement Eqs.~(\ref{551})--(\ref{553}), 
which yields \cite{altarelli6} 
$\Delta P^{(1)}_{gg,off}-{\beta_1 \over 4}=-f \Delta P^{(0)}_{gq}=-2f$.
The form of the matrix in Eq.~(\ref{557}) is quite remarkable, 
in particular the vanishing of the entries in the first 
row. According to Eq.~(\ref{542}) this corresponds to 
\begin{equation}
\Delta P^{(1)}_{qq,off}=0  
\label{558}
\end{equation}
and 
\begin{equation}
\Delta P^{(2)}_{qg,off}=0  \, .
\label{559}
\end{equation}
Note that the vanishing of $\Delta P^{(0)}_{qq}$ and 
$\Delta P^{(0)}_{qg}$ is explicit from Eq.~(\ref{4126}) and the 
vanishing of $\Delta P^{(1)}_{qg}$ has been proven by  
\cite{altarelli2}. In \cite{altarelli2} the entry 
$\Delta P^{(1)}_{gg,off}-{\b_1 \over 4}$ was called $\gamma^{(2)}_{gg}$. 
Eqs.~(\ref{558}) and (\ref{559}) go a step further, saying that in a 
suitable chirality conserving regularisation scheme,  
$\Delta \Sigma$ is a conserved (i.e. $Q^2$--independent) quantity 
beyond the leading order, in fact to any order. 
It is well known that in massless perturbative QCD the 
chiral symmetry is unbroken to any order. 
Eqs.~(\ref{558}) and (\ref{559})  
show that the off--shell scheme 
realizes this fact in the most intuitive way 
by forbidding chirality flip interactions to any order 
of perturbation theory. As we shall see in the following 
this fact is also intimately related to the 
reasonable treatment 
of the ABJ anomaly in the off--shell scheme. 
The anomaly term $-f$ in $C_{off}$, Eq.~(\ref{556}), 
will be shown to correspond  
-- via the consistency requirement Eqs.~(\ref{551})--(\ref{553}) -- 
to a conserved $\Delta \Sigma$ and vice versa. 

Eqs.~(\ref{558}) and (\ref{559}) can be derived, for example, from the 
consistency requirement Eqs.~(\ref{551})--(\ref{553}). To see that 
let us evaluate the products 
$C_{off}E_{off} (\Delta\Sigma_{off} , \Delta\Gamma )^T$ 
and $C_{Kodaira}E_{Kodaira}A_0$ appearing in Eqs.~(\ref{551}) and (\ref{553}): 
\begin{eqnarray} \nonumber 
C_{off}(Q) E_{off}(Q,\mu)
\pmatrix{\Delta\Sigma_{off}(\mu)\cr\Delta\Gamma (\mu) \cr} =
   \biggl \lgroup   (1-{3 \over 2}C_F{\alpha_s (\mu)\over 2\pi}
+{\alpha_s (\mu)-\alpha_s (Q)\over \pi\b_0} 
 (\Delta P^{(1)}_{qq,off} 
\\ \nonumber 
-f \Delta P^{(0)}_{gq}+
{3 \over 2}C_F{\beta_0 \over 2}) \biggr \rgroup
 \Delta\Sigma_{off}(\mu)
 + \biggl \lgroup   -f+c_{\Gamma_{off}}{\alpha_s (\mu)\over 2\pi}
\\
+{\alpha_s (\mu)-\alpha_s (Q)\over \pi\b_0}
 [\Delta P^{(2)}_{qg,off}
-f (\Delta P^{(1)}_{gg}-{\beta_1 \over 4})
-c_{\Gamma_{off}}{\beta_0 \over 2} ] \biggr \rgroup  \Delta\Gamma (\mu)
+O(\alpha_s^2)
\label{560}
\end{eqnarray}
\begin{eqnarray} \nonumber 
C_{Kodaira}  (Q)E_{Kodaira}(Q,\mu)A_0(\mu)=
 \biggl \lgroup   1-{3 \over 2}C_F{\alpha_s (\mu)\over 2\pi}
  +{\alpha_s (\mu)-\alpha_s (Q)\over \pi \b_0}
 (-{3 \over 2}C_F f +
{3 \over 2}C_F {\beta_0 \over 2}) \biggr \rgroup
\\
\times [ \Delta \Sigma_{off}(\mu)-f\Delta\Ga(\mu) ] +O(\alpha_s^2)
 \, . \label{561}
\end{eqnarray}
By comparison one obtains (among other things)  
\begin{equation}
\Delta P^{(1)}_{qq,off}-f\Delta P^{(0)}_{gq}=-{3 \over 2}C_F f \, .
\label{562}   
\end{equation}
Because of the unambiguously determined leading order 
$\Delta P^{(0)}_{gq}={3 \over 2}C_F$,  
the right hand side of (\ref{562}) (=Kodaira's 
anomalous dimension) is saturated by the second term on the 
left hand side, so that 
one arrives at Eq.~(\ref{558}).  
Thus $\Delta \Sigma_{off}(\mu)$ is scale 
independent, i.e. conserved.  

Until now we have discussed only the scheme dependence of the
coefficient of $\Delta g$ in Eq.~(\ref{522}). It turns out
that other quantities differ in the 
on--shell scheme from their corresponding 
counterparts in the off--shell scheme as well.
Therefore, we make a general ansatz
\begin{equation}
C_{on}(Q)=(1,0)+{\alpha_s (Q)\over 2\pi}(C_{\Sigma_{on}},
\/c_{\Gamma_{on}})
\label{563}   
\end{equation}
\begin{equation}
E_{on}(Q,\mu)=1+{\alpha_s (\mu)-\alpha_s (Q)\over \pi\b_0}
\pmatrix{\Delta P^{(1)}_{qq,on}  &\Delta P^{(2)}_{qg,on}\cr
         {3\over 2} C_F       &\Delta P^{(1)}_{gg,on}-{\b_1 \over 4} \cr}
\label{564}   
\end{equation}
for the coefficients and anomalous dimensions in the on--shell
scheme. The entry $\Delta P^{(0)}_{gq}={3\over 2}C_F$ in the anomalous
dimension matrix is scheme independent. Due to the
consistency requirement Eqs.~(\ref{551})--(\ref{553})  for the physical
observable there must be a transformation matrix
\begin{equation}
M(Q)=\pmatrix{1+{\alpha_s (Q)\over 2\pi}m_{11} &-f\cr
                0                    &1\cr}
\label{565}   
\end{equation}
such that
\begin{equation}
C_{on}(Q) = C_{off}(Q)M^{-1}(Q)  
\label{566}   
\end{equation}
\begin{equation}
E_{on}(Q,\mu)=M(Q)E_{off}(Q,\mu)M^{-1}(\mu) 
\label{567}   
\end{equation}
\begin{equation}
\pmatrix{\Delta\Sigma_{on}\cr
          \Delta\Gamma(\mu)\cr}=
M(\mu)\pmatrix{\Delta\Sigma_{off}(\mu)\cr
                \Delta\Gamma(\mu)\cr} \, .       
\label{568}   
\end{equation}
$\Delta\Gamma$ is not redefined at the order at which we are
working; ergo the two entries 0 and 1 in the second row 
of M(Q). 
Eqs.~(\ref{566})--(\ref{568}) can be used to calculate $m_{11}$
as well as the complete set of quantities in $C_{on}$ 
and $E_{on}$. One simply has to work out the matrix expressions 
in Eqs.~(\ref{566})--(\ref{568}). Equivalently, these quantities 
can be derived by using the transformations Eqs.~(\ref{4230}) and (\ref{4231}) 
of Sect. 4.2 for n=1 (with the transition from 
$\Delta g$ to $\Delta\Gamma$ to be carried out). 
Here we do not want to bother with all the details 
of the transition to the on--shell scheme, but concentrate on one particular 
aspect, the non-conservation of $\Delta\Sigma_{on}(\mu)$ which 
becomes explicit in the result \cite{altarelli6}
\begin{equation}
\Delta P^{(1)}_{qq,on}=-{3\over 2}C_F f  
\label{569}   
\end{equation}
which makes clear
that $\Delta\Sigma_{on}(\mu)$ is not a conserved but a scale
dependent quantity (in contrast to $\Delta\Sigma_{off}$).
This result follows from the consistency requirement 
Eq.~(\ref{567}) (by comparison with the off--shell scheme) 
but also from the consistency requirement 
Eqs.~(\ref{552}) and (\ref{553}) (by comparison with Kodaira's result), 
in a similar fashion as $\Delta P^{(1)}_{qq,off}=0$ 
was derived in Eq.~(\ref{558}). 

\begin{figure}
\begin{center}
\epsfig{file=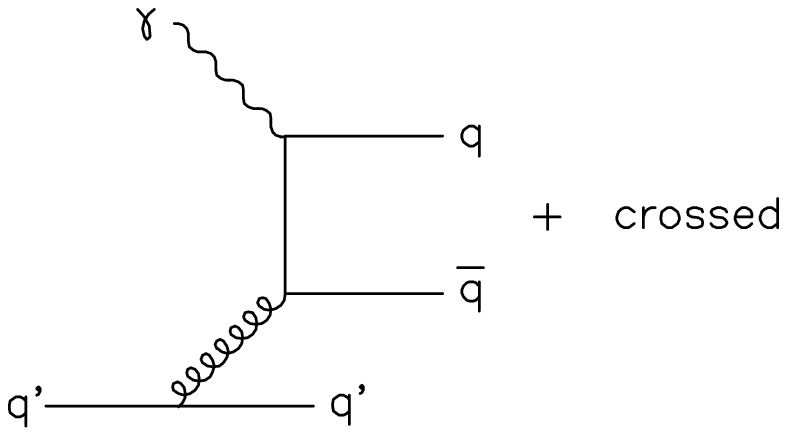,height=4cm}
\vskip 1.cm
\caption{DIS process involving non diagonal transitions  
between quarks of different flavors.}
\label{fig19}
\end{center}
\end{figure}

All of the above considerations where carried out in a 
decent but rather abstract way by imposing the condition of the 
scheme independence of a physical observable. 
An explicit check of the truthfulness of the whole 
approach was made in reference \cite{altarelli6} where 
the relation (\ref{569}) 
was reproven by an explicit two-loop calculation in the 
on--shell scheme (massive quarks). The main point was to 
show that the diagrams in Fig.  \ref{fig19} 
give rise to a nonvanishing anomalous dimension when 
calculated in the scheme with massive quarks, and a 
vanishing anomalous dimension in the off--shell scheme. 
More precisely, the matrix element corresponding to 
Fig.  \ref{fig19} is of the form 
\begin{equation}
ME(Fig. \ref{fig19})=\Delta P^{(1)ND}_{qq} \ln {Q^2 \over \mu^2} +const.
\label{570}   
\end{equation}
after integration over the appropriate phase space \cite{altarelli6}. 
Here ND denotes "non-diagonal" transitions between quarks of 
different flavors q and q' (cf.~Fig.~\ref{fig19}). 
It was shown in  \cite{altarelli6} that 
$\Delta P^{ND(1)}_{qq,on}=-{3\over 2}C_F f$ for any 
masses m and m' of the quarks q and q', whereas in the 
off--shell scheme $\Delta P^{(1)ND}_{qq,off}=0$. 
In other words, the change in $\Delta P^{(1)}_{qq}$ when 
going from the off--shell to the on--shell scheme is entirely 
due to non-diagonal flavor transitions. 
In this way   
a completely consistent picture
arises, in which results of the off--shell scheme can be 
transformed to any other scheme and vice versa. 
Note that diagram Fig. \ref{fig19} can be obtained from 
Fig.  \ref{fig22} by cutting. 

Most important among the other schemes is of course the 
$\overline{MS}$  
scheme, because the two-loop anomalous dimensions  
are now known 
for {\it all} moments \cite{mertig,vogelsang2}. 
The results for the first moment 
were already given in Sect. 4.3. Here we recollect them  
for the convenience of the reader. One has, according to 
Eqs.~(\ref{4321a})--(\ref{4322b}) (see footnote \ref{fn4})    
\begin{equation}
C_{\overline{MS}}(Q)=(1,0)+{\alpha_s (Q)\over 2\pi}
(-{3\over 2}C_F,
\/c_{\Gamma_{\overline{MS}}})
\label{571}   
\end{equation}
\begin{equation}
E_{\overline{MS}} = 1+{\alpha_s (\mu)-\alpha_s (Q)\over \pi\b_0}
\pmatrix{-{3\over 2}C_F f                  &0\cr
        {3\over 2}C_F &\Delta P^{(1)}_{gg,\overline{MS}}-{\b_1 \over 4}\cr} 
\label{572}   
\end{equation}
where $c_{\Gamma_{\overline{MS}}}={3\over 2}C_F f$, which 
will contribute only in NNLO $\alpha_s^2$  
and $\Delta P^{(1)}_{gg,\overline{MS}}-{\b_1 \over 4}=0$ 
can be also obtained from 
consistency by comparison with Kodaira's result, namely from
\begin{eqnarray} \nonumber
\int\limits^1_0 dx g^S_1(x,Q^2) &=& {1\over 9} C_{\overline{MS}} (Q) 
E_{\overline{MS}}
(Q,\mu)\pmatrix{\Delta\Sigma_{\overline{MS}}(\mu)\cr\Delta\Ga (\mu)\cr} 
\\
& & ={1\over 9}C_{\rm Kodaira} (Q)E_{\rm Kodaira} (Q,\mu)A_0(\mu) \quad , 
\label{573}
\end{eqnarray}
in agreement with the results in Eq.~(\ref{519}). It should be noted 
that the entry 0 in Eq.~(\ref{571}) corresponds to a vanishing 
$\Delta C_g$ in Eq.~(\ref{4224}) or (\ref{4315}), i.e. to the fact 
that $\Delta g$ decouples from 
% the singlet contribution $A_0$ to 
$\Gamma_{1,S}^{p,n}$ in the $\overline{MS}$ scheme, as well as in the 
on--shell scheme according to Eq.~(\ref{563}). 

It is now instructive, after these many technical details, 
to summarize the NLO evolution equations for the first moments 
of the flavor singlet quantities in the different schemes. 
In the $\overline{MS}$ and on--shell schemes we have 
\begin{eqnarray} 
{d \over dt} \Delta \Sigma (Q^2)_{\overline{MS},on}&=& 
\left( {\a_s(Q^2) \over 2\pi}\right) ^2 (-2f) 
\Delta \Sigma (Q^2)_{\overline{MS},on} 
\label{574}
\\
{d \over dt} \Delta \Gamma (Q^2)_{\overline{MS},on}&=& 
\left( {\a_s(Q^2) \over 2\pi}\right) ^2 2 \Delta \Sigma (Q^2)_{\overline{MS},on} 
\label{575}
\end{eqnarray}
with the singlet contribution $A_0$ to $\Gamma_1^{p,n}$ being 
given by $A_0(Q^2)=\Delta \Sigma (Q^2)_{\overline{MS},on}$. 
In the off--shell scheme we have obtained 
\begin{equation} 
{d \over dt} \Delta \Sigma (Q^2)_{off}=0
\label{576}
\end{equation}
\be 
{d \over dt} \Delta \Gamma (Q^2)_{off}= 
\left( {\a_s(Q^2) \over 2\pi}\right) ^2 [ 2 \Delta \Sigma_{off}
-2f\Delta \Gamma (Q^2)_{off}] 
\label{577}
\ee
%\begin{eqnarray}
%{d \over dt} \Delta \Sigma (Q^2)_{off}&=&0
%\label{576}
%\\
%{d \over dt} \Delta \Gamma (Q^2)_{off}&=& 
%\left( {\a_s(Q^2) \over 2\pi}\right) ^2 [ 2 \Delta \Sigma_{off}
%-2f\Delta \Gamma (Q^2)_{off}] 
%\label{577}
%\end{eqnarray}
with the singlet contribution now being given by $A_0(Q^2)=\Delta 
\Sigma_{off}-f\Delta \Gamma (Q^2)_{off}$ as was already 
anticipated in Eq.~(\ref{550}). Due to the $Q^2$--independence 
of $\Delta \Sigma_{off}$, Eq.~(\ref{576}), these latter two RG 
equations can be combined into one simple evolution equation for 
the singlet combination $A_0$: 
\begin{equation}
{d \over dt}[\Delta \Sigma-f \Delta \Gamma (Q^2)]_{off}= 
\left( {\a_s(Q^2) \over 2\pi}\right) ^2 (-2f) 
[\Delta \Sigma-f \Delta \Gamma (Q^2)]_{off} \, . 
\label{578}   
\end{equation}
Recalling 
\begin{equation}
{d \alpha_s \over dt}=- \beta_0 {\a_s^2 \over 4\pi} 
- \beta_1 {\a_s^3 \over (4\pi)^2} \quad ,
\label{579}   
\end{equation}
the solutions of the above RG equations are straightforward. In the 
$\overline{MS}$ and in the on--shell scheme, Eqs.~(\ref{574}) and 
(\ref{575}) give 
\be 
\Delta \Sigma(Q^2)_{\overline{MS},on}  =  \left\{ 1-{2f \over \beta_0} 
{\a_s(\mu^2)-\a_s(Q^2) \over \pi}+O(\a_s^2 ) \right\} 
\Delta \Sigma(\mu^2)
_{\overline{MS},on} 
\label{580}
\ee
\be
\Delta \Gamma (Q^2)_{\overline{MS},on}  =  \Delta \Gamma (\mu^2)
_{\overline{MS},on}
-{2f \over \beta_0}                         
[{\a_s(\mu^2)-\a_s(Q^2) \over \pi}+O(\a_s^2 )]\Delta \Sigma(\mu^2)
_{\overline{MS},on}
\label{581}
\ee
with an appropriately chosen input scale $Q_0^2=\mu^2$ and 
$\beta_0=11-{2f \over 3}$. In the off--shell scheme, 
Eqs.~(\ref{576})--(\ref{578}) give 
\begin{equation}
\Delta \Sigma(Q^2)_{off}=\Delta \Sigma(\mu^2)_{off}\equiv \Delta \Sigma = 
const. 
\label{582}        
\end{equation}     
\be 
\Delta \Gamma (Q^2)_{off}=\left\{ 1-{2f \over \beta_0} 
{\a_s(\mu^2)-\a_s(Q^2) \over \pi}+O(\a_s^2 ) \right\} 
 \Delta \Gamma (\mu^2)_{off} 
+{2 \over \beta_0}{\a_s(\mu^2)-\a_s(Q^2) \over \pi}\Delta \Sigma \, .
\label{583}        
\ee
As an illustrative application of these results, let us estimate 
dynamically $\Delta g(Q^2)$ in the off--shell scheme according to 
Eq.~(\ref{583}) \cite{glueck11,kunz,glueck12}. The necessary input 
at $Q^2=\mu^2$, i.e. the boundary condition, 
can be deduced from the unpolarized {\it valence--like} 
gluon and sea input densities (i.e. $xg(x,\mu^2) \sim x^a(1-x)^b$ with 
$a>0$, 
etc.) which are thus integrable, i.e. their n=1 moments $g(\mu^2) 
\equiv \int_0^1 g(x,\mu^2)dx$ etc. exist (only) at $Q^2=\mu^2$ 
with the result $g(\mu^2)\approx 1$ and $s(\mu^2)\approx 0$ at 
$\mu^2 \approx 0.3$ GeV$^2$ \cite{glueck5,glueck9,glueck6}. 
This allows for a parameter--free 
as well as perturbatively stable calculation of structure functions in 
the small--x region ($x \leq 10^{-2}$, at $Q^2>\mu^2$) which is 
entirely based on QCD dynamics and agrees with all present 
measurements obtained at DESY--HERA \cite{glueck6}. Thus, the 
positivity constraints (\ref{4111}) imply 
\begin{equation}
|\Delta g(\mu^2)| \leq g(\mu^2) \approx 1 \quad , \qquad 
|\Delta s(\mu^2)| \leq s(\mu^2) \approx 0 \, .
\label{584}   
\end{equation}
Using therefore $\Delta \Sigma \approx A_8$, with $A_8$ being 
given in (\ref{59}), Eq.~(\ref{583}) yields \cite{glueck12} 
\begin{equation}
-2.6 \lesssim \Delta g(Q^2)_{off} \lesssim 3.9 
\label{585}   
\end{equation}
for $Q^2=10$ $GeV^2$, which is compatible with Eq.~(\ref{525}). 
Here we have used $\beta_0=9$ for $f=3$ light quark flavors 
and ${\a_s(\mu^2) \over 2\pi}=0.108$ and 
${\a_s(10 GeV^2) \over 2\pi}=0.0304$ in NLO. It is also interesting to note 
that on rather general heuristic grounds based on the intrinsic 
bound--state dynamics of the nucleon, counting rules, consistency 
constraints for ${g_- \over g_+}$ and ${\Delta g \over g}$ 
as $x \rightarrow 1$ and $x \rightarrow 0$, etc., one expects \cite{brodsky2} 
the total gluon helicity to be sizeable, $\Delta g \approx 1.2$. 
Note, however, that  
the scale $\mu$ remains undetermined in such considerations, 
in contrast to the RG based result in Eq.~(\ref{585}). 

In the $\overline{MS}$ as well as in the on--shell scheme 
where the full first 
moment $\Delta g$ decouples from $A_0(Q^2)$ and $\Delta \Sigma(Q^2)$ 
is not conserved, cf. Eq.~(\ref{580}), the experimentally 
required suppression  
of $A_8$ in Eq.~(\ref{57}), as required by the constraint (\ref{524}), 
can be achieved either by $\Delta s=\Delta \bar{s} <0$ or by 
$\Delta \bar{u} \approx \Delta \bar{d} <0$ in a $SU(3)_f$ 
broken scenario. 
A detailed $(x,Q^2)$ dependent 
analysis of present polarization data  
will be presented in Sect. 6. Despite 
the fact that $\Delta g$ decouples from $A_0$, the polarized gluon 
density $\delta g(x,Q^2)$ will in future play an important part 
in analyzing presently available data for $x \geq 5 \times 10^{-3}$ which 
give rise to a sizeable $\Delta g$ as well. 

As already mentioned in Sect. 3.4 and to be discussed in more 
detail in Sect. 6, the best information on $\delta g(x,Q^2)$ 
will come from heavy quark production in polarized $\vec{e}\vec{p}$ 
via $\vec\gamma^{\ast}\vec{g} \rightarrow c \bar{c}$ 
(with no light quark contribution present in LO), or 
in polarized $\vec{p}\vec{p}$ collisions at the RHIC 
facility via the dominant $\vec{g}\vec{g} \rightarrow c \bar{c}$ 
subprocess; in the latter case also the production of 
direct photons should offer a good possibility to 
extract information about the magnitude of $\Delta g(Q^2)$. 
Alternatively, $\delta g(x,Q^2)$ could be determined from 
the $Q^2$--dependence of $g_1$ if HERA is run 
with a polarized proton beam.

%% file: k54tex
\subsection{The Bjorken Sum Rule }

Sum rules are relations for moments of the structure functions.
The most important and fundamental sum rule for polarized scattering
is the Bjorken sum rule which was derived in 1966 from SU(6)$\otimes$SU(6) 
current algebra \cite{bjorken}. It describes a relationship between spin 
dependent DIS and the weak coupling constant
defined in neutron $\beta$-decay 
\begin{equation}
\int\limits^1_0 dx [g_1^{ p}(x, Q^2)-g_1^{ n}(x, Q^2)]={1\over 6}
{g_A\over g_V}
\label{586}
\end{equation}
where $g_A$ and $g_V$ are the vector and axial vector couplings
measured in nuclear $\beta$-decay
\begin{equation}
{\cal L}_{\b-\hbox{decay}}=- 
{G_F \over \sqrt{2}}
\cos\theta_c
[
\bar p\gamma_{\mu} 
(1-{g_A\over g_V}\g5 ) 
n ] 
[
\bar e\gamma^{\mu} 
(1-\g5)\nu 
] \, .
\label{587}
\end{equation}
Here $\theta_c$ is the Cabibbo angle and ${g_A\over g_V}=1.2573\pm 0.0028$
in Eq.~(\ref{58}) is known to high precision.

It is not difficult to prove the Bjorken sum rule with the
help of the knowledge which was collected in Sect. 5.1.  
Qualitatively,it arises as follows: ${g_A\over g_V}$ gives the 
strength at which the axialvector transformation from up to 
down quark in the proton takes place. In DIS this is measured 
by $\Delta (u+\bar u) - \Delta (d +\bar d)$. Indeed,
from Eqs.~(\ref{55}), (\ref{530}) and (\ref{587}) one sees that one
simply has to show
\begin{equation}
{}_p\langle PS\vert\bar\psi\ga_\mu\g5\lambda_3\psi\vert PS\rangle_p
= {g_A\over g_V} S_\mu \, .
\label{588}
\end{equation}
One can use the isospin symmetry to apply the Wigner--Eckhart theorem
\begin{equation}
{}_p\langle PS\vert\bar\psi\ga_\mu\g5\lambda_3\psi\vert PS\rangle_p
={}_p\langle PS\vert\bar\psi\ga_\mu\g5\lambda_+\psi\vert PS\rangle_n
\label{589}
\end{equation}
where $\vert\rangle_p$ and $\vert\rangle_n$
denote the proton and neutron states. From Eq.~(\ref{589}) one gets
\begin{equation}
{}_p\langle PS\vert\bar u\ga_\mu\g5 d\vert PS\rangle_n =
{g_A\over g_V} S_\mu
\label{590}
\end{equation}
which completes the proof of the Bjorken sum rule.

This sum rule is so very fundamental because it relies only 
on isospin invariance, i.e. on a $SU(2)_f$ symmetry between 
up-- and down--quarks, cf. Eq.~(\ref{589}). The Bjorken sum 
rule is an asymptotic result which relates low-- and high--$Q^2$ 
quantities. This originates from the fact that, apart from finite 
$O(\a_s)$ corrections, the l.h.s. of Eq.~(\ref{586}) 
$\Gamma_1^p(Q^2)-\Gamma_1^n(Q^2)={1 \over 6} A_3= {1 \over 6} 
(\Delta u +\Delta \bar{u} -\Delta d -\Delta \bar{d})$ is just  
the nonsinglet 
isospin--3 component in Eq.~(\ref{54}) which does not 
renormalize due to the vanishing of $\Delta P_{qq}^{(0)}$ and 
$\Delta P_{NS-}^{(1)}$ in (\ref{519}). Thus it is generally 
believed that the Bjorken sum rule gets only moderate QCD 
corrections. For example, the perturbative QCD corrections are 
given by the nonsinglet Wilson coefficient $\Delta C_{NS}$ in (\ref{520}) 
and known to be small. In addition, there are probably non-perturbative 
higher--twist (HT) corrections , so that  
instead of (\ref{586}) we have in general 
\begin{equation}
\int_0^1 dx [g_1^p(x,Q^2)-g_1^n(x,Q^2)]={1 \over 6}{g_A \over g_V} 
\Delta C_{NS}(Q^2)+{c_{HT} \over Q^2} 
\label{591}
\end{equation}
with \cite{kodaira1,kodaira2,larin2} 
\be 
\Delta C_{NS} =1-{\alpha_s(Q^2) \over \pi}-
3.5833 ({\alpha_s(Q^2) \over \pi})^2
-20.2153 ({\alpha_s(Q^2) \over \pi})^3 \, , 
\label{592} 
\ee
cf. Eq.~(\ref{520}). 
The latter result is obtained for $f=3$ light flavors, and the range 
of $c_{HT}$ has been estimated to be $c_{HT} \approx -0.025$ to 
$+0.03$ GeV$^2$ 
\cite{balitzky,ji}. For recent reviews, see 
\cite{ioffe1,mankiewicz2,hinchliffe} and 
references therein. Additional renormalon contributions 
to the perturbative 3--loop result in Eq.~(\ref{592}) have been studied 
as well but their size seems to be small in the relevant perturbative 
region, $Q^2 \gtrsim 1$ GeV$^2$. For a review, see \cite{ellis5} 
and references therein. Disregarding the small non-perturbative 
contribution for $Q^2 > 1$ GeV$^2$, 
the Bjorken sum rule (\ref{591}) is in reasonable 
agreement with experiments:  According to Table 2, SLAC(E143) finds 
\begin{equation}
\Gamma_1^p(Q^2)-\Gamma_1^n(Q^2)=0.164 \pm 0.017 
\label{593}   
\end{equation}
at $Q^2=3$ GeV$^2$, to be compared with the predicted 
[r.h.s. of (\ref{591})]   
$\Gamma_1^p(Q^2)-\Gamma_1^n(Q^2)=0.187 \pm 0.002$ at the same $Q^2$, 
using ${\a_s \over \pi}=0.076 \pm 0.010$. 
The most recent CERN (SMC) result is \cite{adeva3} 
\begin{equation}
\Gamma_1^p(Q^2)-\Gamma_1^n(Q^2)=0.195 \pm 0.029 
\label{594}   
\end{equation}
at $Q^2=10$ GeV$^2$, where the theoretical prediction is 
$\Gamma_1^p(Q^2)-\Gamma_1^n(Q^2)=0.193 \pm 0.002$, using 
${\a_s \over \pi}=0.061 \pm 0.004$. A more detailed comparison 
between theory and experiment can be found, for example, 
in \cite{voss,ellis6,altarelli10}. 

Finally, it should be kept in mind that the original small EMC 
result for $\Gamma_1^p$ in Eq.~(\ref{321}) implied already dramatic
consequences for $\Gamma_1^n$ {\it prior} to the recent CERN and 
SLAC measurements of the neutron structure function $g_1^n$, 
by assuming the validity of the 'safe' Bjorken sum rule: 
Using Eq.~(\ref{321}) in (\ref{591}) one anticipated 
\begin{equation}
\Gamma_1^n(Q^2 \approx 10 GeV^2)=-0.067 \pm 0.018 
\label{595} 
\end{equation}
which, being in agreement with recent measurements in Table 2, 
is about ten(!) times larger than the naive pre--EMC Ellis--Jaffe 
expectation $\Gamma_{1,EJ}^n \approx -0.008$ based on Eq.~(\ref{515}) 
in conjuction with the Bjorken sum rules. These predictions for 
$\Gamma_1^n$ can be translated into Bjorken--x space, 
\begin{equation}
g_1^n(x,Q^2)=g_1^p(x,Q^2)-{1\over 6}[\delta u_v(x,Q^2)-\delta d_v(x,Q^2)] 
\label{596}
\end{equation}
where $\delta q_v \equiv \delta q -\delta \bar{q}$. The small NLO 
contribution due to $\delta C_q(x)$ in Eq.~(\ref{429}) has been
suppressed. In order to reproduce the strongly negative 
x--integrated result Eq.~(\ref{595}), it can be anticipated 
\cite{glueck13} from Eq.~(\ref{596}) that $g_1^n(x,Q^2)$ has to become 
strongly negative for $x \lesssim 0.1$, in contrast to 
$g_{1,EJ}^n$ which gives rise to an almost vanishing 
$\Gamma_{1,EJ}^n$. It will become transparent from a detailed 
$(x,Q^2)$--dependent analysis of all present data in Sect. 6, 
that such expectations have been confirmed by the 
more recent CERN(SMC) and SLAC measurements.

%% file: k55tex
\subsection{The Drell--Hearn--Gerasimov Sum Rule} 

The DHG sum rule \cite{drell,gerasimov} is a prediction for $\Gamma_1$ at 
$Q^2=0$  
and can be considered as giving qualitative information
on the magnitude of higher twist effects 
in the region between $Q^2=0$ and the present 
high energy data. 
 
It relates the spin-dependent scattering cross section 
of circularly polarized real photons by longitudinally polarized
nucleons N to the anomalous magnetic moment of the nucleon. 
If we define 
\begin{equation}
I_N(Q^2) \equiv 2 { M^2 \over Q^2} \int_0^1 g_1^N(x,Q^2)dx \quad ,  
\label{597}   
\end{equation}
the statement of the DHG sum rule is 
\begin{equation}
I_N(0)=-{\kappa_N^2 \over 4} 
\label{598}   
\end{equation}
where $M$ is the mass of the nucleon and 	
$\kappa_N$ its anomalous magnetic 
moment, $\mu_p=(1+\kappa_p)\mu_B$ and $\mu_n=\kappa_n\mu_B$. 
$\kappa_N$ is defined by the nucleon--photon coupling 
\begin{equation}
\gamma^{\alpha}(i\partial_{\alpha}-e_NA_{\alpha})-{\kappa_N\mu_B \over 2} 
                                  \sigma_{\alpha\beta}F^{\alpha\beta} 
\label{599}   
\end{equation}
with $e_p=+e$ and $e_n=0$ and from experiment one has 
$\kappa_p=1.79$ and $\kappa_n=-1.91$. 

The derivation of the DHG sum rule relies on the relation 
between $g_1$ and the photoabsorption cross sections, 
Eqs.~(\ref{224}) and (\ref{227}), and on the dispersion relation 
for forward Compton scattering. In fact, the spin--flip part of the 
forward Compton scattering amplitude is proportional to 
$\int_0^{\infty} [\sigma_{1/2}-\sigma_{3/2}] {d\nu \over \nu}$ 
and, at low 
$Q^2$, is given by $-{2\pi^2\alpha \over M^2} \kappa_N^2$ 
which is usually referred to as the Low--theorem \cite{low}. 

\begin{figure}
\begin{center}
\epsfig{file=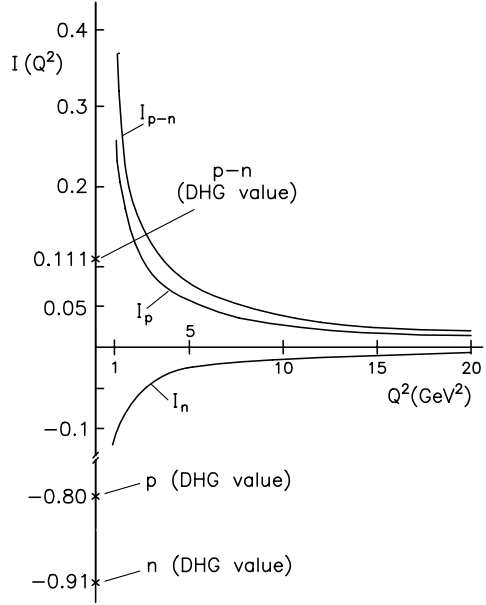,height=8cm}
\vskip 1.cm
\caption{DIS predictions for $I_{p,n}(Q^2)$ according to (\ref{597}) 
using 
$\Gamma_1^p=0.146$ and $\Gamma_1^n=-0.064$. 
These latter numbers 
are 
$Q^2$--independent in LO \protect\cite{glueck2}. 
The DHG 
values refer to (\ref{598}).}
\label{fig20}
\end{center}
\end{figure}

Figure \ref{fig20} compares the result of the DIS data for $I_p(Q^2)$ 
and $I_n(Q^2)$, using the recent $Q^2$--independent LO results 
$\Gamma_1^p=0.146$ and $\Gamma_1^n=-0.064$ \cite{glueck2} in 
Eq.~(\ref{597}), 
with the values at $Q^2=0$ from the DHG sum rule. 
Since perturbative LO and NLO QCD is fully operative 
for $Q^2 \gtrsim 1$ GeV$^2$  \cite{glueck2}, as will be discussed 
in more detail in the next section, the most striking feature 
of this figure is the necessity of a tremendous $Q^2$--dependence 
(variation)  
of $\Gamma_1^p$ and $\Gamma_1^p-\Gamma_1^n$ in the low--$Q^2$ 
region, $Q^2<1$ GeV$^2$: In particular, $\Gamma_1^p$ must change 
sign between $Q^2 \approx 0$ and $Q^2 \approx 1$ GeV$^2$, and this 
must be due to some {\it strong} nonperturbative higher twist 
effect. 
A parametrization (but not an explanation) of this effect 
was suggested by \cite{anselmino2} 
and later on improved by \cite{burkert,soffer1}. 	
It turns out that such effects seem to be 
significantly larger than what one obtains on 
the basis of QCD sum rules. Most recent E143 measurements 
\cite{abe6} at small $Q^2$ confirm these expectations 
as well as the trend depicted in Fig. \ref{fig20}.   		
Reviews on this topic can be found in \cite{ioffe1,bass3}, 
for example.

Experiments are being planned to test the sum rule 
directly \cite{botton,faessler}. 
This is important because, among other things, 
there has been some criticism towards the argument that 
connects the spin structure function integral to the $Q^2=0$ integral 
in Eqs.~(\ref{597}) and (\ref{598}) \cite{ji1}.

%% file: k61tex
\setcounter{equation}{0}
\section{Polarized Parton Densities and Phenomenological Applications}
\subsection{Deep Inelastic Polarized Lepton--Nucleon Scattering}

Nowadays
sophisticated phenomenological models describing the spin structure
functions in LO and NLO QCD are 
used to interpret the experimental data. Before
we discuss these models,
we want to give a short historical overview about some simpler 
(and by now mostly obsolete) 
models that were discussed in the literature in the past, the
Kuti--Weisskopf model \cite{kuti}, the Carlitz--Kaur model 
\cite{carlitz2} and the
Cheng--Fischbach \cite{cheng3,cheng8} model.

When discussing the first moment of $g_1$ in Sect. 5.1 we have
noted that in the SU(6) constituent picture the nucleon consists
of three free quarks and its wavefunction is completely symmetric
in spin and flavor indices.
Squaring the SU(6) wave function one obtains the probabilites for
an up or down quark with spin up or down and, from that, the
SU(6) predictions $\Gamma_1^p={5 \over 54}$ and $\Gamma_1^n=0$ 
according to (\ref{517}).
This prediction is clearly wrong
except in that the proton is expected to have a larger positive
asymmetry than the neutron with a very small asymmetry.

One of the first attempts to describe the nucleon in terms of a
{\it relativistic} picture was the
Kuti--Weisskopf model \cite{kuti}. In this model the nucleon consists
of three valence quarks described by the SU(6) wave function,
allowing however for an x--dependent quark--density distribution,
plus a core of an indefinite number of quark--antiquark pairs
carrying vacuum quantum numbers and zero angular momentum.
The model also postulated the existence of gluons carrying a fraction
of the total momentum, the ratio of gluons to the quark--antiquark
pairs being the only free parameter of the model. At low x the core
dominates while at high x only the valence quarks contribute to the
scattering. The functional form of $g_1$ is given by
\be
g_1(x)={5\over 54}
{\Gamma (\gamma+3(1-\alpha)) \over \Gamma (1-\alpha)
\Gamma (\gamma+2(1-\alpha))}
x^{-\alpha (0)} (1-x)^{-1+\gamma+2(1-\alpha (0))}
\label{600}
\ee
for the proton while for the neutron it is identically zero.
Assuming that the behavior of structure functions near $x=1$ 
is related to the elastic form factor one obtains $-1+\gamma 
+2(1-\alpha (0))=3$. In their original work, Kuti and 
Weisskopf \cite{kuti} considered $\alpha(0)  \simeq 0.5$, 
as suggested by nondiffractive trajectories, which leads to 
an increase of the spin asymmetry at small x. This is in 
contradiction with the present correct view, cf. Sect. 4.4, 
and with experimental data. The normalization factor in 
Eq. (\ref{600}) was chosen to
fulfill the SU(6) predictions for
the first moment sum rules. Therefore, just
as SU(6), the model fails to explain the experimental data.
Here, and in the other following pre--QCD approaches, the 
scale $Q^2$ at which Eq. (\ref{600}) is expected to hold, cannot 
be specified except that $Q^2 \gg M^2$, as usual within the 
framework of the idealized 'impulse approximation'. 

Later attempts \cite{close5} tried to improve the model by taking into
account the difference between constituent and current quarks. The
two are related by a Melosh transformation which is shown to be
equivalent to a rotation in spin space \cite{close5}. The rotation
angle $\theta$ is defined so as to fit the modified SU(6) formula
${g_A \over g_V}={5\over 3}\cos 2\theta$
to the experimental number, yielding $\cos 2\theta \approx 0.75$.
The predictions for the spin asymmetry are
\be
A_1^p=({19\over 15}-{16\over 15}r(x))\cos 2\theta  , 
\qquad \qquad 
A_1^n=({2\over 5r(x)}-{3\over 5})\cos 2\theta
\label{601}
\ee
where $r(x)=F_2^n(x)/F_2^p(x)$ is the ratio of the unpolarized
structure functions.
In this model the prediction for $\Gamma_1^p$ is even somewhat
larger than that of the Ellis--Jaffe sum rule.

In another attempt to satisfy the Bjorken sum rule without
completely abandoning the SU(6) picture of the nucleon,
Babcock, Monsay and Sivers \cite{babcock} suggested to include perturbative
estimates of the $q\bar q$ and gluon sea polarization.
Furthermore, for the valence quarks the parametrization
$\delta u(x) \simeq 0.44u_v(x)$ and $\delta d(x) \simeq -0.35 d_v(x)$ was
proposed. Since the authors oriented their results at the
prediction of the Ellis--Jaffe sum rule, they are not in agreement
with present data.

Most famous among the SU(6) inspired models for the polarized
structure functions is certainly the Carlitz--Kaur model
\cite{carlitz2}. In this model the symmetry is broken by suggesting
that the configuration, in which the non--interacting diquark
system is in an isospin--1 state, is suppressed at high x relative
to the isospin--0 case. This is quantified in terms of two functions,
$I_0(x)$ and $I_1(x)$, describing the valence quark distributions,
where the subscript refers to the isospin of the non--interacting system
and which are obtained from unpolarized DIS data.
At low x, the valence quarks are assumed to lose any memory of the
parent spin orientation through interaction with the gluon sea.
This is described in terms of a factor, called $\sin^2 \theta(x)$,
giving the probability that
the quark will flip its spin through interactions with the sea.
It is given by
\be
\sin^2 \theta(x) \equiv {1\over 2}{H(x)N(x)\over H(x)N(x)+1}
\label{602}
\ee
where $N(x)$ is the density of the gluon sea relative to the
valence quarks and $H(x)$ is the probability of a spin--flipping
interaction between valence quarks and gluons.
The behavior of $N(x)$ for $x\rightarrow 1$ is suggested by
dimensional counting rules and the behavior for $x\rightarrow 0$
by Regge theory, $N(x)\sim (1-x)^2 /\sqrt{x}$.
Assuming the x--independence of $H$,
a measure of the dilution of the valence quark spin due to this
interaction is then given by the spin dilution factor
$\cos 2\theta(x)=H_0 (1-x)^2 /\sqrt{x}$.
The polarized quark densities are given by
\be
\delta u(x)=[u_v(x)-{2\over 3}d_v(x)]\cos 2\theta(x)
\quad \qquad
\delta d(x)=-{1\over 3}d_v(x)\cos 2\theta(x) \/.
\label{6022}
\ee
The only free parameter in the model is $H_0$. It is fixed by the
Bjorken sum rule, $H_0 \approx 0.052$. Correspondingly, the
spin dilution factor can be calculated. It is almost equal to 1
everywhere except at low values of x.
Using Eq. (\ref{6022}), $\Gamma_1^{p,n}$ can be calculated 
and turn out to be essentially identical to the predictions
of the Ellis--Jaffe sum rule. Furthermore, it should be noted
that the d--quark spin density in Eq. (\ref{6022}) does not
satisfy the perturbative QCD requirement
$\delta d(x) \tosim_{x \rightarrow 1} d_v(x)$.

In the Cheng--Fischbach model \cite{cheng3,cheng8}, later refined
by Callaway and Ellis \cite{callaway} and Cheng and Lai \cite{cheng2},
it is simply assumed that the polarized quark distributions are
related to the unpolarized ones via
\be
\delta u_v(x)=\alpha (x)u_v(x)
\quad \qquad
\delta d_v(x)=\beta (x)d_v(x)  \/.
\label{603}
\ee
It is further assumed that $\alpha(x) ,\beta(x)\tosim_{x \rightarrow 1}1$
in order to take account of the argument that the valence quark at x=1
remembers the spin of the parent nucleon. By contrast,
the region near x=0 is expected to be dominated by the sea
so that the spin of the nucleon is no longer reflected by the
valence quarks implying $\alpha(x) ,\beta(x)\tosim_{x \rightarrow 0}0$.
Since $\Delta d$ is negative, the boundary condition
$\beta(x)\tosim_{x \rightarrow 1}1$ implies that $\delta d_v(x)$
changes sign as a function of x. Therefore, $\beta(x)$ was
suggested to be of the form $\beta(x)={x-x_0 \over 1-x_0}x^p$
so that the sign of $\delta d_v(x)$ flips at $x=x_0$.
Furthermore, Cheng and Lai 
\cite{cheng2} have chosen $\alpha (x)=x^{0.26}$.
Roughly, typical values of $x_0$ and p are of the order 0.5.
These valence distributions alone cannot account for the
experimental data which require in addition
a large polarized sea quark and/or
gluon density. In ref. \cite{cheng2} an ansatz of the form
$|\delta s(x)|=x^{\gamma_s} s(x)$, $|\delta g(x)|=x^{\gamma_g} g(x)$ was
made.
Since the data do not really fix $\delta s(x)$ and $\delta g(x)$
it was not possible to determine $\gamma_s$ and $\gamma_g$ from
a fit.

Now we turn to more recent developments.  According to the discussions
and results presented in Sects. 4.1, 4.2 and 5.1, it is straightforward
to perform LO and NLO($\overline{{\rm MS}}$) QCD analyses of all presently
available polarized DIS data on $g_1^{p,n}(x,Q^2)$. 
%, as discussed and
%presented in Sects. 3.1, 2, and 3.  
This affords the knowledge of appropriate
input parton densities $\delta f(x,Q_0^2)$, $f=q,\bar{q},g$, extracted 
(as far as possible) from present measurements at a conveniently chosen
$Q^2=Q_0^2$.  The analyses can be performed either directly in Bjorken--x
space or, more conveniently, in (Mellin) n--moment space where the
RG evolution equations can be solved analytically as in Eqs. 
(\ref{4222})--(\ref{4226}).
During the past decade many LO analyses, partly supplemented by the NLO
gluon anomaly in Eq. 
(\ref{428}), (\ref{4227}) or 
(\ref{522}), have been performed, e.g.\
\cite{altarelli3,glueck13,ross,qiu,
gupta,gupta1,cheng2,chiappetta1,cheng6,
buccella,leader3}, 
and more recently in 
\cite{altarelli5,gehrmann,glueck4,ellis8,florian,bartelski2,
cheng7,ball,bourrely6,florian}. 
With the recently completed calculation of all 2--loop splitting function
(anomalous dimensions) $\delta P_{ij}^{(1)}(x)$, $\, i,j = q,g$\, , in the 
$\overline{\rm MS}$ factorization scheme 
\cite{mertig,vogelsang2}, also fully consistent
NLO($\overline{{\rm MS}}$) analyses became possible  
\cite{glueck2,gehrmann1,ball1,abe7}.

In general, the searched for parton densities $\delta f(x,Q^2)$ have to
satisfy the fundamental positivity constraints (\ref{4111}) at any value of
$x$ and scale $Q^2$, as calculated by the unpolarized and polarized 
evolution equations, within the {\it same} factorization scheme.
It is thus natural to perform polarized NLO analyses in the 
$\overline{\rm MS}$ scheme where practically all unpolarized NLO parton
densities have been analysed and presented.  Furthermore the total
helicities, i.e.\ $n=1$ moments of $\delta f(x,Q^2)$ in Eq. 
(\ref{51}), are
constrained by the sum rules (\ref{58}) and (\ref{59}) 
%Gleichung 6.1
\begin{equation}
\Delta u + \Delta\bar{u} - \Delta d - \Delta\bar{d} = F + D 
     = 1.2573 \pm 0.0028 
\label{61}
\end{equation}
%Gleichung 6.2
\begin{equation}
\Delta u + \Delta\bar{u} + \Delta d + \Delta\bar{d} - 2 (\Delta s + 
  \Delta\bar{s}) = 3 F - D = 0.579 \pm 0.025
\label{62}
\end{equation}
which hold for the flavor SU(3)$_f$ symmetric 'standard' scenario
commonly used.  It should be remembered that the flavor nonsinglet
combinations A$_{3,8}$ in Eqs.
(\ref{55})  and (\ref{56}) which appear in  
(\ref{61}) and (\ref{62}) 
are $Q^2$ independent also in NLO due to $\Delta P_{NS-}^{(1)}=0$
in (\ref{519}).

As has been already discussed at the beginning of Sect. 5.1, there are
serious objections to this latter full SU(3)$_f$ symmetry (mainly due to
$m_{u,d}\ll m_s$) which results in (\ref{62}), in contrast to the unquestioned
isospin SU(2)$_f$ symmetry ($m_u \simeq m_d$) which gives rise to 
(\ref{61}).
A plausible (but extreme) alternative to the full SU(3)$_f$ symmetry is a
'valence' scenario \cite{lipkin1,lipkin2,lichtenstadt} 
where SU(3)$_f$ is (maximally) broken and which is
based on the assumption that the flavor--changing hyperon $\beta$--decay
data fix only the total helicities of valence quarks
$\Delta q_v(Q^2)\equiv \Delta q - \Delta\bar{q}$ :
%Gleichung 6.3
\begin{equation}
\Delta u_v(Q_0^2) - \Delta d_v(Q_0^2) = F + D = 1.2573 \pm 0.0028
\label{63}
\end{equation}
%Gleichung 6.4
\begin{equation}
\Delta u_v(Q_0^2) + \Delta d_v(Q_0^2) = 3F - D = 0.579 \pm 0.025
\label{64}
\end{equation}
at some appropriately chosen input scale $Q^2=Q_0^2$.  Note that
$\Delta q_v(Q^2)$ depends (marginally) on $Q^2$ in NLO due to
$\Delta P_{NS+}^{(1)} \neq 0$ in (\ref{519}).

In the 'standard' SU(3)$_f$ symmetric scenario we need $\Delta s =
\Delta \bar{s} < 0$ in $\Gamma_1^{p,n}$ in Eq. 
(\ref{514}) or (\ref{518}) in
order to comply with recent experiments (cf.\ Table 2), 
i.e.\ in order
to obtain a reduction of the Ellis--Jaffe expectation $\Gamma_{1,EJ}^{p,n}$,
Eq. (\ref{515}), based on $A_3$ and $A_8$ which are entirely fixed by Eqs.\
(\ref{61}) and (\ref{62}), respectively. 
[Remember that $\Delta g(Q^2)$ decouples
from $\Gamma_1^{p,n}$ in (\ref{518}) in NLO($\overline{\rm MS})$ due to
$\Delta C_g =0$.]  In the 'valence' scenario we can do even with 
$\Delta s = \Delta\bar{s}\simeq 0$ since here only the valence contribution
to $A_8$ is fixed (apart from minor $Q^2$ dependent effects in NLO) by
Eq. (\ref{64}), with the entire $A_3$ still being fixed by  
(\ref{63}) due to the
assumption $\Delta\bar{u}=\Delta\bar{d}\equiv \Delta\bar{q}$ (which is
again violated by minor $Q^2$ dependent effects in NLO).  This gives in
LO for $\Gamma_1$ in (\ref{54})  
%Gleichung 6.5
\begin{equation}
\Gamma_1^{p,n}=\pm\frac{1}{12}(F+D)+\frac{5}{36}(3F-D)+\frac{1}{18}
    (10\Delta\bar{q} +\Delta s +\Delta\bar{s})
\label{65}
\end{equation}
and a similar relation holds in NLO \cite{glueck2}.  
Thus, in contrast to Eq. (\ref{514}),
a light polarized sea $\Delta\bar{q}<0$ will account for a reduction of
$\Gamma_1^{p,n}$ {\it even} for the extreme SU(3)$_f$ broken choice
$\Delta s=\Delta\bar{s}=0$!

\begin{figure}
%\vspace{7.0cm}
\begin{center}
\epsfig{file=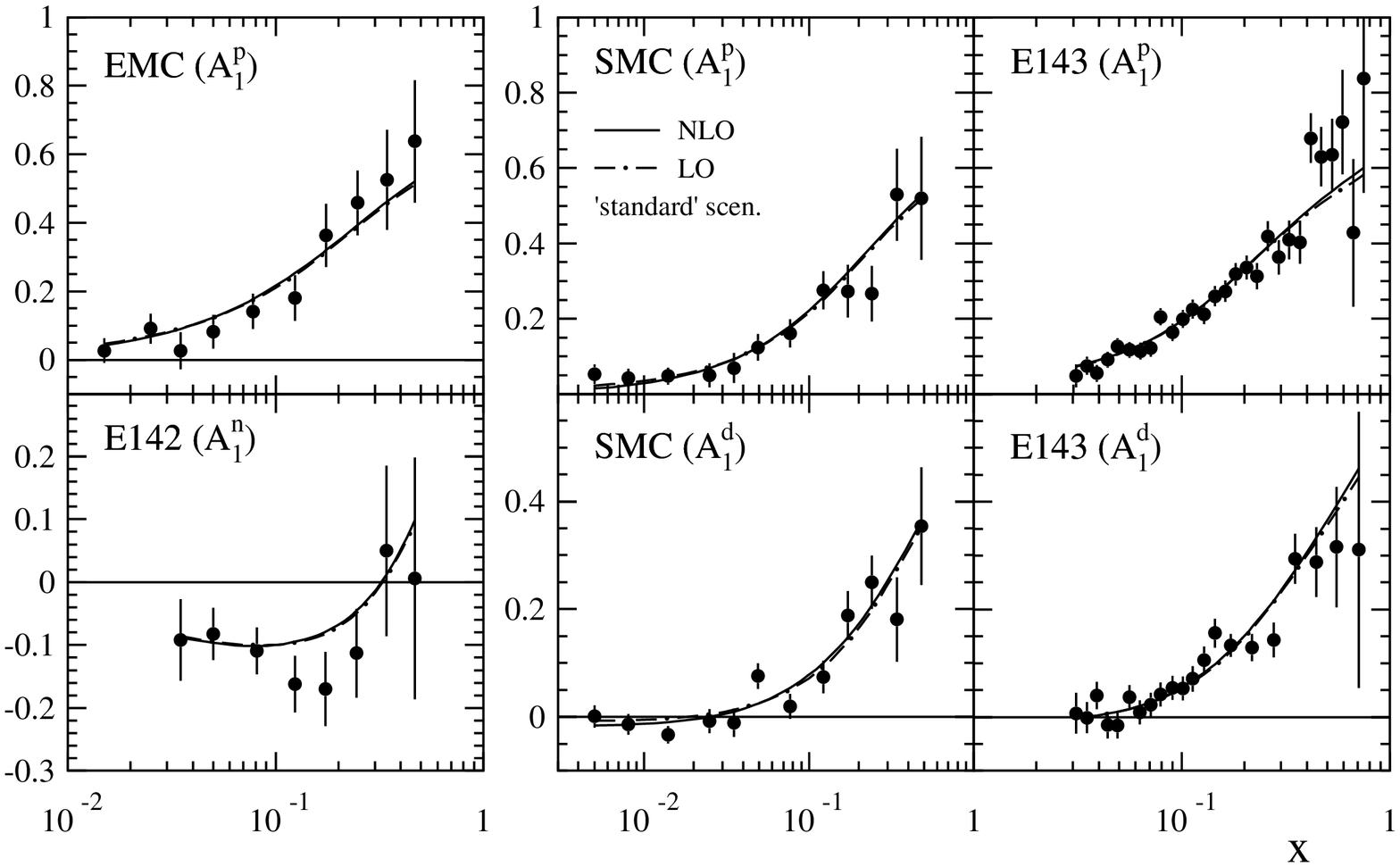,height=7cm,angle=0}
\bigskip
%\parbox{13.0cm}{\caption{
\caption{
Comparison of LO and NLO results for 
$A_1(x,Q^2)$ as obtained \protect\cite{glueck2}
from the fitted inputs at $Q^2=\mu_{LO,NLO}^2$ for the 'standard'
scenario, Eqs. (\ref{61}) and (\ref{62}), with present data 
\protect\cite{papavassiliou,ashman,adeva,adams2,adams,anthony,abe,abe1,abe2,adams6, 
adams7}. 
The $Q^2$ values adopted here correspond to the different values quoted
by the experiments for each data point starting at $Q^2\stackrel{>}{\sim}1$
GeV$^2$ at the lowest available x--bin.}
\label{fig26}
\end{center}
\end{figure}

Turning to the determination of the polarized LO and NLO parton distributions
$\delta f(x,Q^2)$ it is helpful to consider some reasonable theoretical
constraints concerning the sea and gluon densities, in particular in the
relevant small--x region where only rather scarce data exist at present
(in contrast to unpolarized DIS):  Apart from the rather general Regge
constraints in Eq. (\ref{441}) for $x\to 0$, color coherence of the gluon
couplings at $x\simeq 0$, i.e.\ equal partition of the hadron's momentum
among its partons, implies for the gluon and sea densities 
\cite{brodsky2,brodsky1} 
%Gleichung 6.6
\begin{equation}
\frac{\delta f(x,Q_0^2)}{f(x,Q_0^2)} \sim x \quad {\rm{as}} \quad x\to 0\, ,
\label{66}
\end{equation}
and arguments based on helicity retention properties of perturbative 
QCD of valence densities at large--x imply \cite{brodsky2,brodsky1,close6} 
%Gleichung 6.7
\begin{equation}
|\delta q_v(x,Q_0^2)| \sim q_v(x,Q_0^2) \quad {\rm{as}} \quad x\to 1\, .
\label{67}
\end{equation}
The scale $Q_0$ at which these relations are supposed to hold remains
unspecified.  Although not strictly compelling, Eqs. (\ref{66}) 
and (\ref{67})
are expected \cite{brodsky2,brodsky1} 
to hold at some 'intrinsic' bound--state--like scale
($Q_0^2\stackrel{<}{\sim}1$ GeV$^2$, say), but certainly not at much
larger purely perturbative scales $Q_0^2\gg 1$ GeV$^2$.  Despite such
'guidelines', presently available scarce polarization data on $A_1(x,Q^2)
\simeq g_1(x,Q^2)/F_1(x,Q^2)$, Eq. (\ref{227}), constrain the polarized parton
densities rather little, which holds in particular for $\delta g(x,Q^2)$
\cite{glueck2,glueck4,gehrmann1,ball1}.  
The determination of the polarized valence densities
is less ambiguous.  In order to avoid, as far as possible, pure guesses 
for the input densities $\delta f(x,Q_0^2)$, it has been suggested in 
\cite{glueck2,glueck4} to employ the unpolarized valence--like 
input densities $f(x,Q_0^2)$ at $Q_0^2=\mu^2 \simeq 0.3$ GeV$^2$,
properly modified so as to comply with polarized DIS data, with the 
positivity inequalities (\ref{4111}) 
for $Q^2\geq \mu^2$ and with the constraints
(\ref{66}) and (\ref{67}).  Subject to these requirements the following general
ansatz for the LO and NLO polarized parton densities has been employed 
\cite{glueck2,abe7} :
%Gleichung 6.8
\begin{eqnarray}
\delta q_v(x,\mu^2) & = & N_{q_v} x^{a_{q_v}} q_v(x,\mu^2)\nonumber\\
\delta\bar{q}(x,\mu^2) & = & N_{\bar{q}} x^{a_{\bar{q}}}(1-x)^{b_{\bar{q}}}
   \bar{q}(x,\mu^2)\nonumber\\
\delta s(x,\mu^2) & = & \delta\bar{s}(x,\mu^2)=N_s\delta\bar{q}
   (x,\mu^2)\nonumber\\
\delta g(x,\mu^2) & = & N_g x^{a_g} (1-x)^{b_g} g(x,\mu^2)
\label{68}
\end{eqnarray}
where the LO and NLO unpolarized input densities $f(x,\mu^2)$ at
$\mu_{LO}^2=0.23$ GeV$^2$ and $\mu_{NLO}^2=0.34$ GeV$^2$, respectively,
refer to the recent GRV valence--like densities \cite{glueck6}.  
It should be noted that employing valence--like gluon and sea input
densities [i.e.\ $xg(x,\mu^2)\sim x^a, x\bar{q}(x,\mu^2)\sim x^{a'}$   
with $a, a'>0$ as $x\to 0$] allows for a {\it parameter--free} 
calculation of parton densities and DIS structure functions in the 
small--x region ($x\stackrel{<}{\sim}10^{-2}$) at 
$Q^2>\mu^2$ which is entirely based on the QCD dynamics 
\cite{glueck,glueck5,glueck6}.  
The {\it perturbatively stable} LO/NLO predictions turned out to
be in excellent agreement with all DESY--HERA measurements up to now
\cite{abt,ahmed2,aid,derrick1,derrick2,derrick3}.

\begin{figure}
\begin{center}
\epsfig{file=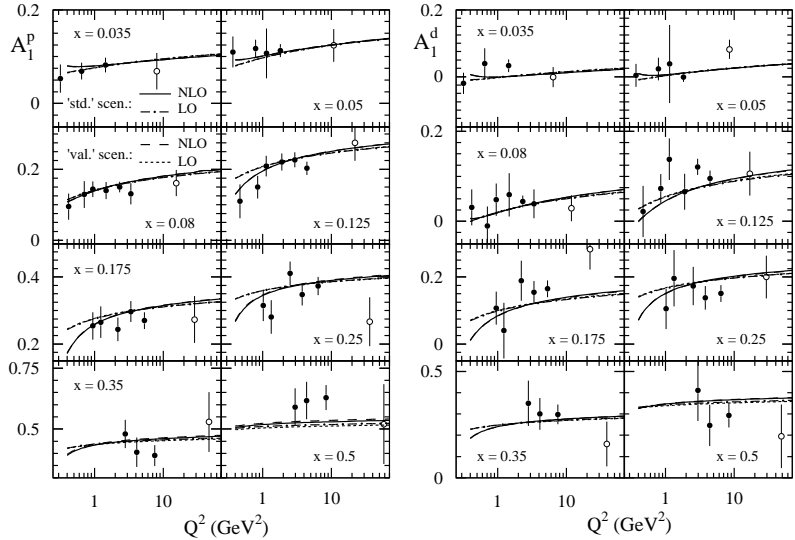,height=8cm,angle=0}
\bigskip
%\parbox{13.0cm}{\caption{
\caption{
The $Q^2$ dependence of $A_1^{p,d}(x,Q^2)$ as
predicted by LO and NLO QCD evolutions \protect\cite{glueck2} 
at various fixed values of
$x$, compared with recent SLAC--E143 
\protect\cite{abe,abe2} 
(solid circles) and SMC data
\protect\cite{adeva,adams,adams2}  (open circles).}
\label{fig27}
\end{center}
\end{figure}

The resulting fit
parameters $N_i$, $a_i$, $b_i$ for the 'standard' and 'valence' scenarios
in LO and NLO can be found in \cite{glueck2} 
where appropriate simple parametrizations
of the rather complicated QCD evolutions have also been given.  The LO 
and NLO results for the asymmetries $A_1^{p,d}(x,Q^2)$ measured up to now,
as discussed in Sect. 3, are presented in Fig. \ref{fig26} 
for the 'standard'
scenario.  The results for the 'valence' scenario are very similar.  
In both cases the LO and NLO results are perturbatively stable and almost
indistinguishable.  The expected $Q^2$ dependence of $A_1(x,Q^2)$ is shown
in Fig.  \ref{fig27} and compared with recent SLAC--E143 and SMC data. [The difference
between the LO and NLO results in the small--$Q^2$ region is mainly due to
different LO ($\mu_{LO}^2=0.23$ GeV$^2$) and NLO ($\mu_{NLO}^2=0.34$ GeV$^2$)
input scales.]  It should be emphasized that $A_1{\simeq}g_1/F_1$
is in general expected to be $Q^2$ {\it dependent} as soon as gluon
and sea densities become relevant, due to the very different polarized and
unpolarized splitting functions $\delta P_{ij}^{(0,1)}(x)$ and 
$P_{ij}^{(0,1)}(x)$, respectively (except for $\delta P_{qq}^{(0)} =
P_{qq}^{(0)}$ which dominates, apart from marginal differences in the 
relevant NLO NS splitting functions, in the large--x region).  Moreover,
the smaller x the stronger becomes 
the dependence of the exactly calculated $A_1(x,Q^2)$ on the 
precise 
form of the input at $Q^2=\mu^2$ \cite{glueck4}.  
For practical purposes, however,
such ambiguities are irrelevant since $A_1\simeq g_1/F_1 \simeq 2x\, 
g_1/F_2\to 0$ as $x\to 0$ is already unmeasurably small (of the order 
$10^{-3}$)
for $x\stackrel{<}{\sim} 10^{-3}$.  Thus the small--x region is unlikely
to be accessible experimentally for $g_1(x,Q^2)$, in contrast to the
situation for the unpolarized $F_{1,2}(x,Q^2)$.  It is furthermore interesting
to note that the approximate asymptotic ($x\to 0$) DLL expression 
(\ref{449})
for $A_1(x,Q^2)$ does not even quantitatively reproduce the exact LO
results for $A_1(x,Q^2)$ for $x\geq 10^{-3}$ \cite{glueck4}.  It is therefore 
misleading to use the simple asymptotic DLL formulae 
(\ref{446})--(\ref{449}) 
for quantitative estimates \cite{bartels,bartels1,ermolaev}.

\begin{figure}
%\vspace{7.0cm}
\begin{center}
\epsfig{file=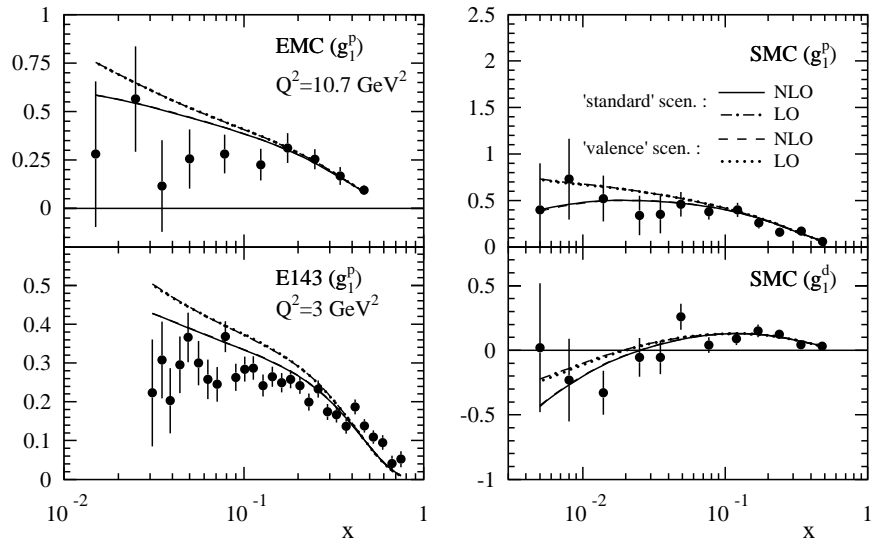,height=9cm,angle=0}
\bigskip
%\parbox{13.0cm}{\caption{
\caption{
Comparison of the 'standard' and 'valence'
LO and NLO results \protect\cite{glueck2} 
with the data for $g_1^{p,d}(x,Q^2)$ 
\protect\cite{papavassiliou,ashman,ashman1,adeva,adams,adams2,anthony,abe,abe1,abe2, 
adams6,adams7,adeva3}.
The SMC data correspond to different $Q^2\stackrel{>}{\sim}1$ GeV$^2$
for $x\geq 0.005$, as do the theoretical results.}
\label{fig28}
\end{center}
\end{figure}

\begin{figure}
%\vspace{7.0cm}
\begin{center}
\epsfig{file=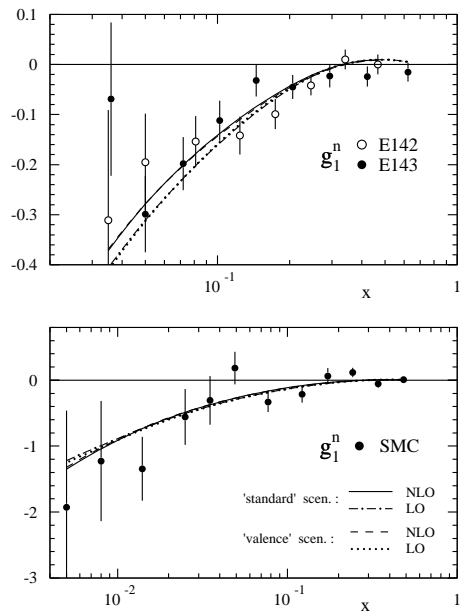,height=10cm,angle=0}
\bigskip
%\parbox{13.0cm}{\caption{
\caption{
Same as in Fig.  \ref{fig28} but for $g_1^n(x,Q^2)$.
The E142 and E143 data \protect\cite{anthony,abe1}  correspond to an average 
$\langle Q^2 \rangle = 2$ and $3$ GeV$^2$, respectively, and the 
theoretical predictions correspond to a fixed $Q^2=3$ GeV$^2$.}
\label{fig29}
\end{center}
\end{figure}

\begin{figure}
%\vspace{7.0cm}
\begin{center}
\epsfig{file=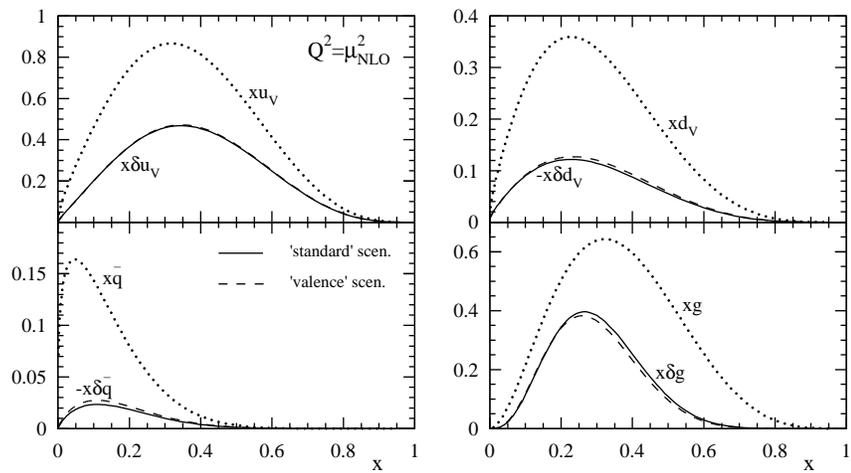,height=8cm,angle=0}
\bigskip
%\parbox{13.0cm}{\caption{
\caption{
Comparison of the fitted 'standard' and
'valence' input NLO($\overline{\rm{MS}}$) densities at $Q^2=\mu_{NLO}^2=
0.34$ GeV$^2$, according to Eq. (\ref{68}) \protect\cite{glueck2}, 
with the unpolarized 
dynamical input densities of \protect\cite{glueck6}.}
\label{fig30}
\end{center}
\end{figure}

\begin{figure}
%\vspace{7.0cm}
\begin{center}
\epsfig{file=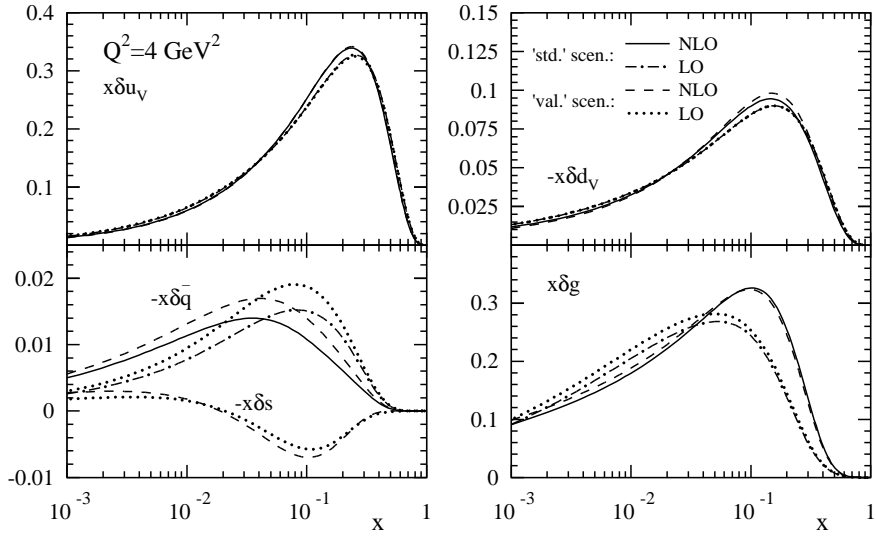,height=9cm,angle=0}
\bigskip
%\parbox{13.0cm}{\caption{
\caption{
The polarized LO and NLO($\overline{\rm{MS}}$) 
densities
at $Q^2=4$ GeV$^2$ as obtained from the input densities at 
$Q^2=\mu_{LO,NLO}^2$ as shown in Fig.  \ref{fig30} for the NLO.  
In the 'standard'
scenario, $\delta s$ coincides with the curves shown for $\delta\bar{q}$
in LO and NLO due to the SU(3)$_f$ symmetric input which is only marginally
broken in NLO for $Q^2>\mu_{NLO}^2$.}
\label{fig31}
\end{center}
\end{figure}

\begin{figure}
%\vspace{7.0cm}
\begin{center}
\epsfig{file=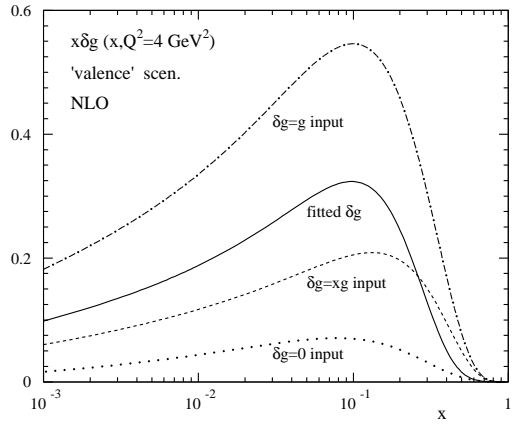,height=7cm,angle=0}
\bigskip
%\parbox{13.0cm}{\caption{
\caption{
The experimentally allowed range of NLO polarized
gluon densities at $Q^2=4$ GeV$^2$ for the 'valence' scenario with 
differently chosen $\delta g(x,\mu_{NLO}^2)$ inputs.  The "fitted
$\delta g$" curve is identical to the one in Fig.  \ref{fig31}.  Very similar
results are obtained if $\delta g(x,\mu_{NLO}^2)$ is varied accordingly
within the 'standard' scenario as well as in an LO analysis 
\protect\cite{glueck2,glueck4}.}
\label{fig32}
\end{center}
\end{figure}

The structure functions $g_1^{p,n}(x,Q^2)$ and $g_1^d$, given by the
relation (\ref{324}), can now be extracted using Eqs. (\ref{227}) and 
(\ref{2210}) where
one usually neglects the subleading contributions proportional to 
$\gamma^2$. These results are shown in Figs. \ref{fig28} and 
\ref{fig29}.  The reason
why the LO results are partly larger by more than about 10\% than the 
NLO ones is mainly due to the LO approximation $R=0$ in Eq. 
(\ref{2211}),
as used in \cite{glueck2}.  
Some of the EMC and E143 asymmetry data 
\cite{papavassiliou,ashman,ashman1,abe,abe2} 
have been analysed by assuming $A_1^p$ to be independent of $Q^2$.
This can be partly responsible \cite{glueck2} 
for these "data" falling somewhat
below the NLO predictions in the small--x region, despite the excellent
fits to $A_1^p$ in Fig.  \ref{fig26}.  The predictions for the NLO parton
distributions at the input scale $Q^2=\mu_{NLO}^2=0.34$ GeV$^2$ in 
Eq.\ (\ref{68}) are shown in Fig.  \ref{fig30} and compared with the reference
unpolarized NLO dynamical input densities of 
\cite{glueck6} which satisfy of
course the positivity requirement (\ref{4111}) 
as is obvious from Eq. (\ref{68}).
The LO predictions are similar \cite{glueck2}.  
It should be noted that the strange
densities correspond to $N_s=1$ in (\ref{68}) for the SU(3)$_f$ symmetric
'standard' scenario, whereas to $N_s=0$ for the SU(3)$_f$ broken
'valence' scenario \cite{glueck2}.  
The corresponding polarized densities at
$Q^2=4$~GeV$^2$, as obtained from these inputs at $Q^2=\mu^2$ for the 
two scenarios in LO and NLO, are shown in Fig.  \ref{fig31}.  It is 
interesting to note that, within the radiative approach 
with its longer 
$Q^2$--evolution "distance" starting at the low input scale $\mu^2$ in
Eq. (\ref{68}), a {\it finite} (negative) strange polarized sea input
$\delta s(x,\mu^2)$ is {\it always} required by present data for
the 'standard' scenario.  This holds true even if one uses (somewhat
inconsistently in NLO) the "off--shell" $\delta\tilde{C}_g$ in 
Eq. (\ref{428}) 
or  (\ref{4227}) which corresponds to $\Delta\tilde{C}_g=\frac{1}{2}$, giving
rise to Eq. (\ref{522}), in contrast to 
$\Delta C_g^{\overline{{\rm MS}}}=0$.
The shape of the polarized gluon densities $\delta g$ presented in 
Figs.\ 30 and 31 is constained rather little by present asymmetry data
\cite{glueck2,glueck4,gehrmann1}:  
Equally agreeable fits can be obtained for a fully 
saturated [inequality  (\ref{4111})] 
gluon input $\delta g(x,\mu^2)=g(x,\mu^2)$
as well as for the less saturated $\delta g(x,\mu^2)=xg(x,\mu^2)$.
A purely dynamical \cite{glueck13} 
input $\delta g(x,\mu^2)=0$ is also compatible
with present data, but such a choice seems to be unlikely in view of
$\delta\bar{q}(x,\mu^2)\neq 0$; it furthermore results in an unphysical
steep \cite{glueck13} 
$\delta g(x,Q^2>\mu^2)$, being mainly concentrated in the 
very small--x region $x<0.01$, as in the corresponding case  
\cite{glueck,glueck16} 
for the unpolarized parton distributions in disagreement with experiment.
The resulting NLO gluon densities $\delta g(x,Q^2)$ at $Q^2=4$ GeV$^2$
which originate from these extreme inputs are compared in 
Fig. \ref{fig32}  with
our 'fitted $\delta g$' curve of Fig. \ref{fig31} obtained for the 'valence'
scenario.  Present data allow even for a partly negative $\delta g(x, 4$
GeV$^2$) \cite{gehrmann1} 
and specific model calculations can accommodate even a
negative $\Delta g(4$ GeV$^2$) \cite{glueck17,jaffe5}.  
It turns out that
$\delta g(x,Q^2)$ is somewhat less ambiguous if only the more global
quantities $\delta q_{NS}$, $\delta\Sigma$ and $\delta g$ are used for
analyzing present data \cite{ball1}, instead of trying to delineate the
individual parton densities.  

\begin{figure}
\begin{center}
\epsfig{file=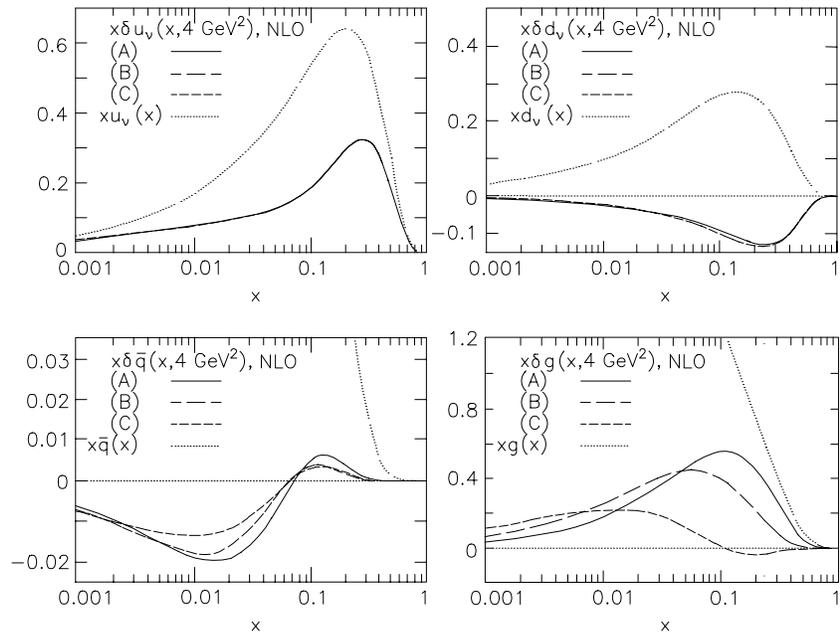,height=11cm,angle=270}
\bigskip
\caption{NLO polarized and unpolarized parton distributions 
at $Q_0^2=4$ GeV$^2$ according to \protect\cite{gehrmann1}}
\label{figstir1}
\end{center}
\end{figure}

An alternative analysis of polarized structure functions has been
performed by Gehrmann and Stirling \cite{gehrmann1}.  
This was done in the same
spirit as the unpolarized analysis \cite{martin1}, 
namely the polarized parton
distributions were chosen to be of the general form
%Gleichung 6.8'
\begin{equation}
x\delta f(x,Q_0^2) =\eta_f\, A_f\, x^{a_f}(1-x)^{b_f} (1+\gamma_f x
   +\rho_f\sqrt{x}) \hspace{\fill} 
\label{68p}
\end{equation}
for $f=u_v,d_v,\bar{q},g$ at the starting scale $Q_0^2=4$ GeV$^2$.
The normalization factors
\begin{equation}
A_f=\left[ \left( 1+\gamma_f\frac{a_f}{a_f+b_f+1}\right)
      \frac{\Gamma(a_f)\Gamma(b_f+1)}{\Gamma(a_f+b_f+1)} + \rho_f
        \frac{\Gamma(a_f+0.5)\Gamma(b_f+1)}{\Gamma(a_f+b_f+1.5)}\right]^{-1}
\label{68ppppp}
\end{equation}
are determined by the condition that the first moments of 
$\delta f(x,Q_0^2)$
are given by $\eta_f$.  The parameters $\eta_f, \, a_f, \, b_f, \, \gamma_f$
and $\rho_f$ for $f=u_v$ and $d_v$ were determined by LO and NLO fits to
recent data.  For the sea, an SU(3)$_f$ symmetric antiquark polarization
was assumed (just as in the 'standard' scenario 
\cite{glueck2} as discussed above,
e.g.\ Eq. (\ref{68}) with $N_s=1$).  For $\delta g(x,Q_0^2)$, which is hardly
constrained by the data [cf.\ Fig. \ref{fig32}], 
three alternative parametrizations
were employed, a hard and a soft distribution (A and B) with the spin aligned
with that of the parent proton and a distribution (called C) with the spin
anti--aligned.  All three choices give equally good descriptions of the
structure function data, but would be relatively easy to discriminate if
data on polarized gluon initiated processes like $\gamma^{(*)}g\to c\bar{c}$
or $qg\to \gamma q$ were available.  In all cases the first moment 
$\eta_g=1.9$ was chosen which is obtained in LO by attributing 
{\it all} the violation of the Ellis--Jaffe sum rule to a large
gluon polarization.  This number is in the same ballpark as if determined
from an NLO analysis of the experimental data ($\Delta g(Q_0^2) = 1.5 \pm
0.5$ \cite{ball1} 
and Eq. (\ref{69}) and Tables 3 and 4 
below), but with a large
error.  A summary of the fitted and chosen parameters together with the
$\chi^2$ of the fit can be found in Table 2 of 
\cite{gehrmann1}.  The NLO polarized 
parton
distributions at $Q_0^2=4$ GeV$^2$ for gluon scenarios A, B and C are
shown in Fig. \ref{figstir1}.  It should be emphasized that present polarization
data can be fitted even with a negative $\delta g$ in the
large--x region, as shown in Fig. \ref{figstir1}, which refers to a more extreme
choice than the ones depicted in Fig. \ref{fig32}.  
Furthermore, the LO GRV 94
\cite{glueck6} and the NLO MRS--A'  
\cite{martin1} parton densities have been chosen as 
reference distributions, which are necessary for comparing with the
positivity constraints  (\ref{4111}), 
and are also shown in Fig. \ref{figstir1}.  The
resulting fits for $g_1^{p,n,d}(x,Q^2)$ are similar
in quality \cite{gehrmann1} as the ones shown in Fig. \ref{fig28} and 
 \ref{fig29}.

Finally let us turn to the first moments (total polarizations) 
$\Delta f(Q^2)$ in Eq. (\ref{51}) of the polarized parton densities 
$\delta f(x,Q^2)$ and the resulting $\Gamma_1^{p,n}(Q^2)$.  It should
be recalled that, in contrast to the LO, the first moments of the NLO
(anti)quark densities do renormalize, i.e.\ are $Q^2$ dependent, due
to the nonvanishing of the 2--loop $\Delta P_{qq}^{(1)}$ and
$\Delta P_{NS+}^{(1)}$ in (\ref{519}).  Let us discuss the two scenarios
in turn:

(i) In the 'standard' scenario the input densities in (\ref{68}), being
constrained by (\ref{61}) and (\ref{62}), imply in LO \cite{glueck2}
%Gleichung (6.9)
\begin{displaymath}
\Delta u_v = 0.9181\, , \quad \quad \Delta d_v = -0.3392
\end{displaymath}
\begin{displaymath}
\Delta\bar{q} = \Delta s = \Delta\bar{s} = -0.0587
\end{displaymath}
\begin{equation}
\Delta g(\mu_{LO}^2) = 0.362\, ,\quad \Delta g(4\,\,{\rm{GeV}}^2) = 1.273\, ,
  \quad \Delta g(10\,\,{\rm{GeV}}^2) = 1.570
\label{69}
\end{equation}
which result in $\Delta\Sigma = 0.227$.  This gives
%Gleichung (6.10)
\begin{equation}
\Gamma_1^p=0.1461\, , \quad \quad \Gamma_1^n = -0.0635
\label{610}
\end{equation}
in reasonable agreement with present data (Table 2).  
The NLO results
are shown in Table 3 which are in even better agreement with 
experiments (Table 2).
%\newpage

\begin{table}[h]
\label{tab3}
\begin{center}
%Tabelle III
%\renewcommand{\arraystretch}{1.5}
\begin{tabular}{|c||c|c|c|c|c|c||c|c|}
\hline
$Q^2$(GeV)$^2$ & $\Delta u_v$ & $\Delta d_v$ & $\Delta\bar{q}$ & 
  $\Delta s = \Delta\bar{s}$ & $\Delta g$ & $\Delta\Sigma$
    & $\Gamma_1^p$ & $\Gamma_1^n$ \\
\hline
$\mu_{NLO}^2$ & 0.9181 & -0.3392 & -0.0660 & -0.0660 & 0.507 & 0.183
    & 0.1136 & -0.0550 \\
 1 & 0.915 & -0.338 & -0.067 & -0.068 & 0.961 & 0.173 & 0.124 & -0.061 \\
 4 & 0.914 & -0.338 & -0.068 & -0.068 & 1.443 & 0.168 & 0.128 & -0.064 \\
10 & 0.914 & -0.338 & -0.068 & -0.069 & 1.737 & 0.166 & 0.130 & -0.065 \\
\hline
\end{tabular}
\bigskip
\caption{First moments of polarized NLO parton densities
$\delta f(x,Q^2)$ and of $g_1^{p,n}(x,Q^2)$ as predicted in the
'standard' scenario \protect\cite{glueck2}.  Note that the marginal differences for
$\Delta\bar{q}$ and $\Delta s$ indicate the typical amount of dynamical
SU(3)$_f$ breaking generated by the RG $Q^2$--evolution to 
$Q^2>\mu_{NLO}^2$.}
\end{center}
\end{table}

%\vspace{0.2cm}

(ii) In the 'valence' scenario the input densities in (\ref{68}), being 
constrained by (\ref{63}) and (\ref{64}), imply in LO \cite{glueck2}
%Gleichung (6.11)
\begin{displaymath}
\Delta u_v = 0.9181\, , \quad\quad \Delta d_v = -0.3392
\end{displaymath}
\begin{displaymath}
\Delta\bar{q} = -0.0712\, , \quad\quad \Delta s = \Delta\bar{s} = 0
\end{displaymath}
\begin{equation}
\Delta g(\mu_{LO}^2) = 0.372\, ,\quad \Delta g(4\,\,{\rm{GeV}}^2) = 1.361\, ,
 \quad \Delta g(10\,\,{\rm{GeV}}^2) = 1.684
\label{611}
\end{equation}
which result in $\Delta\Sigma = 0.294$.  Apart from this maximal SU(3)$_f$
breaking, these results are similar to the 'standard' ones in (\ref{69}) and
yield
%Gleichung (6.12)
\begin{equation}
\Gamma_1^p = 0.1456\, , \quad\quad \Gamma_1^n = -0.0639
\label{612}
\end{equation}
on account of (\ref{65}).  The NLO results are shown in Table 4.
% wenn er eine Zeile doch falsch ausdruckt,liegt das daran,
% dass er bei nur 1-4 Zeilen doch vorstellt!!!
%Tabelle IV
\begin{table}[h]
\label{tab4}
\begin{center}
\begin{tabular}{|c||c|c|c|c|c|c||c|c|}
\hline
$Q^2$(GeV)$^2$ & $\Delta u_v$ & $\Delta d_v$ & $\Delta\bar{q}$ & 
  $\Delta s = \Delta\bar{s}$ & $\Delta g$ & $\Delta\Sigma$
    & $\Gamma_1^p$ & $\Gamma_1^n$ \\
\hline
$\mu_{NLO}^2$ & 0.9181 & -0.3392 & -0.0778  & 0 & 0.496 & 0.268
    & 0.1142 & -0.0544 \\
 1 & 0.915 & -0.338 & -0.080 & -2.5 $\times$ 10$^{-3}$ & 0.982 & 0.252 
    & 0.124 & -0.061 \\
 4 & 0.914 & -0.338 & -0.081 & -3.5 $\times$ 10$^{-3}$ & 1.494 & 0.245 
    & 0.128 & -0.064\\
10 & 0.914 & -0.338 & -0.081 & -3.8 $\times$ 10$^{-3}$ & 1.807 & 0.244 
    & 0.130 & -0.065 \\
\hline
\end{tabular}
\bigskip
\caption{The first moments (total helicities) as in Table 3,
but for the maximally SU(3)$_f$ broken 'valence' scenario 
\protect\cite{glueck2}.}
\end{center}
\end{table}
%\vspace{0.3cm}

These results for the total helicities (first moments) are similar to the
ones observed in other recent LO and NLO analyses 
\cite{gehrmann1,ball1,abe7}.

Due to the similarity of the LO and NLO results in both scenarios, it is
obviously impossible to distinguish experimentally between the 'standard'
(SU(3)$_f$ symmetric) and 'valence' (SU(3)$_f$ maximally broken) scenario. 
In both scenarios the Bjorken sum rule (\ref{586}) manifestly holds due to the
constraints (\ref{61}) and (\ref{63}).  Furthermore, the observed total helicities 
carried by the valence quarks, $\Delta u_v$ and $\Delta d_v$, are compatible
with the ones obtained very recently from semi--inclusive spin asymmetry
measurements \cite{adeva2}, 
cf. Sect. 6.5, which yielded $\Delta u_v = 1.01 \pm 0.24$ and
$\Delta d_v = -0.57 \pm 0.25$ at $\langle Q^2\rangle \simeq 10$ GeV$^2$.
It is also very interesting to note that our optimal fit results shown
above favor a sizeable total gluon helicity $\Delta g(10\,{\rm{GeV}}^2)
\simeq 1.7$, despite the fact that $\Delta g(Q^2)$ decouples from the full
($0\leq x\leq 1$) first moment $\Gamma_1(Q^2)$ in (\ref{518}) in the 
$\overline{\rm{MS}}$ scheme (since $\Delta C_g^{\overline{\rm{MS}}}=0$).  
This implies that for any experimentally relevant analysis (where 
$0.01\stackrel{<}{\sim}x <1$), the NLO $\delta g(x,Q^2)$ in Eq.
 (\ref{429}),   
for example, plays an important role almost regardless of the value
of the full first moment of $\delta C_g(x).$  The importance of 
$\delta g(x,Q^2)$ also holds in LO where $\delta g$ does not directly
appear in $g_1(x,Q^2)$, Eq. 
 (\ref{415}), but enters only via the RG evolution
equations.  

This large $\overline{\rm MS}$ result for $\Delta g(Q^2)$ is also
comparable with the weaker constraint (\ref{585}) obtained in the off--shell
scheme (where $\Delta C_g=-\frac{1}{2}$) which is not too surprising 
since the RG solutions (\ref{581}) and (\ref{583}) differ only to 
${\cal{O}}(\alpha_s)$. Furthermore one can be tempted to reinterpret
our $\overline{\rm MS}$ results for $\Delta\Sigma(Q^2)$ in terms of
$\Delta\Sigma_{off}$ in Eq. (\ref{526}), by assuming that $\Delta g(Q^2)$
in the off--shell scheme is similar to our $\overline{\rm MS}$
results.  This gives
%Gleichung (6.13)
\begin{equation}
\Delta\Sigma_{off} \simeq 0.33\,\, (0.42)
\label{613}
\end{equation}
according to the results for the 'standard' ('valence') scenario in
Table 3 (Table 4).  
Thus the sizeable $\Delta g(Q^2)$ implies a sizeable
amount of total helicity of singlet quark densities, $\Delta\Sigma_{off}$,
which comes close to the naive expectation $\Delta\Sigma_{off} \simeq A_8
\simeq 0.6$ in Eq. (\ref{527}) in contrast to $\Delta\Sigma(Q^2)$ in the 
$\overline{\rm MS}$ scheme.

Finally, it is very interesting to observe that at the low input scales
$Q^2=\mu_{LO,NLO}^2=0.23,\, 0.34$ GeV$^2$ the nucleon's spin is
dominantly carried just by the total helicities of quarks and gluons
%Gleichung (6.14)
\begin{equation}
\frac{1}{2}\Delta\Sigma\left( \mu_{LO[NLO]}^2\right) +
   \Delta g\left(\mu_{LO[NLO]}^2 \right) \simeq 0.5\,\, [0.6]
\label{614}
\end{equation}
according to Eqs. (\ref{69}) and (\ref{611}), 
and Tables 3 and 4.  Thus the
helicity sum rule (\ref{111}) implies that
%Gleichung (6.15)
\begin{equation}
L_z(\mu_{LO,NLO}^2)\simeq 0\, .
\label{615}
\end{equation}
The approximate vanishing of this latter nonperturbative angular momentum,
being built up from the intrinsic $k_T$ carried by partons, is intuitively
expected for low bound--state--like scales (but {\it not} for
$Q^2\gg \mu^2$) implying that the spin of the nucleon is carried solely
by quarks and gluons, Eq. (\ref{614}), i.e.\ there is no 
'spin surprise' whatsoever.  At smaller distances, i.e.\ for $Q^2\gg
\mu_{LO,NLO}^2$, this picture will break down since gluon and $q\bar{q}$
production off the initial partons will increase their $k_T$ which in
turn gives rise to a finite 
orbital angular momentum carried by quarks and gluons,
$L_z(Q^2) = L_q(Q^2) + L_g(Q^2)$ 
\cite{sehgal,jaffe,ratcliffe2}.  Clearly a finite $L_z(Q^2)$
is required to reconcile, for example $\frac{1}{2}\Delta\Sigma(Q^2) + 
\Delta g(Q^2) \simeq 1.8$ at $Q^2=10\,$ GeV$^2$ (Table 3), with the 
sum rule (\ref{111}).  The relevant RG $Q^2$--evolution equations for the quark
and gluon angular momenta $L_q(Q^2)$ and $L_g(Q^2)$ have been written down
recently in LO \cite{ji6}:
%Gleichung (6.16)
\vspace{0.3cm}
\begin{eqnarray}
 \frac{d}{dt} \left( \begin{array}{c}
    L_q(Q^2) \\ L_g(Q^2) \end{array} \right)
& = & \frac{\alpha _s(Q^2)}{2\pi} \left( \begin{array}{cc}
      -\frac{4}{3}C_F & \frac{f}{3} \\
        \frac{4}{3}C_F & -\frac{f}{3} \end{array} \right)
       \left( \begin{array}{c}
         L_q(Q^2) \\ L_g(Q^2) \end{array} \right)\nonumber\\
\rule{0mm}{10mm}
& + &  \frac{\alpha _s(Q^2)}{2\pi} \left( \begin{array}{cc}
       -\frac{2}{3} C_F & \frac{f}{3} \\ 
         -\frac{5}{6}C_F & -\frac{11}{2} \end{array} \right)
        \left( \begin{array}{c}
         \Delta\Sigma (Q^2) \\ \Delta g(Q^2) \end{array} \right)
\label{616}
\end{eqnarray}

\noindent where the second inhomogeneous term was first studied in 
\cite{ratcliffe2} 
with $\Delta\Sigma$ and $\Delta g$ evolving according to (\ref{52}).
The solution of (\ref{616}) is straightforward:
%Gleichung (6.17)
\begin{equation}
L_q(Q^2)=-\frac{1}{2}\Delta\Sigma +\frac{1}{2}\, \frac{3f}{16+3f} +
\left[ L_q(Q_0^2)+\frac{1}{2}\Delta\Sigma -\frac{1}{2}\,\frac{3f}{16+3f}\right]
     L^{2(16+3f)/9\beta_0}
\label{617}
\end{equation}
%Gleichung (6.18)
\begin{equation}
L_g(Q^2)=-\Delta g(Q^2)+\frac{1}{2}\, \frac{16}{16+3f} +
 \left[ L_g(Q_0^2)+\Delta g(Q_0^2)-\frac{1}{2}\, \frac{16}{16+3f}\right]
     L^{2(16+3f)/9\beta_0}
\label{618}
\end{equation}

\noindent with $L\equiv \alpha_s(Q^2)/\alpha_s(Q_0^2)$ and 
$\Delta\Sigma\equiv
\Delta\Sigma(Q_0^2) = \Delta\Sigma(Q^2)$ in LO.  The last Eq. (\ref{618})
demonstrates explicitly that the large gluon helicity $\Delta g(Q^2)$
as obtained above at large $Q^2>\mu^2$ is canceled by an equally large,
but negative, gluon orbital momentum.  Asymptotically ($Q^2\to\infty$)
the solutions (\ref{617}) and (\ref{618}) become particularly simple
%Gleichung (6.19)
\begin{equation} 
L_q + \frac{1}{2}\Delta\Sigma=\frac{1}{2}\, \frac{3f}{16+3f}\, ,\quad
  \quad L_g +\Delta g=\frac{1}{2}\, \frac{16}{16+3f}\, .
\label{619}
\end{equation}
Thus the partition of the nucleon spin between quarks and gluons 
eventually follows the well known partition of the quark and gluon
momenta in the nucleon 
\cite{gross,gross1,georgi}.  If the $Q^2$ evolution is slow,
then Eq. (\ref{619}) predicts that quarks carry less than about 50\%
of the nucleon spin even at low momenta [cf.\ Eqs. (\ref{69}) and 
(\ref{611})].

It should be mentioned that, although $L_z=L_q+L_g$ can be theoretically
formally formulated in a consistent covariant way 
\cite{jaffe,jaffe5,cheng4}, there appears
to be no direct experimental test of the size as well as of the sign
of $L_z(Q^2)$, or more ideally of the separate components $L_q(Q^2)$
and $L_g(Q^2)$.  The possible measurement of azimuthal distributions
has been proposed  
\cite{chou,meng} but these are only sensitive to some
average $\langle k_T^2\rangle$ of rotating constituents in a polarized
nucleon target.

More recently, Ji \cite{ji7,ji8} has suggested to use deeply virtual 
Compton scattering (DVCS) $\gamma^{\ast}(Q^2)p \rightarrow \gamma p'$ 
in the limit of vanishing momentum transfer $t=(p'-p)^2$, in order 
to get direct information about the gauge invariant combinations 
$J_q={1\over 2} \Delta \Sigma +L_q$ and $J_g= \Delta g +L_g$ 
appearing in the spin sum rule (\ref{111}): 
${1\over 2}=J_q(Q^2)+J_g(Q^2)$. In this limit, $J_{q,g}=
{1\over 2}[A_{q,g}(0)+B_{q,g}(0)]$ where $A_{q,g}(t)$ and 
$B_{q,g}(t)$ are the 'Dirac' and 'Pauli' form factors of the quark 
and gluon energy--momentum tensor ( which are analogously defined 
as the well known form factors of the electromagnetic vector current). 
The form factors $A_q(t)$ and $B_q(t)$ are then related via sum rules 
to the $n=2$ moments (Bjorken--x averages) of the structure functions 
of the {\it non}--forward DVCS. Although the extrapolation $t\rightarrow 0$, 
required to obtain $J_{q,g}$, is difficult \cite{ji9,radyushkin,kroll} 
(if at all possible), rough estimates result in a DVCS cross section 
at $-t < 1$ GeV$^2$ above 1 pb at CEBAF and DESY--Hermes kinematics 
\cite{ji9}.

%% file: k62tex
\subsection{Heavy Quark Production in Polarized DIS and in 
Photoproduction}

In the previous sections we have realized, among other 
things, the enormous 
difficulties to extract the polarized gluon distribution 
and in particular its first moment 
from inclusive deep inelastic data. 
These problems have been anticipated several years ago 
by theoretical studies 
\cite{altarelli2,efremov,carlitz,altarelli6}, 
and they are in fact not surprising in view of the 
well known subtleties having occurred in all attempts to 
determine the unpolarized gluon density 
in unpolarized DIS experiments during the past two decades. 

A popular way out of this dilemma is the study of 
semi--inclusive cross sections, and in particular of 
charm production, because the production of heavy quark hadrons  
is triggered in leading order by the photon--gluon fusion 
mechanism [$\gamma^{\ast} (\gamma ) \delta g \rightarrow h \bar h$ 
with $h=b,c$] and is therefore sensitive to the gluon 
density inside the proton, whereas the heavy quark 
content of the proton is usually negligible, if it exists at all, 
at presently 
available $Q^2$--values. 
Due to its prominent decay mechanism, $J/\psi$ production is  
the most distinct among the charmed events. 
In contrast to open charm production one faces here, however, 
the additional model dependence for bound--state production, 
such as the 'duality model' 
\cite{fritzsch3,halzen,jones,glueck18,glueck19} 
(nowadays also called 'color 
evaporation' model) and the color--singlet model 
\cite{chang,berger,baier}. In the duality or 'soft color' 
treatment of color quantum numbers the cross section for 
bound charm production is given by 
\begin{equation}
\sigma_{onium}={1\over 9} \int_{2m_c}^{2m_D} dm 
{d\sigma_{c\bar c} \over dm}
\label{60001}
\end{equation}
where $d\sigma_{c\bar c}$ is computed perturbatively in LO and NLO 
due to $\gamma^{\ast}  g \rightarrow c \bar c$, etc. 
(or due to $q\bar q \rightarrow c\bar c$, $gg\rightarrow
c\bar c$, etc. for hadronically produced quarkonia), since here 
the color singlet property of the $J/\psi$ is ignored at the 
perturbative stage of the calculation. The subsequent hadronization of 
the color singlet state is then assumed to be characterized 
by multiple (nonperturbtive) soft gluon emissions, i.e. 
the treatment of color is, on the average, statistical with the factor
${1\over 9}$ representing the statistical probability that 
the $3\times \bar 3$ charm pair is asymptotically in a color 
singlet state. Alternatively, in the color--singlet model  
the color singlet property of the produced onium states 
($J/\psi$,  etc.) is enforced already at short distances, 
$\Delta x \sim m_{\psi}^{-1}$, by the emission of a perturbative (octet) 
gluon off the produced charm quark. It is this latter assumption 
which casts doubt on the color--singlet model since it does not seem 
logical to enforce perturbatively the color--singlet property of the onia 
at short distances, given that there remains practically an infinite 
time for soft gluons to readjust the color of the $c\bar c$ pair 
before it appears as an asymptotic $J/\psi$ or, alternatively, 
$D\bar D$ state. In other words, it is hard to imagine 
that a color singlet state formed at a range $m_{\psi}^{-1}$, 
automatically survives to form a $J/\psi$. Indeed, the duality 
('color evaporation') treatment has received renewed attention 
and appears to be the favorite mechanism of heavy quarkonia 
production \cite{amundsen,eboli,schuler,gavai,schuler1} 
(or a variant of it which differs in its nonrelativistic 
treatment of the nonperturbative long--distance part of the 
$c\bar c$ matrix elements which obey simple 'velocity--scaling' 
laws with respect to the relative velocity $\beta$ of the $c\bar c$; 
this allows for a systematic expansion in $\alpha_s (2m_c)$ and 
$\beta$ \cite{bodwin1}.) The reason for this revival is that 
some data on the production of $\psi$-- and $\Upsilon$--states 
disagree with the simple minded color--singlet model predictions; 
occasionally by well over one order of magnitude as in the 
case of $\psi '$ production at the Fermilab Tevatron \cite{braaten1}. 
Thus it seems to be more appropriate to study and delineate the 
relevant quarkonia production mechanism using unpolarized 
reactions first, instead of using a particular model to get 
access to $\delta g(x, Q^2 \approx m_{J/\psi}^2)$ via polarized 
deep inelastic (or photon) production of $J/\psi$'s 
\cite{glueck10,guillet,kalyniak,godbole,sridhar}. 
This statement is even more true for 
(diffractive) elastic $J/\psi$ production
where some very speculative 
models exist \cite{brodsky3,ryskin,levin}. They are 
based on a two--gluon exchange but it is not clear whether 
the square of the polarized gluon density   
$\delta g(x, Q^2 \approx m_{J/\psi}^2)$ or some independent 
two--gluon correlation function appears in the cross section 
formula.

Therefore, from the theoretical point of view 
a much cleaner signal for the gluon density in heavy quark 
production is open charm production, although 
experimentally it has worse statistics due to the  
difficulties in identifying D--mesons. 
Instead of the deep inelastic process one may as well look 
at photoproduction, because the
mass of the charm quark forces the process to take 
place in the perturbative regime. The advantage 
of photoproduction over DIS is its larger cross section. 
Several fixed target experiments, HERMES at DESY \cite{amarian}, 
COMPASS at CERN \cite{baum2} and a new 
facility at SLAC \cite{arnold2} are 
being developed to measure the polarized gluon 
distribution via photoproduction.  

\begin{figure}
\begin{center}
\epsfig{file=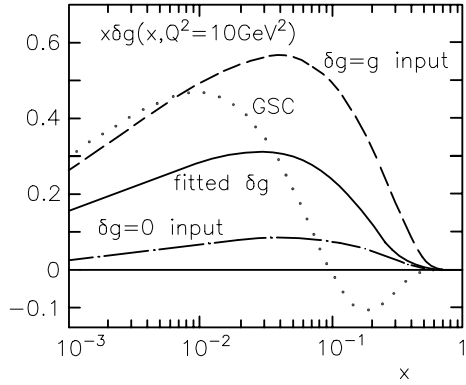,height=5cm}
\vskip 0.5cm
\caption{Polarized gluon densities at $Q^2=10$GeV$^2$ ($\approx 4m_c^2$) 
of the four LO sets used in this subsection. The dotted curves refers 
to set C of \protect\cite{gehrmann1} whereas the other densities are taken 
from \protect\cite{glueck4} as described in Sect. 6.1. }          
\label{fig34}                            
\end{center}                                
\end{figure}

\begin{figure}
\begin{center}
\epsfig{file=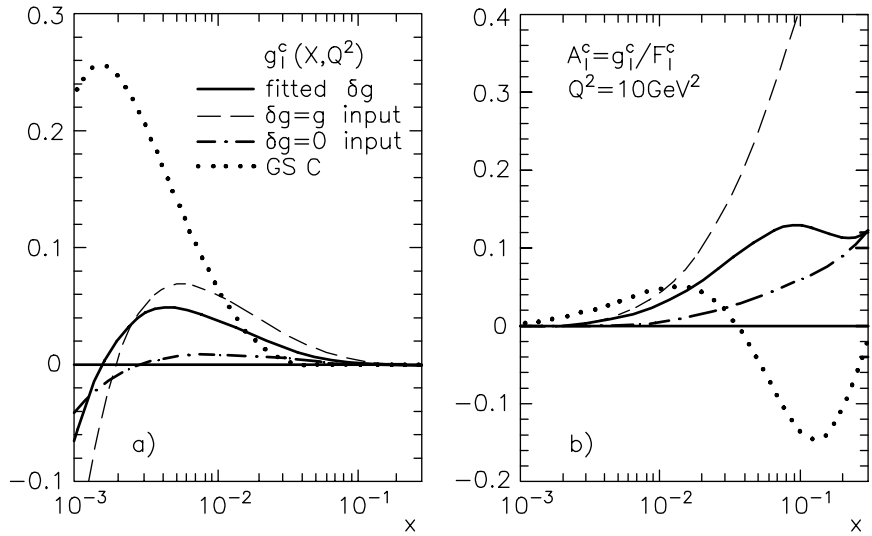,height=7cm}
\vskip 0.5cm
\caption{Charm contribution $g_1^c$ to $g_1$ at $Q^2=10$GeV$^2$ 
for the four gluon distributions of Fig. \ref{fig34}, 
calculated according to Eq. (\ref{65n1}) using $\mu_F=2m_c$ with 
$m_c=1.5$GeV \protect\cite{stratmann1}. }
\label{fig35}                            
\end{center}                                
\end{figure}

In leading order the inclusive 
polarized deep 
inelastic open charm production cross section is given by, 
using Eq. (\ref{221}) or (\ref{2414})  
\begin{equation}
{d^2 \Delta \sigma_c \over dx dy}=
{d^2  \over dx dy}
{1\over 2}
(\sigma_{\Rightarrow}^{\leftarrow}-\sigma_{\Leftarrow}^{\leftarrow})
\approx 
{4\pi \alpha^2 \over Q^2} 
(2-y) g_1^c(x,Q^2) 
\label{65n0}
\end{equation}
where 
\cite{watson,glueck8} 
\begin{equation}
g_1^c(x,Q^2)={\alpha_s (\mu_F^2) \over 9\pi}\int_{(1+{4m_c^2 \over Q^2})x}^1 
{dx' \over x'} \delta g(x',\mu_F^2) \delta C_c({x \over x'},Q^2) 
\label{65n1}
\end{equation}
is the charm contribution to the polarized structure 
function $g_1$ and where  
\begin{equation}
\delta C_c(z,Q^2)=(2z-1)\ln {1+\beta \over 1-\beta}+(3-4z)\beta 
\label{65n2}
\end{equation}
is the partonic matrix element due to $\gamma^{\ast} \delta g 
\rightarrow c \bar c$. Note that the renormalization scale of 
$\alpha_s$ in (\ref{65n1}) has been set equal to the factorization 
scale $\mu_F$ appearing in $\delta g$ and that one usually 
takes $\mu_F=2m_c$. Furthermore,   
$\beta =\sqrt{1-{4m_c^2 \over \hat s}}$ where 
$\hat s=(p+q)^2=Q^2{1-z \over z}$ is the Mandelstam 
variable of the subprocess. 
By combining these formulae with the unpolarized cross 
section one can obtain 
the polarization asymmetry $A^c = {d \Delta \sigma^c \over d \sigma^c}$. 
If one plugs in the drastically differing LO polarized gluon 
densities of \cite{glueck2} and the oscillating GS--C density of 
\cite{gehrmann1} which are for convenience compared to each other in Fig. 
\ref{fig34},  
one obtains the results for $g_1^c(x,Q^2)$ and the deep--inelastic 
charm asymmetry $A^c=g_1^c /F_1^c$ shown in Fig. \ref{fig35} 
at $Q^2=10$GeV$^2$ \cite{glueck8,stratmann1}. 
(Note that the corresponding NLO densities are shown in Figs. 
\ref{fig32} and \ref{figstir1} at $Q^2=4$ GeV$^2$.)
It should further 
be noted that the dashed curve in Fig. \ref{fig35}a corresponds, 
at $x \approx 0.01$, to about 10\% of the full $g_1^p$ which 
implies that a fairly accurate high statistics experiment 
would be required in order to extract $\delta g(x,\mu_F^2)$ from 
$g_1^c$. The size of $A_1^c$ in Fig. \ref{fig35}b in the 
large--x region, $x > 0.01$, is deceptive since here the 
individual $g_1^c$ and $F_1^c={F_2^c-F_L^c \over 2x}$ 
\cite{glueck25} rapidly 
decrease as a result of the threshold condition $\beta \geq 0$.
Realistically, $A_1^c$ is of the order of 5\% at $x \simeq 0.01$ 
where $g_1^c$ is also sizable as shown in Fig. \ref{fig35}. 
For $x \lesssim 0.005$, which could be reached at the HERA 
$\vec e \vec p$ collider, the situation is even less 
favorable, since $g(x,\mu_F^2)$ becomes much larger than  
$\delta g(x,\mu_F^2)$ as $x \rightarrow 0$ and $A_1^c$ is 
correspondingly small. Furthermore, the contribution $\delta C_c$ 
in (\ref{65n2}) from the polarized subprocess 
$\gamma^{\ast}  g \rightarrow c \bar c$ changes sign towards the 
HERA small--x region, so that $A_1^c$ is further suppressed, 
cf. Fig. \ref{fig35}, and becomes probably unmeasurable 
below $x \simeq 0.005$. It should be noted that this latter 
oscillation of $\delta C_c$ causes the strong increase of 
$g_1^c$ in the region of very small x, as shown by the dotted curve 
in Fig. \ref{fig35}a, via the convolution with the peculiarly 
oscillating polarized gluon density GS--C of Fig. \ref{fig34}. 
The relevant asymmetry $A_1^c$ in Fig. \ref{fig35}b remains 
negligible due to the enormously increasing unpolarized gluon 
density for $x\rightarrow 0$.   

%relatively large gluon contribution 
%of \cite{altarelli3}, one gets asymmetries 
%of the order 0.1 in a fixed target experiment such as COMPASS. 
%Under HERA conditions the situation is not as favourable, 
%as will be discussed in detail below for charm photoproduction. 
%First of all, the bulk of the HERA data would come from the 
%small--x region, where $g(x)$ is much larger that $\delta g(x)$ 
%and the asymmetry $A^c$ correspondingly small. Secondly, the 
%matrix element for the polarized subprocess $\gamma^{\ast}g 
%\rightarrow c \bar{c}$ changes sign towards the HERA region, 
%so that $A^c$ is further suppressed.  

\begin{figure}
\begin{center}
\epsfig{file=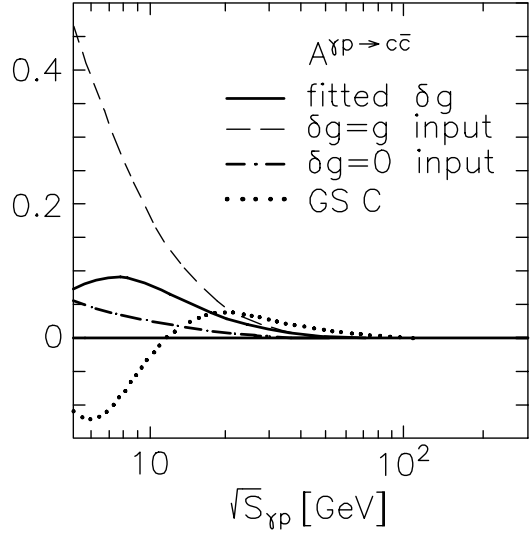,height=7cm}
\vskip 0.5cm
\caption{Longitudinal spin asymmetry for the total charm 
photoproduction cross section calculated according to Eq. 
(\ref{65n3}), using $\mu_F=2m_c$ with 
$m_c=1.5$GeV, for the four polarized gluon 
densities shown in Fig. \ref{fig34} \protect\cite{stratmann1}. }          
\label{figac}                            
\end{center}                                
\end{figure}

We now turn to the case of {\it photo}production of charm. 
It is straightforward to obtain from the above expressions 
(\ref{65n0}) -- (\ref{65n2}) 
the inclusive open charm photoproduction cross section 
by taking the simultaneous limits $Q^2 \rightarrow 0$ and 
$z \rightarrow 0$ while keeping ${Q^2 \over z} \approx \hat s$ 
fixed: 
\begin{equation}
\Delta \sigma^c_{\gamma p} (s_{\gamma})=
{4 \pi \alpha \alpha_s(\mu_F^2) \over 9 s_{\gamma}}
\int_{{4m_c^2 \over s_{\gamma}}}^1 
{dx' \over x'} \delta g(x', \mu_F^2)
(3\beta -\ln {1+\beta \over 1-\beta})  
\label{65n3}
\end{equation}     
where $\beta=\sqrt{1-{4m_c^2 \over \hat s}}$ and $\hat s=x' s_{\gamma}$. 
This integrated cross section depends only on the 
total proton--photon energy $s_{\gamma}=(P+q)^2$ which for 
a fixed target experiment is given by $s_{\gamma}=2ME_{\gamma}$ 
where $E_{\gamma}$ is the photon energy. By varying the 
photon energy it is in principle possible to explore the 
x--dependence of $\delta g$. Very high photon energies 
correspond to small values of x. However, as we shall see 
later, it is not trivial to obtain the first moment 
of $\delta g$ from the cross section Eq. (\ref{65n3}). 
The expected polarization asymmetry is given as a function of 
$\sqrt{s_{\gamma}}$ 
in Fig. \ref{figac} for various possible forms of $\delta g(x,\mu_F^2)$. 
It is clearly seen that it becomes rapidly smaller towards higher 
energies due to the two reasons disussed above [oscillation 
of the parton matrix element and singular behavior of $g(x,4m_c^2)$]. 
Thus the {\it total} charm cross section and the asymmetry become 
unmeasurably small at HERA energies, $\sqrt{s_{\gamma}} \sim 200$GeV, 
to be reached with a possible future doubly polarized $\vec e \vec p$ 
collider \cite{stratmann1}. On the other hand, 
it seems more feasible that 
the total asymmetry in $\gamma p \rightarrow c\bar c$ can be 
measured at smaller energies, $\sqrt{s_{\gamma}} \lesssim 20$ GeV, 
where future polarized fixed--target experiments like 
COMPASS \cite{baum2} will be performed. 

It should be noted that the choice of scale $\mu_F$ 
in $\delta g$ in Eqs. (\ref{65n1}) and (\ref{65n3}) is not 
certain: probably $2m_c$ is a reasonable choice but it 
might as well be $\sqrt{s_{\gamma}}$ or any number in 
between. This uncertainty 
reflects our ignorance about the magnitude of the 
higher order correction and could be resolved if a 
higher order calculation of these cross sections will be 
performed. 
The same statement holds true for the argument 
of $\alpha_s$. Therefore, in the equations presented below 
the scale of $\delta g$ and $\alpha_s$ will be chosen 
to be more general, $\mu_F$ and $\mu_R$, respectively.  
It turns out that the variation of the cross sections when one varies 
$\mu_F$ and $\mu_R$ is larger ($\sim 20$ \%) than that of the asymmetries 
($\leq 5$ \%) 
so that one may speculate that the higher order corrections to 
$A^c$ are small. However, to prove this conjecture a higher order 
calculation is necessary.  

Equation (\ref{65n3}) was obtained after integration over the 
charm quark production angle $\hat \theta$ (in the gluon--photon cms). 
If one is interested in the 
$p_T$ distribution or wants to introduce a $p_T$--cut, it is 
appropriate to keep the $\hat \theta$ dependence in the 
fully differential cross section \cite{frixione,stratmann1}
\be
{d^2 \Delta \sigma^c_{\gamma p} \over dx' d \cos  \hat \theta} 
= {e_c^2 \alpha_s(\mu_R^2) \over 16\hat s}
  \delta g(x',\mu_F^2)   \beta    
\bigl\{ 2 {\hat t^2+ \hat u^2 -2m_c^2\hat s \over \hat t\hat u}
+4m_c^2{\hat t^3+\hat u^3 \over \hat t^2\hat u^2} \bigr\}
\label{65n4}
\ee
where $\hat s=x' s_{\gamma}$, 
$\hat t=-{\hat s\over 2}(1-\beta \cos \hat \theta)$ and 
$\hat u=-{\hat s\over 2}(1+\beta \cos \hat \theta)$. 
It is possible to make a transformation to the transverse 
momentum of the charmed quark by using 
$p_T^2=({\hat s \over 4}-m_c^2)\sin^2\hat \theta$: 
\be
\Delta \sigma^c_{\gamma p} (p_{Tcut})=
{e_c^2 \alpha_s(\mu_R) \over 16 s_{\gamma}}
\int_{{4m_c^2 \over s_{\gamma}}}^1 {dx' \over x'}
\delta g(x',\mu_F^2) {\beta \over {\hat s \over 4}-m_c^2}   
\int_{p_{Tcut}^2}^{{\hat s \over 4}-m_c^2}
{d p_T^2 \over \sqrt{1-{p_T^2 \over {\hat s \over 4}-m_c^2}}}
\biggl[ 2m_c^2{\hat s \over \hat t\hat u}-{\hat t \over \hat u}
                                    -{\hat u \over \hat t} 
-2m_c^2({\hat t \over \hat u^2}+{\hat u \over \hat t^2}) \biggr] \/.  
\label{65n5}
\ee
There are several good reasons to study the $p_T$ distribution. 
First of all and in general, it gives more information than the 
inclusive cross section. Secondly and in particular, it can be 
shown that the integrated photoproduction cross section 
Eq. (\ref{65n3}) as well 
as the corresponding 
DIS charm production cross section in (\ref{65n0}) are not sensitive 
to the first moment of $\delta g$. The sensitivity is strongly 
increased, however, if a $p_T$--cut of the order of 
$p_T \geq 1$ GeV is introduced, see below. 
Last but not least, 
it is experimentally reasonable 
and often necessary to introduce a $p_T$--cut. 

Let us dwell on the first moment discussion for a moment. 
It is true that the first moment is only one among an infinite 
set of moments and the most interesting quantity to know is the full 
x--dependence of $\delta g(x,Q^2)$ at a given $Q^2$. 
However, as was shown in Sect. 5, 
the first moment $\Delta g$ certainly has its 
significance, firstly because it enters the fundamental spin 
sum rule (\ref{111}) and secondly because it gives the contribution 
within the proton to the $\gamma_5$ anomaly, 
$<PS|\bar q \gamma_{\mu} \gamma_5 q|PS>=( \Delta q-{\alpha_s \over 2\pi}
 \Delta g ) S_\mu $,  
within the proton.  
In massless DIS it is straightforward to find out what the 
contribution of the first moment to the cross section is. 
One can apply the convolution theorem (\ref{4123})  
to see that the 
contribution of $\Delta g$ is given by the first moment of the 
parton matrix element, i.e. by the $n=1$ gluonic Wilson coefficient, 
c.f. Eq. (\ref{4224}). 
If masses are involved, like $m_c$, the  
answer to this question is somewhat more subtle. Since the cross 
section is not any more a convolution of the standard form, 
(\ref{4114}), one can formally write the integrals in 
(\ref{65n1}) and (\ref{65n3}) as 
$\int_{\xi}^1 {dx' \over x'} \delta g(x',\mu_F^2) H({\xi \over x'},Q^2)$ 
%$\sigma (a,e)=\int_a^1 {dw \over w} \delta g(w) H({a\over w},e)$ 
where $\xi =(1+{4m_c^2 \over Q^2})x$ for DIS charm production 
and $\xi ={4m_c^2 \over S_{\gamma}}$ for photoproduction ($Q^2=0$) 
of charm. Now one can apply the 
convolution theorem by integrating this expression over $\xi$ 
and the first moment $\int_0^1 dzH(z,Q^2)$ 
gives essentially the contribution from $\Delta g(\mu_F^2)$. 
By integrating Eqs. (\ref{65n2}) and (\ref{65n3}) it turns out 
that both for the inclusive charm DIS and photoproduction 
the corresponding 
quantities $\int_0^1 dzH(z,Q^2)$ identically vanish 
\cite{watson,glueck8,glueck10}. 
This can be traced back to the 
small--$p_T$ behavior of the (perturbative) 
partonic cross section (Wilson coefficient) for 
$\gamma^{\ast} g \rightarrow c \bar c $
which cancels the contribution of the large--$p_T$ region
in $\int_0^1 dzH(z,Q^2)$ \cite{mankiewicz,vogelsang1}.
It is 
not really a surprise in view of the structure of the anomaly 
in massive QCD [cf. the discussion after Eq. (\ref{539}) and 
the Appendix of \cite{lampe1}]. 
Since the integrals $\int_0^1 dzz^{n-1}H(z,Q^2)$ keep being small 
in a neighborhood of n=1 one may conclude from this that 
these cross sections are not suited for determining the first 
moment of $\delta g$. Fortunately, the situation changes 
if one includes a $p_T$--cut of greater than 1 GeV. 
In that case the sensitivity 
to $\delta g(x,\mu_F^2)$ is reestablished because the  
small--$p_T$ behavior of the matrix element for 
$\gamma g \rightarrow c \bar c $ 
does not cancel the contribution of the large--$p_T$ region 
any more \cite{lampe4}. 

\begin{figure}
\begin{center}
\epsfig{file=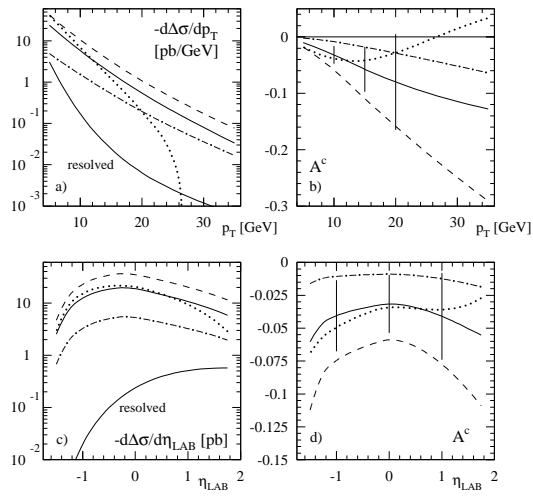,height=7.5cm} 
% BEI DIESER FIGUR WIRD AB 8CM AUF EINE SEPARATE SEITE GEDRUCKT!!!!
\vskip 0.5cm
\caption{$p_T$-- and $\eta_{LAB}$--dependence of the (negative) polarized
charm-photoproduction
cross section and asymmetry in $ep$-collisions at HERA             
\protect\cite{stratmann1}. The considered kinematic regions are 
$-1 \leq \eta_{LAB} \leq 2$ for the $p_T$--distribution and  
$p_T>8$ GeV for the $\eta_{LAB}$--distribution. 
The various curves correspond to the various polarized gluon densities 
in Fig. \ref{fig34}. For comparison, the 'resolved' contribution 
to the cross section, calculated with the 'fitted' $\delta g$ in 
Fig. \ref{fig34} and the 'maximally' saturated set of polarized photonic 
parton densities, is shown by the lower solid curves.}          
\label{figsv3}                            
\end{center}                                
\end{figure}

The formulae presented in Eqs. (\ref{65n3})--(\ref{65n5}) 
were obtained for 
strictly real photons $Q^2=0$. This is a reasonable 
approximation for the projected fixed target experiment 
(photoproduction) 
but may be improved, if one is interested in operating 
the HERA ep collider also with polarized 
high energy protons. In that case 
the Weizs\"acker--Williams approximation may be introduced 
\cite{frixione,stratmann1} to account for the tail of the 
photon propagator. The Weizs\"acker--Williams approximation 
is also advantageous because tagging of the electron, needed 
for the extraction of the cross section at fixed photon energy 
would reduce the cross section too strongly. 
On the basis of the Weizs\"acker--Williams approximation one may go 
on to include a possible resolved photon contribution to the 
cross section described by polarized photon structure functions 
$\delta f^{\gamma}(x,\mu_F^2)$ with f=q,g 
\cite{glueck20,glueck21,glueck22,stratmann5}. 
These quantities are completely unmeasured 
so far and could, in the 'maximal' scenario, contribute up to 20\% of 
the cross section \cite{stratmann1}. 
The polarized 
lowest order cross section for producing a charm quark with 
transverse momentum $p_T$ and cms--rapidity $\eta$ then is 
\be 
\frac{d^2 \Delta \sigma^c_{ep}}{dp_T d\eta} = 2 p_T
\sum_{f^e,f} \int_{\frac{\rho e^{-\eta}}
{1-\rho e^{\eta}}}^1 d x_e x_e \delta f^e (x_e,\mu_F^2) 
x_p \delta f (x_p,\mu_F^2)
\frac{1}{x_e - \rho e^{-\eta}} \frac{d\Delta \hat{\sigma}}{d\hat{t}} 
\label{wqc}
\ee
where  $\rho \equiv m_T/\sqrt{s}$ with $m_T \equiv \sqrt{p_T^2+m_c^2}$ and
$x_p \equiv x_e \rho e^{\eta}/(x_e - \rho e^{-\eta})$.
The sum runs over all relevant parton species. $\delta f^e$ are 
effective polarized parton densities in the longitudinally 
polarized electron defined by 
\begin{equation}  \label{elec}
\delta f^e (x_e,\mu_F^2) = \int_{x_e}^1 \frac{dy}{y} \Delta P_{\gamma/e} (y)
\delta f^{\gamma} (x_{\gamma}=\frac{x_e}{y},\mu_F^2) \;
\end{equation}
where $\Delta P_{\gamma/e}$ is the polarized Weizs\"{a}cker-Williams
spectrum   
\begin{equation}  \label{weiz}
\Delta P_{\gamma/e} (y) = \frac{\alpha}{2\pi} \left[
\frac{1-(1-y)^2}{y} \right] \ln \frac{Q^2_{max} (1-y)}{m_e^2 y^2} \; ,
\end{equation}
and the same cuts as in the unpolarized case should be used, 
$Q^2_{max}=4$ GeV$^2$ and the $y$-cuts $0.2 \leq y \leq 0.85$ 
\cite{derrick}. The cross section Eq. (\ref{wqc}) can be 
transformed to the more relevant HERA laboratory frame by a simple boost
which implies 
$\eta \equiv \eta_{cms} = \eta_{LAB} -\frac{1}{2} \ln (E_p/E_e)$,
where we have counted positive rapidity in the proton
forward direction. The spin-dependent differential LO subprocess cross
sections $d\Delta \hat{\sigma}/d\hat{t}$ for the resolved processes   
$gg \rightarrow c\bar{c}$ and $q\bar{q} \rightarrow c\bar{c}$ with $m_c
\neq 0$ can be found in \cite{contogouris,karliner}.  
The dominant direct ('unresolved') contribution derives from 
$\delta f^{\gamma}(x_{\gamma},\mu_F^2) \equiv \delta (1-x_{\gamma})$ in 
(\ref{elec}) with the corresponding polarized cross sections 
for the direct subprocess $\gamma g \rightarrow c\bar c$ 
being readily obtained from that for $g g \rightarrow c\bar c$ by dropping 
the non--abelian (s--channel) part and multiplying by 
$2N_c e_c^2 \alpha / \alpha_s (\mu_F^2)$ where $e_c=2/3$. 
Note that the resolved photon contributions are relevant mainly 
for (real) photoproduction and that they are appreciable only at 
very high energies 
$\sqrt{s_{\gamma}} \geq 100$ GeV. 
Furthermore, there are experimental techniques to separate 
the resolved part from the direct photon contribution, see e.g. 
\cite{forshaw}. 
Figure \ref{figsv3} shows results for the $p_T$ and $\eta_{LAB}$ 
distributions obtained for the four different polarized gluon  
densities in Fig. \ref{fig34} 
for $E_p=820$ GeV and $E_e=27$ GeV \cite{stratmann1}. 
The curve denoted by 'resolved' is an estimated upper limit for 
the resolved photon contribution. It is negligibly small unless 
$p_T$ becomes very small. Also shown are the corresponding 
asymmetries $A^c$ which are much larger than for the total cross 
section if one goes to $p_T$ of about 10-20 GeV. Furthermore, 
the asymmetries are sensitive to the size and shape of the
polarized gluon distribution used. Included in the asymmetry 
plots are the expected
statistical errors $\delta A^c$ at HERA which can be estimated from
\begin{equation}  \label{aerr}
\delta A = \frac{1}{P_e P_p \sqrt{{\cal L} \sigma \epsilon}} \; ,
\end{equation} 
where $P_e$, $P_p$ are the beam polarizations, ${\cal L}$ is the integrated
luminosity and $\epsilon$ the charm detection efficiency, estimated to be 
$P_e * P_p=0.5$, ${\cal L}=100$pb$^{-1}$ and $\epsilon=0.15$.

%% file: k622tex
\subsection{Heavy Quark Production in Hadronic Collisions} 

{\it Hadronic} heavy quark production proceeds via the (so far 
available) LO subprocesses 
\begin{equation}
\delta g \, \delta g \rightarrow h\bar h \qquad , 
\qquad \delta q \, \delta \bar q \rightarrow h\bar h \qquad , 
\label{t637}
\end{equation}
and appears to be a very sensitive and presumably the 
most realistic test of $\delta g(x,\mu_F^2)$, since 
$\delta g$ enters 'quadratically' and the $\delta q \delta \bar q$
contribution is small \cite{cortes,doncheski,contogouris,karliner}. 
Here the polarized $\vec p \vec p$ RHIC collider ($\sqrt{s}=50-500$GeV) 
with high luminosity (${\cal L} \gtrsim 10^{32}$ cm$^{-2}$ s$^{-1}$) will 
play a decisive role \cite{bunce,yokosawa}. The differential 
cross sections for the subprocesses in Eq. (\ref{t637}) are 
given by \cite{contogouris} 
\begin{equation}
{d \over d \hat t } \Delta \hat \sigma^{gg\rightarrow h\bar h}
\equiv {d \over d \hat t } {1\over 2} 
(\hat \sigma^{gg\rightarrow h\bar h}_{++} - 
\hat \sigma^{gg\rightarrow h\bar h}_{+-} )
={\pi \alpha_s^2 \over 8 \hat s ^2} 
({3\over \hat s^2}-{4 \over 3 \tilde t \tilde u})
[\tilde t^2 +\tilde u^2 -{2m_h^2\hat s \over \tilde t \tilde u} 
          (\tilde t^2 +\tilde u^2)] 
\label{t638} 
\end{equation}
\begin{equation}
{d \over d \hat t } \Delta \hat \sigma^{q\bar q\rightarrow h\bar h}
= -{d \over d \hat t } \hat \sigma^{q\bar q\rightarrow h\bar h}
=-{\pi \alpha_s^2 \over 3 \hat s ^2} {4\over 3} 
{\tilde t^2 +\tilde u^2 +2m_h^2\hat s \over \hat s^2} 
\label{t639} 
\end{equation}
where $\pm$ refers to the helicity of the incoming partons, 
$\alpha_s=\alpha_s(\mu_F^2)$ and $\tilde t \equiv \hat t -m_h^2$, 
$\tilde u \equiv \hat u -m_h^2$, i.e. $\hat s +\tilde t 
+\tilde u=0$. By integrating with respect to $\tilde t$ the 
total cross sections are then easily obtained : 
\begin{equation}
\Delta \hat \sigma^{gg\rightarrow h\bar h} (\hat s,\mu_F^2)
={\pi \alpha_s^2 \over 16 \hat s } 
[2(3\beta^2-{17\over 3})\ln {1+\beta \over 1-\beta}
          +5\beta (5-\beta^2)]
\label{t640} 
\end{equation}
\begin{equation}
\Delta \hat \sigma^{q\bar q\rightarrow h\bar h}(\hat s,\mu_F^2)
= - \hat \sigma^{q\bar q\rightarrow h\bar h}(\hat s,\mu_F^2)
=-{\pi \alpha_s^2 \over 9 \hat s } {4\over 3} \beta (3-\beta^2) 
\label{t641} 
\end{equation}
with $\beta^2=1-4m_h^2/\hat s$. The unpolarized differential and 
total cross sections for $gg\rightarrow h\bar h$ are well known 
\cite{glueck18}, with the latter being given by 
\begin{equation}
\hat \sigma^{gg\rightarrow h\bar h} (\hat s,\mu_F^2)
\equiv {1\over 2}
(\hat \sigma^{gg\rightarrow h\bar h}_{++} + 
\hat \sigma^{gg\rightarrow h\bar h}_{+-} )
={\pi \alpha_s^2 \over 16 \hat s } 
[(11-6\beta^2+{1\over 3}\beta^4)\ln {1+\beta \over 1-\beta}
          +{1\over 3}\beta (31\beta^2-59)] \quad .
\label{t642} 
\end{equation}
A NLO analysis of the polarized cross section $\Delta \hat \sigma^{ij}$ 
for heavy quark production is unfortunately still missing. 

\begin{figure}
\begin{center}
\epsfig{file=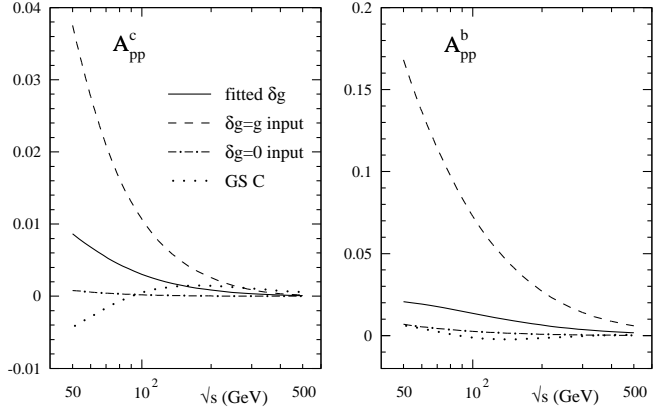,height=6.0cm}
\bigskip
\caption{Asymmetries for the total charm and bottom 
hadroproduction, according to Eq. (\ref{t645}), at RHIC energies 
using $m_c=1.5$ GeV and $m_b=4.5$ GeV, for the four polarized 
gluon densities shown in Fig. \ref{fig34}.}  
\label{fig37}                            
\end{center}                                
\end{figure}

For not too small polarized gluon densities, 
the total polarized cross section for heavy quark production 
($4m_h^2 \leq \hat s \leq s$)  
is dominated by the gg--initiated subprocess, because  
the contribution from the subprocess 
$q\bar q \rightarrow h\bar h$ is marginal 
for large energies ($\sqrt s \gtrsim 50$ GeV). 
Although the latter can be easily implemented \cite{contogouris}, 
we give here for simplicity only the result for the former 
cross section :
\begin{eqnarray} \nonumber
\Delta \sigma^h_{pp} (s) &=& 
\int_{4m_h^2}^s d \hat s \int dx_1 dx_2 \delta g(x_1,\mu_F^2)
\delta g(x_2,\mu_F^2) \Delta \hat \sigma^{gg\rightarrow h\bar h}
(\hat s,\mu_F^2) \delta(\hat s -x_1 x_2 s) \\ \nonumber 
&=&\int_{{4m_h^2\over s}}^1 dx_1 \int_{{4m_h^2\over s x_1}}^1dx_2 
\delta g(x_1,\mu_F^2) \delta g(x_2,\mu_F^2) 
\Delta \hat \sigma^{gg\rightarrow h\bar h}
(x_1 x_2 s,\mu_F^2) \\ 
&=&\int_{{4m_h^2\over s}}^1 {d\tau \over \tau} 
\Delta \hat \sigma^{gg\rightarrow h\bar h} (\tau s,\mu_F^2) 
\Phi_{gg}(\tau ,\mu_F^2)
\label{t643} 
\end{eqnarray}
where the polarized gluon luminosity (flux) is given by 
\begin{equation}
\Phi_{gg}(\tau ,\mu_F^2)
=\tau \int_{\tau}^1 {dx_1 \over x_1} 
\delta g(x_1,\mu_F^2) \delta g({\tau \over x_1},\mu_F^2) \, . 
\label{t644} 
\end{equation}
The relevant spin--spin asymmetry is defined by 
\begin{equation}
A_{pp}^h(s)
={\Delta \sigma_{pp}^h(s) \over \sigma_{pp}^h(s)} 
\label{t645} 
\end{equation}
where the unpolarized cross section is analogous to the 
polarized cross section Eq. (\ref{t643}), with $g(x,\mu_F^2)$ appearing
instead of $\delta g(x,\mu_F^2)$ and the unpolarized partonic 
cross section (\ref{t642}) has to be used instead of 
$\Delta \hat \sigma$. The typical scale to 
be used in
(\ref{t643}) is again $\mu_F \approx 2m_h$. For the production of heavy 
quarkonia ($J/\psi$, $\Upsilon$, etc.) 
one proceeds in the same way except that in 
Eq. (\ref{t643}) the region of integration has to be taken 
as $4m_c^2 \leq \hat s \leq 4m_D^2$ or 
$4m_b^2 \leq \hat s \leq 4m_B^2$, instead of 
$4m_h^2 \leq \hat s \leq s$. However, one faces here the additional 
bound--state model dependence as discussed in the previous 
section \cite{cortes,doncheski}. The expected asymmetries 
(\ref{t645}) for 
total charm and bottom production 
at typical RHIC energies are shown in Fig. \ref{fig37} for 
the various possible forms of $\delta g(x,\mu_F^2)$
shown in Fig. \ref{fig34}. Although the asymmetries are 
very sensitive to the polarized gluon density 
$\delta g(x,4m_h^2)$, they become relatively small at top 
RHIC energies, with $A_{pp}^b$ being almost an order 
of magnitude larger than $A_{pp}^c$. It seems that realistic 
measurements of $\delta g$ will be possible preferably at medium 
RHIC energies, $\sqrt s \lesssim 100$ GeV, or at future HERA--$\vec N$ 
energies, $\sqrt{s} \approx 40$ GeV (for a recent review see 
\cite{korotkov}, and references therein). It should be noted, 
however, that the introduction of a $p_T$--cut, as discussed 
in the previous section for charm photoproduction 
\cite{stratmann1,lampe4}, could result in sizeably larger asymmetries 
at high energies.

%% file: k63tex
\subsection{High $p_T$ Jets in High Energy Lepton Nucleon Collisions} 

A possible way to distinguish the spin--dependent parton distributions 
is to study jet production at HERA operated in the polarized mode 
\cite{carlitz,altarelli3,glueck8}. 
The idea is based on the fact that the parton subprocesses 
$\gamma^{\ast}q \rightarrow gq$ and 
$\gamma^{\ast}g \rightarrow q\bar q$ lead to a different 
large--$p_T$ behavior so that the contributions from 
$\delta q(x,\mu_F^2)$ and from $\delta g(x,\mu_F^2)$ might be distinguished. 
The undetermined QCD renormalization/factorization 
scale $\mu_F$ is taken to be 
$\mu_F^2=Q^2$, although a choice like $\mu_F=p_T$ is equally 
feasible. Unfortunately, at CERN (SMC), SLAC and DESY (Hermes) 
jet cross sections are difficult to measure because the 
signatures are minute at small $\sqrt s$. 
Furthermore, it turns 
out that the polarization asymmetries 
for jet production even at HERA ($\sqrt s \approx 300$ GeV) 
are not too large, and  
in view of the statistical error obtained for luminosities 
${\cal L} \sim 100 pb^{-1}$, 
cf. Eq. (\ref{aerr}) with $\epsilon \sim 1$, one should 
not be too optimistic that $\delta g(x,\mu_F^2)$ can be determined 
from such measurements. Nevertheless, an overview 
of the physics issues will be given in the following. 

The result for $<g_1^p>$ when written differentially  
in $\lambda ={4p_T^2 \over Q^2}$ is 
\begin{equation}
{d \over d\lambda}\int_0^1 dx g_1^p(x,Q^2,\lambda)= 
{\alpha_s(\mu_R^2) \over 4\pi} \sum_q e_q^2 
\bigl[ C_F M_q(\lambda)(\Delta q+\Delta \bar q)(\mu_F^2) 
+T_R M_g(\lambda)\Delta g(\mu_F^2) \bigr]
\label{6j1}
\end{equation}
with $C_F=4/3$ and $T_R=1/2$, and 
where the functions $M_q(\lambda)$ and $M_g(\lambda)$ are given by 
\begin{equation}
M_q(\lambda)={{11 \over 8} \lambda +{13   \over 4  } +{ 2  \over \lambda } 
                   \over (1+ \lambda)^{3/2}    } 
         \ln {\sqrt{1+\lambda}+1 \over \sqrt{1+\lambda}-1} 
         -{12+11\lambda \over 4\lambda (1+\lambda)} 
\label{6j2} 
\end{equation}
\begin{equation}                                          
M_g(\lambda)={\lambda -2 \over 2 (1+\lambda)^2} 
       -{\lambda (4+\lambda) \over 4 (1+\lambda)^{5/2}  }
        \ln {\sqrt{1+\lambda}+1 \over \sqrt{1+\lambda}-1}  \, .
\label{6j3} 
\end{equation}
As usual, one takes $\mu_R^2=\mu_F^2=Q^2$. 
The important point to observe is \cite{altarelli3} 
that $M_g(\lambda)$ gives 
a negative contribution which, for a large positive 
$\Delta g \sim 3$ would be the dominant contribution in Eq. (\ref{6j1}). 
However, this derivation is strongly idealized, because the 
negative $M_g$ gets a large contribution from the small--x 
region where measurements become difficult due to the large 
unpolarized cross section. Imposing a cut $x \geq x_0 $ on the 
x--integration \cite{glueck8}, typically $x_0=Q^2/2ME_l$ with 
$x_0 \leq x \leq (1+\lambda)^{-1}$, the negative signal from $M_g$ 
is drastically reduced so that it is doubtful that this effect can be 
used to determine $\Delta g$. 

\begin{figure}
\begin{center}
\epsfig{file=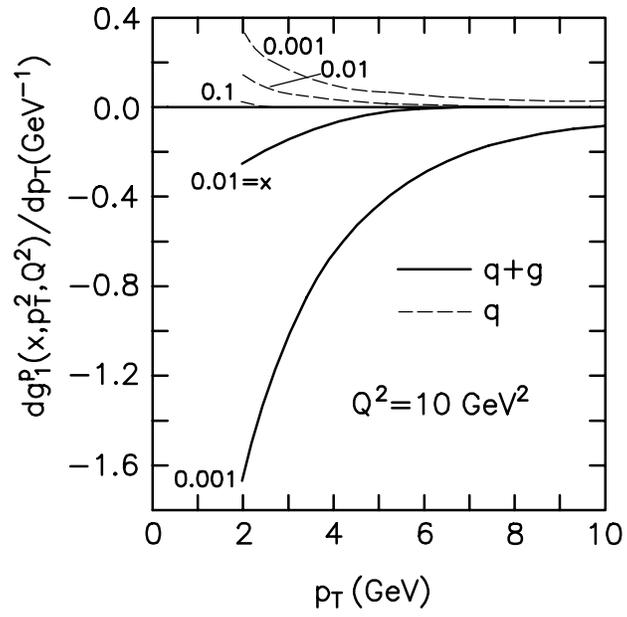,height=8cm}
\vskip 0.5cm
\caption{$p_T$ depencence of $g_1^p(x,p_T^2,Q^2)$ for various values of x 
at $Q^2=10$ GeV$^2$ \protect\cite{glueck8}.}
\label{figxt}                            
\end{center}                                
\end{figure}

Instead of using a cut one can directly verify this finding by a study 
of the distribution ${dg_1^p(x,Q^2,\lambda) \over d\lambda}$ as 
a function of x and $\lambda$ \cite{glueck8}. The result is shown 
in Fig. \ref{figxt}. It can be seen that only for $x \leq 0.001$ 
a clear negative signal in the perturbatively safe region 
$p_T \gtrsim 4$ GeV develops. This feature has nothing to do 
with the particular $\delta g(x,Q^2)$ and $\delta q(x,Q^2)$ 
chosen for this plot, 
but is a consequence of the structure of the perturbative 
parton matrix element. 

So far in this subsection  
we have discussed the determination of the first moment 
$\Delta g$ . Now let us come to the directly measurable 
x--dependent densities $\delta g(x,\mu_F^2)$ etc. 
In general, the cross section for polarized deep inelastic electron 
proton scattering with several partons
in the final state 
\begin{equation}
e^-_{\Rightarrow}(l) + p_{\Rightarrow}(P) \rightarrow  e^-(l^\prime)+
\mbox{remnant}(p_r) +
\mbox{parton} \,\,1 (p_1) +
\ldots
+\mbox{parton}\,\, n (p_n)
\label{eqn621}
\end{equation}
is generically given by 
\begin{equation}
d\Delta \sigma^{had}_{n-\mbox{jet}} = \sum_f
\int dx_f \,\, \delta f(x_f,\mu_F^2)\,\,\, 
d \Delta \hat{\sigma}^f(p=x_f P,
\alpha_s(\mu_R^2), \mu_R^2, \mu_F^2)
\label{eqn622}
\end{equation}
where the sum runs over incident partons $f=q,\bar{q},g$ which carry
a fraction $x_f$ of the proton momentum. 
$\Delta \hat{\sigma}^f$ denotes the partonic cross section from which
collinear
initial state singularities are factorized out
(in a next-to-leading order calculation) 
and included in the scale dependent parton densities
$\delta f$. A natural scale for multi--jet production in DIS 
may be 
$ \mu_R^2 = \mu_F^2 = 1/4\;(\sum_j \,p_T^B(j))^2$, 
with the average $p_T^{B}(j)$ being defined 
in the Breit frame by $2\,E_j^2(1-\cos\theta_{jP})$, 
where the subscripts $j$ and $P$
denote the jet and proton. 
Results about this cross section 
including HO effects can be found in \cite{mirkes,feltesse}. 
If the jets are defined in
a modified JADE scheme, the theoretical uncertainties 
of the two-jet cross section can be very large 
due to higher order effects. 
These uncertainties are smaller for the cone algorithm,
which is defined in the laboratory frame.
In this algorithm  the distance 
$\Delta R=\sqrt{(\Delta\eta)^2+(\Delta\phi)^2}$ between two partons
decides whether they should be recombined to a single jet. Here the
variables
are the pseudo-rapidity $\eta$ and the azimuthal angle $\phi$. 
Partons with 
$\Delta R<1$ are recombined.
Furthermore, a cut on the jet transverse momenta of $p_T(j)>5$~GeV in the
lab frame and in the Breit frame was imposed.
Using the polarized parton densities (set A) of \cite{gehrmann} 
it was found \cite{feltesse} that the LO 
polarized dijet cross section
$\Delta{\sigma}(2-\mbox{jet})$ is $-45$ pb 
at HERA energies ($\sqrt s \approx 300$ GeV).
This negative result for the polarized dijet cross section is entirely due
to the boson-gluon fusion process, which is negative
for $x \lesssim 0.025$ and its  contribution to the total
polarized dijet cross section
is -53 pb. The contribution from the quark initiated subprocess
is positive over the whole kinematical range and contributes with
8 pb to the resulting dijet cross section.
With these numbers one obtains a rather small 
average asymmetry of $\approx -0.015$. 
To obtain these numbers, polarizations of 70\%  for both
the electron and the proton  beams have been assumed. 
Note, however, that the shape of the
polarized gluon density
is hardly (or even not at all)  constrained by currently
available DIS data, in particular for small $x$.
Alternative parametrizations of the polarized gluon distributions in
the small--$x$ region,
which are still consistent with all present data \cite{gehrmann,glueck2},
can lead to
asymmetries 
which are a factor two larger. 
Although the dijet events are in principle 
sensitive to this lower $x$ range, the resulting numbers for the 
asymmetries are in general small because of the dominance of 
the unpolarized gluon 
density in the small--$x$ region. 

Reducing the proton beam energy to 410 GeV, instead of the nominal
820 GeV, does not improve the signal,
although the mean value of $y$ is higher.
The asymmetry signal increases only for a few points around $x>0.1$,
since a lower incident energy probes slightly higher values of $x$ 
\cite{mirkes,feltesse}.

One may also have a look at large--$p_T$ jet photoproduction 
\cite{stratmann1}. In that case 
the generic cross section formula for the production of a single jet
with transverse momentum $p_T$ and rapidity $\eta$ is similar to
that in (\ref{wqc}), the sum now running over all properly symmetrized
$2\rightarrow 2$ subprocesses for the direct ($\gamma b\rightarrow cd$)
and resolved ($ab\rightarrow cd$) cases. 
The corresponding differential helicity-dependent LO subprocess
cross sections can be found in \cite{babcock} for the case that 
only light flavors are involved. 
One may neglect the 
charm content of the nucleon and consider charm only contributing 
as a final state via 
$\gamma g \rightarrow c\bar{c}$ (for the direct part) and $gg \rightarrow
c\bar{c}$, $q\bar{q} \rightarrow c\bar{c}$ (for the resolved part). 

Just as for the case of heavy quark photoproduction it turns out 
that the resolved photon contributions are at most subdominant 
(in the 'maximal' scenario studied by \cite{stratmann1}), 
if not negligible. As these authors have found, the 'direct' 
cross sections are fairly large over the whole rapidity 
range, and also as a funtion of $p_T$, and sensitive to the shape
and size of $\delta g(x,p_T^2)$ with, unfortunately, not too sizeable
asymmetries as compared to the statistical errors for 
${\cal L}=100$ pb$^{-1}$.
A measurement of $\delta g$ thus appears to be possible under the imposed
conditions only if luminosities exceeding 100 pb$^{-1}$ can be reached.

%% file: k64tex
\subsection{Semi--Inclusive Polarization Asymmetries} 

A straightforward idea to study polarized DIS in more detail is 
to analyze semi--inclusive cross sections.   
Heavy quark production as discussed in Sect. 6.2 and 6.3 is an example 
for this. Another possibility is to tag for light quark hadrons, 
like the pions, kaons or protons, in order to obtain information on the spin 
flavor content of the nucleon, i.e. consider the process 
\begin{equation}
\vec e +\vec p \rightarrow e'+H+X
\, . 
\label{6400}
\end{equation}
The charged hadrons H most suited for this analysis are mainly 
$\pi^{\pm}$ and $K^{\pm}$ but also $p$ and $\bar p$ may be 
interesting \cite{adeva2}. Recently, $\pi^0$ production has 
been used as well for extracting $\delta s(x,Q^2)$ 
\cite{avakian}.  
In particular one can 
compare cross sections of hadrons with positive and negative 
electric charge to obtain additional information. 
In the unpolarized case and within the framework 
of the LO--QCD quark parton model the cross section is given by 
\begin{equation}         
{1 \over \sigma} {d\sigma^{\pm} \over dz} =
{\sum_{q,H} e_q^2 q(x,Q^2) D_q^H(z,Q^2) \over \sum_{q} e_q^2 q(x,Q^2)} 
\label{6401}
\end{equation}           
where $\sigma$ is the inclusive cross section and one is considering 
the production of hadrons $H^{\pm}$. The fragmentation function 
$D_q^H(z,Q^2)$ represents the probability that a struck quark 
with flavor q fragments into a hadron H carrying fractional 
momentum z of the parent (=struck) quark q (see, for example, 
\cite{altarelli1,reya}). In NLO the cross section in (\ref{6401}) 
does not factorize in x and z anymore, but instead one has more 
complicated double convolution integrals over parton densities, 
fragmentation functions and Wilson coefficients 
\cite{altarelli9,furmanski}. 

In the polarized case one may define 
asymmetries $A_{1}^{\pm}$, similar 
to the inclusive asymmetry Eq. (\ref{224}), 
namely 
\begin{equation}         
A_1^{\pm}=\frac{\sigma_{1/2}^{\pm}-\sigma_{3/2}^{\pm}}
{\sigma_{1/2}^{\pm}+\sigma_{3/2}^{\pm}}
\, .
\label{6499}
\end{equation}           
In analogy to (\ref{6401})  
these semi--inclusive 
asymmetries in LO--QCD are given by  
\begin{equation}         
A_1^{\pm}(x,z,Q^2)={\sum_{q,H} e_q^2 \delta q(x,Q^2) D_q^H(z,Q^2) \over 
\sum_{q,H} e_q^2 q(x,Q^2) D_q^H(z,Q^2) } \, .
\label{6402}
\end{equation}
This result holds 
under the (to some extent questionable)  
assumption that the fragmentation functions do not depend 
on the quark helicity \cite{frankfurt,close4}.  
The idea then is to take the $q(x,Q^2)$ and the $D_q^H(z,Q^2)$  
from other (unpolarized) experimental data and use the 
measurement of $A_1^{\pm}$ to determine the $\delta q(x,Q^2)$. 
By using different targets (proton, deuteron and $^3$He) one 
measures different linear combinations of the $\delta q(x,Q^2)$ 
according to Eq. (\ref{6402}). For example, the deuteron 
cross section is considered the sum of the proton and 
neutron cross section, corrected by the D--state factor 
$1-{3\over 2}w_D$, as will be discussed in Sect. 6.10.  
Because of isospin symmetry the asymmetries 
for proton and neutron are related 
by exchange of up and down quarks and antiquarks. 
To increase the statistics, the LO relation (\ref{6402}) 
is sometimes integrated over the measured z--domain 
\cite{adeva2}, i.e. $D_q^H(Q^2)=\int_{0.2}^1 dz D_q^H(z,Q^2)$ 
is used in (\ref{6402}) instead of $D_q^H(z,Q^2)$. 

One may take higher order QCD corrections to Eq. (\ref{6402}) 
into account. The NLO corrections to the denominator 
of the asymmetry are well known \cite{altarelli9,furmanski}, and the 
NLO corrections to the numerator have been calculated too 
\cite{chiappetta,florian2}. 
In the $\overline{MS}$ scheme, where the anomalous gluon 
contribution is hidden in the definition of the singlet 
quarks $\Delta \Sigma$, the NLO corrections turn out to be much smaller 
than the present experimental accuracy. Furthermore, 
present measurements \cite{adeva2} do not involve the small--x 
region $x \lesssim 0.005$.   

If one considers the production of the 6 charged hadrons 
$\pi^{\pm}$, $K^{\pm}$, $p$ and $\bar p$ from 3 quarks 
and 3 antiquarks, there are 
in principle 36 independent fragmentation functions. 
Among these the 
fragmentation functions of 
strange quarks into pions can be neglected. 
The fragmentation functions of 
non--strange quarks into pions 
can be obtained, for example, from EMC measurements 
\cite{arneodo} 
by using 
charge conjugation and isospin symmetry. 
The number of independent 
fragmentation functions can be further reduced by assumptions 
like $D_d^{(K^-=\bar u s)}=D_{\bar d}^{(K^-=\bar u s)}$ and 
$D_s^{(K^-=\bar u s)}=D_{ d}^{(\pi^-=\bar u d)}$ for unfavored 
and favored fragmentations, respectively, so that finally 
the number of independent quark fragmentation functions is 6. 
In addition there are the three gluon fragmentation functions 
$D_g^H(z,Q^2)$ which in LO enter only via the evolution 
equations, whereas in NLO they enter the cross sections 
and asymmetries (\ref{6402}) directly.

The quark fragmentation functions 
together with 
the unpolarized parton 
densities and a measurement of the spin asymmetries of 
proton and deuteron and/or $^3$He serve as input to  
determine all the valence and 
sea quark densities $\delta q_v(x,Q^2) =\delta q(x,Q^2)-\delta \bar q(x,Q^2)$ 
and $\delta q_{sea}(x,Q^2) =2\delta \bar q(x,Q^2)$ (q=u,d,s). 
If one 
measures the spin asymmetries in certain x--bins, there is 
for each x--bin 
a system of linear equations for the 6 unknown spin 
distributions. 
The weight of the strange quark distributions 
in these equations is marginal, so that 
they cannot be determined. Recently, however, asymmetries for 
$\pi^0={1\over 2}(\pi^+ +\pi^-)$ production off a $^3He$ target 
have been measured \cite{avakian} which allow the extraction of 
$\delta s \over s$ according to a suggestion of $\cite{frankfurt}$. 
Although the statistics is still inferior, there is an indication 
that $\delta s(x,Q^2)$ turns negative for $x \lesssim 0.1$ and 
vanishes at larger values of x as theoretically anticipated. 
The statistical errors of present polarized experiments 
are so large that the power of the method can be   
increased by the additional assumption $\delta \bar u(x,Q^2) 
= \delta \bar d(x,Q^2) \equiv \delta \bar q(x,Q^2)$. 
Finally one is left with 3 unknown 
functions $\delta u_v(x,Q^2)$, $\delta d_v(x,Q^2)$ and $\delta \bar q(x,Q^2)$.  
These have been determined in a recent analysis by the SMC 
collaboration \cite{adeva2} for 12 x--bins between 
0.005 and 0.48 and assuming $Q^2$--independence of the asymmetries. 
Their results show that at the present stage 
this method is not really accurate enough for a quantitative 
analysis. Typical errors are about 50\% or larger. However, 
in future this method will be certainly very fruitful to 
discriminate between the polarized up and down and sea quark 
contribution. A check on the consistency of the 
procedure is possible by using the relation 
\begin{equation}
g_1^p(x,Q^2)-g_1^n(x,Q^2)={1 \over 6} [\delta u_v(x,Q^2)-\delta d_v(x,Q^2)] 
+O(\delta \bar u-\delta \bar d)+(HO-QCD) 
\, .
\label{6404}
\end{equation}
The l.h.s. of this equation is obtained from inclusive 
data while the r.h.s. can be obtained from the semi--inclusive 
asymmetries. In addition the assumptions $\delta \bar u(x,Q^2)
= \delta \bar d(x,Q^2)$ and the smallness of the HO corrections can 
be checked. 

An alternative procedure, 
in case of pions, has been suggested by 
\cite{frankfurt}.  
These authors consider the asymmetry 
\begin{equation}
A_{mixed}=\frac{(\sigma_{1/2}^+ -\sigma_{1/2}^-)
-(\sigma_{3/2}^+ -\sigma_{3/2}^-)} 
{(\sigma_{1/2}^+ -\sigma_{1/2}^-)+(\sigma_{3/2}^+ -\sigma_{3/2}^-)} 
\label{6498}
\end{equation}
where 'mixed' refers to the combination $\pi^+ - \pi^-$ of 
$\pi^{\pm}$ production, for example, and 
which has 
a simple expression in terms of up-- and down--quark densities, 
\begin{equation}
A_{mixed}=\frac{4 \delta u_v -\eta \delta d_v}{4 u_v -\eta d_v} \, .
\label{6497}
\end{equation}
For the combination $\pi^+ - \pi^-$ the fragmentation functions 
cancel and thus $\eta =1$. If, however, the hadrons are not identified, 
as in the SMC experiment \cite{adeva2}, the factor $\eta$ has to be 
evaluated by averaging the relevant fragmentation functions and is 
found to be about 0.5 for $z > 0.2$. This method, however, is not as 
simple as it looks, because the spectrometer acceptance is different 
for positive and negative 
hadrons and the ratio of these 
acceptances does not cancel in the asymmetries Eq. (\ref{6498}). 
Still one can measure this ratio and correct for the 
acceptance difference. Finally, the quark distributions 
$\delta u_v$ and $\delta d_v$ obtained this way should be in agreement 
with the distributions obtained from $A_1^{\pm}$. 

There is the possibility to obtain information on 
$\delta s$ by looking at processes with fast kaons ($K^-=\bar u s$) in 
the final state \cite{close4}. These have a high probability to contain the 
initial struck quark, i.e. a fast kaon can be a signal for a 
s or $\bar u$ quark struck by the photon. Observation of the 
corresponding polarization asymmetry $A_{1,fast}^{K^-}$ is a 
signature of the existence of a polarized s resp. $\bar u$ sea. 
However, there are strong systematic uncertainties in such an 
experiment because some contribution will arise from $s \bar s$ 
pairs created in the photon--gluon fusion process.  

Polarization asymmetries for semi--inclusive pion production 
have been measured from doubly longitudinally polarized 
(anti)proton--proton collisions at the Fermilab SPF 
\cite{adams1} resulting in $A_{LL}^{\pi^0} \approx 0$ 
for $1 \lesssim p_T^{\pi^0} \lesssim 4$ GeV at 
$E_{beam}^{p}=200$ GeV, i.e. $\sqrt s =20$ GeV. It should 
be pointed out that this result does not necessarily imply a 
vanishing \cite{adams1,ramsey} gluon polarization $\Delta g$, 
but is equally consistent with a large $\Delta g \approx 3-6$ 
\cite{vogelsang4}. A clean distinction between a large and 
a small $\Delta g$ scenario could be achieved, if it were 
possible to perform such a semi--inclusive experiment at, say, 
$\sqrt s \approx 100$ GeV with $p_T^{\pi^0} \gtrsim 5$ GeV 
\cite{vogelsang4}.

%% file: k65tex
\subsection{Information from Elastic Neutrino--Proton Scattering} 

(Quasi)--elastic neutrino--proton scattering ($\nu p \rightarrow \nu p$) 
is mediated by the 
exchange of the Z boson. Since the parity violating 
Z--quark coupling involves $\gamma_{\mu} \gamma_5$, the 
{\it unpolarized} cross 
section will depend on the proton matrix element 
of the axial vector current. As will be shown below 
this offers in particular 
the possibility to measure a combination of the strange quark matrix
$\left\langle p | \bar s \gamma_{\mu} \gamma_5 s |p\right\rangle $ 
and the anomalous gluon component.  
Theoretical aspects of this process have been reviewed 
by \cite{kaplan} and more recently by \cite{bilenky}. 
A very readable experimental paper on the subject is 
\cite{ahrens}. New neutrino experiments
like CHORUS, NOMAD, ICARUS, MINOS, COSMOS, etc. \cite{morales} 
aimed to search for neutrino oscillations
are taking data or are under preparation and could 
also be used to get information
on the NC neutrino
(and antineutrino) elastic scattering on protons.

The one-nucleon matrix element
of the hadronic neutral current
has the form
\begin{equation}
\left\langle
p'
\left|
J^Z_{\mu}
\right|
p 
\right\rangle
=
\bar{u}(p')
\left[
\gamma_{\mu}
F_V^Z(Q^2)
+
\frac{i}{2M}
\,
\sigma_{\mu\nu}
q^{\nu}
F_M^Z(Q^2)
+
\gamma_{\mu}
\gamma_5
G_A^Z(Q^2)
\right]
u(p)  
\;.   
\label{e651}
\end{equation}
As will be explained now, 
$F_V^Z$, $F_M^Z$ and $G_A^Z$ can be written as linear 
combinations of normalized matrix elements ('form factors') 
of the following $U_3 \times U_3$ quark currents : 
\begin{equation}
V_{\mu}^a=\bar q \gamma_{\mu} T^a q 
\qquad
V_{\mu}^0={1 \over 3} \bar q \gamma_{\mu} q 
\label{e652}
\end{equation}
\begin{equation}
A_{\mu}^a=\bar q \gamma_{\mu} \gamma_5 T^a q 
\qquad
A_{\mu}^0={1 \over 3} \bar q \gamma_{\mu} \gamma_5 q  \, .   
\label{e653}
\end{equation}
The normalized form factors are defined according to 
\begin{equation}
\left\langle N(p') \left| V_{\mu}^{0,8} \right| N(p) \right\rangle
= \bar{u}(p') \left[ F_1^{0,8}(Q^2)\gamma_{\mu} 
 +F_2^{0,8}(Q^2)\frac{i\sigma_{\mu\nu} q^{\nu}}{2M} \right] u(p)
\label{e655}
\end{equation}
\begin{equation}
\left\langle N(p') \left| V_{\mu}^{3} \right| N(p) \right\rangle
= \bar{u}(p') \left[ F_1^{3}(Q^2)\gamma_{\mu}
 +F_2^{3}(Q^2)\frac{i\sigma_{\mu\nu} q^{\nu}}{2M} \right] 
 \tau_3 u(p)
\label{e656}
\end{equation}
\begin{equation}
\left\langle N(p') \left| A_{\mu}^{0,8} \right| N(p) \right\rangle
= G_1^{0,8}(Q^2)\bar{u}(p') \gamma_{\mu}\gamma_5 u(p)
\label{e657}
\end{equation}
\begin{equation}
\left\langle N(p') \left| V_{\mu}^{3} \right| N(p) \right\rangle
=  G_1^{3}(Q^2) \bar{u}(p')\gamma_{\mu}\gamma_5 
 \tau_3 u(p)  \, .
\label{e658}
\end{equation}
The four 
vector form factors $F_{1,2}^{3,8}$ can be measured in electromagnetic 
scattering. At $Q^2=0$ they are given in terms 
of the anomalous magnetic moments $\kappa_N$ of the 
nucleons by
\begin{equation}
F_1^3(0)={1 \over 2}  
\qquad 
F_2^3(0)={1 \over 2}(\kappa_p-\kappa_n)
\label{e659}
\end{equation}
\begin{equation}
F_1^8(0)={1 \over 2} \sqrt{3}  
\qquad 
F_2^8(0)={1 \over 2}\sqrt{3} (\kappa_p+\kappa_n)   \, .
\label{e6510}  
\end{equation}
The $Q^2 \rightarrow 0$ limit of $F_1^0$ is the baryon number,  
\begin{equation}
F_1^0(0)=1
\label{e6511}  \, .        
\end{equation}
The nonsinglet axial vector form factors $G_1^{3,8}$ at zero 
momentum transfer are related to the constants F and D introduced 
in hyperon semileptonic decays, cf. Eqs. (\ref{510}) and (\ref{511}), 
\begin{equation}
G_1^3(0)={1 \over 2}(F+D)={1 \over 2}{g_A \over g_V} 
\qquad
G_1^8(0)={1 \over \sqrt{12}}(3F-D)
\label{e6512}  \, .   
\end{equation}
One is left with two singlet form factors undetermined at 
$Q^2 \rightarrow 0$, $F_2^0(0)$ and $G_1^0(0)$. $F_2^0(0)$ 
is the 'anomalous baryon number magnetic moment'. $G_1^0(0)$ 
is the singlet axial current form factor relevant to spin physics. 
Both $F_2^0$ and $G_1^0$ can be measured in elastic neutral 
current scattering. 
In the naive parton model $G_1^0(0)$ can be expressed by 
the singlet polarized quark density $\Delta \Sigma$ introduced 
in Sect. 5. As discussed in detail in Sect. 5, higher order 
QCD corrections may introduce, within certain schemes, 
an anomalous gluon contribution. 

Since the hadronic neutral current is a linear combination of the 
$U_3 \times U_3$ currents $V_{\mu}^a$ and $A_{\mu}^a$, 
\begin{equation}
J^Z_{\mu}=\sum_{a=0,3,8}(v_a V_{\mu}^a +a_a A_{\mu}^a) 
\label{e6513}
 \, , \end{equation} 
its matrix element Eq. (\ref{e651}) and thus the form 
factors $F_V^Z$, $F_M^Z$ and $G_A^Z$ appearing on the r.h.s. 
of Eq. (\ref{e651}) can be given as linear combinations    
of the normalized form factors $F_{1,2}^{0,3,8}$ and $G_1^{0,3,8}$, 
\begin{equation}
F_V^Z=\sum_{a=0,3,8}v_a F_1^a 
\qquad
F_M^Z=\sum_{a=0,3,8}v_a F_2^a
\qquad
G_A^Z=\sum_{a=0,3,8}a_a G_1^a
\label{e6514}
 \, . \end{equation} 
To lowest order the coefficients $v_a$ and $a_a$ are 
determined by the vector and axial vector couplings of the 
Z to the quarks and given by 
\begin{equation}
v_0=-{1 \over 2}   
\qquad 
v_3=1-2 \sin ^2 \theta_W
\qquad 
v_8={1 \over \sqrt{3}} (1-2 \sin ^2 \theta_W) 
\label{e6515}
\end{equation}
\begin{equation}
a_0={1 \over 2}
\qquad 
a_3=-1
\qquad 
a_8=-{1 \over \sqrt{3}} \, .
\label{e6516}
\end{equation}

The impact of these results, and in particular of the last relation in  
Eq. (\ref{e6514}), on spin physics is as follows: In the naive 
(i.e. non--QCD) parton model there is no $Q^2$ dependence and one can identify 
the axial vector form factors with combinations of first 
moments of polarized parton densities, for $f=3$ active flavors,  
\begin{equation}
G_1^0 ={1\over 3} [ \Delta (u+\bar u) +\Delta (d+\bar d) +\Delta (s+\bar s)]
\quad
G_1^3 ={1\over 2} [\Delta (u+\bar u) -\Delta (d+\bar d)]
\quad
G_1^8 ={1\over \sqrt{12} } [ \Delta (u+\bar u) +
      \Delta (d+\bar d) -2\Delta (s+\bar s) ]
\label{e657790}
 \, . 
\end{equation}
If these results are combined to $G_A^Z$ in Eq. (\ref{e6514}), one finds that 
$G_A^Z$ measures the following combination of first moments   
\begin{equation}
G_A^Z =-{1\over 2 } [ \Delta (u+\bar u) -\Delta (d+\bar d) 
      -\Delta (s+\bar s) ]
\label{e657788890}
 \, . 
\end{equation}
This is an important result because it shows that (quasi)--elastic 
neutral current processes allow to determine a linear combination 
of first
moments of polarized parton densities, which is independent and different 
of what is measured in polarized deep inelastic electroproduction. 

Higher order QCD, QED and quark mass 
corrections to this result have been estimated by  
\cite{kaplan} to be small. 
There is some effect from the renormalization group 
running of $a_0$ between the scale $<$ 1 GeV at which 
the hadronic matrix elements are defined and the scale $m_Z$ 
at which the couplings are given. 
The effect of this running can be summarized as effectivly replacing
$a_0={1 \over 2}$ by $a_0 \approx 0.48$. 
The running of the other couplings can be neglected to a good 
approximation. 

The discussion so far seems to imply that elastic neutrino scattering 
is an extremely elegant method to determine the spin of the proton. 
Unfortunately, the above discussion is not the whole story. The point is 
that to fit neutrino--hadron scattering data one needs the 
form factors at $Q^2 \neq 0$ and there is a significant $Q^2$--dependence 
due to the finite extension of the proton. 
The exact $Q^2$--dependence of the form factors being unknown,  
a dipole approximation of the form 
$\sim (1+Q^2/M_D^2)^{-2}$ is usually applied. In 
these expressions new phenomenological 
parameters $M_D$ (different for each form factor) appear, which have to 
be determined experimentally. 
This uncertainty in the $Q^2$--dependence 
reduces the potential of the method very much. 
Recently, a method has been suggested by \cite{bilenky} 
to reduce this sensitivity on the poorly known (non--strange) 
axial form factor  
and increase the accuracy 
and sensitivity to the strange axial form factor 
($\Delta s$) by considering a certain 
combination of neutral and charged current neutrino and antineutrino 
cross sections, namely the asymmetry 
\begin{equation}
A_{N}(Q^2)
=
\frac{
\left( \frac { d \sigma }{ d Q^2 } \right)_{\nu N}^{NC}
-
\left( \frac { d \sigma }{ d Q^2 } \right)_{\bar{\nu} N}^{NC}
}{
\left( \frac { d \sigma }{ d Q^2 } \right)_{\nu n}^{CC}
-
\left( \frac { d \sigma }{ d Q^2 } \right)_{\bar{\nu} p}^{CC}
}
\; .
\label{e6518}
\end{equation}
In principle, it should be possible to obtain information
about the proton spin 
from the investigation of the
quasi--elastic CC processes
$\nu_\mu + n \rightarrow \mu^{-} + p$ and 
$\bar\nu_\mu + p \rightarrow \mu^{+} + n$. 
However,
the existing CC data
are not accurate enough to compete with the 
NC elastic scattering data \cite{garvey}.
These and other suggestions are awaiting the experimental tests. 
For example, a measurement of $G_1^0$ in 
elastic scattering of unpolarized electrons off unpolarized protons 
might be feasible, 
because the forward backward asymmetry of the elctrons in the cms 
is determined by   
the $\gamma$--Z interference terms and these in turn are 
proportional to $G_1^0$.

%% file: k66tex
\subsection{The OPE and QCD Parton Model for $g_3$ and $g_{4+5}$}

In Sect. 2.4 the 'kinematics' of polarized charged and neutral 
current processes has been introduced and the structure functions 
$g_{3,4,5}$ have been defined. 
In this section we want to discuss the parton model 
and the phenomenological consequences 
of polarization 
effects involving  
charged and neutral currents. 
Further information can be found in the literature 
\cite{craigie,lampe,vogelsang3,jenkins,ravishankar,mathews,ji2,anselmino}. 
The first reference \cite{craigie} is an old review and 
summarizes the theoretical articles  
\cite{nash,derman,ahmed1,kaur,cahn} from before 
the startup of the CERN experiments.  
Just as for $g_1$ the naive parton 
model can be applied as a first approximation, 
whenever longitudinally polarized
leptons probe longitudinally polarized protons ($P_{\mu}=MS_{\mu}$) 
at high 
energies (much larger than $\Lambda_{QCD}$). 
Under this condition the relevant structure 
functions in the hadronic tensor Eq. (\ref{244}) are $g_3$ and $g_{4+5} := 
g_4 +g_5$. 

Let us first
consider neutrino nucleon scattering $\nu p \rightarrow l^- X$ 
which proceeds via $W^+$ exchange. 
(Analogously, for antineutrino scattering, $\bar{\nu}N \rightarrow l^+X$ 
refers to the charged $W^-$ current.)   
The results can also be taken over rather directly to the 
charged current reactions $l^+ p \rightarrow \bar{\nu} X$. 
Neutrinos couple 
to d-type quarks and to $\bar{u}$-type antiquarks  
so that 
the densities of these partons will appear in the structure 
functions. 
In principle, one has also to take into account the proper 
CKM mixing occuring at the charged current vertex. However, 
if one considers the contributions of 4 flavors (u,d,s and c), 
one always encounters the factor $\cos^2 \theta_c +\sin^2 \theta_c =1$. 
One can get the parton model expressions 
for $g_3$ and $g_{4+5}$ by an explicit calculation of the 
lowest order processes and by comparing it to the general 
form of the hadron tensor (\ref{241}) 
\begin{equation}
g_1^{\nu N}(x,Q^2)=\d d(x,Q^2) +\d s(x,Q^2) 
+ \d \bar{u}(x,Q^2) + \d \bar{c}(x,Q^2)   
\label{661}
\end{equation}
\begin{equation}
g_3^{\nu N}(x,Q^2)=-[\d d(x,Q^2) +\d s(x,Q^2) 
- \d \bar{u}(x,Q^2) - \d \bar{c}(x,Q^2)] 
\label{662}
\end{equation}
\begin{equation}
g_{4+5}^{\nu N}(x,Q^2)=2xg_3^{\nu N}(x,Q^2) 
\label{663}
\end{equation} 
where the index $\nu N$ always implies $W^+$ exchange. 
One obtains the corresponding formulae for $\bar \nu N$ ($W^-$ exchange)  
by the flavor interchanges $d \leftrightarrow u$ and 
$s \leftrightarrow c$.
The contribution of $g_{4-5}^{\nu N}$ to the cross section 
vanishes in the framework 
of the QCD improved parton model.  

Some remarks are in order. Due  
to the charge conjugation property of $\gamma_5$ the 
antiquarks always appear with an opposite sign in $g_{3}$ and $g_{4+5}$ 
as compared to $g_1$. As discussed in Sect. 5, the sum 
\begin{equation}
\int^1_0 dx (g_1^{\nu N}+g_1^{\bar{\nu} N})(x,Q^2)= 
\Delta \Sigma -f {\alpha_s \over 2 \pi} \Delta g(Q^2)  
\label{664}
\end{equation}
is a measure of the axial vector singlet current matrix element 
and its measurement would be a nice way to verify the 
value of $\Delta \Sigma -f {\alpha_s \over 2 \pi} \Delta g$ 
without recurrence to the low energy determination of the 
matrix elements $A_3$ and $A_8$. In contrast to $g_1$, the 
structure functions $g_3$ and $g_{4+5}$ measure solely 
nonsinglet combinations of parton densities (just as $F_3$ 
in unpolarized DIS), so that they do not get a 
contribution from the gluon densitiy. 
They can yield informations of the following type : 
For example, by scattering on an isoscalar target one could find out , 
how large the polarized strange quark sea is: 
\begin{equation}
(g_3^{\nu N}-g_3^{\bar{\nu} N})_p(x,Q^2)+
(g_3^{\nu N}-g_3^{\bar{\nu} N})_n(x,Q^2)= 
2(\d c +\d \bar{c} -\d s -\d \bar{s})(x,Q^2) \; .
\label{665}
\end{equation}
Other combinations have been studied by \cite{mathews} 
and \cite{anselmino}. Furthermore, it should be mentioned that 
the treatment of the charm quark contributions to the above 
structure functions in terms of a massless intrinsic density 
$\delta c(x,Q^2)=\delta \bar c(x,Q^2)$ is controversial. 
It is more appropriate to calculate the  
heavy quark contributions to neutral and charged current 
processes perturbatively 
via the subprocess $W^+ g \rightarrow c\bar s$ \cite{vogelsang3}, 
for example.  

Just as $F_1$ and $F_2$ in unpolarized DIS, 
$g_3$ and $g_{4+5}$ are 
related by a Callan-Gross like 
relation, Eq. (\ref{663}), originally derived by 
\cite{dicus}. This relation will be violated beyond the 
leading order of QCD, except in the case of neutral currents 
where $g_3$ and $g_{4+5}$ do not receive a gluonic contribution 
to $O(\alpha_s)$ \cite{lampe,vogelsang3}.     
Further new relations analogous to the Wandzura--Wilczek 
relation for $g_2$ (c.f. \cite{wandzura} and Sect. 6.9) 
have been recently derived from a study of the twist--2 and 
twist--3 OPE \cite{bluemlein3,bluemlein4}. 

The close relationship between unpolarized and polarized 	
DIS in the limit $Q^2 \rightarrow \infty$ , in which 
only longitudinal polarization survives, 
becomes very transparent within the OPE. In the high energy 
limit one can stick to the leading twist-2 operators, and 
the OPE for the hadronic tensor reads 
\begin{eqnarray}  \nonumber 
W_{\mn}^{\nu N} = 
i~\ve_{\mn\lambda\si}q^\lambda
\sum\limits_n ({2\over Q^2})^n q_{\mu_1}\ldots q_{\mu_{n-1}}
\sum\limits_i (R_i^{\si\mu_1\ldots\mu_{n-1}}E^n_i+ 
               Q_i^{\si\mu_1\ldots\mu_{n-1}}C^n_i) 
& & \\ \nonumber
     +\Bigl(-g_{\mn}+{q_\mu q_\nu\over q^2}\Bigr)
\sum\limits_n ({2\over Q^2})^n
q_\sigma q_{\mu_1}\ldots q_{\mu_{n-1}}\sum\limits_i
 (R_i^{\si\mu_1\ldots\mu_{n-1}}V^n_i+ 
  Q_i^{\si\mu_1\ldots\mu_{n-1}}X^n_i) 
& & \\
    +\Bigl(-g_{\mu\si}+{q_\mu q_\si\over q^2}\Bigr)
\Bigl(-g_{\nu\mu_1}+{q_\nu q_{\mu_1}\over q^2}\Bigr) 2\sum\limits_n
({2\over Q^2})^{n-1}q_{\mu_2}\ldots q_{\mu_{n-1}}\sum\limits_i
  (R_i^{\si\mu_1\ldots\mu_{n-1}} W^n_i
   +Q_i^{\si\mu_1\ldots\mu_{n-1}} Y^n_i) & &
\label{666}
\end{eqnarray}
where $R_i$ are the operators relevant for polarized 
scattering (Eqs. (\ref{433}) and (\ref{434})) 
and $Q_i$ are the corresponding operators 
for unpolarized scattering 
(essentially the same as Eqs. (\ref{433}) and (\ref{434}) 
but without a $\gamma_5$). 
$E^n_i, C^n_i, V^n_i, X^n_i, W^n_i$ 
and $Y^n_i$ are the corresponding Wilson 
coefficients. For example, $E^n_i$ is related to the structure 
function $g_1$, $C^n_i$ to the structure function $F_3$ etc.
More precisely, one has \cite{ravishankar}  
\begin{equation}
\int\limits^1_0 dx x^{n-1}F_1(x,Q^2)=\phantom{-}\sum\limits_i b^i_n
X^n_i (Q^2/\mu^2,\alpha_s)\quad n=1,3,5,\ldots 
\label{667}
\end{equation}
\begin{equation}
\int\limits^1_0 dx x^{n-2}F_2(x,Q^2)=2 \phantom{-}\sum\limits_i b^i_n
Y^n_i (Q^2/\mu^2,\alpha_s)\quad n=1,3,5,\ldots 
\label{668}
\end{equation}
\begin{equation}
\int\limits^1_0 dx x^{n-1}F_3(x,Q^2)=\phantom{-}\ha\sum\limits_i b^i_n
C^n_i (Q^2/\mu^2,\alpha_s)\quad n=1,3,5,\ldots 
\label{669}
\end{equation}
\begin{equation}
\int\limits^1_0 dx x^{n-1}g_1(x,Q^2)=\phantom{-}\ha\sum\limits_i a^i_n
E^n_i (Q^2/\mu^2,\alpha_s)\quad n=1,3,5,\ldots
\label{6610}
\end{equation}
\begin{equation}
\int\limits^1_0 dx x^{n-1}g_3(x,Q^2)=\phantom{-}\sum\limits_i a^i_n
V^n_i (Q^2/\mu^2,\alpha_s)\quad n=1,3,5,\ldots
\label{6611}
\end{equation}
\begin{equation}
\int\limits^1_0 dx x^{n-2}g_{4+5}(x,Q^2)=2 \phantom{-}\sum\limits_i a^i_n
W^n_i (Q^2/\mu^2,\alpha_s)\quad n=1,3,5,\ldots
\label{6612}
\end{equation}
where the $a^i_n (b^i_n)$ are the matrix elements for a (un)polarized 
proton, cf. Eq. (\ref{435}).  
From these equations the Callan-Gross relations are apparent, 
because in leading order all Wilson coefficients are equal to 1. 
For more details we refer the reader to 
\cite{ravishankar,bluemlein3,bluemlein4}.

Let us now turn to neutral current exchange where the description 
is similar, although somewhat complicated by the $\gamma Z$-
interference. In LO the parton model predictions for the structure 
functions are 
\begin{equation}
g_1^{\gamma Z}(x,Q^2)= \sum\limits_q e_q v_q 
(\d q+ \d \bar{q})(x,Q^2) 
\label{6613}
\end{equation}
\begin{equation}
g_3^{\gamma Z}(x,Q^2)= \sum\limits_q e_q a_q 
(\d q- \d \bar{q})(x,Q^2) 
\label{6614}
\end{equation}
\begin{equation}
g_1^{Z Z}(x,Q^2)= {1 \over 2} \sum\limits_q (v_q^2+a_q^2) 
(\d q+ \d \bar{q})(x,Q^2) 
\label{6615}
\end{equation}
\begin{equation}
g_3^{Z Z}(x,Q^2)= \sum\limits_q v_q a_q 
(\d q- \d \bar{q})(x,Q^2) 
\label{6616}
\end{equation}
where $v_q$ and $a_q$ are the vector and axialvector coupling 
of the quark flavor q to the $Z^0$, 
$v_u={1 \over 2}-{4 \over 3}sin^2 \theta_W$,
$a_u={1 \over 2}$,$v_d=-{1 \over 2}+{2 \over 3}sin^2 \theta_W$ and 
$a_d=-{1 \over 2}$, cf. Sect. 2.4. Furthermore, one has 
$g_{4+5}^{\gamma Z,ZZ}=2xg_3^{\gamma Z,ZZ}$ and 
$g_1^{\gamma}$ as before, cf.  Eq. (\ref{415}). 
NLO contributions can be found in 
\cite{florian1,lampe,mathews,vogelsang3,stratmann}.
For comparison let us include here the corresponding LO relations 
for the unpolarized structure functions, namely 
$F_1^{\gamma Z,ZZ}=\sum\limits_q (e_q v_q,v_q^2+a_q^2) ( q+  \bar{q})$,  
$F_3^{\gamma Z,ZZ}=2\sum\limits_q (e_q a_q,v_q a_q)( q-  \bar{q})$ 
and $F_2^{\gamma Z,ZZ}=2xF_1^{\gamma Z,ZZ}$. 
For the charged current processes $l^+ N \rightarrow \bar{\nu}X$ 
one can take over the expressions for $g_1^{\nu N}$ etc. 
from above, Eqs. 
(\ref{661})--(\ref{665}).

%% file: k67tex
\subsection{Single Spin Asymmetries and Handedness}

Single spin asymmetries $A_L={\sigma_{\Rightarrow}-\sigma_{\Leftarrow} 
\over \sigma_{\Rightarrow}+\sigma_{\Leftarrow}}$ are asymmetries that 
arise if only one of the external particles is 
longitudinally (or transversally) polarized. 
In deep inelastic lepton nucleon scattering at $Q^2 << m_Z^2$ 
all single spin asymmetries vanish -- at least within the framework  
of the (twist--2) parton model -- and the same holds true for many processes 
at RHIC, like direct photon, heavy quark or jet production. 
Still, there are circumstances under which nonvanishing 
single spin asymmetries arise. For instance, many authors have 
tried to derive single spin asymmetries from higher twist effects 
like the intrinsic transverse momentum of the partons within 
hadrons \cite{sivers1,sivers,collins1,boros,efremov8,qiu1,qiu2,lampe2,
efremov7,teryaev,artru1,anselmino3,anselmino4,anselmino5,korotkov}. 
Such considerations have been triggered by several experiments 
which have 
shown that single spin asymmetries can indeed be large 
both in semi--inclusive $pp^\uparrow \rightarrow \pi X$ 
\cite{adams9,adams8,adams4} and exclusive $pp^\uparrow \rightarrow pp$ 
\cite{antille,cameron,crabb} reactions.  
These ideas go beyond
the perturbative QCD (twist--2) 
parton model and are more difficult to test than the 
parton model inspired predictions which usually 
lead to double spin asymmetries. 
Another example of single spin asymmetries is the possibility to 
determine the spin of final state particles, like the 
handedness of jets. All these ideas and 
approaches will be discussed at the end of this section. 
We shall start the discussion with a leading, twist--2 source 
of single spin asymmetries which is due to weak interactions, i.e. 
the presence of parity violating couplings in processes in which 
$W^{\pm}$ and $Z^0$ are involved. In such processes the $\gamma_5$ from the 
polarized particle and the $\gamma_5$ from the axial vector coupling 
combine in the matrix element to give a nonvanishing 
single spin asymmetry even within the ordinary parton model. 
This point will be discussed first in this section 
because it can lead to drastic effects 
and, at RHIC, is an independent way to determine 
the polarized parton densities. 
%There is another, less well known 
%source of single spin asymmetries due to the intrinsic transverse 
%momentum of the partons within hadrons. For deep inelastic 
%scattering this has been studied by \cite{lampe2} and for proton scattering 
%in a different approach by \cite{anselmino3}. These ideas go beyond    
%the ordinary parton model in which partons carry only longitudinal 
%momentum and, in addition, are more difficult to test than the single 
%spin asymmetries due to weak interactions, to which we come now. 

\begin{figure}
\begin{center}
\epsfig{file=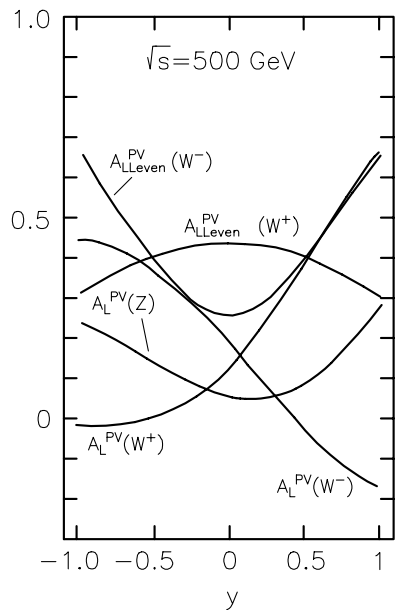,height=8cm}
\bigskip
\caption{The parity violating spin asymmetries for $W^{\pm}$ 
and $Z^0$ production 
as a function of y under RHIC conditions 
\protect\cite{bourrely5}. The curves correspond 
to a suggested choice of the sea quark polarizations, cf. ref. 
\protect\cite{bourrely2}. }
\label{figwz}
\end{center}
\end{figure}

\begin{figure}
\begin{center}
\epsfig{file=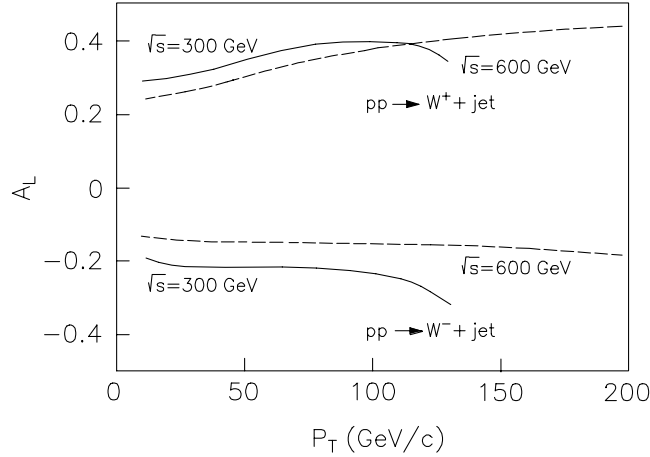,height=6cm}
\bigskip
\caption{The parity violating single spin asymmetries 
in $pp \rightarrow W^{\pm}$ + jet     
at y=0 as a function of $p_T$ under RHIC conditions. The results 
are taken from ref.  
\protect\cite{bourrely1}. }                                                     
\label{figwzj}
\end{center}  
\end{figure}

Let us recall that in unpolarized proton scattering the production of W's 
via $pp \rightarrow W^{\pm}X$ is sensitive to the form of the antiquark 
distributions, because the dominant contribution to the cross 
section comes from the quark--antiquark fusion reactions 
$u\bar d\rightarrow W^+$ and $\bar u d\rightarrow W^-$. 
The same holds true for polarized scattering where at least one  
of the protons is polarized. The corresponding single spin 
asymmetry is given by 
\begin{equation}
A_L^{PV}={\sigma_{\Rightarrow}-\sigma_{\Leftarrow}
\over \sigma_{\Rightarrow}+\sigma_{\Leftarrow}}=
{\delta u(x_1,m_W^2) \bar d(x_2,m_W^2) -(u \leftrightarrow \bar d) 
\over u(x_1,m_W^2)  \bar d(x_2,m_W^2)+(u \leftrightarrow \bar d)} 
\, .
\label{m70}              
\end{equation}  
Note that for kinematical reasons $A_L^{PV}$ depends only on 
$y=\ln {x_1 \over x_2}$ because one has $x_1={m_W \over \sqrt{S}} 
e^y$ and $x_2={m_W \over \sqrt{S}}e^{-y}$.          
Eq. (\ref{m70}) is a crude LO approximation but it shows how 
sensitive one is to the polarized antiquark densities. 
A phenomenological analysis on the basis of this formula 
has been presented by \cite{bourrely5,soffer2} 
for the RHIC machine parameters 
chosen to be $\sqrt{s}=0.5$ TeV, $L=2 \times 10^{32}$ cm$^{-2}$ 
sec$^{-1}$ and a polarization of 70\%. The resulting single 
spin asymmetries are shown in 
Fig. \ref{figwz} as a function of y. In ref. \cite{bourrely1} 
the analysis has been extended to processes 
like $pp\rightarrow W^{\pm}$ + jet + X. 
The corresponding single spin asymmetries are shown in Fig. \ref{figwzj} 
as a function of $p_T$.   
Later on, a Monte Carlo study including collinear gluon bremsstrahlung 
has been carried out by \cite{saalfeld}. 
The full NLO--QCD calculation has been performed by \cite{weber1}.

In principle, there are two other parity violating asymmetries which can 
be measured if both protons are polarized, namely 
\begin{equation}
A_{LL even}^{PV}(y)=\frac{\si^{\Leftarrow}_{\Leftarrow}
-\sigma^{\Rightarrow}_{\Rightarrow}}
          {\si^{\Leftarrow}_{\Leftarrow} +\si^{\Rightarrow}_{\Rightarrow}}
\label{m71}                                  
\end{equation}
and
\begin{equation}
A_{LL odd}^{PV}(y)=\frac{\si^{\Leftarrow}_{\Rightarrow}
-\sigma^{\Rightarrow}_{\Leftarrow}}
          {\si^{\Leftarrow}_{\Rightarrow} +\si^{\Rightarrow}_{\Leftarrow}}
\label{m72}                                  
\end{equation}
Note that if parity is conserved one has $\sigma_{a,b}=\sigma_{-a,-b}$ 
and $\sigma_{a}=\sigma_{-a}$ 
so that all these asymmetries, including 
$A_L^{PV}$, vanish.  
In the parity violating $W^\pm$ production process 
they are given in leading order by \cite{bourrely5} 
\begin{equation}
A_{LL even}^{PV}(y)={[\delta u(x_1,m_W^2) \bar d(x_2,m_W^2) -
u(x_1,m_W^2)\delta \bar d(x_2,m_W^2)] - [u \leftrightarrow
\bar d ]
\over [u(x_1,m_W^2) \bar d(x_2,m_W^2) -
\delta u(x_1,m_W^2)\delta \bar d(x_2,m_W^2)]+[u \leftrightarrow \bar d]}
\label{m73}                                  
\end{equation}
and
\begin{equation}
A_{LL odd}^{PV}(y)={[u(x_1,m_W^2) \delta \bar d(x_2,m_W^2) +
\delta u(x_1,m_W^2) \bar d(x_2,m_W^2)] - [u \leftrightarrow
\bar d ]
\over [u(x_1,m_W^2) \bar d(x_2,m_W^2) +
\delta u(x_1,m_W^2)\delta \bar d(x_2,m_W^2)]+[u \leftrightarrow \bar d ]}
\label{m74}     
\end{equation}                               
with the property $A_{LL even}^{PV}(y)=A_{LL even}^{PV}(-y)$ and 
$A_{LL odd}^{PV}(y)=-A_{LL odd}^{PV}(-y)$   
and again sensitive to the polarized sea. 

Results for these asymmetries are included in Fig. \ref{figwz}, 
obtained under the assumptions of ref. 
\cite{bourrely5}. These authors 
have suggested that one should make the approximation 
$\delta u(x_1) \delta \bar d(x_2) << u(x_1)\bar d(x_2)$. 
In that case the three asymmetries 
are not independent quantities any more but $A_{LL even}^{PV}(y)$ and 
$A_{LL odd}^{PV}(y)$ can be expressed in terms of $A_L^{PV}(y)$ and 
$A_L^{PV}(-y)$. Whether this approximation is reasonable or 
not will be shown in future. In any case, among the three, $A_L^{PV}$ 
is certainly the most interesting one because it will have the smallest 
statistical error. This is true despite the fact that the single 
spin asymmetry will be smaller in absolute magnitude than 
the double spin asymmetries by roughly a factor of 2.  

Similar considerations as for W--production apply for Z--production 
at RHIC. One may also consider the Z--contribution to the Drell-Yan 
process $pp \rightarrow \gamma^{\ast}, Z^{\ast} \rightarrow \mu^+ 
 \mu^-$ which is relevant for large invariant masses of 
the $\mu^+ \mu^-$ pair. 
There have been studies of large--$p_T$ W 
\cite{bourrely1} (and also large--$p_T$ $Z^{\ast}$ 
\cite{leader2,leader1} and 
large--$p_T$ jet production) induced 
by the higher order processes $q_i \bar q_j\rightarrow Wg$ and 
$q_ig\rightarrow q_jW$, where $i \neq j$ denotes the quark flavor. 
Let us discuss now qualitatively the significance of these 
processes for the determination of the polarized gluon density 
$\delta g(x,\mu_F^2)$ where $\mu_F$ may be roughly chosen 
as $\mu_F^2  \approx (p_T^2 + m_W^2)/4$.  
If the beams are polarized and the initial partons are carrying     
a given helicity, one has for the Compton--type scattering process 
$q_i(h)g(\lambda)\rightarrow q_jW$ the parton level cross section 
\begin{equation}
{d \hat \sigma \over d \hat t}={\pi \alpha \alpha_s \over 12 
\sin \theta_W^2 \hat s^3 \hat u} 
(h-1)(c_2\lambda + c_1(1-\lambda)) 
\label{m75}     
\end{equation}                                
where 
$c_1=(\hat s - m_W^2)^2 +(\hat u - m_W^2)^2$ and 
$c_2=2(\hat u - m_W^2)^2$. 
For $\bar q_i(h)g(\lambda)\rightarrow \bar q_jW$ the same formula 
holds but with $h\rightarrow -h$ and $\lambda \rightarrow -\lambda$. 
The resulting parton level single spin asymmetries are
$\hat a_L^{PV}=1$ for polarized quarks and $\hat a_L^{PV}=1-{c_2 \over c_1}$ 
for polarized gluons. Since the last quantity is rather small on 
average \cite{bourrely1}, 
the single hadron helicity asymmetry will be dominated 
by polarized quarks and is not very sensitive to $\delta g$. 

If it were possible to determine the polarization of the outgoing
photon or jet, one would have nonvanishing single spin asymmetries 
in processes without parity violation. Namely, one could study 
processes like $q +\vec g \rightarrow q +\vec \gamma$ or 
$q +\vec g \rightarrow \vec q +g$ etc. either at RHIC 
or one could use the energetic unpolarized proton beams of the 
Tevatron $\bar p p$ or HERA $ep$ colliders ($E_p \approx 1$ TeV)  
to be scattered off a polarized fixed proton target. 
In the former case one would have to measure the circular 
polarization of the final state photon. This is quite difficult. 
It could be done either by selective absorbtion using a polarized detector 
\cite{craigie1} 
or by making use of the fact that in high energy photon induced 
showers the longitudinal polarization is conserved to a good degree 
of accuracy \cite{mathews1,glueck15}. 
In the latter 
case an unpolarized quark collides with a polarized gluon and 
one would have to measure the polarization of the outgoing quark 
jet. It has been speculated \cite{nachtmann,efremov3,dalitz,efremov4} 
that information about the polarization 
of the initiating parton, i.e. the outgoing quark, 
can be obtained from the 'handedness' 
of a jet. As will be discussed below, present experimental data 
show unfortunately so far no evidence for this concept to be 
relevant. Here handedness is defined as follows: consider the 
triple product of vectors $S:=\vec t \cdot (\vec k_1 \times \vec k_2)$ 
where $\vec t$ is a unit vector along the jet axis, and $\vec k_1$ 
and $\vec k_2$ are the momenta of two particles in the jet chosen by 
some definite prescription, e.g. the two fastest particles. The 
jet is defined as left(right)--handed if $S$ is negative(positive). 
For an ensemble of jets the handedness is defined as the asymmetry 
in the number of left-- and right--handed jets, 
$H:={ N_{S<0}-N_{S>0} \over N_{S<0}+N_{S>0}}$. It can then be 
asserted that $H=\alpha P$ where P is the average polarization 
of the underlying partons in the ensemble of jets and $\alpha$ is 
the 'analyzing power' of the handedness method. 

In order to obtain information about $\delta g(x,\mu_F^2)$ from 
processes like  
$p + \vec p \rightarrow \vec{jet} + X$ one needs to know the 
value of $\alpha$. It has been attempted \cite{abe4} to determine $\alpha$ 
from jet production at SLD where a polarized $Z$--boson decays 
into two jets originating from a $q\bar q$ pair. 
The quark and antiquark in a $Z^0$ decay have opposite helicities. 
The SM predicts $P_{u,c} \approx 0.67$ and
$P_{d,s,b} \approx 0.94$ for the average polarizations of quark 
flavors 
so that the quarks are produced predominantly left-handed and
the antiquarks predominantly right-handed.
In order to observe a net polarization in an ensemble of jets from $Z^0$
decays it is necessary to distinguish quark jets from antiquark
jets.
This separation can be achieved
at SLC where $Z^0$ bosons are produced in
collisions
of highly longitudinally polarized electrons with unpolarized positrons.
In this case the SM predicts a large
difference in polar angle distributions between quarks and antiquarks. 
Unfortunately, no evidence for handedness was found by the SLD 
collaboration \cite{abe4}. This negative result seems to imply 
that the connection 
between the polarization of the hard partons before the 
fragmentation and the 
orientation of the final state hadrons is very loose and washed 
out by confinement effects. 

Instead of producing and delineating a polarized photon, Drell--Yan 
dilepton production has been suggested as a probe for polarized 
densities : here either the polarization of one of the final 
leptons in $p \vec p \rightarrow \gamma^{\ast}X \rightarrow \mu^+ 
\vec \mu^- X$ has to be measured \cite{contogouris4}, or 
the angular distribution of the produced lepton pair 
\cite{carlitz1} as a polarimeter for the virtual intermediate 
photon. 

Similarly, instead of producing a polarized jet (quark), 
semi--inclusive polarized single--particle (e.g. $\vec \Lambda$ ) 
production has been suggested and analyzed as well. Either 
$e^+\vec e^- \rightarrow \vec \Lambda X$ \cite{burkhardt} or 
$l \vec p \rightarrow l \vec \Lambda X$ \cite{jaffe6,lu,lu1,kochelev}, 
where in the latter case a longitudinally polarized lepton beam 
instead of a polarized nucleon target could do as well 
as a sensitive probe of $\delta g(x,Q^2)$. 
Here the (time--like) polarized fragmentation functions 
$\delta D_f^{\Lambda}(z,,Q^2)$, $f=q, \bar q,g$ also enter, 
which are defined in analogy to the space--like polarized 
parton densities. Theoretically these processes have been studied 
rather thoroughly during the past few years, not only in LO, but 
in NLO as well \cite{florian3,stratmann4}.  

Finally, we want to come back to higher twist contributions as a 
source of single (transverse) spin asymmetries $A_N$. 
Let us first discuss semi--inclusive reactions such as 
$pp^\uparrow \rightarrow \pi X$ and $pp^\uparrow \rightarrow \gamma X$. 
At present, there are basically 3 sources for a nonvanishing $A_N$: 
%Several models and theoretical analyses suggest possible higher
%twist effects: there might be twist--3 

(i) dynamical contributions, i.e. 'hard' partonic twist--3 
scattering effects, which  
%which are usually called 'hard scattering' effects. 
result from a short distance part 
calculable in perturbative QCD combined with a long distance part  
%with slightly modified Feynman 
%rules, combined with a long distance part 
related to quark--gluon correlations \cite{efremov8,qiu1,qiu2,teryaev};  
%There might also be

(ii) intrinsic $k_T$ effects 
%, both in the quark fragmentation process
in parton distribution functions which, being 
%. The latter are not by
%themselves higher twist contributions -- they are rather
non--perturbative
universal nucleon properties, give rise to twist--3 contributions
when convoluted with the hard partonic scattering cross sections.
Such contributions are usually referred to as 'Sivers effect' 
\cite{sivers1,sivers,anselmino3,mulders,mulders1,kotzinian}; 

(iii) intrinsic $k_T$ effects in the parton fragmentation functions 
which is known as 'Collins' or
'sheared jet' effect \cite{collins1,mulders,mulders1,artru1}.  

\begin{figure}
\begin{center}
\epsfig{file=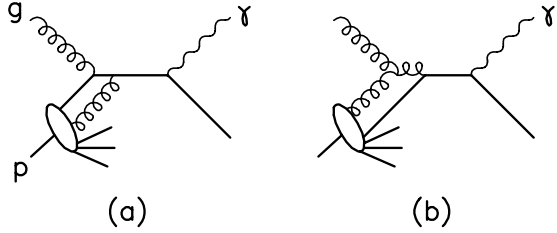,height=3cm}
\vskip 0.5cm
\caption{Sample twist--3 quark--gluon correlation diagrams 
corresponding to fermionic (a) and gluonic (b) pole contributions 
which give rise to a nonvanishing single spin asymmetry in hadronic 
direct--photon production. }
\label{fig40}
\end{center}
\end{figure}

A typical example for dynamical quark--gluon correlation  
contributions (i) to direct--$\gamma$ production is depicted 
in Fig. \ref{fig40}. They give  
rise to $A_N^{\gamma} \neq 0$ 
and correspond to fermionic (quark) pole 
\cite{efremov8} and/or gluonic pole \cite{qiu1,qiu2,ehrnsperger1} 
dominance in the calculation of the twist--3 partonic scattering 
cross sections. Clearly, single spin asymmetries for direct--$\gamma$ 
production are not 'contaminated' by possible intrinsic $k_T$ effects 
in fragmentation functions (iii), but $k_T$ effects in parton 
densities (ii) could also be relevant besides (i). In the latter 
case (ii)
one expects that the number of quarks with longitudinal momentum
fraction $x$ and transverse intrinsic motion $\vec k_T$ depends, 
on account of soft gluon interactions between initial state partons, 
%Namely, one may take into account 
%soft gluon interactions between initial state partons 
%which are most certainly present for hadron--hadron
%interactions. In that case, one can expect that the number of quarks with
%longitudinal momentum fraction $x$ and transverse intrinsic motion
on the transverse spin direction of the parent nucleon,
so that the 'quark distribution analysing power'
$N_q(\vec k_T) \equiv
(f_{q/N^\uparrow}(x, \vec k_T) - f_{q/N^\downarrow}(x, \vec k_T)) /
(f_{q/N^\uparrow}(x, \vec k_T) + f_{q/N^\downarrow}(x, \vec k_T))
$ can be different from zero.
Similar soft interactions in the final (fragmentation) state 
may give rise to the 'Collins effect' (iii);  
it simply amounts to say that the number
of hadrons $h$ (say, pions) resulting from the fragmentation
of a transversely polarized quark, 
with longitudinal momentum fraction $z$ and transverse 
momentum $\vec k_T$, depends on the 
quark spin orientation. That is, one expects the 'quark
fragmentation analysing power' $A_q(\vec k_T) \equiv
(D_{h/q^\uparrow}(z, \vec k_T) - D_{h/q^\downarrow}(z, \vec k_T)) /
(D_{h/q^\uparrow}(z, \vec k_T) + D_{h/q^\downarrow}(z, \vec k_T))
$ to be different from zero,
where, by parity invariance, the quark spin should be orthogonal
to the $q-h$ plane. Notice also that time reversal invariance
does not forbid such a quantity to be nonvanishing because of the
(necessary) soft interactions of the fragmenting quark with
external strong fields, i.e. because of final state interactions.
This idea has been applied \cite{artru1} to the computation of the single
spin asymmetries observed in $pp^\uparrow \to \pi X$ 
\cite{adams9,adams8,adams4}.  
%A similar effect for the distribution functions has also been 
%suggested \cite{anselmino3} to explain the single
%spin asymmetries observed in $pp^\uparrow \to \pi X$ \cite{adams8}.
As mentioned above, both $A_q(\vec k_T)$ and $N_q(\vec k_T)$
are leading twist quantities which, when convoluted
with the elementary cross-sections and integrated over, give
twist-3 contributions to the single spin asymmetries.

Each of the above mechanisms might be present and important
in understanding twist--3 contributions.
It is then important to
study possible ways of disentangling these different contributions
in order to be able to assess the importance of each of them.
We discuss now single spin asymmetries 
for various processes $AB^\uparrow \to CX$. 
To obtain a complete picture one 
needs
to consider nucleon--nucleon interactions together with other
processes,
like lepton--nucleon scattering which might add valuable information too.
For each of them one should discuss the possible sources of higher twist
contributions, distinguishing, according to the above discussion,
between those originating from the hard scattering and those
originating
either from the quark fragmentation or distribution analysing power 
\cite{anselmino5,anselmino4,korotkov}: 
\begin{itemize}
\item 
$pN^\uparrow \to hX$:
all kinds of higher twist contributions may be present;
this asymmetry {\it alone} could not help in evaluating the relative
importance of the different terms; 
\item
$pN^\uparrow \to \gamma X, \> pN^\uparrow \to \mu^+\mu^-
X,
\> pN^\uparrow \to jets + X$:
no fragmentation process is involved and one remains with possible
sources of non--zero single spin asymmetries in the hard scattering 
(i) or the quark distribution analysing power due to an intrinsic 
$k_T$ (ii);
\item
$lN^\uparrow \to hX$:
a single spin asymmetry can originate either from
hard scattering (i) or from $k_T$ effects in the fragmentation
function (iii),
but not in the distribution functions, as soft initial state
interactions
are suppressed by powers of $\alpha_s$; 
\item
$lN^\uparrow \to \gamma X, \> \gamma N^\uparrow \to \gamma X, \>
lN^\uparrow \to \mu^+\mu^- X, \> lN^\uparrow \to jets + X$:
a single spin asymmetry in any of these processes
may only be due to higher-twist dynamical hard scattering effects (i).
\end{itemize}
In the following we want to concentrate on some specific processes, 
in order to see to what extent valuable information can be obtained 
from measuring single spin asymmetries at RHIC and at the future 
HERA--$\vec N$ (where we specifically refer to the 'phase I' 
program, i.e. to polarized fixed target experiments 
with an unpolarized beam; 'phase II' refers to a 
polarized proton beam as well \cite{korotkov}). One example 
is semi--inclusive $\pi^{\pm ,0}$ production by 
$p^\uparrow p$ collisions which is known to exhibit 
surprisingly large single spin asymmetries at 200 GeV.  
This was measured a few years ago
by the E704 Collaboration using a transversely
polarized beam \cite{adams9,adams8,adams4}. 
For any kind of pions the asymmetry 
shows a
considerable rise above
$x_F > 0.3$, i.e. in the fragmentation region of
the polarized nucleon. It is positive ($> 20$\% ) for both
$\pi^+$ and $\pi^0$ mesons, while it has the opposite sign for $\pi^-$
mesons. 
The charged pion data were taken in the $0.2 < p_T < 2$ GeV range 
and it was found that the asymmetry is larger for $p_T > 1$ GeV  
than for $p_T < 1$ GeV. 
Theoretical approaches as to the interpretation of these data 
have been put forward in \cite{sivers1,sivers,boros,
efremov7,artru1,anselmino3}. The authors of Refs. 
\cite{anselmino4,korotkov} have examined to what accuracy this 
asymmetry could be measured at HERA--$\vec N$. 
They have shown that it would be possible to quantitatively 
determine the $p_T$ dependence of the asymmetry up to 
$p_T$ values of about 10 GeV.  

Another example is to measure the single transverse spin asymmetry 
in inclusive direct photon production, $p p^{\uparrow} \to \gamma
X$. Since this process 
proceeds without fragmentation, i.e. the photon carries directly the
information from the hard scattering process, it 
measures
a combination of initial $k_T$ effects and hard scattering twist--3
processes \cite{efremov8,qiu1,qiu2,ehrnsperger1}. 
The first and only results up to now were obtained by the E704
Collaboration \cite{adams3}
showing an asymmetry compatible with zero within
large errors for $2.5 < p_T <3.1$ GeV in the central region
$ | x_F | < 0.15$.   
Again, the authors of Refs.
\cite{anselmino4,korotkov} have examined to what accuracy this
asymmetry could be measured at HERA--$\vec N$.
The contributions of the 
gluon--Compton scattering
($qg \rightarrow \gamma q$) and quark--antiquark annihilation
($q \bar q \rightarrow \gamma g$) were compared to the
background photons that originate mainly from $\pi^0$ and
$\eta$ decays.  
It turns out that a good sensitivity $\delta A_N^\gamma$ of about 0.05 
can be maintained up to $p_T < $ 8 GeV.
For increasing transverse momentum the annihilation subprocess and the
background photons are becoming less essential. 

Thirdly, the single spin asymmetry in Drell--Yan production,
$pp^{\uparrow}\rightarrow l \bar lX$, at small
transverse momenta was calculated \cite{hammon} in the framework
of twist--3 perturbative QCD at HERA-$\vec N$ energies. The resulting 
asymmetry
does not exceed 2\% 
and depends strongly on the kinematical domain. 
It should be noted that asymmetries of the 
size of a few percent represent the canonical order of
magnitude for single spin asymmetries induced by 
twist--3 perturbative QCD effects, and that one may expect 
larger values only under very special circumstances. 
Note also, that such 
asymmetries on the few percent level are
difficult to measure, even with sufficiently small statistical errors,
since the systematic errors originating mainly from beam and target
polarization measurements constitute a severe limit.
 
Recently, it has been suggested \cite{lampe2} that deep 
inelastic Compton 
scattering $lp \rightarrow l \gamma X$, where either the 
incoming lepton or proton are polarized, might lead to 
appreciable and well--measurable 
single spin asymmetries in the small--x region. 
The idea is that intrinsic $k_T$ and 
absorptive effects of the strong interactions lead to a complex phase
in the vertex between the incoming quark and the prompt photon. 
This phase, in turn, induces a single spin asymmetry.   

Finally, it should be remembered that 
large spin effects in proton--proton elastic scattering,
$pp^{\uparrow}\rightarrow p p$, have been discovered many years ago 
\cite{antille,cameron,crabb} with proton beam energies of 24 
and 28 GeV.
The single spin asymmetry
was found significantly different from zero for transverse momenta 
of the outgoing protons 
larger than 2 GeV. 
The transverse single spin asymmetry in elastic $pp$ scattering
at HERA-$\vec{N}$ and RHIC energies has been calculated in a
dynamical model that leads to spin--dependent pomeron couplings
\cite{goloskokov}. The predicted asymmetry is about 0.1 for
$p_{T}^2=4 - 5$ GeV$^2$
with a projected statistical error of $0.01 - 0.02$
for HERA-$\vec{N}$, i.e.
a significant measurement of the asymmetry can be performed
to test the spin dependence of elastic $pp$ scattering at high energies. 
Although we do not have any detailed quantitative theoretical 
understanding of elastic single spin asymmetries for the time 
being, it should be emphasized that, apart from possible 
helicity nonconserving effects occuring in the hadronic wave function, 
one expects at least qualitatively a {\it non}vanishing elastic 
single spin asymmetry due to degenerate multiple Regge exchanges which 
give rise to different phases in different helicity amplitudes 
(see, e.g., \cite{sivers2} for a recent review).

%% file: k68tex
\subsection{Structure Functions in 
DIS from Polarized Hadrons and Nuclei of Arbitrary Spin}

Shortly after the measurement of $g_1^p$ at CERN 
there were several 
proposals to investigate spin effects in other nuclear 
targets. Most prominent among them are of course the 
deuteron (with spin J=1) and Helium-3 (J=${1 \over 2}$), 
because they are used to determine $g_1^n$, 
but there are other 
potentially polarizable nuclei as well, like 
$^7Li(J={3 \over 2})$, $^{10}B(J=3)$, $^{27}Al(J={5 \over 2})$, 
etc. In those higher spin configurations additional 
structure functions arise which are in principle 
measureable, although they are typically smaller 
than the unpolarized structure functions, because 
polarization dependent effects arise only from the 
unpaired nucleons in the nucleus and are therefore suppressed as 1/A.  
Furthermore, 
it should be noted that all these new structure functions 
would vanish for free nucleons. Thus 
such experiments offer the possibility to confirm the spin 
structure of nucleons, and at the same time to obtain new 
information on nuclear binding effects. 

Among the new structure functions defined by \cite{jaffe2,hoodbhoy} 
there are some which have 
a simple description within the parton model 
(like $g_1$ for the nucleons) and there are others which 
vanish in the parton model (i.e., appear only in higher orders 
of QCD like $F_L$ or are of higher twist like $g_2$ for the nucleons). 
Those with a parton model interpretation 
can be described in terms of parton densities 
$q_{\pm}^{JM}(x,Q^2)$ which give the probability 
to find a quark with momentum fraction 
x and helicity $\pm {1 \over 2}$ 
in a spin--J target with spin M along the z--axis, at some scale $Q^2$.  
Measuring the cross section for an unpolarized target 
determines the averaged quark distribution 
\begin{equation} 
q^J(x,Q^2)= {1 \over 2J+1} \sum_{M=-J}^J q^{JM}(x,Q^2) \, .
\label{a11}
\end{equation} 
Because of QCD parity invariance one has 
$q_{\pm}^{JM}=q_{\mp}^{J,-M}$ 
so that only  2J+2 (2J) of the 4J+2 (4J) polarized parton densities 
are independent. (The numbers in brackets hold for half integer J.) 
It is appropriate to introduce combinations 
\footnote{Instead of Eqs.~(\ref{a8}) and (\ref{a12}) it is sometimes more 
convenient to define "multipole" 
structure functions  
$[JL]:=\sum_{M=-J}^J (-1)^{J-M} (J \, M \, J \, -M |L \, 0) q_+^{JM}$ 
where $(J \, M \, J \, -M |L \, 0)$ are the Clebsch--Gordan coefficients. 
These function can be measured using (un)polarized leptons 
for odd (even) L. 
They are particularly useful if the relative motion of nucleons inside the 
nucleus is to be considered. 
Furthermore it should be noted 
that the first $L-1$ moments of these functions vanish. 
\label{fnas}} 
\begin{equation}
F_1^{JM}={1\over 2}(q_+^{JM} + q_-^{JM})  
\label{a8}
\end{equation}
\begin{equation}
g_1^{JM}={1\over 2}(q_+^{JM} - q_-^{JM})  \, ,
\label{a12}
\end{equation}
because in the Bjorken limit the cross sections on targets 
with spin (JM) probed with helicity $\pm {1\over 2}$ 
electrons can be written as 
\begin{equation}
{d {1\over 2}(\sigma_+^{JM}+\sigma_-^{JM}) \over dx dy} = 
{8\pi \alpha^2 {\cal M} E \over Q^4} [y^2 x+2x(1-y)] F_1^{JM}(x,Q^2) 
\label{a13} 
\end{equation}
and
\begin{equation}
{d {1\over 2}(\sigma_+^{JM}-\sigma_-^{JM}) \over dx dy} = 
{8\pi \alpha^2 {\cal M} E \over Q^4} y(2-y)xg_1^{JM}(x,Q^2) \, . 
\label{a14} 
\end{equation}
with ${\cal M}$ denoting the target mass. 

Up to now we have introduced 2J+2 (2J) quark densities 
for (half)integer J corresponding to 2J+2 (2J) naive parton model  
structure functions. In addition there are  
4J (4J+1) functions which get contributions either from 
higher twist operators or from HO QCD so that one has 
altogether 
6J+2 (6J+1) independent 
structure functions for integer (half integer) spin, as will be 
shown after Eq. (\ref{a17}).  
For example, for $J={1\over 2}$ there are $F_1$, $F_2-2xF_1$, $g_1$ and $g_2$, 
among them $F_1$ and $g_1$ of leading and $F_2-2xF_1$ of higher order 
in QCD, and $g_2$ being a higher twist contribtuion. 
For J=1 one has $F_1$, $F_2-2xF_1$, $b_1$, $b_2-2xb_1$, $g_1$, $g_2$, 
$b_3$ and $b_4$.    
In the naive quark parton approximation the only nonzero  
structure functions are $F_1$, $g_1$, $b_1$ and $b_3$. 
For example, $b_1$ can be measured by scattering 
an unpolarized electron beam from a polarized deuteron target with the 
target spin directed parallel to the direction of the incident electron beam
and arranged in each of its $m_I=+1,0,-1$ substates,   
\begin{equation}
b_1(x,Q^2)= \sum_q e_q^2 [F_1^{1,+1}(x,Q^2) +
F_1^{1,-1}(x,Q^2)-2F_1^{1,0}(x,Q^2)]  \/.
\label{a66}
\end{equation} 
In general all $b_i$ can be measured using unpolarized lepton beams. 
Furthermore, they all arise from nuclear 
binding effects and are not present for a system of free nucleons. 
One has $b_1=0$ if the spin--1 target consists of two spin ${1 \over 2}$  
constituents at rest, or in relative s-wave, and thus one 
expects $b_1(deuteron) \approx 0$, because the relative motion 
of the nucleons in 
the deuteron is nonrelativistic. 
In contrast one expects a large $b_1^{\rho}$ because 
the quarks in the $\rho$--meson move relativistically 
\cite{jaffe2,hoodbhoy}. 
Furthermore, an interesting sum rule for $b_1$ has been derived 
which is related to the electric quadrupole moment of the spin--one 
target and to its polarized sea--quark content \cite{close7}.

It is possible to write down the general hadron tensor for 
a spin--1 target
\begin{eqnarray} \nonumber 
W_{\mu \nu}=-F_1 g_{\mu \nu}+F_2 {P_{\mu } P_{\nu} \over P \cdot q} 
-b_1 r_{\mu \nu}+{b_2 \over 6} (s_{\mu \nu}+t_{\mu \nu}+u_{\mu \nu}) 
+{b_3 \over 2} (s_{\mu \nu}-u_{\mu \nu})
+{b_4 \over 2} (s_{\mu \nu}-t_{\mu \nu}) 
\\
+{ig_1 \over P \cdot q} \epsilon_{\mu \nu \lambda \sigma}q^{\lambda}s^{\sigma} 
+{ig_2 \over (P \cdot q)^2} \epsilon_{\mu \nu \lambda \sigma}q^{\lambda} 
(P \cdot q s^{\sigma} -s \cdot q P^{\sigma})  
\label{a16}
\end{eqnarray}
and to obtain the cross section from it.
In Eq. (\ref{a16}) we have defined 
$r_{\mu \nu}={g_{\mu \nu} \over (P \cdot q)^2}
(q \cdot {\cal E}^{\ast}q \cdot {\cal E}- 
{\kappa \over 3}(P \cdot q)^2)$ with $\kappa =1+{4x^2{\cal M}^2 \over Q^2}$, 
$s_{\mu \nu}={2P_{\mu } P_{\nu} \over (P \cdot q)^3}
(q \cdot {\cal E}^{\ast}q \cdot {\cal E}- 
{\kappa \over 3}(P \cdot q)^2)$, 
$t_{\mu \nu}={1 \over 2(P \cdot q)^2}(q \cdot {\cal E}^{\ast}P_{\mu }
{\cal E}_{\nu} 
+q \cdot {\cal E}^{\ast}P_{\nu }{\cal E}_{\mu}
+q \cdot {\cal E}P_{\mu }{\cal E}_{\nu}^{\ast} 
+q \cdot {\cal E}P_{\nu }{\cal E}_{\mu}^{\ast}
-{4\over 3}P \cdot q P_{\mu } P_{\nu})$ 
and 
% {1 \over P \cdot q} IN u_{\mu \nu} IST KORREKT!!!!!!!!!!!!!!!!!!!
$u_{\mu \nu}={1 \over P \cdot q}({\cal E}_{\mu}^{\ast}
{\cal E}_{\nu}+
{\cal E}_{\nu}^{\ast}{\cal E}_{\mu}+{2 \over 3} {\cal M}^2  g_{\mu \nu}
-{2 \over 3} 
P_{\mu }P_{\nu})$.  
${\cal E}_{\mu}$ is the polarization 4--vector 
of the spin--1 target, 
i.e. it fulfills $P \cdot {\cal E} =0$ and ${\cal E}\cdot {\cal E} =-{\cal M}^2$ 
and $s^{\sigma}\equiv {-i\over {\cal M}^2}\epsilon^{\sigma \alpha \beta \tau} 
{\cal E}^{\ast}_{\alpha} {\cal E}_{\beta}P_{\tau}$ . 

Instead of this cumbersome approach one may gain 
more physical insight by studying the various amplitudes 
appearing in the forward Compton helicity matrix elements. 
Namely, just as for the spin--${1 \over 2}$ target the optical theorem 
relates the hadron tensor 
for arbitrary spin to the imaginary part of the forward Compton 
scattering amplitude which in turn can be expressed in 
terms of helicity amplitudes. Let $A_{mM,m'M'}$ denote the 
imaginary part of the forward Compton helicity amplitude 
for $\gamma_m +target_M \rightarrow \gamma_{m'} +target_{M'}$, 
\begin{equation}
A_{mM,m'M'}^J = \epsilon_{m'}^{\ast \mu}W_{\mu \nu}^{JM'M}\epsilon_m^{\nu} 
\label{a17}
\end{equation}
where $\epsilon_m^{\nu}$, $m=\pm 1,0$ are the photon polarization vectors. 
{\it All} the structure functions like $b_1$, $b_4$ etc. discussed before can 
be expressed as linear combinations of these amplitudes. 
The $A^J_{mM,m'M'}$ are easily enumerated. Angular momentum 
conservation requires that the total helicity is conserved, 
$m+M=m'+M'$, which leaves 18J+1 independent helicity amplitudes. 
Time reversal invariance requires $A_{mM,m'M'}^J=A_{m'M',mM}^J$ 
which leaves 12J+2 independent amplitudes. Parity invariance 
requires $A_{mM,m'M'}^J=A_{-m-M,-m'-M'}^J$ so that one finally has 
the 6J+2 (6J+1) independent amplitudes, as mentioned before. 
Among them are diagonal transverse amplitudes 
 $A_{\pm M,\pm M}^J$ 
which in the Bjorken limit 
correspond to the quark densities $q_{\pm}^{JM}$ 
introduced above: $A_{\pm M,\pm M}^J=q_{\pm}^{JM}$. 
%and $A_{+ H,+ H}^J$, J+1(J+1/2) diagonale longitudinal 
%amplitdues $A_{0  H,0 H}^J$ for $H \geq 0$, 

The amplitudes naturally arise when the hadron tensors 
$W_{\mu \nu}^{JM'M}$ are multiplied 
with the lepton tensor to form the cross section. 
To work that out in detail one has to expand the leptontensor 
Eq. (\ref{242}) in the basis of virtual photon helicity eigenstates 
\begin{eqnarray}  \nonumber 
L_{\mu \nu}={2Q^2 \over \kappa y^2} 
\biggl\{ \lambda^2 \epsilon_0^{\mu} \epsilon_0^{\nu} 
+{1\over 2}(\lambda^2+\kappa y^2)(\epsilon_+^{\mu \ast} \epsilon_+^{\nu } 
             +\epsilon_-^{\mu \ast} \epsilon_-^{\nu }) 
-{\lambda^2 \over 2}(\epsilon_+^{\mu \ast} \epsilon_-^{\nu }e^{2i\phi} 
 +\epsilon_-^{\mu \ast} \epsilon_+^{\nu }e^{-2i\phi})
\\  \nonumber
+\lambda (1- {y\over 2})(\epsilon_-^{\mu \ast}\epsilon_0^{\nu}e^{-i\phi} 
+\epsilon_0^{\mu}\epsilon_-^{\nu }e^{i\phi}
-\epsilon_+^{\mu \ast}\epsilon_0^{\nu}e^{i\phi}
-\epsilon_0^{\mu}\epsilon_+^{\nu }e^{-i\phi}) \biggr\} 
\\
+{2Q^2 \over y\sqrt{\kappa}} \biggl\{ (1- {y\over 2})
(\epsilon_+^{\mu \ast} \epsilon_+^{\nu }- 
\epsilon_-^{\mu \ast} \epsilon_-^{\nu }) 
-{\lambda \over 2}
(\epsilon_-^{\mu\ast}\epsilon_0^{\nu}e^{-i\phi} 
+\epsilon_0^{\mu}\epsilon_-^{\nu }e^{i\phi}
+\epsilon_+^{\mu \ast}\epsilon_0^{\nu}e^{i\phi}
+\epsilon_0^{\mu}\epsilon_+^{\nu }e^{-i\phi}) \biggr\}
\label{a127}
\end{eqnarray}
where $\lambda^2=2(1-y)-{y^2 \over 2}(\kappa -1)$ 
and $\phi $ is the 
azimuthal angle (measured with respect to the x--axis in the 
xy--plane) of the final lepton. The photon momentum is used as 
the spin quantization axis. 
Note that the first term ($\sim {2Q^2 \over \kappa y^2}$) 
in Eq.~(\ref{a127}) is spin--independent 
and the last term ($\sim {2Q^2 \over y\sqrt{\kappa}}$) 
is the spin dependent term.   

The resulting expressions for the cross sections in terms of 
the helicity amplitudes are 
\begin{eqnarray} \nonumber  
{d  \Sigma^{J\lambda ' \lambda} \over dxdyd\phi}=
{8\pi \alpha^2MEx \over 2\pi Q^4\kappa} 
\biggl\{  \lambda^2 (A_{0M,0M'}^J-{1\over 2} A_{+M,-M'}^Je^{-2i\phi} 
-{1\over 2} A_{-M,+M'}^Je^{2i\phi}) 
\\ \nonumber
+{1\over 2}(\lambda^2+\kappa y^2)(A_{+M,+M'}^J+A_{-M,-M'}^J)
\\ 
+(1- {y\over 2})\lambda (A_{-M,0M'}^Je^{i\phi}+A_{0M,-M'}^Je^{-i\phi}
-A_{+M,0M'}^Je^{-i\phi}-A_{0M,+M'}^Je^{i\phi}) \biggr\}  
\label{a613}
\end{eqnarray}
\begin{eqnarray} \nonumber 
{d \Delta \Sigma^{J\lambda ' \lambda} \over dxdyd\phi}=
{8\pi \alpha^2 MEx \over 2\pi Q^4 \sqrt \kappa}
\biggl\{ y(1- {y\over 2}) (A_{+M,+M'}^J-A_{-M,-M'}^J)
-{1\over 2}y\lambda (A_{-M,0M'}^Je^{i\phi} 
+A_{0M,-M'}^Je^{-i\phi}
\\ 
+A_{+M,0M'}^Je^{-i\phi}+A_{0M,+M'}^Je^{i\phi}) \biggr\}
\label{a614}
\end{eqnarray}
These cross sections are not yet in a useful form because the helicities 
M and M' are defined w.r.t. the virtual photon direction    
which changes event by event. It is better to define cross sections 
for targets with definite helicites $\lambda$ and $\lambda '$ w.r.t. the 
incident beam direction. To transform between the 2 frames 
one has to perform an Euler rotation. The state $|J\lambda >$ can 
be written in terms of $|JM>$ as 
\begin{equation}
|J\lambda >=\sum_M e^{i(\lambda -M)\phi} d_{M\lambda}^J(\beta) |JM>
\label{a615}
\end{equation}
where $d_{M\lambda}^J(\beta)$ is the Wigner rotation matrix and 
$(\beta)$ the angle between $\vec q$ and the incoming lepton 
direction $\vec k$. 
The cross sections for targets with definite polarizations 
in the lab frame can therefore be obtained from Eqs. (\ref{a613}) 
and (\ref{a614}) as 
\begin{equation} 
{d  \sigma^{J\lambda ' \lambda} \over dxdyd\phi}= 
\sum_{M,M'} {d  \Sigma^{J\lambda ' \lambda} \over dxdyd\phi }
d_{M\lambda}^J(\beta)d_{M'\lambda '}^J(\beta)
e^{i(\lambda -\lambda '+M'-M)\phi}
\label{a616}
\end{equation}
\begin{equation}
{d \Delta \sigma^{J\lambda ' \lambda} \over dxdyd\phi}= 
\sum_{M,M'} {d \Delta \Sigma^{J\lambda ' \lambda} \over dxdyd\phi}
d_{M\lambda}^J(\beta)d_{M'\lambda '}^J(\beta)
e^{i(\lambda -\lambda '+M'-M)\phi}
\label{a617}
\end{equation}
For more details see  \cite{hoodbhoy,jaffe2}. 
In those references models have been studied, like the bag model, 
to predict some of the 'new' nuclear structure functions 
$F_1^{JM}$ and $g_1^{JM}$.

%% file: k69tex
\subsection{Nuclear Bound State Effects}

In the previous section 
only the appearance of new structure functions in higher 
spin nuclei has been discussed. Another, probably even more 
important question is how the 'old' functions like $g_1$ are modified 
by nuclear effects, i.e. how much e.g. $g_1^D$ deviates from 
$g_1^p+g_1^n$. This is a very important question because $g_1^n$ 
cannot be determined directly but only through a measurement of $g_1^D$ 
or $g_1(^3He)$. Let us start with the discussion of deuterium, for 
which a number of studies exists. 
The simplest and firmest theoretical approach to the deuteron is 
to approximate it as a free neutron and proton with 
polarization $1-{3 \over 2} w_D$, where $w_D$ is the deuteron 
d--state probability. 
All other nuclear corrections are essentially small because 
they are suppressed 
by powers of ${\vec{p}^2 \over M^2}$ where $\vec{p}$ is the 3--momentum 
of the nucleon in the restframe of the nucleus. 
$w_D$ can be calculated in NN potential 
models to be $w_D \approx 0.050 \pm 0.010$ \cite{buck,arnold1}. 
There is a relatively large theoretical error in the 
prediction of $w_D$, because 
depending on which of the phenomenological potentials one uses 
one gets different answers. This theoretical error is by 
far the dominant source of uncertainty in the deuteron 
analysis, much larger than the $O({\vec{p}^2 \over M^2})$ effects 
mentioned above. This is true except for the large--x region 
($x \geq 0.8$) where large nuclear effects are known to be 
present for unpolarized scattering and expected for polarized 
scattering as well. It is not clear whether these effects 
are spin independent, in the sense that they drop out in the 
asymmetries $\sim {g_1 \over F_1}$, or not. 

The fact that $w_D$ is not precisely known may become a problem 
for future precision measurements of $g_1^n$ because a variation 
of 2\% in $w_D$ corresponds to an error of roughly 10\% in $g_1^n$. 
This is also the order of magnitude which other nuclear effects 
may have on the determination of $g_1^n$. To incorporate these 
effects there has been a sequence of papers 
(\cite{melnitchouk,melnitchouk1} and references therein, but see 
also the work of \cite{woloshyn} to be discussed below)  
working in the impulse approximation, i.e. neglecting final state 
interactions. There may be final state interactions and related 
effects like final state pion exchange or nuclear shadowing, but 
these effects are usually assumed to be small for light nuclei 
and to some extent spin independent.  

At first the so--called convolution model was applied 
to include nuclear binding and relativistic effects, and afterwards 
the effects of off-mass-shellness of bound nucleons were studied. 
In the convolution model the free nucleon structure function is 
convoluted with the light cone momentum distribution of nucleons 
in the nucleus \cite{roberts}
\begin{equation} 
g_1^D(x,Q^2)=\sum_{N=n,p} \int_x^1 {dy \over y} \delta f_{N/D}(y)
g_1^N({x \over y},Q^2) \,  .
\label{a618}
\end{equation} 
This formula holds in the Bjorken limit (infinite $Q^2$ and $\nu$). For 
finite $Q^2$ there is a spectral representation which generalizes 
Eq. (\ref{a618}) \cite{melnitchouk,melnitchouk1,thomas}.  
It was found \cite{tokarev} that the convolution approach gives 
a result very close to the description with a constant factor 
$1-{3 \over 2} w_D$, at least for $x \leq 0.7$. Above 0.7 
there are appreciable corrections to this factor which can become as large 
as 10\%.  The simple behaviour for $x \leq 0.7$ is not modified when 
off-mass-shell corrections \cite{melnitchouk,melnitchouk1} are included 
but there are additional contributions at $x \geq 0.7$ 
(of up to 5\%) which should 
be taken into account. 
Although the nuclear effects are within the error bars of presently 
available data, they will be required in the upcoming 
high statistics E154 and HERMES experiments. 
  
Now coming to $^3He$, as a first approximation one may say that 
$g_1(^3He)$ directly gives $g_1^n$ because the spins of the protons 
compensate each other due to the Pauli principle. 
This relies on the assumption that all nucleons are in a S wave. 
However, such a cancellation
does not occur
if other components
of the three body wave function are considered. 
In references \cite{woloshyn,atti,atti1}  
the question has been quantitatively discussed as to whether
and to what extent the extraction of $g_1^n$
from the asymmetry of the process
$ {^3\vec{\rm He}}({\vec e}, e')$X could be
hindered by nuclear effects arising
from small wave function components of $^3$He, as well as from
Fermi motion and binding
effects on DIS.
The basic nuclear ingredient
used in \cite{atti,atti1} is
the spin dependent spectral
function of $^3$He, which allows one to take into account
at the same time Fermi motion and binding corrections. 
Of particular relevance are the up and down quark spectral 
functions $P^N_{\sigma \sigma ' M}$, 
$N_{3+}\equiv P^N_{\ha \ha \ha}$ and $N_{3-}\equiv 
P^N_{-\ha -\ha \ha}$,  
because the integral of their difference determines the 
nucleon polarizations $P_N^{(\pm)}$, $N=p,n$ \cite{atti,atti1}.  
These functions enter the so called spectral representation of 
$g_1(^3 He)$ as a function of $g_1^N$. Note that in the Bjorken limit this 
spectral representation goes over into Eq. (\ref{a618}). 

\begin{figure}
\begin{center}
\epsfig{file=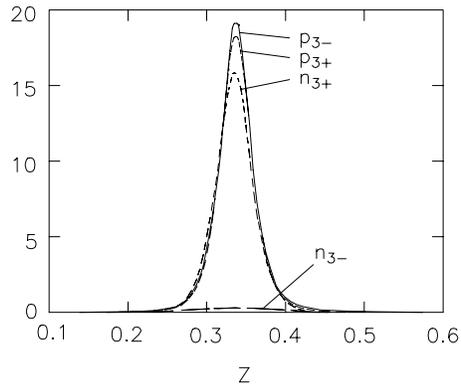,height=5cm}
\bigskip
\caption{Nucleon momentum distributions for spins
parallel and antiparallel to a $^3$He target, with z being 
the momentum fraction of the nucleon in the target, according to 
\protect\cite{woloshyn}.}
\label{figw1}
\end{center}
\end{figure}

In ref. \cite{woloshyn} the $\vec {^3{\rm He}}$
asymmetry has been calculated taking into
account $S'$ and $D$ waves but considering only Fermi motion and omitting
$Q^2$ dependent terms. Woloshyn used $^3$He wave functions 
\cite{afnan} calculated 
using the momentum space Faddeev equation, which are known to 
give a good 
description of the spin averaged quasi--elastic data at medium 
energies.  
The extent to which the proton contribution to the spin 
asymmetry of $^3$He can be neglected, can be envisaged 
by a look at the proton momentum distributions for spins 
parallel and antiparallel to the target, cf. Fig. \ref{figw1} 
where they are plotted against the light cone momentum fraction. 
The spin dependence is seen to be quite small. 
Also shown in Fig. \ref{figw1} are the corresponding curves for 
the neutron. The neutron curves answer  
the important question, how large the probability is of 
having a neutron with spin antiparallel to the nuclear spin. 
One sees that $n_{3-}$  is very small as compared to $n_{3+}$ but somewhat    
broader since it comes from components in the $^3$He wave function 
with a larger high momentum tail than the dominant S--wave. 

\begin{figure}
\begin{center}
\epsfig{file=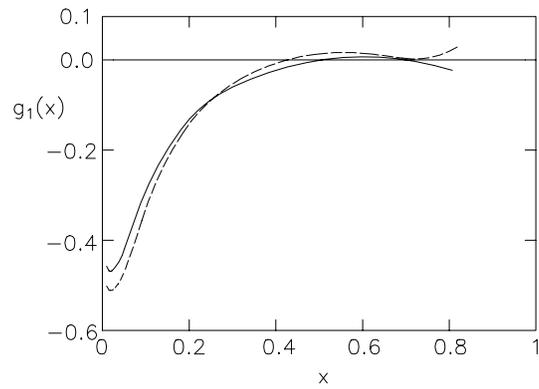,height=5cm}
\bigskip
\caption{$g_1(x)$ for neutron (dashed) and $^3$He (solid curve) 
according to \protect\cite{woloshyn}.}
\label{figw2}
\end{center}
\end{figure}

Fig. \ref{figw2}  shows typical results for $g_1(^3He)(x,Q^2)$ as 
compared to $g_1^n(x,Q^2)$ at some unknown (small) value of 
$Q^2$. The curves are given only up to $x \approx 0.9$ 
because above this value the calculation becomes unreliable. In fact, 
nuclear effects are expected to become much larger as $x\rightarrow 1$. 

It is possible to summarize the 
effects of higher wave functions $S'$ and $D$ in $^3$He 
by means of the following 
procedure: 
In a pure $S$ wave state the nucleon polarisations are given by 
$P_n^{(+)}=1$, $P_n^{(-)}=0$ and
$P_p^{(+)}=P_p^{(-)}={1\over2}$, whereas for a three--body wave function
containing $S$, $S'$ and $D$ waves, one has
\begin{equation}
P_n^{(\pm)}={1\over 2} \pm {1 \over 2} \mp \Delta, \label{pnxxx}
\end{equation}
\begin{equation}
P_p^{(\pm)}={1 \over 2}                     
\mp \Delta'~~, \label{ppxxx}                   
\end{equation}
where $\Delta={1 \over 3} [P_{S'}+2P_D]$ and
$\Delta'={1 \over 6}[P_D-P_{S'}]$.          
From
world calculations on the three body system one obtains,
in correspondence with the experimental value of the 
binding energy of $^3$He,                   
$\Delta=0.07 \pm 0.01$ and $\Delta'=        
0.014\pm 0.002$ \cite{friar}.                 
Thus if the $S'$ and $D$ waves are          
considered and                              
Fermi motion and binding effects            
are disregarded, one can write              
\begin{equation}
g_1(^3He)(x,Q^2)  =  2p_p g_1^p(x,Q^2) + p_n g_1^n(x,Q^2)\label{gmod2} 
\end{equation}
\begin{equation}
A_{\vec {^3{\rm He}}}  =  2 f_p p_p A_{\vec p} + f_n p_n A_{\vec n}
\label{mod2}                                
\end{equation}
where $f_{p(n)}(x,Q^2)=                     
F_2^{p(n)}(x,Q^2)/[2F_2^p(x,Q^2)+F_2^n(x,Q^2)]$
is the proton (neutron) dilution factor,    
$A_{{\vec p}({\vec n})}(x,Q^2)=g_1^{p(n)}(x,Q^2)/F_1^{p(n)}(x,Q^2)$
is the proton (neutron) asymmetry           
and $p_{p(n)}$ are the effective nucleon polarizations 
\begin{equation}
p_p  =  P_p^{(+)}-P_p^{(-)}=-0.028{\pm} 0.004
\label{polp} 
\end{equation}
\begin{equation}
p_n  =  P_n^{(+)}-P_n^{(-)}=0.86 {\pm} 0.02 \label{pol}
\end{equation}
The above values correspond to Eqs. (\ref{pnxxx}) and (\ref{ppxxx}), while
the spin dependent spectral functions of 
\cite{woloshyn,atti,atti1} yield 
$p_p=-0.030$ and $p_n=0.88$.

For future precision 
studies one will have to go even 
beyond those approximations, because there 
may be modifications due to the presence of the nuclear medium, like 
nuclear swelling and binding effects 
on the parton densities. These effects have recently 
been studied by \cite{indumathi} by depleting the parton densities of 
\cite{glueck2} at small $Q^2$ and evolving them according to the 
AP evolution equations. It turns out that these effects are far below 
the present experimental accuracy, but may be of some relevance for 
future precision measurements, in 
particular at small $x \leq 0.03$.

%% file: k699tex
\subsection{Direct Photons and Related Processes in Proton Collisions 
      using Polarized Beams}
%--68.tex-----}

The most interesting prospect for polarized high energetic 
proton experiments is the possibility to determine $\d g(x,\mu_F^2)$ 
from the process $\vec p\vec p \rightarrow \gamma X$ with a high energetic
photon in the final state 
\cite{einhorn,bourrely2,berger1,gupta,cheng2,contogouris1,gordon,glueck14,
bourrely1,bunce,mathews1,guellenstern,soffer2}.   
In unpolarized scattering hard photons are known to be a clean 
probe of the gluon distribution, because they can be directly detected,
without undergoing fragmentation.
Note that for physics beyond the standard model direct photon 
production is claimed to be a signal to probe compositeness. The 
observation of spin asymmetries in this reaction could help to 
disentangle the structure of new contact interactions 
\cite{bourrely2}. 

\begin{figure}
\begin{center}
\epsfig{file=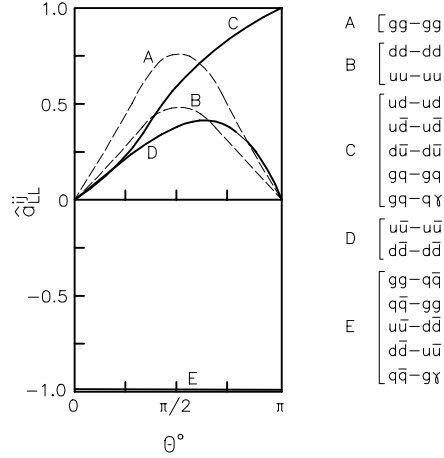,height=6cm}
\bigskip
\caption{Parton level asymmetries, including direct photon production 
processes as
a function of the scattering angle in the parton-cms 
\protect\cite{bourrely3}.}
\label{gamas}
\end{center}
\end{figure}

On the parton level the process is induced in lowest order by 
the annihilation of light quarks $q \bar{q} \rightarrow \gamma g$ and 
by the Compton scattering 
$gq \rightarrow \gamma q$ , and is strongly dependent on the 
magnitude of  $\d g(x,\mu_F^2)$ with $\mu_F \sim p_T^{\gamma}$. 
It may also be possible to determine  $\d g(x,\mu_F^2)$ from high-$p_T$ 
jet production, but this process has a large background of 
quark-initiated events $qq \rightarrow partons$, and is therefore 
less sensitive to $\d g(x,\mu_F^2)$. 
In Table 5 and Fig. \ref{gamas} the parton level asymmetries for all the 
possible partonic $2 \rightarrow 2$ processes 
including direct photon production are shown as 
a function of the scattering angle in the parton-cms. 
The figure clearly shows that the photon processes 
are expected to give   
potentially large effects. 

\begin{table}                             
\label{tab1267}                              
\begin{center}                            
\begin{tabular}{|c|c|c|}                  
\hline
 & ${d\hat \sigma_{ij} \over d\hat t}$ & $\hat a_{LL}^{ij}$ \\  
\hline
$qq \rightarrow qq$ & 
${4\over 9} \bigl( {\hat s^2 +\hat u^2 \over \hat t^2} +
{\hat s^2+\hat t^2\over \hat u^2}
-{2\over 3}{\hat s^2\over \hat t\hat u}\bigr)$ & 
${(\hat s^2-\hat u^2)/\hat t^2+(\hat s^2-\hat t^2)/\hat u^2 
-{2\over 3}\hat s^2/\hat t\hat u \over 
(\hat s^2+\hat u^2)/\hat t^2+(\hat s^2+\hat t^2)/\hat u^2 
-{2\over 3}\hat s^2/\hat t\hat u}$  \\
$qq' \rightarrow qq'$ &
${4\over 9} {\hat s^2 +\hat u^2 \over \hat t^2}$ &
${\hat s^2-\hat u^2\over \hat s^2+\hat u^2}$ \\
$q\bar q \rightarrow q'\bar q'$ &
${4\over 9} {\hat t^2 +\hat u^2 \over \hat s^2}$ &
$-1$ \\
$q\bar q \rightarrow q\bar q$ &
${4\over 9}\bigl( {\hat t^2 +\hat u^2 \over \hat s^2}+ 
{\hat s^2 +\hat u^2 \over \hat t^2}- 
{2\over 3}{\hat u^2\over \hat s\hat t}\bigr)$ &
${(\hat s^2-\hat u^2)/\hat t^2-(\hat t^2+\hat u^2)/\hat s^2
+{2\over 3}\hat u^2/\hat s\hat t \over
(\hat s^2+\hat u^2)/\hat t^2+(\hat t^2+\hat u^2)/\hat s^2
-{2\over 3}\hat u^2/\hat s\hat t}$ \\
$q\bar q \rightarrow gg$ &
${32\over 27} {\hat t^2 +\hat u^2 \over \hat t\hat u} 
-{8\over 3}{\hat t^2 +\hat u^2 \over \hat s^2}$ &
$-1$ \\
$qg \rightarrow qg$ &
${\hat s^2 +\hat u^2 \over \hat t^2}-
{4\over 9}{\hat s^2 +\hat u^2 \over \hat s\hat u}$ &
${\hat s^2-\hat u^2\over \hat s^2+\hat u^2}$ \\
$gg\rightarrow q\bar q$ &
${1\over 6} {\hat t^2 +\hat u^2 \over \hat t\hat u}-
{3\over 8}{\hat t^2 +\hat u^2 \over \hat s^2}$ &
$-1$ \\
$gg\rightarrow gg$ &
${9\over 2}\bigl( 3-{\hat s\hat u\over \hat t^2}
-{\hat s\hat t\over \hat u^2}-{\hat t\hat u\over \hat s^2}\bigr)$ &
${-3+2\hat s^2/\hat t\hat u+\hat t\hat u/\hat s^2 \over 
3-\hat s\hat u/\hat t^2-\hat s\hat t/\hat u^2-\hat t\hat u/\hat s^2}$ \\
$qg\rightarrow q\gamma$ &
$-{1\over 3}\bigl( {\hat s\over \hat u}+{\hat u\over \hat s}\bigr)$ &
${\hat s^2-\hat u^2\over \hat s^2+\hat u^2}$  \\
$q\bar q\rightarrow \gamma g$ &
${8\over 9}\bigl( {\hat t\over \hat u}+{\hat u\over \hat t}\bigr)$ &
$-1$ \\
\hline
\end{tabular}
\bigskip
\caption{Tree--level partonic cross sections
and asymmetries $\hat a_{LL}^{ij}=d\Delta \hat \sigma_{ij} 
/ d \hat \sigma_{ij}$, including direct
photon production processes \protect\cite{babcock,bourrely3}.
A factor of $\pi \alpha_s^2/\hat s^2$ has been factored out of
the jet cross sections and $\pi e_Q^2\alpha \alpha_s/\hat s^2$
has been factored out of the single--photon cross sections. }
\end{center}
\end{table}

From a theoretical point of
view it would also be interesting to study 
the production of high-$p_T$ muon pairs in Drell--Yan processes 
involving off-shell photons,  
because they offer the possibility 
to obtain additional information,namely about the polarization 
of the final state. This information can be obtained, for example,
from the angular distribution of the leptons 
\cite{cheng2,leader1,leader2}. For an interesting application 
of HO QCD to this process see \cite{carlitz1}.
The Drell--Yan process will be discussed later on in this 
subsection. 

\begin{figure}
\begin{center}
\epsfig{file=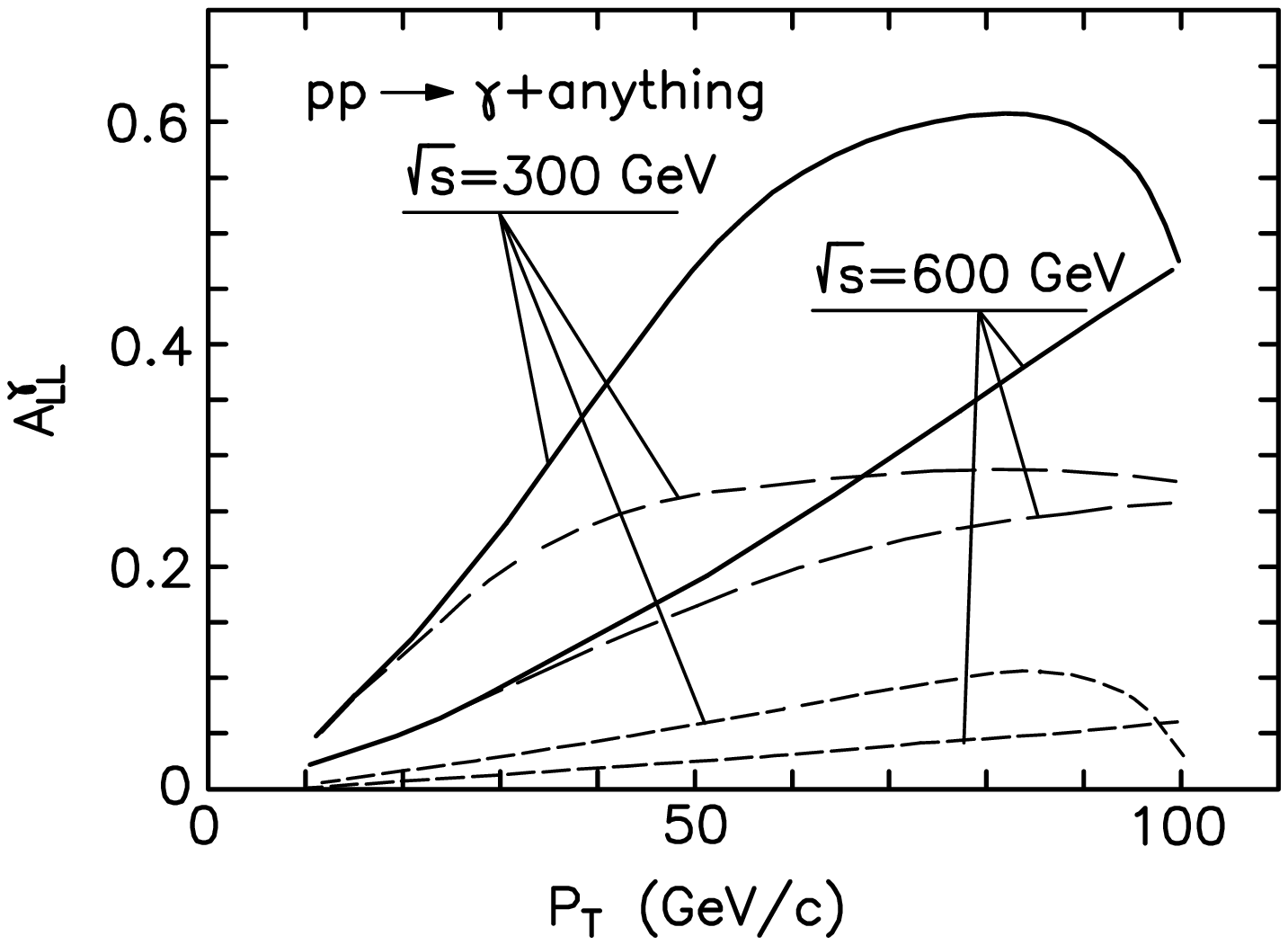,height=7cm}
\bigskip
%\caption{$A_{LL}^{\gamma}$ at 300 GeV pp beam energy 
%with a large $\delta g(x,\mu_F^2)$  
%as a function of $p_T$ at $\theta_{cm} =45^o$ (upper solid 
%curve) and $\theta_{cm} =90^o$ (dashed curve). The lower short-dashed 
%curve corresponds to a small $\delta g(x,\mu_F^2)$ at $\theta_{cm} =90^o$. 
%The factorization scale was chosen to be $\mu_F=p_T$. 
%Analogous curves for $\sqrt{s}=$ 600 GeV are also included 
%\protect\cite{bourrely1}.  }
\label{gamq}
\end{center}
\end{figure}

The observable to be measured in direct-$\gamma$ production 
is the inclusive double spin 
asymmetry 
\begin{equation}
A^{\gamma}_{LL}={\sigma_{\Leftarrow}^{\Leftarrow} 
+\sigma_{\Rightarrow}^{\Rightarrow}
-\sigma_{\Rightarrow}^{\Leftarrow} 
-\sigma_{\Leftarrow}^{\Rightarrow}
\over
\sigma_{\Leftarrow}^{\Leftarrow} 
+\sigma_{\Rightarrow}^{\Rightarrow}
+\sigma_{\Rightarrow}^{\Leftarrow}
+\sigma_{\Leftarrow}^{\Rightarrow}}
\label{681}
\end{equation}
which reduces to 
\begin{equation}
A^{\gamma}_{LL}={\sigma_{\Leftarrow}^{\Leftarrow}-
\sigma_{\Rightarrow}^{\Leftarrow} \over 
\sigma_{\Leftarrow}^{\Leftarrow}+
\sigma_{\Rightarrow}^{\Leftarrow}} 
\label{682}
\end{equation}
if parity is conserved. In Eqs. ~(\ref{681}) and (\ref{682}) it has been 
assumed that both protons are longitudinally polarized.  
More explicitly, this asymmetry is given by 
\begin{equation}
A_{LL}^{\gamma} d \sigma = 
\sum\limits_{i,j=q,\bar q,g} {1 \over 1+ \delta_{ij}} \int dx d x' 
\biggl\{ \d f_i(x,\mu_F^2) \d f_j'(x',\mu_F^2) 
\hat a_{LL}^{ij} d \hat \sigma_{ij} +(i \leftrightarrow j) \biggr\}
\label{683}
\end{equation}
where $d \hat \sigma_{ij}$ and $d \sigma$ are the parton and hadron 
level cross sections for unpolarized direct-$\gamma$ production, 
\begin{equation}
d \sigma = 
\sum\limits_{i,j=q,\bar q,g} {1 \over 1+ \delta_{ij}} \int dx d x'
\biggl\{  f_i(x,\mu_F^2)  f_j'(x',\mu_F^2)  
d \hat \sigma_{ij} +(i \leftrightarrow j) \biggr\}
\label{684}
\end{equation}
The prime refers to the second proton and $f_i$ and $f_j'$ 
are the parton densitites in the two protons. 
$\hat a_{LL}^{ij}$ are the subprocess asymmetries, which along with the  
$d \hat \sigma_{ij}$ can be calculated in perturbative QCD, 
cf. Table 5 and Fig. \ref{gamas}. 
The product $d \Delta \hat \sigma_{ij}=\hat a_{LL}^{ij} d \hat \sigma_{ij}$ 
is much larger for 
the Compton subprocess than for the annilation subprocess.  
This can be deduced from the explicit form of the parton 
level cross sections and asymmetries in 
Table 5 and Fig. \ref{gamas}. 
%${d (\Delta ) \sigma_{ij} \over d\hat t}= 
%{(\Delta ) |M|^2(ij \rightarrow \gamma X) \over 16 \pi \hat s^2}$ 
%with  
%\begin{eqnarray} \nonumber
%(\Delta ) |M|^2(q \bar q \rightarrow \gamma g)&=& \pm {8 \over 9}
%({\hat u \over \hat t} + {\hat t \over \hat u})   \\ \nonumber 
%(\Delta ) |M|^2(q  g \rightarrow \gamma q)&=&-{1 \over 3}  
%({\hat s \over \hat t} \pm {\hat t \over \hat s})   \\    
%(\Delta ) |M|^2(g \bar q \rightarrow \gamma q)&=&-{1 \over 3}  
%({\hat s \over \hat u} \pm {\hat u \over \hat s})
%\label{685}
%\end{eqnarray}
%where $\hat s$, $\hat t$ and $\hat u$ are the Mandelstam 
%variables for the parton subprocess and the lower signs 
%stands for the polarized matrix element. 
The Compton subprocess has a positive $\hat a_{LL}^{i=g,j=q} $ and leads 
to a positive contribution to $A_{LL}^{\gamma}$ which is proportional 
to $\d g$. For the annihilation subprocess one has 
a negative $\hat a_{LL}^{ij} $. Since the Compton subprocess dominates 
on the parton level, one has usually a much smaller asymmetry 
$A_{LL}^{\gamma}$ (positive or negative) if $\d g(x,\mu_F^2)$ is small, 
in contrast to a large  
$\d g(x,\mu_F^2)$. In the latter case one encounters 
large positive values of $A_{LL}^{\gamma}$ up to 50 percent. 
In Fig. \ref{gamq} $A_{LL}^{\gamma}$ 
is shown as a function of the photon-$p_T$ 
for the two scenarios of large and small $\d g$. 
It is clearly seen that a measurement of $A_{LL}^{\gamma}$ 
provides a valuable probe of $\delta g(x,\mu_F^2)$. 

Higher order QCD corrections to the process 
$\vec p\vec p \rightarrow \gamma X$ involving polarized proton beams 
have been calculated by \cite{contogouris1,gordon}.  
A particular feature is the appearance of the process 
$gg \rightarrow \gamma q \bar{q}$ whose contribution to the
spin asymmetry is of the order of $[\d g]^2$. 
The calculation has been performed in the framework of 
dimensional regularization, in which case special care is needed 
for the treatment of $\gamma_5$. 
In ref. \cite{contogouris1} the method 
of dimensional reduction was used and it was shown 
how to circumvent the problems with $\gamma_5$.  
In ref. \cite{gordon} the 't Hooft-Veltman scheme was used 
and the dependence of the cross section 
on isolation cuts has been examined 
carefully. Isolation cuts are needed in order to single 
out isolated direct hard photons from the background 
\cite{gordon2,gordon1}. 
  
\begin{figure}
\begin{center}
\epsfig{file=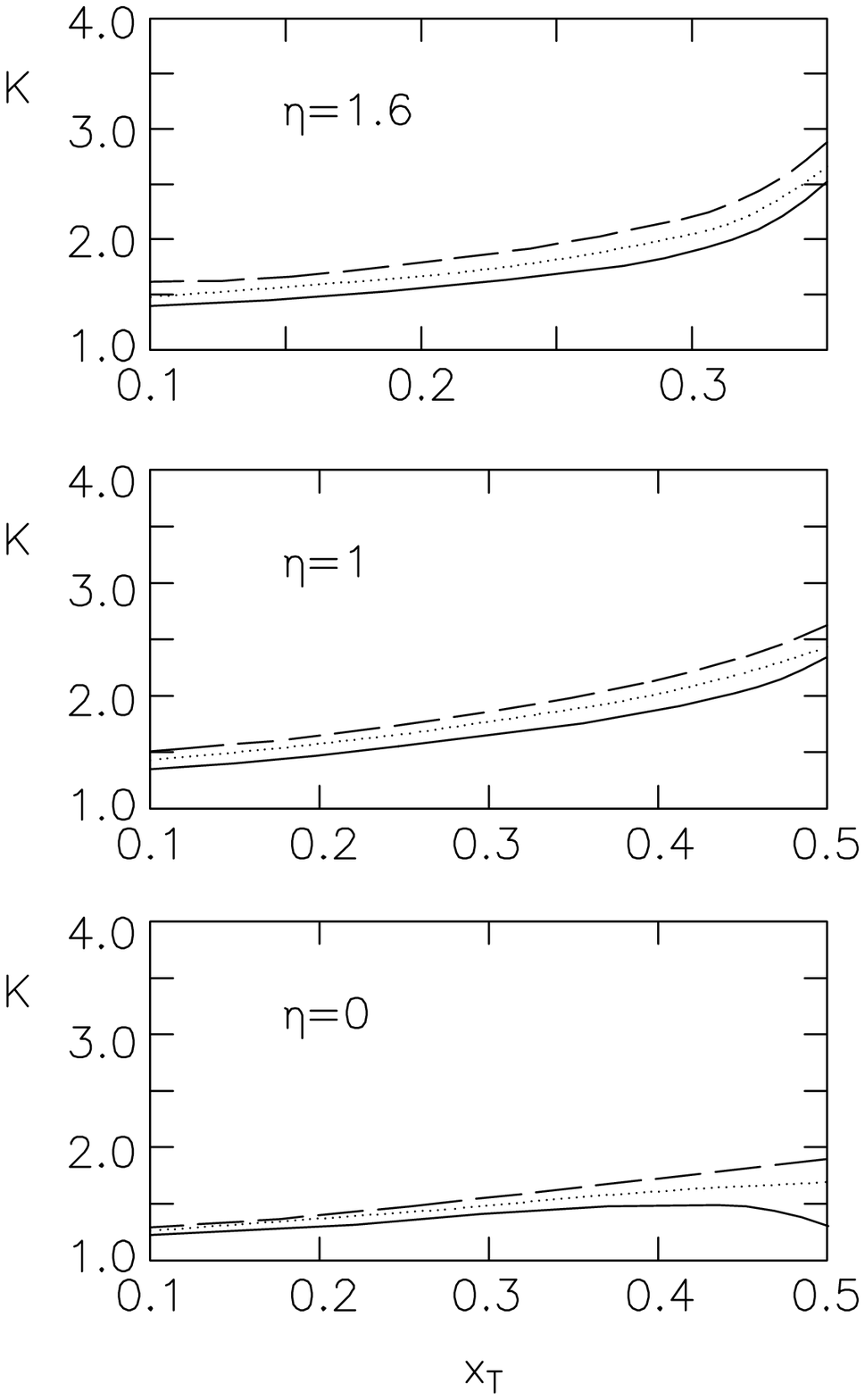,height=9cm}
\bigskip
%\caption{K--factor for direct $\gamma$--production in 
%$\vec p \vec p$--collisions as a function of $x_T$ \protect\cite{contogouris1}. 
%The following parameters have been used: $\Lambda_{\overline{MS}}=200$ 
%MeV with 4 flavors and the parton distributions of ref. 
%\protect\cite{altarelli3} (set 1) with $\mu_F=p_T$.  
%The three curves correspond to $\sqrt{s}=$38(dashed), 100(dotted) and 
%500(solid) GeV.}
\label{ficon2}
\end{center}
\end{figure}

The results of the calculation of ref. \cite{contogouris1} 
are shown in Fig. \ref{ficon2} 
in the form of the K-factor for the polarized cross section 
defined as the ratio of $K={LO+HO \over LO}$. 
The K--factors are shown as a function of $x_T={2p_T \over \sqrt{s}}$  
for various values of the rapidity $\eta$ and $\sqrt{s}$. 
Note that for the $p_T$ distribution one has the 
simple formula 
\begin{equation}
{d (\Delta )\sigma \over d p_T^2}=
\sum\limits_{i,j=q,\bar q,g} {1 \over 1+ \delta_{ij}} 
\int_{{4p_T^2\over S}}^1 dx (\delta )f_i(x,p_T^2) 
\int_{{4p_T^2\over xS}}^1 dx' (\delta ) f_j'(x',p_T^2)
{d (\Delta )\hat \sigma_{ij \rightarrow \gamma X} \over d p_T^2}
\label{686}
\end{equation}
where ${d (\Delta )\hat \sigma \over d p_T^2}= 
{d (\Delta )\hat \sigma \over d \hat t} 
{\hat s \over \hat t -\hat u} $ and 
${d (\Delta )\hat \sigma \over d \hat t}$ was given in Table 5. 
One finds in Fig. \ref{ficon2}  
that in all cases the HO corrections are positive 
and quite large, in particular at large $x_T$.
Due to the quite large K-factors there is some ambiguity 
of the results concerning the unknown higher orders. 
However, this ambiguity is much smaller than the dependence 
on the input parton densities 
%-- as is clear from Fig. \ref{ficon2}  
%where the solid lines corresponds to three different 
%parametrizations of the polarized parton densities, 
as has been discussed in \cite{contogouris1} with 
a rather moderate gluon density as input. 
It should be kept in mind, however, that  
strictly speaking, there is a very subtle interplay between 
the magnitude of the K--factor in various schemes and the 
correct choice of the HO parton densities, the value of $\Lambda$,etc.  
Similar results have been obtained, using more recent polarized 
parton densities, for prompt photon production at a future fixed 
target HERA--$\vec N$ (phase II) experiment ($\sqrt{s}=39$ GeV) 
\cite{gordon2} and at the RHIC collider ($\sqrt{s}=50-500$ GeV) 
\cite{gordon1}. 

Instead of direct photon production one may consider 
production of W's or Z's balanced by a jet. 
These processes are very closely related because 
they involve analogous diagrams. The heavy vector bosons 
show the additional feature that spin asymmetries arise 
even if only one of the proton beams is polarized 
('single spin asymmetries'). This 
is due to the axialvector couplings involved and 
essentially a parity violating effect, as already 
discussed in Sect. 6.8. The corresponding 
single spin asymmetry 
\begin{equation}
A_L^{W,Z} 
={
\sigma_{\Rightarrow}-\sigma_{\Leftarrow} \over
\sigma_{\Rightarrow}+\sigma_{\Leftarrow}
}
\label{687}
\end{equation}
involves products of polarized and 
unpolarized parton densities and would in principle be 
sensitive to $\d g(x,m_W^2)$. It turns out, however, that the QCD matrix 
element for 'Compton' scattering of a polarized gluon on an 
unpolarized quark and associated production of a $W^{\pm}$ is very small, 
so that the single spin asymmetries are not very suitable for 
the determination of $\d g(x,m_W^2)$. They $are$, however, suitable for the 
determination of the various flavor contributions $\d u(x,m_W^2)$ 
and $\d d(x,m_W^2)$ to the proton spin. 
Recent analyses of these effects can be found in 
\cite{saalfeld} and \cite{bourrely2,bourrely5,soffer2} 
where it is shown that the process $p\vec p \rightarrow 
W^{\pm}X$ at RHIC can lead to asymmetries around 50\% 
with a large effect in the asymmetry for 
$p\vec p \rightarrow W^{-}X$ if $\delta \bar u (x,m_W^2)$ 
is small. 

\begin{figure}
\begin{center}
\epsfig{file=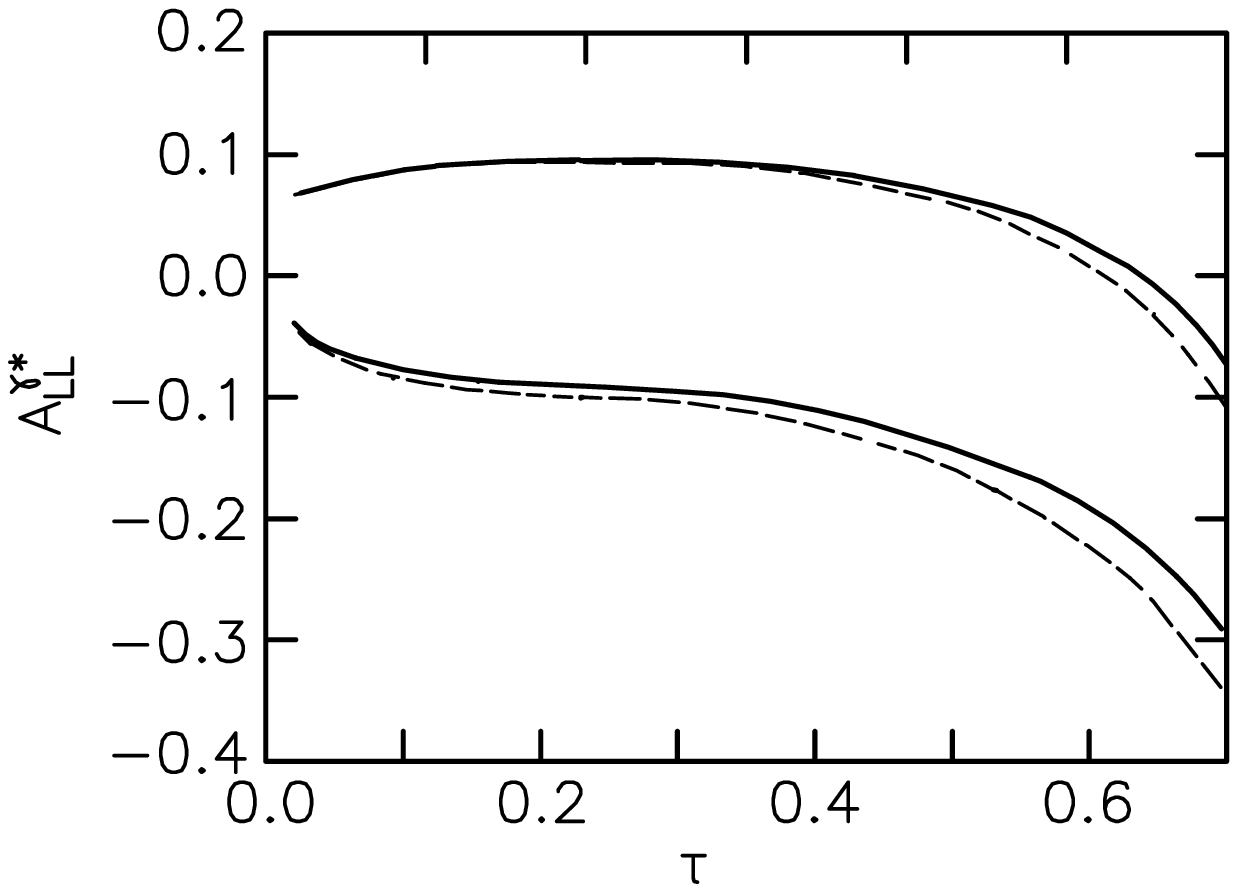,height=5cm}
\bigskip
%\caption{Predicted Drell--Yan spin asymmetry as a function 
%of $\tau =Q^2/s$ at $\sqrt{s} =100$ GeV for large positive and 
%large negative $\delta s$ \protect\cite{cheng2}. Dashed curves are the 
%asymmetries without gluon contributions. }
\label{figdry}
\end{center}
\end{figure}

Let us now come to the Drell--Yan process 
\cite{ratcliffe,gupta,cheng2,bourrely1,mathews1,weber,kamal,gehrmann2}. 
We shall discuss here only the case of longitudinal polarization. 
The NLO corrections to the Drell--Yan process in transversely polarized 
hadron--hadron collisions have been analyzed in 
\cite{contogouris3,vogelsang5}. 
The longitudinal spin 
asymmetry $A_{LL}^{\gamma^{\ast}}=
{ d \Delta \sigma / d m_{ll}^2 \over d \sigma / d m_{ll}^2 }$  
is defined in analogy to 
$A_{LL}^{\gamma}$ but depends on the invariant mass of the 
produced lepton pair $m_{ll}^2$. The polarized and 
unpolarized cross sections are given by 
\begin{eqnarray} \nonumber 
{d (\Delta )\sigma \over d m_{ll}^2} &=&(-) {4\pi \alpha^2 \over 9 s m_{ll}^2} 
\int {dx_1 \over x_1} {dx_2 \over x_2} 
\biggl\{ \sum_{q=u,d,s} e_q^2 (\delta )q(x_1,m_{ll}^2) 
(\delta )\bar q(x_2,m_{ll}^2) 
[\delta (1-z)+{\alpha_s(m_{ll}^2) \over 2\pi} 
\theta (1-z) (\delta )c_{q \bar q}] 
\\ \nonumber 
& & +{\alpha_s(m_{ll}^2) \over 2\pi} \theta (1-z) (\delta )c_{q g}(z)  
(\delta )g(x_2,m_{ll}^2)  
\sum_{q=u,d,s} e_q^2 [(\delta )q(x_1,m_{ll}^2) 
+(\delta )\bar q(x_1,m_{ll}^2)]
\biggr\} 
\\ & & 
+\left( 1\leftrightarrow 2 \right) 
\label{688}
\end{eqnarray}
% FORMEL IST AUS CHENG+LAI 
where $z={m_{ll}^2 \over x_1 x_2 s}$ and the coefficient 
functions $(\delta )c_{q \bar q}(z)$ 
and $(\delta )c_{q g}(z)$ for the relevant subprocesses 
are in the $\overline{MS}$--scheme given by \cite{furmanski,ratcliffe,
weber,mathews,kamal,gehrmann2}
\begin{eqnarray}
c_{q \bar q}(z)&=&C_F \biggl\{ ({2\over 3}\pi^2-8)\delta (1-z) 
+4(1+z^2) ({\ln (1-z) \over 1-z})_+ -2{1+z^2\over 1-z}\ln z \biggr\}
\label{6891}
\\
c_{q g}(z)&=&T_R[(2z^2-2z+1)\ln {(1-z)^2\over z} -{7\over 2}z^2+3z+{1\over 2}]
\\
\delta c_{q \bar q}(z)&=&-c_{q \bar q}(z)
\\
\delta c_{q g}(z)&=&-T_R[(2z-1)\ln {(1-z)^2\over z} 
-{3\over 2}z^2-z+{5\over 2}]   \,  .
\label{689}   
\end{eqnarray}
It should be noted that the dominant $O(\alpha_s)$ contribution 
in Eq. (\ref{688}) comes from the $\delta$--function term in 
Eq. (\ref{6891}). It gives a large effect for the cross sections 
but not in the double spin asymmetry $A_{LL}^{\gamma^{\ast}}$.  
The asymmetry is 
shown in Fig. \ref{figdry} as a function of 
$\tau = {m_{ll}^2 \over  s}$ for different parametrizations of  
$\delta g(x,m_{ll}^2)$ and $\delta s(x,m_{ll}^2)$. 
The asymmetries in Fig. \ref{figdry} are of the same order of magnitude 
as for the direct photon process. 
In principle one could expect
that both the $\delta g$ and the $\delta s$ term contribute 
about the same 
to the asymmetry. However, due to the smallness 
of $\delta c_{q g}(z)$ it turns out that the sea polarization plays 
a stronger role in $A_{LL}^{\gamma^{\ast}}$ than the gluon polarization. 
This is in contrast to the direct photon process which depends strongly 
on the polarization of the gluons. 

Figure \ref{figdry} also shows that the Drell-Yan process could be very 
valuable to determine the strange sea polarization in the proton. 
A measurement of the sign of $A_{LL}^{\gamma^{\ast}}$ would 
already give information about the sign of $\delta s$. 
Unfortunately, 
the Drell-Yan process has a larger background than 
the direct photon production so that the expected statistical 
errors at RHIC are larger.

Finally we want to add some remarks about the prospects 
of determining the polarized parton densities from jet production 
in hadronic collisions, 
in particular   
at RHIC \cite{babcock,ranft,einhorn,bourrely2,bourrely1,kunszt,
chiappetta1,ramsey,doncheski1,soffer2}. 
Quite in general many features of unpolarized jet production can be 
applied directly to 
the polarized case. For example, at low $p_T$ the main contribution 
comes from gluon--gluon scattering so that this region is 
particularly suited for attempts to determine the (polarized) 
gluon density. Secondly, event rates and statistics are much 
larger than in direct photon production. A drawback is that 
the signatures 
are less clear as will be discussed below. 

\begin{figure}
\begin{center}
\epsfig{file=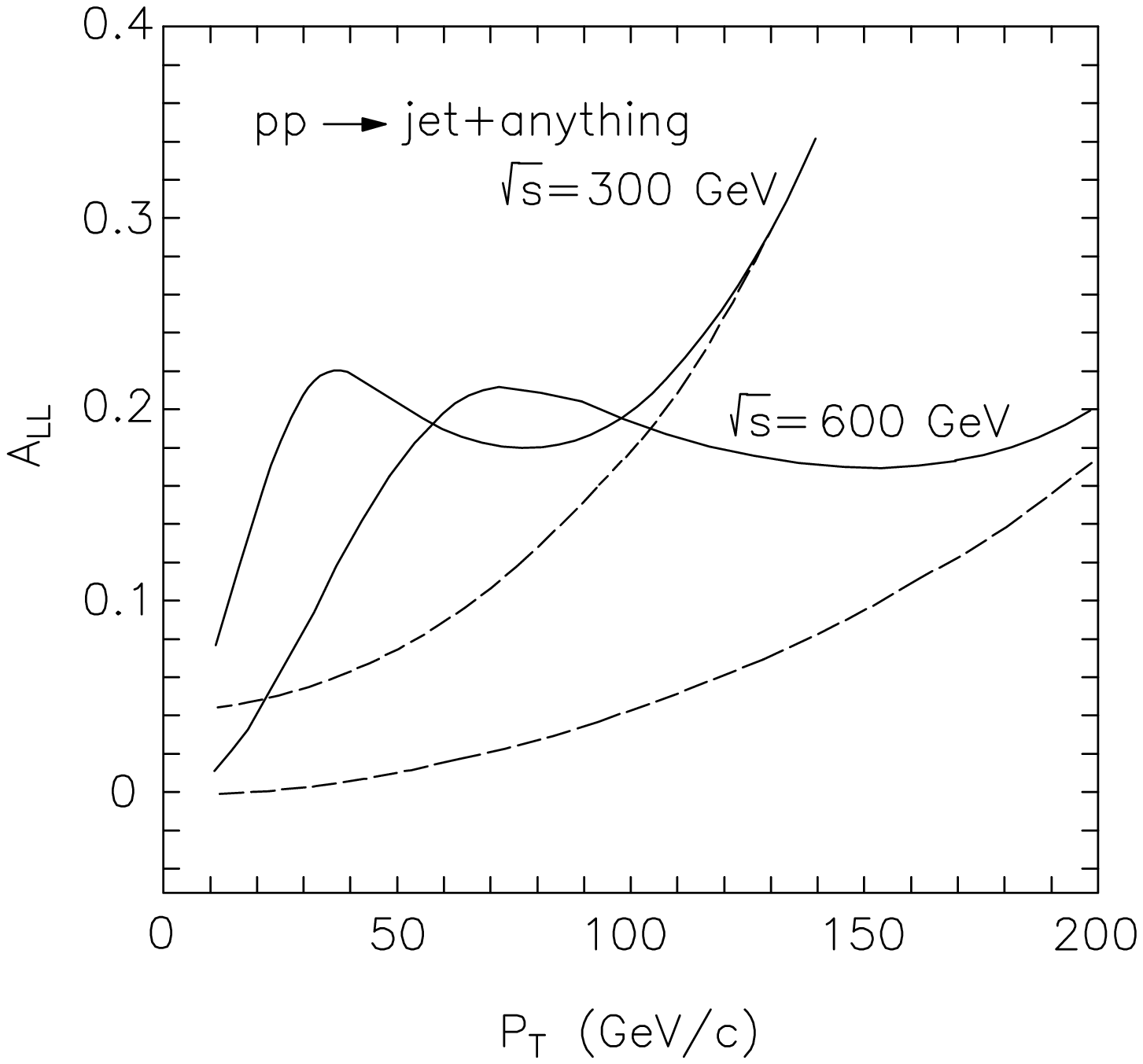,height=6cm}
\bigskip
%\caption{Predicted spin asymmetries for single jet production 
%as a function of the jet--$p_T$ for a large (solid curve) and small 
%(dashed curve) first moment $\Delta g(p_T^2)$ 
%and two values of the RHIC beam energy \protect\cite{bourrely1}. }
\label{asjet}
\end{center}
\end{figure}

One has in principle the possibility to analyze single--jet 
production $\vec p \vec p \rightarrow JX$ as well as di--jet production
$\vec p \vec p \rightarrow J_1 J_2X$. The latter is more 
difficult but not impossible to analyse in a high
luminosity machine like RHIC. The single jet spin asymmetry 
can be calculated as a function of rapidity and transverse 
momentum of the jet from the ratio of the polarized and the 
unpolarized cross sections. These are given by 
\be 
E{d(\Delta )\sigma \over d^3p}={1\over \pi} \sum_{i,j} 
{1\over 1+\delta_{ij}} \int_{x_{min}}^1 
dx_1 {2x_1x_2 \over 2x_1-x_T e^y} 
[ (\delta )f_i(x_1,\mu_F^2) 
(\delta )f_j(x_2,\mu_F^2) {d (\Delta )\hat \sigma_{ij} \over d\hat t} 
+(i \leftrightarrow j) ]
\label{6810} 
\ee
where $x_T={2p_T\over \sqrt{s}}$, $x_{min}={x_Te^y \over 2-x_Te^{-y}}$ 
and $x_2={x_1x_Te^{-y} \over  2x_1-x_Te^y}$. The explicit 
expressions for the parton cross sections can be found in 
Table 5 \cite{babcock,bourrely3} and the corresponding 
parton level spin asymmetries are shown in Fig. \ref{gamas}. 
For all the dominant subprocesses the corresponding asymmetries
$\hat a_{LL}^{ij}$ are positive, except for $q\bar q$ annihilation. 
This leads to positive values of the single
jet double spin asymmetry $A^{jet}_{LL} 
={d\Delta \sigma /dp_T \over d\sigma /dp_T}$ 
in the low
$p_T$--range where the gluons dominate.
The parton level spin asymmetries 
combine with suitable parton densities to give the 
proton asymmetries. These are shown in Fig. \ref{asjet} 
as a function of $p_T$ at two values of the beam energy 
for a $\delta g(x,\mu_F^2)$, $\mu_F \sim p_T$, 
with a small and with a large first 
moment $\Delta g(\mu_F^2)$ \cite{bourrely1}. 
The significance of the small and 
intermediate $p_T$ regime for 
the determination of $\delta g$ is clearly exhibited. 
If $\Delta g(p_T^2)$ is large one can easily have asymmetries of 
about 20\% in the small $p_T$ range. 
The relatively large RHIC luminosity 
${\cal L} \sim 10^{32}$ cm$^{-2}$ s$^{-1}$ 
in combination with the large jet cross sections make the 
signal in principle large and will lead to small statistical 
errors \cite{soffer2}. 
The drawback of this method to determine $\delta g(x,p_T^2)$ 
is the large background of soft events in the small $p_T$ region. 

\begin{figure}
\begin{center}
\epsfig{file=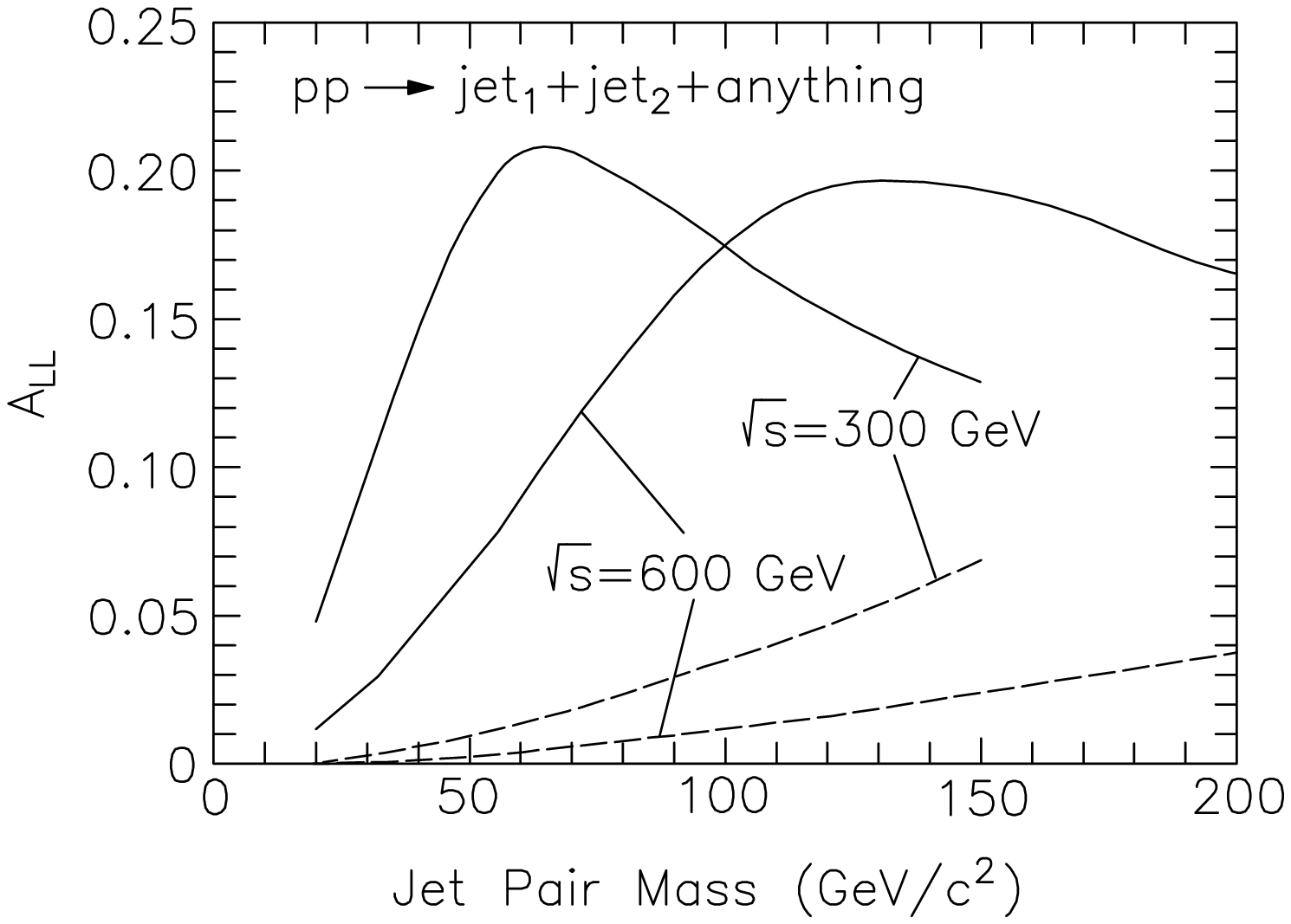,height=6cm}
\bigskip
%\caption{Predicted asymmetries $A_{LL}$ for dijet production 
%as a function of the jet pair 
%mass for two values of the proton energy and two choices of the polarized 
%gluon density, a small $\delta g(x,M_{JJ}^2)$ (dashed curve) and a 
%large one (solid curve) \protect\cite{bourrely1}.}
\label{figpair}
\end{center}
\end{figure}

Let us now discuss 2--jet production. An observable to study is the 
distribution in the 2--jet invariant mass, $M_{JJ}=2p_T$cosh($y^{\ast}$), 
where $y^{\ast}={1\over 2}(y_1-y_2)$. It is given by 
\be 
{d(\Delta )\sigma \over dM_{JJ}}={M_{JJ}^3\over 2s} \sum_{i,j} 
\int_{-Y}^{+Y} dy_1 \int_{y_m}^{y_M} dy_2 
{(\delta )f_i(x_1,\mu_F^2)(\delta )f_j(x_2,\mu_F^2) \over cosh (y^{\ast})} 
 {d (\Delta ) \hat \sigma_{ij} \over d\hat t}
\label{6811}
\ee
where $x_{1,2}=\tau e^{\pm {1\over 2}(y_1+y_2)}$,  
$y_m=$max$(-Y,\log \tau -y_1)$ and 
$y_M=$min$(-Y,-\log \tau -y_1)$. $Y \sim 1$ is a cutoff 
on the jet rapidity and $\tau$ is defined as 
$\tau=\sqrt{{4p_T^2 \over S}} cosh^2 (y^{\ast})$. 
Numerical results for these distributions have been presented 
by \cite{chiappetta1} and \cite{bourrely1}. They are shown 
in Fig. \ref{figpair} as a function of the jet pair mass for 
two values of the proton energy and two choices of the polarized 
gluon density, a small $\delta g(x,M_{JJ}^2)$ (dashed curve) and a 
large one (solid curve).  
At intermediate values of the jet pair mass $M_{JJ} \sim 100$ GeV,  
they are even more 
sensitive to the magnitude of the polarized gluon density than 
the single jet $p_T$ distribution but somwhat more difficult 
to measure.    
One may also examine di--jet production and higher order effects 
\cite{doncheski1}.

%% file: k6999tex
\subsection{Spin--dependent Structure Functions and Parton Densities 
of the Polarized Photon}
% 6.12

Structure functions $F_i^{\gamma}(x,Q^2)$ and parton densities 
$f^{\gamma}(x,Q^2)$, with $f=q,\bar q,g$, of unpolarized, i.e. 
helicity averaged, photons are theoretically well known (see, e.g. 
\cite{glueck30,glueck1} and references therein) and experimentally rather 
well studied (see, e.g. 
\cite{berger2,erdmann} and references therein). In contrast to the 
(un)polarized  
hadronic parton densities studied so far, these densities obey 
{\it in}homogeneous evolution equations where the inhomogeneous 
LO and NLO terms $k_{i=q,g}^{(0,1)}(x)$ derive from the pointlike 
splitting of the photon into quarks and gluons which can be calculated 
from first QED principles. In LO, $\gamma \rightarrow q\bar q$ gives 
rise to $k_q^{(0)}$, and $k_g^{(0)}=0$. They are the so-called driving 
terms which uniquely fix the 'pointlike' (inhomogeneous) solution of 
the evolution equations for the parton densities of the photon, once 
a specific input scale $Q_0^2$ for $\alpha_s(Q_0^2)$ has been 
chosen. This is in contrast to the conventional 'hadronic' 
(homogeneous) part of the general solution, which derives from the 
common homogeneous evolution equations and which requires  
some nonperturbative hadronic (vector--meson--dominance oriented) 
input $f_{had}^{\gamma}(x,Q_0^2)$. 

In complete analogy to the helicity averaged case $f^{\gamma}=
f_+^{\gamma}+f_-^{\gamma}$, the spin--dependent parton densities 
of a longitudinally (more precisely, circularly) polarized photon 
are defined as \cite{irving,hassan,xu,manohar1}
\begin{equation}
\delta f^{\gamma}(x,Q^2)=f_+^{\gamma}(x,Q^2)-f_-^{\gamma}(x,Q^2) 
\label{68881}
\end{equation}
with $f_+^{\gamma}$ ($f_-^{\gamma}$) denoting the parton densities in 
the photon aligned (anti--aligned) with its helicity. 
As in Eq. (\ref{4111}), they satisfy the general positivity 
constraints 
\begin{equation}
|\delta f^{\gamma}(x,Q^2)|\leq f^{\gamma}(x,Q^2)\/.
\label{68882}   
\end{equation}
Similarly to Eq. (\ref{415}), the polarized photon structure 
function is given by 
\begin{equation}
g_1^{\gamma}(x,Q^2)={1\over 2}\sum_q e^2_q 
[\delta q^{\gamma}(x,Q^2)+\delta \bar q^{\gamma}(x,Q^2) ] 
+O(\alpha_s,\alpha)
\label{68883}
\end{equation}
where the NLO (2--loop) contributions \cite{stratmann5} 
have been suppressed. Note that $\delta q^{\gamma}=\delta \bar q^{\gamma}$ 
and $\delta q^{\gamma}=O({\alpha \over \alpha_s})$ in LO. 
In Bjorken x--space, these photonic parton densitites obey the 
following inhomogeneous LO evolution equations: 
\begin{equation}
\frac{d}{dt} \delta q_{NS}^{\gamma}(x,Q^2)=
\frac{\alpha}{2\pi} \delta k_{NS}^{(0)}(x)+
\frac{\alpha_s(Q^2)}{2\pi}\delta
P^{(0)}_{NS}\otimes\delta q^{\gamma}_{NS}
\label{68884}
\end{equation}
\begin{equation}
\frac{d}{dt}{\delta\Sigma^{\gamma}(x,Q^2) \choose \delta g^{\gamma}(x,Q^2)}=
\frac{\alpha}{2\pi} 
{\delta k_q^{(0)} \choose 0}  + 
\frac{\alpha_s (Q^2)}{2\pi}{\delta P^{(0)}_{qq} \quad 2f\delta P^{(0)}_{qg}
                      \choose\delta P^{(0)}_{gq} \quad \delta P^{(0)}_{gg}}
\otimes {\delta\Sigma^{\gamma}\choose\delta g^{\gamma}}
\label{68885}
\end{equation}
which are a straightforward generalization of Eqs. (\ref{4112}) and 
(\ref{4119}), taking into account the 'pointlike' photon splitting 
$\gamma \rightarrow q$ which gives rise to the inhomogeneous terms: 
$\delta k_q^{(0)}=\delta P^{(0)}_{q\gamma}$ can be obtained, apart 
from obvious charge factors, from $\delta P^{(0)}_{qg}$ in 
(\ref{4121}) by multiplying it with $2fN_c/T_R$: 
\begin{equation}
f(e_q^2-<e^2>)^{-1}\delta k_{NS}^{(0)}=<e^2>^{-1}\delta k_q^{(0)}=6f(2x-1)
\label{68886}
\end{equation}
where $<e^2>\equiv f^{-1}\sum_q e_q^2$. Furthermore, 
$\delta k_g^{(0)}=\delta P^{(0)}_{g\gamma}=0$ has been used in 
(\ref{68885}). The LO equations (\ref{68883})--(\ref{68885}) 
can be straightforwardly extended to NLO \cite{stratmann5} where 
the $O(\alpha \alpha_s)$ terms $\delta k_q^{(1)}$ and $\delta k_g^{(1)}$ 
derive from the $C_FT_f$ pieces of $2f\delta P^{(1)}_{qg}$ and  
$\delta P^{(1)}_{gg}$ in (\ref{p1qg}) and (\ref{p1gg}), 
respectively, multiplied by $fN_c/T_f$. 

The evolution equations (\ref{68884}) and (\ref{68885}) are most 
conveniently solved in Mellin n--moment space (cf. Eqs. 
(\ref{4122}) and (\ref{4123})) where the solutions can be given 
analytically and one can easily keep track of the contributions 
stemming form different powers of $\alpha_s$ in order to avoid terms 
beyond the order considered. The inversion to Bjorken x--space is 
again straightforward with the help of (\ref{4136}). The general 
solution decomposes into 
\begin{equation}
\delta f^{\gamma,n}(Q^2)=\delta f_{PL}^{\gamma,n}(Q^2)+
\delta f_{had}^{\gamma,n}(Q^2)
\label{68887}
\end{equation}
with the 'pointlike' (inhomogeneous) solution being given by 
\begin{equation}
\delta q^{\gamma,n}_{NS,PL}(Q^2)={4\pi\over\alpha_s(Q^2)} 
[1-L^{1-\frac{2}{\beta_0}\delta P_{NS}^{(0)n} }] 
{1\over 1-\frac{2}{\beta_0}\delta P_{NS}^{(0)n} } 
{\alpha\over2\pi\beta_0}\delta k_{NS}^{(0)n} 
\label{68888}
\end{equation} 
\begin{equation}
{\delta\Sigma^{\gamma,n}_{PL} (Q^2)\choose
\delta g^{\gamma,n}_{PL} (Q^2)}=
{4\pi\over\alpha_s(Q^2)} 
[1-L^{1-\frac{2}{\beta_0}\delta \hat P^{(0)n}  }]
{1\over 1-\frac{2}{\beta_0}\delta \hat P^{(0)n} }
{\alpha\over2\pi\beta_0} 
{\delta k^{(0)n}_{q} \choose 0}
\label{68889}
\end{equation}
and the 'hadronic' (homogeneous) solution given by
\begin{equation}
\delta q^{\gamma,n}_{NS,had}(Q^2)=L^{-\frac{2}{\beta_0}
\delta P_{NS}^{(0)n}}
\delta q^{\gamma,n}_{NS,had}(Q_0^2)
\label{688810}
\end{equation}
\begin{equation}
{\delta\Sigma^{\gamma,n}_{had} (Q^2)\choose
\delta g^{\gamma,n}_{had} (Q^2)}=L^{-\frac{2}{\beta_0}
\delta\hat P^{(0)n}}{\delta\Sigma^{\gamma,n}_{had} (Q^2_0)
\choose \delta g^{\gamma,n}_{had}(Q^2_0)}
\label{688811}
\end{equation}
where $L(Q^2)\equiv\alpha_s(Q^2)/\alpha_s(Q^2_0)$. 
Note that the 'hadronic' solutions are formally identical to the 
usual ones in Eqs. (\ref{4127}) and (\ref{4128}) since they are 
derived from the homogeneous part of the evolution equations 
(\ref{68884}) and (\ref{68885}). The structure of these LO 
solutions can be straightforwardly extended to NLO \cite{glueck1,stratmann5} 
by using the techniques discussed in Sect. 4.2 being based on the 
(2--loop) evolution matrix in Eq. (\ref{4218}). 

\begin{figure}
\begin{center}
\epsfig{file=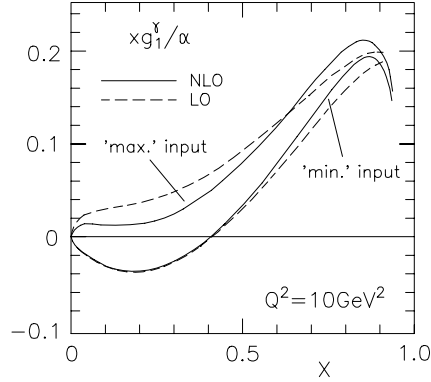,height=5cm}
\vskip 0.5cm
\caption{LO and NLO expectations for $g_1^{\gamma}$ according to the 
'minimal' and 'maximal' inputs in Eqs. (\ref{688812}) and (\ref{688813}), 
respectively \protect\cite{stratmann5}. The results shown correspond to 
f=3 flavors. }
\label{fig48}
\end{center}
\end{figure}

The new ingredient of these solutions is the 'pointlike' component 
which is driven by the inhomogeneous (photon splitting) terms 
$\delta k_i^{(0)}$ in Eqs. (\ref{68888}) and (\ref{68889}), i.e. 
they uniquely determine the 'pointlike' parton densities in the 
photon once an appropriate input scale $Q_0^2$ has been specified. 
This is in contrast to the hadronic components in (\ref{688810})  
and (\ref{688811}) which require, as usual, also the specification 
of the input densities $\delta f_{had}^{\gamma}(x,Q_0^2)$. 
In general one expects the 'pointlike' $\delta f_{PL}^{\gamma}(x,Q^2)$ 
to be dominant in the large--x region since the $\delta k_i^{(0)}$ 
in (\ref{68888}) and (\ref{68889}) increase as $x\rightarrow 1$ 
according to (\ref{68886}). This is in contrast to the 'hadronic' 
$\delta f_{had}^{\gamma}(x,Q^2)$ where the (VMD oriented) input 
$\delta f_{had}^{\gamma}(x,Q_0^2)\sim (1-x)^a$ as $x\rightarrow 1$.
Such expectations have been well established in the case of 
unpolarized photons \cite{glueck1,glueck31,berger2,erdmann}. For polarized 
photons, several model calculations have been performed for 
estimating $\delta f^{\gamma}(x,Q^2)$ in LO \cite{irving,hassan,xu}. 
More recently, in order to obtain a somewhat more realistic estimate 
for the theoretical uncertainties in the experimentally entirely 
unknown $\delta f^{\gamma}$, two very different scenarios were 
considered in \cite{glueck20,glueck21}: The 'minimal scenario' is 
characterized by the input 
\begin{equation}
\delta f^{\gamma}_{had}(x,\mu^2)=0 \,  ,
\label{688812}
\end{equation}
whereas the 'maximal scenario' is defined by the other extreme 
input 
\begin{equation}
\delta f^{\gamma}_{had}(x,\mu^2)=f^{\gamma}_{had}(x,\mu^2) \,  ,
\label{688813}
\end{equation}
with $Q_0^2\equiv \mu^2=\mu^2_{LO}=0.25$ GeV$^2$ and the unpolarized LO GRV 
photon densities $f^{\gamma}_{had}(x,\mu^2)$ have been taken from 
\cite{glueck31}. It should be mentioned that the range of such VMD 
inspired inputs can be further restricted \cite{glueck21} by using 
the sum rule 
\begin{equation}
\int_0^1\delta q^{\gamma}(x,Q^2)dx=0 
\label{688814}                          
\end{equation}
which derives from the vanishing of the first moment of $g_1^{\gamma}$ due 
to the conservation of the electromagnetic current (gauge invariance) 
\cite{efremov6,bass2,narison1,freund}. Note that the 'minimal' 
hadronic input (\ref{688812}) satisfies this sum rule automatically. 
Such inputs have also been implemented in NLO \cite{stratmann5} 
and some typical expectations for $g_1^{\gamma}$ are shown 
in Fig. \ref{fig48}. The strong increase in the large--x region 
is typical for the 'pointlike' $\delta q^{\gamma}_{PL}$ as discussed above.
The 'minimal' scenario is 'pointlike' throughout the entire x--region 
due to the vanishing hadronic input in (\ref{688812}). The possible 
importance of the hadronic components in Eqs. (\ref{688810}) and 
(\ref{688811}) is illustrated by the 'maximal' scenario results in 
Fig. \ref{fig48} which involve an additional hadronic component 
due to the input (\ref{688813}). 

Several suggestions have been proposed to measure $\delta f^{\gamma}(x,Q^2)$ 
at polarized $ep$ \cite{glueck32,glueck20} and $e^+e^-$ colliders 
\cite{glueck22}.  
It has been discussed already in Sect. 6.2 that these resolved photon 
contributions could amount up to 20\% of the total charm production 
cross section at future $\vec e\vec p$ collisions \cite{stratmann1}. 
The situation improves for photoproduction of jets at future 
$\vec e\vec p$ colliders ($\sqrt{s_{\gamma p}} \approx 200$ GeV) 
where the much larger size of the resolved photon contributions 
to single--inclusive jet and, in particular, dijet production 
\cite{stratmann1} could give rise to experimentally testable signatures 
of the parton densities of a polarized photon in the not too 
distant future.

%% file: k71tex
\setcounter{equation}{0}
\section{Nonperturbative Approaches to the Proton Spin}

In Sect. 5 it was shown that the apparent smallness of 
the flavor singlet axial vector current matrix element 
is due either to a negative sea polarization or to a 
positive gluon polarization. However, no theoretical 
prediction for the value of this matrix element was 
presented. There are several nonperturbative approaches 
which try to remedy this situation \cite{veneziano,shore,cheng4,birkel} 
and lattice calculations have been attempted as well 
\cite{mandula,goeckeler,dong,fukugita}.   
The well--established isotriplet Goldberger--Treiman (GT) relation 
\be
g_A^3(0)={\sqrt{2}f_{\pi} \over 2M}g_{\pi_3 NN} \, , \qquad 
f_{\pi}=132 MeV \, ,
\label{701} 
\ee
which fixes the axial coupling $g_A^3(0)=g_A=1.2573\pm 0.0028$ 
in terms of the strong 
coupling constant $g_{\pi_3 NN}$, has 
inspired \cite{veneziano,shore} to generalize it to the isosinglet  
$U_A(1)$ case to see if one can learn something about the 
magnitude of $g_A^0 S_{\mu}\equiv \langle  PS\vert \bar \psi \gamma_{\mu} 
\gamma_5 \psi \vert PS\rangle$. \footnote{Note that in Sect. 5 a 
different notation was used, namely 
$A_0S_{\mu}=\Delta \Sigma_{on}S_{\mu}=\langle  PS\vert \bar \psi \gamma_{\mu}
\gamma_5 \psi \vert PS\rangle$ and $g_A^3(0)=g_A/g_V$. 
Further, in Sect. 6.6 an isotriplet form factor $G_1^3(Q^2)$ 
has been introduced which is normalized  
differently than $g_A^3(Q^2)$, namely  
$G_1^3(Q^2)={1\over 2}g_A^3(Q^2)$. }  

Many discussions on the isosinglet GT relation 
were mainly motivated by the desire to understand why the 
$g_A^0$ inferred from the EMC experiment is so small 
($g_A^0 \sim 0.15$, cf. Table 3) 
\cite{cheng9,fritzsch,fritzsch1,fritzsch2,
schechter,bartelski1,veneziano,shore,cohen,hatsuda,
ji5,birse,chao,efremov1,ellwanger1,ellwanger2,
wakamatsu,liu,shore1,narison,efremov5}. 
At first sight, the $U_A(1)$ GT relation seems not to be in the 
right ballpark, as the naive SU(6) quark--model prediction yields far
too large a value $g_A^0 \approx 0.80$, 
because  
$g_{\eta_0 NN}={\sqrt{6} \over 5}g_{\pi_3 NN}$. 
It was then suggested 
that the 'physical' $\eta_0$--nucleon coupling is composed 
of a bare coupling $g_{\eta_0 NN}^{(0)}$ and a coupling between 
the ghost field and the nucleon. In QCD 
the ghost field $G\equiv \partial_{\mu}K^{\mu}$, cf. (\ref{538}), 
is necessary 
to solve the $U_A(1)$ problem. It mixes with the bare $\eta_0$ 
and is allowed to have a direct 
$U_A(1)$ invariant interaction $g_{GNN}$ with the proton. 
Unfortunately, the definition of this coupling is not free of 
ambiguities. For example, it is sometimes assumed in the literature 
to be the 
coupling between the glueball and the nucleon. 
Furthermore, unlike in the case of $g_A^3(0)$, the predictive 
power of the $U_A(1)$ GT relation is partly lost by the mixing. 
In chirally invariant schemes (like the off--shell regularization 
discussed in Sect. 5) one may at least identify the 
term ${\sqrt{n_f}f_{\pi} \over 2M}g_{\eta_0 NN}$ with the total 
quark spin inside the proton $\Delta \Sigma_{off}$, cf. Sect. 5 \cite{shore}. 

It is interesting to obtain a scheme independent relation. 
For the isotriplet GT relation this is possible because it holds 
irrespective of the light quark masses. For $m_{\pi}^2 \neq 0$ it 
may be derived from PCAC, while in the chiral limit 
$g_A^3(q^2)$ can be related to an isotriplet form factor 
which in turn receives nonvanishing pion pole contributions. 
By the same token it can be shown in the OZI limit that the isosinglet 
$U_A(1)$ GT relation \cite{shore,shore1}
\be
g_A^0(0)={\sqrt{3}f_{\pi} \over 2M}g_{\eta_0 NN}^{(0)} 
\label{702}
\ee
remains totally scheme and mass independent. 
%To see this explicitly, let us define the form factors 
%$g_A^0(q^2)$ and $f_A^0(q^2)$ by 
%\be
%<N(p')\vert J_{\mu}^5\vert N(p)\rangle= 
%\bar u (p') (g_A^0(q^2) \gamma_{\mu} \gamma_5 +
%f_A^0(q^2) q_{\mu}\gamma_5 )u(p)
%\label{703}
%\ee
In Eq. (\ref{702}) $g_{\eta_0 NN}^{(0)}$ is a bare unphysical 
coupling between $\eta_0$ and the nucleon. The $\eta_0$ is not 
a physical meson but constructed from the mass eigenstates via 
\cite{schechter,bartelski1,cheng4} 
\be
\pmatrix{\pi_3 \cr \eta_8 \cr \eta_0 \cr} =
\pmatrix{1 & \theta_1 \cos \theta_3 +\theta_2 \cos \theta_3 & 
             \theta_1 \sin \theta_3 -\theta_2 \cos \theta_3 \cr 
        -\theta_1 & \cos \theta_3 & \sin \theta_3 \cr 
        \theta_2 & -\sin \theta_3 & \cos \theta_3 \cr} 
\pmatrix{\pi_0 \cr \eta \cr \eta ' \cr} 
\label{703}
\ee
where $\theta_1=-0.016$, $\theta_2=-0.0085$ and $\theta_3=-18.5^o$ 
are the mixing angles of the $\pi$--$\eta$ system \cite{efremov5}. 
Consequently,  
the complete(=isotriplet+octet+singlet) GT relation in terms 
of physical coupling constants reads 
\be
g_A^3(0)={\sqrt{2}f_{\pi} \over 2M}g_{\pi_3 NN}=
{\sqrt{2}f_{\pi} \over 2M} [g_{\pi NN} \pm 
g_{\eta ' NN} (\theta_1 \sin \theta_3 -\theta_2 \cos \theta_3) 
\pm g_{\eta  NN}(\theta_1 \cos \theta_3 +\theta_2 \sin \theta_3)] 
\label{704}
\ee
\be
g_A^8(0)={\sqrt{6}f_{\pi} \over 2M}g_{\eta_8 NN}=
{\sqrt{6}f_{\pi} \over 2M} (g_{\eta NN} \cos \theta_3 + 
g_{\eta ' NN} \sin \theta_3 \mp 
g_{\pi  NN}\theta_1 )
\label{705}
\ee
\be
g_A^0(0)={\sqrt{3}f_{\pi} \over 2M}g_{\eta_0 NN}^{(0)}=
{\sqrt{3}f_{\pi} \over 2M} (g_{\eta ' NN} \cos \theta_3 -  
g_{\eta  NN} \sin \theta_3 \pm                     
g_{\pi  NN}\theta_2 ) +...
\label{706}
\ee
where the upper sign is for the proton and the lower sign 
for the neutron and the ellipses in the last relation are related 
to the ghost coupling to be discussed below. 

Although the isosinglet GT relation is scheme invariant, 
the interpretation of the $\eta_0$ field is scheme dependent. 
When the SU(6) quark model is applied to the coupling 
$g_{\eta_0 NN}^{(0)}$ the predicted $g_A^0$ is too large, 
$g_A^0=0.80$. This can be resolved by applying the SU(6) model 
to the physical coupling $g_{\eta_0 NN}$ rather than 
to $g_{\eta_0 NN}^{(0)}$. One receives a two--component 
expression \cite{shore}
\be
g_A^0(0)={\sqrt{3}f_{\pi} \over 2M}(g_{\eta_0 NN}- 
{1 \over 2\sqrt{3}} m^2_{\eta_0} f_{\pi} g_{GNN}) 
\label{707}                                                
\ee                    
where $g_{GNN}$ is the coupling defined in the effective 
QCD nucleon Lagrangian 
\be
{\cal L}=...+{g_{GNN}\over 4M} \partial^\mu G Tr(\bar N 
\gamma_{\mu} \gamma_5 N) +{\sqrt{3}\over f_{\pi}} 
(\partial K) \eta_0 +...
\label{708}
\ee
where $\partial K={1\over \sqrt{3}} m^2_{\eta_0}f_{\pi}\eta_0 
+{1\over 12}g_{GNN}m^2_{\eta_0}f_{\pi}\partial^\mu Tr(\bar N
\gamma_{\mu} \gamma_5 N)$ 
is the anomaly with matrix element 
$\langle N\vert \partial K\vert N\rangle={1\over 3}\langle N
\vert \partial^\mu 
J^5_\mu \vert N\rangle ={1\over \sqrt{3}}f_{\pi} 
g_{\eta_0 NN}^{(0)}={2\over 3}M g_A^0(0)$. 

In order to arrive at a small EMC--like $g_A^0$ 
one may now argue that there is 
a cancellation between the two terms 
of Eq. (\ref{707}). This is reminiscent 
to the situation in the perturbative framework where a 
cancellation between $\Delta \Sigma$ and $\Delta g$ was proposed 
to explain the spin EMC effect. However, there is no 
scheme independent  
identification between the various terms of the perturbative 
and nonperturbative approach. 

Some authors have questioned \cite{efremov5} the U(1) GT relation 
(\ref{707}) but it 
seems that it survives as long one takes proper care of all the 
mixing effects. Still it should be stressed that until now no 
direct experimental confirmation exists, and in this sense
it is one of the speculations from the realms of nonperturbative 
QCD. The point is that both sides of the relation are difficult 
to grab. The righthand side (i.e. the low energy couplings) 
cannot really be determined from low energy data, 
because $g_{GNN}$ is not known 
\footnote{
There is a conjecture, though no proof, that $g_{GNN}$ is given by 
the quark sea contribution ('disconnected insertion'), 
$-{m_{\eta_0}^2 f_{\pi}^2 \over 4M}g_{G NN}= \Delta (u_s+\bar u) 
+\Delta (d_s+\bar d) +\Delta (s+\bar s) 
\approx 3 \Delta (s+\bar s)$ \cite{cheng4}. 
Furthermore, $g_{GNN}$ could in principle be obtained from 
low energy baryon--baryon scattering in which an additional SU(3) 
singlet contact interaction arises from the ghost interaction  
but this is very difficult to measure \cite{schechter}. } .  
Even the $\eta NN$ couplings 
are only known within large errors, cf. the discussion in \cite{cheng4}. 
For example, a determination of $g_{\eta ' NN}$ using F, D and 
$\theta_3$ values gives $g_{\eta ' NN}=3.4$ \cite{cheng5} whereas 
an estimate of the $\eta ' \rightarrow 2\gamma$ decay rate through 
the baryon triangle contributions yields $g_{\eta ' NN}=6.3$ 
\cite{bagchi}. The analysis of the NN potential gives 
$g_{\eta ' NN}=7.3$ \cite{dumbrajs} 
while the forward NN scattering analyzed by 
dispersion relations gives $g_{\eta ' NN}<3$ \cite{brein}.  
On the other hand, 
the only chance to determine $g_{GNN}$ is from the polarized 
DIS data via the U(1) GT relation. However, there 
is no direct identification between $g_{GNN}$ and an observable 
defined in high energy scattering. Therefore, the whole issue 
remains somehow speculative and undecided. 

Another attempt to obtain information on the 
proton spin matrix elements 
from low energy meson properties 
was recently made by Birkel and Fritzsch \cite{birkel}. 
They used the masses and properties of the axial vector mesons with 
quantum numbers $J^{PC}=1^{++}$. The spectrum consists of the 
%isovector meson $a_1(1260)$ and the 
isoscalar 
mesons $f_1(1285)$, $f_1(1420)$ and $f_1(1510)$. 
After correcting for the mixing with the isovector $a_1(1260)$ and  
the strange isodoublet $K_1$, the $f_1$ mesons can be written 
as a linear combination of the three axial vector states 
$\vert N\rangle ={1\over \sqrt{2}} \vert \bar u u +\bar d d\rangle$, 
$\vert S\rangle =\vert \bar s s\rangle$ and an exotic gluonic meson state  
$\vert G\rangle$. The latter interpretation arises because within 
the SU(3) multiplets one expects only two isoscalar mesons $f_i$. 
A relatively large mixing is found between the N, S and G state, 
and this is due largely to the effect of the anomaly and 
reminiscent of what happens in the case of pseudoscalar mesons 
\cite{birkel}. On the basis of this analysis one may use the 
idea of 'axial vector' dominance to get information on the 
proton spin matrix element. The basic relation is 
\be
\langle P \vert \bar q \gamma_{\mu} \gamma_5 q \vert P\rangle =\sum_A
{\langle 0 \vert \bar q \gamma_{\mu} \gamma_5 q \vert A
\rangle \langle AP\vert P\rangle 
\over m_A^2-k^2} \vert_{k^2=0} 
\label{709}
\ee
where $\langle 0 \vert \bar q \gamma_{\mu} \gamma_5 q \vert A\rangle$ 
denotes the transition element of the axial vector current between 
the vacuum and the axial vector meson A, and $\langle AP\vert P\rangle$ 
describes the coupling of the axial vector meson to the proton. 
The 4--momentum transfer is $k$. 
Using Eq. (\ref{709}) one can relate $\Delta (u+\bar u)$, 
$\Delta (d+\bar d)$ and $\Delta (s+\bar s)$ 
to the corresponding couplings of axial vector mesons. 
Without the gluonic state, a relatively large value of the 
flavor singlet quark spin contribution 
$\Delta \Sigma \approx 0.52$ is obtained.    
Inclusion of the gluonic state $\langle GP\vert P\rangle \neq 0$ leads 
to numbers of the order $\Delta \Sigma \approx 0.25$ in accordance 
with DIS data. However, 
just as in the U(1) Goldberger Treiman model there is 
a free parameter which has to be fixed, namely the coupling 
$g_{GP}$ to the proton defined by $\langle GP\vert P\rangle =ig_{GP} 
\bar u(P)\gamma_{\nu} \gamma_5 u(P)\epsilon^{\nu}$ where 
$\epsilon^{\nu}$ is the polarization vector of the gluonic state. 
It is hard to determine $g_{GP}$ experimentally. The above number  
$\Delta \Sigma \approx 0.25$ corresponds to the choice 
$g_{GP}=19$.  

A more general method to obtain nonperturbative results on the 
proton spin matrix elements is lattice gauge theory. 
After the 1987 EMC spin surprise, several attempts were made 
to compute $\Delta g$ and $\langle PS\vert \bar \psi \gamma_\mu \g5 
\psi \vert PS\rangle$ using lattice QCD. 
A first direct calculation of $\Delta \Sigma$ was 
made in \cite{mandula} but without final results. Successful lattice 
computations in the quenched approximation were published 
recently \cite{dong,fukugita,goeckeler}. A more ambitious program of computing the 
polarized structure functions $g_1$ and $g_2$ is also feasible 
and encouraging early results were reported in \cite{goeckeler}. 
In refs. \cite{dong,fukugita}, 
the scheme and gauge invariant matrix elements 
$\langle PS\vert \bar \psi \gamma_\mu \g5 \psi \vert PS\rangle$ were 
calculated. 
The results are shown in Table 6 and compared to 
the experimental data, although one should keep in mind 
that the computation of sea quark contributions might 
be questionable in a quenched calculation. 
Varying the quark masses it was found in \cite{fukugita}
that a considerable amount of the sea contributions is 
mass independent and therefore must be induced by gluons 
through the ABJ anomaly. This is in accord with arguments 
based on perturbative QCD and presented in Sect. 5. 
However, for a direct lattice computation of $\Delta g$ 
one would need gauge configurations on a sizeable lattice 
not available so 
far. It is hoped that lattice results for $\Delta g$ 
will be available in the near future. 
The most recent improvement in this field is the implementation 
of an improved action by the DESY/HLRZ collaboration 
\cite{best}, i.e. of a systematic procedure for the removal 
of all terms linear in the lattice spacing $a$ from the 
lattice observables \cite{symanzik,luescher} 
which reduces the cutoff errors order by 
order in $a$, yielding a 
better extrapolation towards the continuum limit \cite{best}. 
This $O(a)$ improved lattice theory yields, for example, 
$\Delta u_v=0.841(52)$ and $\Delta d_v=-0.245(15)$, in reasonable 
agreement with DIS experiments (cf. Sect. 6.1 and ref. \cite{abe7}).     

\begin{table}
\label{tab1778}    
\begin{center}
\begin{tabular}{|c|c|c|c|}
\hline
 & \cite{dong} & \cite{fukugita} & Experiment \\
\hline
$g_A^0$ & 0.25(12) & 0.18(10) &  0.22(6) \\
$g_A^3$ & 1.20(10) & 0.985(25) & 1.2573(28) \\
$g_A^8$ & 0.61(13) & -- & 0.579(25) \\
$\Delta (u+\bar u)$ & 0.79(11) & 0.638(54) & 0.80(3) \\ 
$\Delta (d+\bar d)$ & --0.42(11) & --0.347(46) & --0.46(3) \\ 
$\Delta (s+\bar s)$ & --0.12(1) & --0.109(30) & --0.12(3) \\ 
F & 0.45(6) & 0.382(18) & 0.459(8) \\ 
D & 0.75(11) & 0.607(14) & 0.798(8) \\ 
\hline
\end{tabular}
\bigskip
\caption{Axial couplings and quark spin content of the proton 
from lattice calculations according to \protect\cite{cheng4}. Note that 
the 'experimental' singlet results are model and scheme dependent 
and the stated numbers refer to an average of recent LO and 
NLO ($\overline{MS}$) analyses. In the off-shell scheme one 
obtains \protect\cite{altarelli10}, for example, $g_A^0=0.45(9)$; 
see, however, ref. \protect\cite{abe7}. }
\end{center}  
\end{table}

There are also attempts to explain the smallness of $g_A^0$, 
i.e. of $\Delta \Sigma$, by invoking the Skyrme model 
\cite{brodsky,ellis1,ryzak,clement,bernard}. Here one argues 
\cite{brodsky,ellis1} that 
the quark singlet contribution $\Delta \Sigma$ is suppressed by 
$1/N_c$ while $\Delta (q+\bar q)=O(1)$ for each separate flavor, and 
a similar suppression should hold for $\Delta g(Q^2)$. 
Besides the fact that the precise relation 
between the Skyrme model and QCD is 
somewhat unclear, in particular in connection with $\Delta g$, 
this explanation has been questioned within the Skyrme model 
itself \cite{ryzak,clement,bernard,jaffe}. 

Alternatively, QCD sum rules \cite{belyaev,chiu} result in 
a similarly small total singlet spin contribution $g_A^0 =0.1-0.2$ 
\cite{gupta2,henley}. Again, a distinction between the individual 
$\Delta \Sigma$ and $\Delta g$ contributions cannot be obtained. 

Most promising appears to be the chiral soliton approach towards 
the structure of the nucleon within the effective chiral theory 
\cite{diakonov,diakonov1,weigel} which allows for the calculation 
of the {\it full} x--dependence of the structure functions and the 
parton densities from first principles, in contrast to just their 
n--th moments as obtained in the nonperturbative approaches 
discussed so far. The relevance and influence of the instanton 
vacuum on low--energy QCD observables has been emphasized 
in particular by Diakonov et al. \cite{diakonov,diakonov1} 
who calculated unpolarized (spin--averaged) and polarized valence and sea 
input densities from first principles at the typical scale 
which is set by the inverse average instanton size $\bar \rho$, 
i.e. $Q_0^2 \approx \bar \rho^{-2} \approx 0.36$ GeV$^2$. The 
instanton size remains the only free parameter in the calculation, 
and its inverse serves as an UV cutoff of the nonrenormalizable 
effective chiral field theory. What makes this approach quite 
promising is the fact that it predicts \cite{diakonov,diakonov1}, 
besides the valence densities, a {\it valence}--like input 
(unpolarized) sea density in the small--x region at 
$Q_0^2=0.3-0.4$ GeV$^2$, which forms the basic ingredient 
for understanding and predicting all small--x unpolarized 
DIS HERA--data \cite{abt,ahmed2,aid,derrick1,derrick2,derrick3} 
from first principles, i.e. pure (parameter--free) 
QCD dynamics \cite{glueck1,glueck9,glueck6}. 
So far, the polarized sea and the (un)polarized gluon input 
densities have not been calculated. It is in particular the 
valence--like gluon densities \cite{glueck6,glueck2}, being 
$1/N_c$ 'suppressed', which have to come out sizeable at 
$Q_0^2=0.3-0.5$ GeV$^2$. Otherwise the chiral soliton approach 
does not refer to a perturbative (twist--2) input scale 
reachable by perturbative RG evolutions, but instead would refer to 
some nonperturbative input quark scale which cannot be 
reached by perturbative evolutions. Nevertheless, for the time 
being, the chiral soliton approach appears to be a realistic 
model of nonperturbative QCD which might eventually link, 
from first principles, the confining regime to the perturbative sector.  

It should be stressed that only future (dedicated) experiments 
can ultimately decide about the physical reality of the various 
theoretical ideas and scenarios discussed so far. It should be 
kept in mind, however, that all realistic experiments are of course 
sensitive to the explicit x--dependence of the polarized 
parton densities $\delta f(x,Q^2)$ and in most cases are not 
directly related to their first moments $\Delta f(Q^2)$.

%% file: k81tex
\setcounter{equation}{0}
\section{Transverse Polarization}
\subsection{The Structure Function $g_2$}

For pure photon exchange the
complete polarization part of the hadron tensor is antisymmetric
and given by, cf. (\ref{211}),  
\begin{equation}
W_{\mu\nu}^A=i 
{M \over Pq}
\ve_{\mu\nu\rs}q^\r \bigl\{ S^\si g_1(x,Q^2)+
(S^\si -{Sq\over Pq}P^\si)g_2(x,Q^2) \bigr\}  \, . 
\label{811}
\end{equation}
For longitudinal polarization and 
at $Q^2$ much larger than $M^2$ the $g_1$--piece gives 
the dominant contribution, with $g_2$ being suppressed by 
a factor $x^2M^2/Q^2$ according to (\ref{221}). However, 
for a nucleon transversely polarized with respect to the 
beam direction, $W_{\mu\nu}^A$ is proportional
to ${xM \over Q} (g_1+g_2)$,
so that $g_1$ and $g_2$ enter with equal coefficients but the
whole contribution is down by a factor ${xM \over Q}$ with 
respect to $g_1$ in the longitudinally polarized case. 
This can again be derived from the inclusive, fully differential cross section 
Eq. (\ref{221}). 
Therefore $g_2$ being measured at SLAC and DESY 
will have much less accuracy than $g_1$. 
Furthermore, it is really the combination $g_T \equiv g_1+g_2$ 
which is the 'transverse spin structure function' 
although, for obvious reasons, one usually refers just to $g_2$ 
\cite{jaffe1}. 

Since $g_2$ is related to a transverse polarization, it is not 
easy to find a partonic interpretation \cite{altarelli1,jaffe8,jaffe4}: 
In a transversely polarized nucleon, the quark spin operator 
projected along the nucleon spin, 
$\Sigma_T=\gamma_0 \gamma_5 S\llap{/} _T$ with 
$S\llap{/} _T \sim \gamma_1$, does not commute with the 
free quark Hamiltonian $H_0=\alpha_z p_z$ and thus there exists 
no energy eigenstate $\vert p_z\rangle$ such that 
$\Sigma_T \vert p_z\rangle=\lambda_T \vert p_z\rangle$. Therefore, 
$\Sigma_T$ is a 'bad' operator and depends on the dynamics. 
Nevertheless, a transverse--spin {\it average} for quarks can still be 
defined in the nucleon and it is just $g_T\equiv g_1+g_2$ which is 
sensitive to the quark--gluon interactions -- a clear sign that 
no simple parton interpretation can be 
made for it \cite{jaffe1,jaffe7}. This is in contrast to the 
longitudinally polarized nucleon, where the quark helicity 
operator $\Sigma_{\vert \vert}=\gamma_0 \gamma_5 S\llap{/} _{||}$, 
with $S\llap{/} _{||} \sim \gamma_3$, commutes with $H_0$ and 
thus $g_1(x)$ measures directly the quark helicity distribution. 
It should be further noted, that $g_2$ vanishes 
\cite{altarelli1,jaffe7} for a free (massless or massive) quark, i.e.      
for a pointlike nucleon, and thus $g_2$ cannot be expressed as 
an incoherent sum over free on--shell partons. The partons must 
be interacting and/or virtual in order to contribute to $g_2$. 
Therefore, $g_2$ will serve as a unique probe of 'higher twist' 
(twist$\equiv$dimension--spin$=$3) as well.  

\begin{figure}
\begin{center}
\epsfig{file=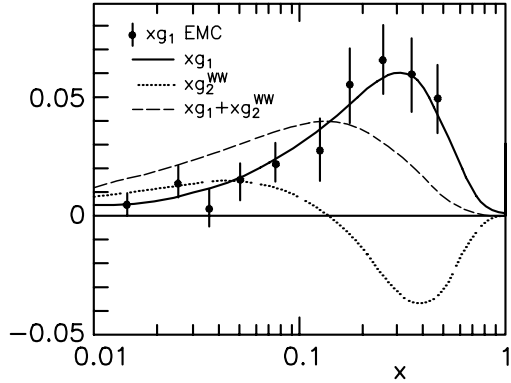,height=5cm}
\bigskip
\caption{Twist--2 expectations for the structure function 
$g_2^{WW}(x,Q^2=10$ GeV$^2)$,  
using a parametrization of the EMC data (solid curve) \protect\cite{ashman,ashman1}  
on $g_1^p(x,Q^2=10$ GeV$^2)$, 
according to the Wandzura--Wilczek relation (\ref{8117}). }
\label{figexpectg2}
\end{center}
\end{figure}

Regardless of the difficulties with a partonic interpretation, 
$g_2$ itself consists of a twist--2 ($g_2^{WW}$) and a 
twist--3 ($\bar g_2$) contribution,  
\begin{equation}
g_2=g_2^{WW}+\bar g_2
\label{vvxy}
\end{equation}
both of which can a priori contribute the same order of 
magnitude. 
The twist--2 contribution $g_2^{WW}$ is the so--called 
'Wandzura-Wilczek' piece \cite{wandzura} which 
will be discussed first: 
In leading twist--2, the same operators in Eqs. (\ref{433}) and (\ref{434}) 
contribute to $g_1$ and $g_2$; therefore one has, through the optical
theorem \cite{jaffe1,jaffe7}, 
\begin{equation}
\int\limits^1_0 dx x^{n-1}g_2^{WW}(x,Q^2)=-\phantom{-}\ha {n-1 \over n}
\sum\limits_i M^n_i
E^n_i (Q^2/\mu^2,\alpha_s)\quad , n=1,3,5,\ldots
\label{812}
\end{equation}
By comparing this to the corresponding equation for $g_1$, (\ref{436}), 
one obtains \cite{wandzura}
\begin{equation}
\int\limits^1_0 dx x^{n-1}[{n-1 \over n} 
g_1(x,Q^2)+ g_2^{WW}(x,Q^2)]^{twist-2} =0  \, .
\label{812a}
\end{equation}
This can be inverted to Bjorken--x space 
to give 
\begin{equation}
g_2^{WW}(x,Q^2)=-g_1(x,Q^2)+ \int_x^1 {dy \over y}g_1(y,Q^2) 
\label{8117}
\end{equation}
which is the so--called  Wandzura-Wilczek 
relation \cite{wandzura}. This twist--2 
expectation for $g_2$ and the present 
experimental results on $g_1$ are shown in Fig. \ref{figexpectg2}. 
Note that the twist--2 Wandzura-Wilczek piece $g_2^{WW}$ of $g_2$ 
obeys automatically the so--called  
Burkhardt--Cottingham sum rule \cite{burkhardt1}, 
$\int_0^1 g_2^{WW}(x,Q^2)dx=0$, which follows from Eq. (\ref{812a})  
or (\ref{8117}) and to which we shall return below. 

It is a unique feature of $g_2$  
that the higher twist term $\bar g_2$ in (\ref{vvxy}) 
is not suppressed by inverse powers of $Q^2$ and thus 
could in principle be equally important as the 
twist--2 contributions $g_2^{WW}$ discussed so far. 
It should, however, be kept in mind that $\bar g_2$ vanishes 
for ultrarelativistic on--shell quarks where $S^\sigma \sim 
P^\sigma$, i.e. in this case there are not enough four--vectors 
to form the required antisymmetric combination in (\ref{811}). 
Furthermore, one might naively expect $\bar g_2$ to be small because 
non--relativistic corrections, being of the order $m_q/M$ or 
$m_q/\Lambda$, are small for light quarks.  
Future experiments will 
prove to what extent this is actually the case. 
%In the literature 
%the theoretical opinions about this question are split. 
%On the one hand, 
A considerable twist--3 contribution, however, might be 
due to the off--shellness of interacting quarks with virtuality 
$k^2$ where $k^2/\Lambda^2$ is not small 
\cite{shuryak,altarelli4,altarelli7,jaffe1,jaffe7}. On the 
other hand, within the covariant relativistic parton model approach 
it has been argued \cite{jackson} 
that the quark virtuality $k^2$ is expected to 
be not sizeable and therefore deviations from the Wandzura--Wilczek relation 
(\ref{8117}) should be small. 
In the following we shall explain this argument, because it allows 
to get some physical insight into the properties of $g_2$. 
As a straightforward generalization of the unpolarized case \cite{glueck38}, 
the covariant parton model expression for the antisymmetric part 
(\ref{811}) of the 
hadron tensor is given by a convolution over the struck parton's 
momentum in the 'hand bag' diagram, 
\begin{equation}
W_{\mu\nu}^A(q,P,S)=
\sum_{s=\pm} \int d^4kf_s(P,k,S)w_{\mu\nu ,s}^A(q,k)\delta [(k+q)^2-m^2]
\label{8cc1}
\end{equation}
where $f_s$ is some Lorentz--invariant hadronic wave function. 
The antisymmetric tensor for the parton--photon interaction 
is given by 
\begin{equation}
w_{\mu\nu ,\pm}^A= 
{1\over 4} tr [(1\mp \gamma_5 n\llap{/} )(k\llap{/} +m)\gamma_{\mu} 
             (k\llap{/} +q\llap{/} +m)\gamma_{\nu} ] -
           (\mu \leftrightarrow \nu  ) 
=\pm im \epsilon_{\mu \nu \alpha \beta}q^{\alpha }n^{\beta}
\label{8cc2}
\end{equation}
with $n^\beta$ being the off--shell parton spin vector, 
$k \cdot n =0$, $n^2=-1$. 
Defining 
\begin{equation}                                                        
\tilde f(P\cdot k,k^2)=-{M\over k\cdot S} 
{m\over \sqrt{k^2-M^2k^4/(P\cdot k)^2}} (f_+-f_-)  \quad , 
\label{8cc3}
\end{equation} 
a comparison of Eq. (\ref{8cc1}) and (\ref{811}) yields \cite{jackson} 
\begin{equation}
g_1(x)={\pi x\over 8} \int dk^2dk^2_T
\tilde f(x+{k^2+k_T^2 \over xM^2},k^2)(1-{k^2+k_T^2 \over x^2M^2})
 (1-{2k^2\over x^2M^2+k^2+k_T^2})
\label{8cc4}
\end{equation} 
\begin{equation}
g_T(x)\equiv g_1(x)+g_2(x)={\pi x\over 8} \int dk^2dk^2_T
\tilde f(x+{k^2+k_T^2 \over xM^2},k^2){k_T^2 \over x^2M^2}
\label{8cc5}
\end{equation}
for a longitudinally and transversely polarized photon, 
respectively. The last equation, (\ref{8cc5}), immediately proves that a 
finite transverse polarization result $g_T$ can arise only for a
non--vanishing parton transverse momentum $k_T^2 \neq 0$. Therefore,
the EMC observation of a non--vanishing value of $\int_0^1 g_1(x)dx$ 
is direct evidence for $k_T^2 \neq 0$ (via the Burkhardt--Cottingham 
sum rule  $\int_0^1 g_2(x)dx=0$, to be discussed below). 
Furthermore, the second factor in parentheses in the integrand 
of Eq. (\ref{8cc4}) proportional to $2k^2$ violates the 
Wandzura--Wilczek sum rule and explicitly describes higher--twist 
corrections ($\bar g_2$) 
to it. In the absence of this term one easily derives 
from (\ref{8cc4}) and (\ref{8cc5}) 
\begin{equation}
{d\over dx} [g_1(x)+g_2(x)]=-{1\over x}g_1(x)
\label{8cc6}
\end{equation}
and hence the Wandzura--Wilczek sum rule (\ref{8117}). There seems to be not 
much room for a sizeable $k^2\neq 0$ since the first term 
in parentheses in the integrand of Eq. (\ref{8cc4}) gives rise 
to a zero for $g_1(x)$ if $k^2+k_T^2 \approx x^2M^2$. The lack of 
experimental evidence for such a zero suggests that $k^2+k_T^2 << 
x^2M^2$. Ignoring the possibility that $k^2<0$ contributes 
appreciably, the second factor in parentheses in (\ref{8cc4}) 
is essentially unity, giving thus only small corrections to the 
Wandzura--Wilczek sum rule. It should be noted that $g_T(x)$ 
in Eq. (\ref{8cc5}) gets no additional off--mass--shell correction, 
besides from the one entailed in $\tilde f$, and that a 
measurement of $g_1$ gives a direct estimate \cite{jackson} 
of the mean intrinsic $k_T$ of partons as a function of x, 
in complete analogy with the unpolarized case \cite{glueck38}.   

Unfortunately, it must be noted 
that the scale $Q^2$, at which these results are supposed 
to hold, remains undetermined within the covariant parton model 
\cite{glueck38}. Therefore, these results have only a qualitative rather 
than a quantitative character. Furthermore, and more importantly,    
the virtuality $k^2$ alone is not a reliable measure 
for the importance of all possible twist--3 contributions to $g_2$, 
because of the appearance of additional twist--3 operators 
which describe quark--gluon correlations and explicitly 
quark--mass--dependent operators. 
These operators will now be discussed in some detail:   
In the OPE approach to $g_2$ the 
higher twist terms are determined by a tower of 
operators whose number increases with increasing moments. 
Therefore, Eqs. (\ref{812})--(\ref{8117}) are incomplete 
due to the neglect of $\bar g_2$ in (\ref{vvxy}), and 
in principle significant modifications of Fig. 
\ref{figexpectg2} might be expected.  
%Writing $g_2=g_2^{twist-2}+\bar g_2$, the
The modifications $\bar g_2$ due to higher twist  
can be described in terms of matrix elements of
the following twist--3 operators 
\cite{kodaira1,shuryak,jaffe7,kodaira5,kodaira3}  
%$$\displaylines{ 
%\hfill T^{\si\mu_1\ldots\mu_{n-1}}_{\ell~~i} = {i^{n-4}\over 4}
%\bar\Psi (0)D^{\mu_1}\ldots D^{\mu_{\ell -1}}G^{\si\mu_\ell}
%D^{\mu_{\ell +1}}\ldots
%D^{\mu_{n-2}}\gamma^{\mu_{n-1}}\gamma_5\lambda_i\Psi(0)\hfill\cr
%\hfill S^{\si\mu_1\ldots\mu_{n-1}}_\ell = {i^{n-3}\over 4}
%\bar\Psi (0)D^{\mu_1}\ldots D^{\mu_{\ell -1}}\tilde G^{\si\mu_\ell}
%D^{\mu_{\ell +1}}\ldots             
%D^{\mu_{n-2}}\gamma^{\mu_{n-1}}\lambda_i\Psi(0)\hfill\cr
%\hfill i=0,\ldots, 9\qquad \ell =1,\ldots, n-2\hfill\cr}
%$$                                  
%where $\tilde G$ is the dual field strength.
%Only $n-2$ of these operators are independent. One may
%take, for example, the $n-2$ independent combinations
%$$                                  
%W_k=-T_{\ell =k}+T_{\ell =n-1-k}+S_{\ell =k}+S_{\ell =n-1-k}
%$$                                  
%In addition there is a twist--3 quark mass dependent operator
%$$                                  
%O_{m_q}^{\si\mu_1\ldots\mu_{n-1}}={i^n\over 4}\bar\Psi (0)
%m_q\(\gamma^\si \gamma^{\mu_1}\)\gamma^5 D^{\mu_2}\ldots
%D^{\mu_{n-1}}\Psi (0)               
%$$                                  
%and, for the flavor singlet, twist--3 operators from gluon fields
%$$                         
%U_\ell^{\si\mu_1\ldots\mu_{n-1}}=& i^{n-1}
%G^{\a\mu_1}D^{\mu_2}\ldots D^{\mu_{\ell -1}}
%\tilde G^{\si\mu_\ell}D^{\mu_{\ell +1}}\ldots
%D^{\mu_{n-2}}G_\a^{\mu_{n-1}}
%$$    
%$$
%V_\ell^{\si\mu_1\ldots\mu_{n-1}}=& i^{n-3}
%\tilde G^{\a\mu_1}D^{\mu_2}\ldots D^{\mu_{\ell -1}}
%G^{\si\mu_\ell}D^{\mu_{\ell +1}}\ldots
%D^{\mu_{n-2}}G_\a^{\mu_{n-1}}   
%$$ 
\begin{eqnarray} \nonumber 
  R_F^{\sigma\mu_{1}\cdots \mu_{n-1}} &=&
         \frac{i^{n-1}}{n} [ (n-1) \overline{\psi}\gamma_5
       \gamma^{\sigma}D^{\{\mu_1} \cdots D^{\mu_{n-1}\}}\psi
\\ & &      
      - \sum_{l=1}^{n-1} \overline{\psi} \gamma_5
       \gamma^{\mu_l }D^{\{\sigma} D^{\mu_1} \cdots D^{\mu_{l-1}}
            D^{\mu_{l+1}} \cdots D^{\mu_{n-1}\}}
                             \psi ] ,
                                            \label{op3}\\
  R_{m_q}^{\sigma\mu_{1}\cdots \mu_{n-1}} &=&
          i^{n-2}  \overline{\psi} m_q \gamma_5
       \gamma^{\sigma}D^{\{\mu_1} \cdots D^{\mu_{n-2}}
        \gamma ^{\mu_{n-1}\}} \psi ,
                                             \label{op4} \\
  R_k^{\sigma\mu_{1}\cdots \mu_{n-1}} &=& \frac{1}{2n}
              \left( V_k - V_{n-1-k} + U_k + U_{n-1-k} \right) ,
                                                  \label{op5}
\end{eqnarray}
where $m_q$ in (\ref{op4}) represents the quark mass (matrix) 
and $\{ \}$ means symmetrization over the Lorentz indices; 
furthermore, the flavor structure ($\lambda_a$) for the 
quark fields $\psi$ has been suppressed for simplicity  
and the appropriate subtraction 
of trace--terms is always implied in order to render 
the resulting operators traceless, i.e. of definite spin.
The operators in (\ref{op5}) contain explicitly 
the gluon field strength $G_{\mu\nu}$ and its dual tensor
$\widetilde{G}_{\mu \nu}={1\over
2}\varepsilon_{\mu\nu\alpha\beta}
G^{\alpha\beta}$ and are given by
\begin{eqnarray}
    V_k &=& i^n g S \overline{\psi}\gamma_5
       D^{\mu_1} \cdots G^{\sigma \mu_k } \cdots D^{\mu_{n-2}}
        \gamma ^{\mu_{n-1}} \psi , \nonumber \\
    U_k &=& i^{n-3} g S \overline{\psi}
       D^{\mu_1} \cdots \widetilde{G}^{\sigma \mu_k } \cdots
             D^{\mu_{n-2}} \gamma ^{\mu_{n-1}} \psi , \nonumber
\end{eqnarray}
where $S$ means symmetrization over $\mu_i$ and $g$ is the QCD
coupling constant.
It is a well-known fact 
\cite{shuryak,altarelli4,altarelli7,jaffe1,jaffe7,kodaira5,kodaira3} 
that these operators (\ref{op3})-(\ref{op5}) are not
independent and related through the 'equation of motion' operator 
\begin{equation}
   R_{eq}^{\sigma\mu_{1}\cdots \mu_{n-1}}
    =   i^{n-2} \frac{n-1}{2n} S [ \overline{\psi} \gamma_5
             \gamma^{\sigma} D^{\mu_1} \cdots D^{\mu_{n-2}}
               \gamma ^{\mu_{n-1}} (i\not{\!\!D} - m_q )\psi 
    + \overline{\psi} (i\not{\!\!D} - m_q )
             \gamma_5 \gamma^{\sigma} D^{\mu_1} \cdots D^{\mu_{n-2}}
                 \gamma ^{\mu_{n-1}} \psi ] 
\end{equation}
Making use of the identities $D_\mu={1\over2}\{\gamma_\mu,\not{\!\!\!D}\}$
and $[D_\mu,D_\nu]=g G_{\mu\nu}$ 
one can obtain the following relation for the twist--3 operators,
\begin{equation}
     R_F^{\sigma\mu_{1}\cdots \mu_{n-1}} =
       \frac{n-1}{n} R_{m_q}^{\sigma\mu_{1}\cdots \mu_{n-1}}
         + \sum_{k=1}^{n-2} (n-1-k)
            R_k^{\sigma\mu_{1}\cdots \mu_{n-1}} +
              R_{eq}^{\sigma\mu_{1}\cdots \mu_{n-1}} .
\label{op6}
\end{equation}
Leaving aside the $Q^2$ dependence for the time being,   
the moments of $g_2$ are given by  
\begin{equation}
\int\limits^1_0 dx x^{n-1}g_2(x,Q^2)=\phantom{-}\ha {n-1 \over n}
(d_n-a_n) \,   , \, 
\quad n=3,5,7,\ldots     
\label{8120}
\end{equation}
which represents the generalization of the pure twist--2 
relation (\ref{812}) and 
where the contribution from the twist--2 operators is summarized 
in $a_n=\sum\limits_i M^n_i E_i^n$ and $d_n$ is the matrix element 
of the sum of all twist--3 operators contributing to the 
n-th moment of $g_2$. 
Note that this formula is only true if one formally keeps $R_F$ in 
the operator basis because the matrix element $d_n$ is defined by 
\begin{equation} 
  \langle P,S | R_F^{\sigma\mu_{1}\cdots \mu_{n-1}} |P,S \rangle
       = -  \frac{n-1}{n} d_n ( S^{\sigma}P^{\mu_1} - S^{\mu_1}P^{\sigma})
                    P^{\mu_2} \cdots P^{\mu_{n-1}}   \,  .
\label{rtu1}
\end{equation} 
If one eliminates $R_F$ from the basis via Eq. (\ref{op6}), the 
matrix elements of the operators $R_{m_q}$, $R_k$ and $R_{eq}$ will 
appear. 
Note further that there is {\it no} relation (\ref{8120}) for 
the first moment n=1, because there is no twist--3 operator for n=1. 
This is in contrast to $g_1$ whose first moment is fixed in the 
operator product expansion by the matrix element of the 
axial vector singlet current (cf. Sect. 5). 

Combining Eq. (\ref{8120}) with the analogous pure twist--2 relation 
(\ref{436}) for $g_1$, 
\begin{equation} 
\int_0^1 dx x^{n-1}g_1(x,Q^2)={1\over 2} a_n   \,  , 
\label{rtu2}
\end{equation} 
it has become customary to extract the pure twist--3 matrix element 
$d_n$: 
\begin{equation} 
d_n(Q^2)=2\int_0^1 dx x^{n-1}[g_1(x,Q^2)
+{n\over n-1} g_2(x,Q^2)]=2{n\over n-1}
\int_0^1 dx x^{n-1} \bar g_2(x,Q^2)  
\label{rtu3}
\end{equation} 
where the latter equality follows from Eqs. (\ref{vvxy}) and 
(\ref{812a}). Being pure twist--3, $d_n(Q^2)$ is a direct probe 
of non--partonic effects such as quark--gluon correlations. In   
other words, it is a direct measure of deviations from the 
(twist--2) Wandzura--Wilczek relation (\ref{8117}). 
Several experimental 
attempts at CERN (SMC \cite{adams5}) and SLAC (E143 
\cite{abe3}, E154 \cite{abe8}) to observe such deviations 
by measuring $g_2^{p,n,d}(x,Q^2)$ via $A_T$ in (\ref{226a}) 
did not result in any statistically relevant twist--3 
contribution ($\bar g_2$) to $g_2$, i.e. present data 
are in agreement with the twist--2 Wandzura-Wilczek prediction 
derived from  (\ref{8117}), $g_2(x,Q^2)\approx g_2^{WW}(x,Q^2)$, 
at presently attainable values of $Q^2$. Qualitatively, the 
observed tendency is that $g_2^p(x,Q^2)$ is positive in the 
region of smaller $x$ and negative in the region of larger 
$x$ values, in agreement with the twist--2 Wandzura-Wilczek 
expectations \cite{abe3} (cf. Fig. \ref{figexpectg2}). 
More specifically, present 
measurements imply, for example, for 
the third $n=3$ moment $d_3(Q^2\approx 5GeV^2)$ 
in Eq. (\ref{rtu3}) the following results \cite{abe3,abe8} :  
\begin{eqnarray} \nonumber 
d_3^p&=&(5.4\pm 5.0)\times 10^{-3}  \qquad (E143)  
\\ \nonumber 
d_3^d&=&(3.9\pm 9.2)\times 10^{-3}  \qquad (E143) 
\\
d_3^n&=&(-10\pm 15)\times 10^{-3}   \qquad (E154, SLAC \, average)  \, . 
\label{rtu4}
\end{eqnarray}
Comparing these results with bag model expectations 
\cite{stratmann2,ji,song2,song1}, 
$d_3^p\approx (6 \, $to$ \, 18) \times 10^{-3}$,  
$d_3^d\approx (3 \, $to$ \, 7)\times 10^{-3}$ and 
$d_3^n\approx (-2.5 \, $to$ \, 0.3)\times 10^{-3}$ or with those 
obtained from QCD sum rules \cite{balitzky,stein}, 
$d_3^p\approx (-9 \, $to$ \, 0)\times 10^{-3}$,  
$d_3^d\approx (-22 \, $to$ \, -8)\times 10^{-3}$ and 
$d_3^n\approx (-40 \, $to$ \, -15)\times 10^{-3}$, it becomes clear 
that for the time being there is no possibility to distinguish 
between different models.  

As a side remark, 
it should be noted that 
the matrix elements $d_n$ appear also in higher twist corrections 
to $g_1$. For example, the first moment of $g_1$ has an expansion 
\cite{ehrnsperger,ji}
\begin{equation}
\int\limits^1_0 dx g_1(x,Q^2)=\phantom{-}\ha a_1 
+{M^2 \over 9Q^2}(a_3+4d_3+4f_3) +O({M^4 \over Q^4})  \, . 
\label{8121}
\end{equation}
The higher twist corrections in this result are not completely 
fixed by $a_3$ and $d_3$, but there is another matrix element $f_3$  
of a twist--4 operator involved, defined by 
\begin{equation}
\sum\limits_i e_q^2 \langle PS | 
\bar{q} g \tilde{G}_{\alpha \beta}  \gamma^{\beta} q 
|PS \rangle =2M^2 f_3 S_{\alpha}   \, . 
\label{8122}
\end{equation}
This twist--4  
matrix element $f_3$ 
has been estimated \cite{balitzky,ehrnsperger,ji11}, partly   
from a QCD sum rule approach, with similar uncertainties 
as the above estimates of $d_3$. 

\begin{figure}
\begin{center}
\epsfig{file=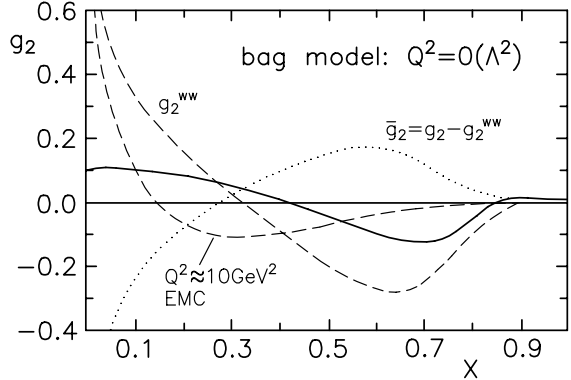,height=5cm}
\bigskip
\caption{Predictions for $g_2$ \protect\cite{stratmann2} 
of an extended (one--gluon exchange) 
version of the MIT bag model \protect\cite{schreiber} 
using a bag radius R=0.7 fm and spin--singlet and 
--triplet intermediate diquark masses of 0.56 and 
0.8 GeV. The EMC prediction for $g_2^{WW}$ of 
Fig. \ref{figexpectg2} is also shown. }
\label{bagm}
\end{center}
\end{figure}

There have been attempts to calculate also $g_2(x,Q^2)$ in the bag model 
\cite{jaffe7,stratmann2,song2,song1}. The 
bag model does not contain gluon fields explicitly, but the
boundary of the bag-confined quarks simulates the binding
effect coming from quark-gluon and gluon-gluon interactions. 
Hence, the structure function $g_2$ calculated in the bag
model includes higher twist effects and measures possibly 
large twist--3 matrix elements comparable in size to the twist--2 ones. 
Indeed, sizeable departures 
from the Wandzura--Wilczek relation 
\cite{stratmann2} have been noted in an extended 
version of the MIT bag model \cite{schreiber}. 
They are mainly induced by the 
non--covariance of the relativistic bag model, which originates 
from the implementation of the bag boundary in the equation of 
motion. This is in spite of the fact that the average parton virtuality 
in the bag is small, from which one might erroneously conclude 
that $\bar g_2 \approx 0$ according to the covariant parton model 
discussed above. The predictions for $g_2$ \cite{stratmann2} 
are shown in Fig. \ref{bagm}. It should, however, be emphasized that 
the results of such bound--state models are expected to hold at some 
non--perturbative bound--state scale, typically $Q^2 \sim \Lambda^2$. 
Strictly speaking, it is therefore not even possible to use these 
predictions as an input for an evolution to a larger scale 
$Q^2 \gtrsim 1$ GeV$^2$, unless one arbitrarily chooses the 
bag bound--state scale $Q^2 >> \Lambda^2$, from where a perturbative 
evolution could be started. In order to demonstrate the importance 
of different scale effects, the EMC prediction for the twist--2 $g_2^{WW}$ 
at $Q^2 \approx 10$ GeV$^2$ is taken from Fig. \ref{figexpectg2} 
and shown in Fig. \ref{bagm} as well. The difference between the 
bag and EMC prediction is indeed very large, indicating that 
expectations from bound--state models are not very relevant for 
actual deep inelastic measurements. This is not very surprising 
since the bag predictions for $g_1$ \cite{jaffe7} and for unpolarized 
structure functions \cite{schreiber} 
also disagree with actual deep inelastic measurements.    
Therefore, the large bag model prediction for $\bar g_2$ appears 
to be not too relevant for future experiments. In fact, other 
models \cite{mankiewicz3,
mankiewicz4,schaefer} based on the light--cone quark model 
\cite{dziembowski} 
give substantially different predictions for $g_2(x)$ than 
Fig. \ref{bagm}. According to these models, 
$g_2$ and $\bar g_2$ could  
also become strongly negative (of the order of $-1$) for 
$x < 0.1$. Therefore, it should be reemphasized that it is 
very important to check experimentally first 
the Wandzura--Wilczek relation (\ref{8117}) and then to 
extract $\bar g_2\equiv g_2 -g_2^{WW}$, despite the fact 
that present experiments \cite{abe3,abe8,adams5} are, within 
large errors, consistent with a vanishing $\bar g_2(x,Q^2)$.  

%The twist--3 component of 
%$g_2$ can be further decomposed into a quark transverse 
%polarization part and a twist--3 part coming
%from the quark--gluon interaction.
We have already noted, that for the first moment of $g_2$ there is no 
twist--3 operator within the light--cone OPE. 
This is a hint, though no proof, of the so-called 
'Burkhardt-Cottingham (BC) sum rule' \cite{burkhardt1} 
\begin{equation}
\int^1_0 dx g_2(x,Q^2)=0 \quad .
\label{bcsr}
\end{equation}
This relation 
can be derived as a superconvergence relation based on 
Regge asymptotics (see \cite{jaffe1} for a review).  
Recently it has been shown by an explicit calculation, 
that it is fulfilled in higher order QCD \cite{altarelli8,kodaira4}. 
The (more general but also more questionable) 
Regge proof was derived by considering 
the asymptotic behavior of the virtual Compton helicity 
amplitude $A_2$ related to $g_2$, 
\begin{equation}
g_2(x,Q^2)={\nu^2 \over 2 \pi M^4}Im A_2(q^2,\nu) 
\label{8123}
\end{equation}
where $\nu\equiv P\cdot q$. 
The first moment of $g_2$ is given in terms of $A_2$ as 
\begin{equation}
\int^1_0 dx g_2(x,Q^2)={Q^2 \over 2\pi M^4}\int_{-q^2/2}^{\infty}
d \nu^\prime Im A_2(q^2,\nu^\prime)  \, . 
\label{8124}
\end{equation}
Combining Cauchy's theorem with crossing symmetry one finds the 
following dispersion relation for $A_2$: 
\begin{equation}
A_2(q^2,\nu)={2 \over \pi} \nu \int_{-q^2/2}^{\infty} 
{d \nu^\prime \over \nu^{\prime 2}- \nu^2}Im A_2(q^2,\nu^\prime) \, .  
\label{8125}
\end{equation}
Burkhardt and Cottingham argue that all known Regge 
singularities contributing to $A_2$ have an 
intercept $\alpha(0)$ less than zero. This is equivalent to 
the statement
that $A_2$ falls off to zero stronger than ${1 \over \nu}$ 
as $\nu \rightarrow \infty$ (more precisely $\sim \nu^{\alpha(0)-1}$ 
modulo logarithms). 
This implies that one can take 
the limit $\nu \rightarrow \infty$ under the $\nu^\prime$ 
dispersion integral and obtains 
\begin{equation}
A_2(q^2,\nu) \approx -{2 \over \pi \nu} \int_{-q^2/2}^{\infty}
d \nu^\prime Im A_2(q^2,\nu^\prime) 
\label{8126}
\end{equation}
which is compatible with the rapid Regge fall-off only if 
the integral vanishes. Thus (\ref{8124}) implies the BC sum rule 
(\ref{bcsr}). 

The derivation raises some questions. For example, Heimann \cite{heimann} 
%and similarly Ioffe,Khoze and Lipatov have  
has 
argued that there might be contributions from multi-pomeron 
cuts which invalidate the assumption of intercept less than zero 
and would imply a highly singular behavior 
$g_2(x) \rightarrow x^{-2}$ as $x \rightarrow 0$ 
so that the first moment integral would not even converge. 
On the other hand, Regge cuts with branch points at 
$\alpha (0) \geq 0$ or specific nonpolynomial residue $J=0$ 
fixed poles in Compton amplitudes \cite{cheng10,jaffe1} 
could invalidate the BC sum rule by terms of order at most $1/Q^2$.  
%The latter authors have suggested that the BC sum rule might 
It is, therefore, possible that the BC sum rule might 
be restored asymptotically, as $Q^2 \rightarrow \infty$,  
because the residues of the Pomeron cuts fall off at 
$Q^2 \rightarrow \infty$. Apart from such an 'exotic' situation,  
the BC sum rule (\ref{bcsr}) appears to be very robust and is 
most probably true \cite{jaffe1,jaffe7}.  
Furthermore, 
in the framework of perturbative QCD there is no indication 
of a violation of the BC sum rule \cite{altarelli8,kodaira4} 
so that we believe that it is probably valid at 
sufficiently high $Q^2$. 
It should be emphasized again, that the light--cone OPE per se 
is {\it non}--committal about the BC sum rule since the twist--3 
operators with mixed symmetry in (\ref{op3})--(\ref{op5}) need 
at least two (antisymmetrized) indices ($n > 1$) which implies 
the validity of Eq. (\ref{8120}) only for $n > 1$. We would 
need the $n\rightarrow 1$ continuation of the twist--3 
matrix element $d_n$: if $d_n$ is less singular than 
$(n-1)^{-1}$ as $n\rightarrow 1$, the BC sum rule (\ref{bcsr})   
would hold, as one might naively expect; it would not hold 
if $d_n\sim (n-1)^{-1}$ for $n\rightarrow 1$ (but on the other 
hand we know that the twist--2 matrix element $a_1$ in 
(\ref{rtu2}) is not singular because the $n=1$ moment of 
$g_1$ is finite). It is clearly important to check 
the validity of the BC sum rule experimentally -- as far as 
possible. Present measurements at an average $Q^2$ of 3 to 5 
GeV$^2$ imply \cite{abe3,abe8} 
\begin{equation}
\int_{0.03}^1 dx g_2^p(x,Q^2)=-0.013\pm 0.028, \quad 
\int_{0.014}^1 dx g_2^n(x,Q^2)=0.06\pm 0.15, 
\label{81177}
\end{equation}
which are consistent with (\ref{bcsr}) but certainly not 
conclusive. It would nevertheless be very interesting if the 
BC sum rule were not true, because this would imply a 
non--conventional behavior of twist--3 operators 
($d_n\sim (n-1)^{-1}$) or the importance of long range 
effects \cite{jaffe7,mankiewicz4,mankiewicz3,schaefer}. 
 
Another interesting, but less problematic sum rule concerns the 
valence content of $g_1$ and $g_2$ \cite{efremov2,efremov9} 
\begin{equation}
\int\limits_0^1 dx x [g_1(x,Q^2)+2g_2(x,Q^2)]^{valence}=0
\label{81173}
\end{equation}
which amounts 
to the vanishing of all twist--3 contributions to the second 
($n=2$) moment of $g_2$. Apart from the fact that this sum rule can be 
independently derived from the OPE \cite{bluemlein4}, there 
is also a wealth of similar sum rules between structure 
functions in the electroweak sector (see \cite{bluemlein3,bluemlein4} 
and references therein) which are unfortunately beyond experimental 
reach for the time being.     
%Note that the 'Wandzura Wilczek' twist--2 part of $g_2$ 
%fulfils the BC sum rule. 
%Therefore one may say that the present experimental situation 
%supports the BC sum rule. 

Let us finally come to the logarithmic scaling violations 
to Eq. (\ref{8120}), in particular the $Q^2$--evolution 
of $\bar g_2(x,Q^2)$ is theoretically unknown and in 
general an unsolved intricate problem, because   
the number of independent twist--3 operators increases with $n$ 
\cite{antoniadis,bukhvostov,bukhvostov1,ratcliffe3}. 
Thus the dimension of the anomalous dimension matrix 
becomes larger with increasing $n$ and therefore there 
exist no Altarelli--Parisi--type evolution equations. 
Anomalous dimensions and coefficient functions for $g_2$ 
have been calculated, see \cite{ji10,kodaira3} and earlier 
references quoted therein, but for principal reasons a 
satisfactory physical model to predict $\bar g_2(x,Q^2)$ 
does not exist. There have been attempts \cite{ali}, to 
which we shall return below,  
to construct practical approximations to the $Q^2$ evolution 
in the large--$n$ limit ($x\rightarrow 1$) as well as in the 
limit $N_c\rightarrow \infty$. 
Unfortunately, the very large--x region and probably also 
the large $N_c$ limit are  
experimentally not very relevant. The latter just gives 
an indication that the effect of the $Q^2$ evolution of 
$\bar g_2(x,Q^2)$ might be large throughout the whole x--region 
\cite{stratmann2}. Furthermore, a parton inspired   
picture involves two--parton 'correlation functions' 
$C(x_1,x_2)$,  
where two partons (with Bjorken $x_1$ and $x_2$)   
split from the proton at the same time and interact with the 
photon (see, e.g., \cite{mankiewicz3,teryaev}). 
However, these functions are as 
undetermined as the matrix elements of the higher twist operators.

At the end of this section we want to discuss 
%It is perhaps enlightening to study 
the operator mixing and its 
effects on the $Q^2$--evolution and the coefficient functions for the 
simple case of $n=3$.  
The coefficient functions at the tree level and the anomalous dimensions 
for the operators depend separately 
upon the choice of the independent operator basis.  
But the $Q^2$ evolution of the moments of $g_2$ does 
of course not depend on it. 
In the case $n=3$ there are four operators 
with the constraint 
\begin{equation}
  R_F = \textstyle{\frac{2}{3}} R_m + R_1 + R_{eq},
\label{n3}
\end{equation}
where the Lorentz indices of operators are omitted.
One may choose the operators $R_F, R_m, {\rm and}R_1$ as independent
operators
and eliminate the 'EOM' operator $R_{eq}$.
One then 
gets the following renormalization matrix for the composite operators 
\begin{equation}
\bmat{c}
R_F \\ R_1 \\ R_m \\ R_{eq1}
\emat_{\hspace{-0.1cm}R}
=
\bmat{cccc}
Z_{11} & Z_{12} & Z_{13} & Z_{14} \\
Z_{21} & Z_{22} & Z_{23} & Z_{24} \\
0 & 0 & Z_{33} & 0 \\
0 & 0 & 0 & Z_{44}
\emat
\bmat{c}
R_F \\ R_1 \\ R_m \\ R_{eq1}
\emat_{\hspace{-0.1cm}B}
\label{n3z1}
\end{equation}
where the 
$Z_{ij}$ can be calculated in dimensional regularization and are of form 
$Z_{ij} = \delta_{ij} +
{1\over\varepsilon}{{g^2}\over{16\pi^2}}z_{ij}$ 
with $D=4-2\varepsilon$. 
A straightforward but tedious calculation gives \cite{kodaira5,kodaira3}
\begin{equation}
\ba{ll}
z_{11}={7\over 6}C_F+{3\over 8}N_c, &
z_{12}=-{3\over 2}C_F+{21\over 8}N_c, \\
&\\
z_{13}=3C_F-{1\over 4}N_c, & z_{14}=-{3\over 8}N_c, \\
&\\
z_{21}={1\over 6}C_F-{1\over 8}N_c, &
z_{22}=-{1\over 2}C_F+{25\over 8}N_c, \\
&\\
z_{23}=-{1\over 3}C_F+{1\over 12}N_c, &
z_{24}={1\over 8}N_c, \\
&\\
z_{33}=6C_F, & z_{44}=0 .
\label{zn3}
\ea
\end{equation}
$R_{eq1}$ is a gauge non-invariant 'equation of motion' operator 
\begin{equation}
  R_{eq1}^{\sigma\mu_{1} \mu_{2}} =i \textstyle{1\over 3} S
  [ \overline{\psi} \gamma_5 \gamma^{\sigma} \partial ^{\mu_1}
  \gamma ^{\mu_{2}} (i\not{\!\!D} - m_q )\psi + \overline{\psi}
  (i\not{\!\!D} - m_q ) \gamma_5 \gamma^{\sigma} \partial ^{\mu_1}
  \gamma ^{\mu_{2}} \psi ] 
\label{zn33}
\end{equation}
which comes into play when renormalizing the operators 
in (\ref{n3}). 
Although this operator is gauge non-invariant,
it may be chosen to appear in the operator basis
because it vanishes by the equation of motion.
The above results in Eq. (\ref{zn33}) satisfy the equalities 
\cite{kodaira5} 
\begin{equation}
z_{11}+z_{12} = z_{21}+z_{22}, \quad 
    {2\over 3}z_{11}+z_{13} = {2\over 3}z_{21}+z_{23}+{2\over 3}z_{33}, 
\quad 
  z_{13}-{2\over 3}z_{12} = z_{23}-{2\over 3}z_{22}+{2\over 3}z_{33} 
         \, . 
\label{zrel}
\end{equation}
%\begin{equation}
%\ba{ll}
%z_{11}+z_{12} = z_{21}+z_{22}, &
%    {2\over 3}z_{11}+z_{13} = {2\over 3}z_{21}+z_{23}+{2\over 3}z_{33}, \\
%&\\
%  z_{13}-{2\over 3}z_{12} = z_{23}-{2\over 3}z_{22}+{2\over 3}z_{33}.&
%\ea
%\label{zrel}
%\end{equation}
What happens if one chooses $R_1$, $R_m$, $R_{eq}$ and $R_{eq1}$,
and eliminates $R_F$?
Using (\ref{n3}) and the relations (\ref{zrel})
one gets the renormalization matrix
\begin{equation}
\bmat{c}
R_1 \\ R_m \\ R_{eq} \\ R_{eq1}
\emat_{\hspace{-0.1cm}R}
=
\bmat{cccc}
Z_{21}+Z_{22} & {2\over 3}Z_{21}+Z_{23} & Z_{21} & Z_{24} \\
0 & Z_{33} & 0 & 0 \\
0 & 0 & Z_{11}-Z_{21} & Z_{14}-Z_{24} \\
0 & 0 & 0 & Z_{44}
\emat
\bmat{c}
R_1 \\ R_m \\ R_{eq} \\ R_{eq1}
\emat_{\hspace{-0.1cm}B} 
\end{equation}
This choice of basis was adopted, for example, by \cite{ali} 
in their approximate calculation of anomalous dimensions 
in the large $N_c$ limit:  
\begin{equation}
\bar \gamma_{NS}^n
=4N_c ( \sum_{j=1}^n {1\over j}
-{1\over 4}-{1\over 2n})
\label{zu77}
\end{equation}
valid for large $n$, i.e. for large values of x close to 1. 
It should be noted that this anomalous dimension differs 
substantially from the one naively obtained by ignoring 
the operator mixing as was done in the early days 
\cite{hey,sasaki,ahmed,ahmed1,kodaira1,kodaira2,antoniadis}. 
In that approximation the evolution equation for the moments 
of the twist--3 contribution $\bar g_2$ to $g_2$ in 
Eq. (\ref{vvxy}) reads 
\begin{equation}
\int\limits^1_0 dx x^{n-1}\bar g_2(x,Q^2)=
\biggl \lgroup  { \alpha_s(Q^2) \over \alpha_s(Q_0^2) } \biggr \rgroup
^{\bar \gamma_{NS}^n /2\beta_0} 
\int\limits^1_0 dx x^{n-1}\bar g_2(x,Q_0^2)
\label{812uu}
\end{equation}
with $\beta_0$ given in (\ref{4113}). This $Q^2$ evolution 
has been quantitatively studied in \cite{stratmann2} and 
used by \cite{abe3} to compare bound--state model predictions,  
e.g. in Fig. \ref{bagm}, with experimental results, such 
as the ones in (\ref{rtu4}), at larger values of $Q^2$.   

%% file: k82tex
\subsection{Transverse Chiral--Odd ('Transversity') 
            Structure Functions}

In analogy to the unpolarized and polarized structure functions $F_1$
and $g_1$, the
`transversity' structure function 
\cite{ralston,kodaira1,bukhvostov1,artru,cortes1,jaffe8,jaffe4,ratcliffe1} 
is, in LO, formally given by (cf.~Eq.\ (4.5))
%Gl. (8.39)
\begin{equation}
 h_1(x,Q^2)=\frac{1}{2}\sum_qe_q^2 \left[\delta_Tq(x,Q^2)+
                 \delta_T\bar{q}(x,Q^2)\right]
\end{equation}
where, similarly to Eqs.\ (4.1) and (4.2),
%Gl. (8.40)
\begin{equation}
 \delta_T\!\! \stackrel{(-)}{q}(x,Q^2)\equiv
      {\stackrel{(-)}{q}}\,^{\uparrow}
        (x,Q^2)-{\stackrel{(-)}{q}}\,^{\downarrow}(x,Q^2)
\end{equation} 
describes the (anti)quark `transversity' distribution with 
${\stackrel{(-)}{q}}\,^{\uparrow}\, ({\stackrel{(-)}{q}}\,^{\downarrow})$
being the probability of finding a (anti)quark in a 
transversely polarized proton with spin parallel 
(antiparallel) to the proton spin. The transverse polarization of a 
quark entering an interaction kernel is obtained by using 
$u(p,s)\,\bar{u}(p,s)=-p\!\!\!/ \,s\!\!\!/\,\gamma_5$ for its spinor 
$u(p,s)$ with $s\cdot p=0$. The $\delta_T\!\!\stackrel{(-)}{q}$ are related 
to the tensor current $\bar{q}\,i\sigma^{\mu\nu}\gamma_5q$ which is 
chiral (and charge conjugation) {\it {odd}}, i.e.\ measure
correlations between left-- and right--handed quarks, 
$q_L \leftrightarrow q_R$, induced for example by nonperturbative
condensates $\langle \bar{q}_L\, q_R\rangle$ in the nucleon.  Thus, 
unlike in the familiar case of the $q(x,Q^2)$ and $\delta q(x,Q^2)$,
there is no gluonic transversity density at leading twist 
\cite{jaffe9,ji12,soffer3}.  
Together, $q,\, \delta q$ and $\delta_Tq$ provide a complete
description of quark momentum and spin at leading twist as can be
seen generically from a spin--density matrix representation in the 
quark-- and nucleon--helicity basis,
%Gl. (8.41)
\begin{equation}
{\cal{F}}(x,Q^2)=\frac{1}{2}q(x,Q^2)I\otimes I +\frac{1}{2}\delta q(x,Q^2)
  \sigma_3\otimes\sigma_3 +\frac{1}{2}\delta_T q(x,Q^2)
   (\sigma_+\otimes\sigma_-\, +\, \sigma_-\otimes\sigma_+)\, .
\end{equation}
Thus the $\delta_T\!\!\stackrel{(-)}{q}\!\!(x,Q^2)$ are leading twist--2
densities and {\it complete} the twist--2 sector of nucleonic
parton distributions, and are therefore in principle as interesting
as the familiar $f(x,Q^2)$ and $\delta f(x,Q^2)$.  Unfortunately, the
$\delta_T\!\!\stackrel{(-)}{q}$ densities are experimentally more
difficult to access and entirely unknown so far.  (Although the name
`transversity' is fairly universal, the notation is not.  In addition
to $\delta_T q$, notations such as $h_1^q,\, h_T,\, \Delta_Tq,\, 
\Delta_1q,\, \delta q$ etc.\ are common as well 
\cite{ralston,cortes1,jaffe8,jaffe4,artru}).

It should be remembered that the common unpolarized and longitudinally
polarized quark densities $q$ and $\delta q$, respectively, considered
up to now are related to the matrix elements of vector and axial--vector
currents, $\bar{q}\gamma_{\mu}q$ and $\bar{q}\gamma_{\mu}\gamma_5q=
\bar{q}_L\gamma_{\mu}{\gamma_5}q_L+\bar{q}_R\gamma_{\mu}\gamma_5q_R$,
respectively, which preserve chirality, i.e.\ are chiral--{\it {even}}
$(q_L\to q_L,\, q_R\to q_R)$ in contrast to the
chiral--odd $\delta_T q$. Thus, the transverse spin structure function
$g_T=g_1+g_2$ (or $g_2$) of the previous Sect.\ 8.1, which preserves
chirality, must not be confused with $h_1$ which flips chirality.  
We have seen that for $g_2$, arising from transversally polarized nucleons, 
the cross section picks up a factor of $\frac{M}{\sqrt{Q^2}}$ because in 
these processes the nucleon helicity changes but the quark chirality does 
not change in the hard scattering subprocess 
\cite{jaffe8,jaffe4,bunce}. 
Therefore it is obviously not possible to measure the chiral--odd 
transversity structure function $h_1(x,Q^2)$ in usual $ep$ inclusive
DIS which, apart from small quark mass corrections, gives always rise
to chiral--even transitions ($\gamma^*q_L\to q_L,$ etc.), i.e.\ to
$F_{1,2}$ and $g_{1,2}$.  The chiral--even transitions are illustrated
in Fig.\ 52 where the quark lines leaving and entering the nucleon are
of a single chirality.  This is so because the photons and gluons
participating in the hard scattering process have a vectorlike interaction
with the massless quark.  Thus in lowest order only two independent
quark--nucleon amplitudes enter the description of DIS:  the sum in 
Fig.\ 53 over chirality gives the unpolarized structure function 
$F_1$ (or $F_2$), whereas the difference gives the structure function
$g_1$ of longitudinal polarization. 

\begin{figure}
\begin{center}
\epsfig{file=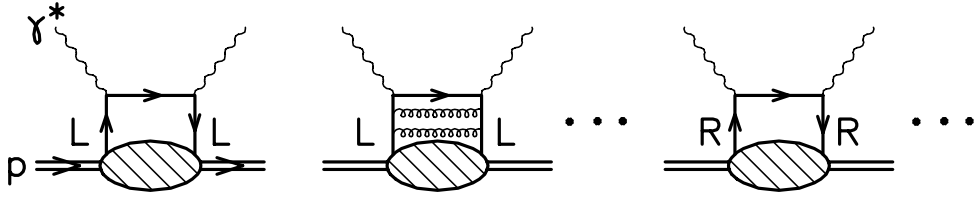,height=2.5cm}
\bigskip
\caption{Chiral (even) structure of deep inelastic
scattering.}
\label{figeven}
\end{center}
\end{figure}

\begin{figure}
\begin{center}
\epsfig{file=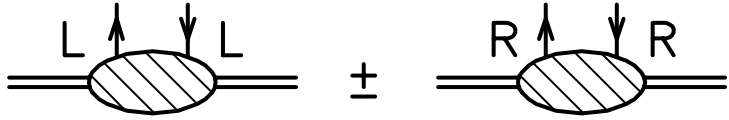,height=1.5cm}
\bigskip
\caption{Left-- and right--handed chiral even quark
distributions as measured in DIS whose sum and difference give
$F_1(x,Q^2)$ and $g_1(x,Q^2)$, respectively.}
\label{figeme}
\end{center}
\end{figure}

On the other hand, in hadronic collisions, the chirality of the quark 
lines leaving and entering a given nucleon need not be the same, as
illustrated in Fig.\ 54.  This is due to nonperturbative condensates
$\langle\bar{q}_L q_R\rangle\neq 0$ which break the chiral symmetry
of the QCD vacuum as well as of the nucleon structure due to bound
state effects 
\cite{jaffe8,jaffe4,bunce}.  The corresponding new leading twist--2
structure function is referred to as $h_1(x,Q^2)$ in (8.39).  Its 
interpretation is obscure in the chiral basis, but is revealed by using 
the transverse projection operators $P_{\Uparrow\Downarrow}=\frac{1}{2}
(1\pm \gamma_5 S\!\!\!/_T)$ instead of $P_{R,L}=\frac{1}{2}(1\pm\gamma_5)$,
c.f.\ Eq.~(1.3).  Here the transversely polarized nucleon and quarks are
classified as eigenstates of the transversely projected Pauli--Lubanski
spin  operator $\gamma_5 S\!\!\!/_T$ (which is the more familiar form
proportional to $W_{\mu}=-\frac{1}{2}\varepsilon_{\mu\nu\rho\sigma}
J^{\nu\rho}P^{\sigma})$, i.e.\ $\gamma_5 S\!\!\!/_T u(P_z,S_T)=\pm
u(P_z,S_T)$.  In contrast to the transverse quark spin operator 
$\gamma_0\gamma_5 S\!\!\!/_T$ relevant for $g_2$ in the previous 
Sect.\ 8.1, the Pauli--Lubanski operator $\gamma_5 S\!\!\!/_T$ commutes
with the free quark Hamiltonian $H_0=\alpha_z p_z$, and therefore 
$h_1(x,Q^2)$ can be interpreted in terms of the quark--parton model
as done in (8.39).

\begin{figure}
\begin{center}
\epsfig{file=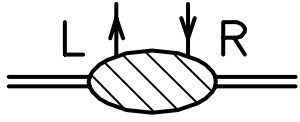,height=1.5cm}
\bigskip
\caption{Chiral (odd) flip distribution which 
gives the
`transversity' structure function $h_1(x,Q^2)$.}
\label{figodd}
\end{center}
\end{figure}

There are further subleading (twist--3) transversity distributions
\cite{jaffe8,jaffe4,mulders}
such as $h_2$ which is the chiral--odd analog of $g_2$.
There is, furthermore, a twist--3 chiral--odd scalar distribution
$e(x,Q^2)$ related to the unit operator, i.e.\ measuring, via its 
first moment,
the nucleon $\sigma$--term.  More explicitly, the complete set of 
structure functions are defined by the light cone Fourier transform
of nucleon matrix elements of quark bilocal operators as follows:
%Gl. 8.42 - 8.45
\begin{eqnarray}
\int\frac{d\lambda}{2\pi}\, e^{i\lambda x} \langle PS|\bar{q}(0)
  \gamma_{\mu}q(\lambda n)|PS\rangle & = & 2[f^q_1(x,Q^2)p_{\mu}
      + M^2f^q_4(x,Q^2)n_{\mu}] \\
\int\frac{d\lambda}{2\pi}\, e^{i\lambda x} \langle PS|\bar{q}(0)
   \gamma_{\mu}\gamma_5 q(\lambda n)|PS\rangle & = & 2M[g^q_1(x,Q^2)p_{\mu}S
       \cdot n + g^q_T(x,Q^2)S_{T\mu}\nonumber \\
          & & +\,M^2g^q_3(x,Q^2)n_{\mu}S\cdot n] \\
\int\frac{d\lambda}{2\pi} \, e^{i\lambda x} \langle PS|\bar{q}(0)q
         (\lambda n)|PS \rangle & = & 2 Me^q(x,Q^2) \\
\int\frac{d\lambda}{2\pi} \, e^{i\lambda x} \langle PS|\bar{q}(0)
         i\sigma_{\mu\nu}\gamma_5 q(\lambda n)|PS \rangle & = & 2[h^q_1
          (x,Q^2)(S_{T\mu}p_{\nu}-S_{T\nu}p_{\mu})\nonumber\\ 
            & & +\, h^q_L(x,Q^2)M^2(p_{\mu}n_{\nu}-p_{\nu}n_{\mu})
                        S\cdot n\nonumber \\
            & &  +\, h^q_3(x,Q^2)M^2(S_{T\mu}n_{\nu}-S_{T\nu}n_{\mu})]
\end{eqnarray}
which holds at some factorization scale $Q^2,\,h^q_L\equiv\frac{1}{2}\,
h^q_1+h^q_2$, and in order to follow more closely the original notation 
\cite{jaffe8,jaffe4} we have used $f^q_1 \equiv q$, 
$g^q_1 \equiv \delta q$, $h^q_1 \equiv \delta_T q$, etc.  
Two light--like null--vectors 
$(p^2=n^2=0)$ have been introduced via the relation $P_{\mu}=p_{\mu}+
\frac{M^2}{2}\,n_{\mu}$ with $p\cdot n=1$, and the nucleon spin vector
is decomposed as $S_{\mu}=S\cdot n\, p_{\mu}+S\cdot p\, n_{\mu} + S_{T\mu}$.
The twist--4 contributions $f_4^q, \, g_3^q$ and $h_3^q$ 
will not be considered
in the following.

The $\delta_T\!\!\stackrel{(-)}{q}$ and the longitudinal 
$\delta\!\!\stackrel{(-)}{q}$ densities are not entirely independent of
each other since one clearly has $\stackrel{(-)}{q}_+ + \stackrel{(-)}{q}_-
={\stackrel{(-)}{q}}\,^{\uparrow} + {\stackrel{(-)}{q}}\,^{\downarrow}$ by
rotational invariance, which implies the general positivity constraints
\cite{jaffe,jaffe4} 
%Gl. 8.46
\begin{equation}
|\delta_T\!\!\stackrel{(-)}{q}(x,Q^2)| \leq\, \stackrel{(-)}{q}(x,Q^2)
\end{equation}
in complete analogy to (4.11).  A second inequality has been derived by
Soffer \cite{soffer}, 
%Gl. 8.47
\begin{equation}
|\delta_T\!\!\stackrel{(-)}{q}(x,Q^2)| \leq\, \frac{1}{2}
  \left[ \stackrel{(-)}{q}(x,Q^2) + \delta\!\!\stackrel{(-)}{q}
    (x,Q^2)\right] = \stackrel{(-)}{q}_+(x,Q^2)\, ,
\end{equation}
which has its origin in the positivity properties of helicity amplitudes.
This latter inequality can also be derived in the context of the parton
model  \cite{soffer,sivers3,goldstein}. 

Being a twist-2 quantity, where only fermions contribute, $h_1(x,Q^2)$,
i.e.\ $\delta\!\!\stackrel{(-)}{q}_T(x,Q^2)$ obey a simple nonsinglet--type
Altarelli--Parisi evolution equation.  In LO it is similar to (4.12) and
reads \cite{artru} 
%Gl. 8.48
\begin{equation}
\frac{d}{dt}\delta_T\!\!\stackrel{(-)}{q}(x,Q^2)=\frac{\alpha_s(Q^2)}{2\pi}\, 
  \delta_T P_{qq}^{(0)} \otimes \delta_T\!\!\stackrel{(-)}{q}
\end{equation}
where $t=\ell n\frac{Q^2}{Q_0^2}$ and
%Gl. 8.49
\begin{equation}
\delta_T P_{qq}^{(0)}(x) = C_F\left[ \frac{2x}{(1-x)_+} + \frac{3}{2}\,
  \delta(1-x) \right]
\end{equation}
with $C_F=\frac{4}{3}$.  Alternatively, in Mellin $n$--moment space this
RG evolution equation becomes similar to (4.24) and reads
%Gl. 8.50
\begin{equation}
\frac{d}{dt}\, \delta_T\!{\stackrel{(-)}{q}}\,^n(Q^2)
         = \frac{\alpha_s(Q^2)}
   {2\pi}\, \delta_T P_{qq}^{(0)n}\, \delta_T\!{\stackrel{(-)}
                {q}}\,^n(Q^2)
\end{equation}
where the $n$--th moment is defined by Eq.\ (4.22) and the $n$--th moment
of (8.49) is given by
%Gl. 8.51
\begin{equation}
\delta_T P_{qq}^{(0)n} = C_F\left[\frac{3}{2}-2S_1(n)\right]
\end{equation}
with $S_1(n)$ being defined right after Eq.\ (4.26).  The solution of
(8.50) is straightforward and is formally identical to the one in 
(4.27).  Recently, the calculation of the NLO 2--loop splitting functions
$\delta_TP_{qq\pm}^{(1)}(x)$ has been completed in the ${\overline{\rm MS}}$
factorization scheme 
\cite{kumano,haya,vogelsang6} 
for the flavor combinations
$\delta_T q_{\pm}\equiv \delta_T q\pm\delta_T\bar{q}$.  The LO evolutions
of the twist--3 distributions $e(x,Q^2$) and $h_L(x,Q^2)$ in (8.44) and
(8.45) have been studied as well recently 
\cite{koike,koike1,balitzky1,belitsky}.  

Since the corresponding evolution kernels are entirely different for the
$\stackrel{(-)}{q}\!\!\!(x,Q^2)$, 
\mbox{$\delta\!\!\stackrel{(-)}{q}\!\!(x,Q^2)$}
and $\delta_T\!\!\stackrel{(-)}{q}\!\!(x,Q^2)$ densities, the question 
immediately arises whether the Soffer inequality (8.47) is maintained when 
QCD is
applied \cite{goldstein,kamal1,barone}.  
It turns out, \mbox{however}, that this 
inequality
is preserved by LO \cite{barone} and NLO 
\cite{vogelsang6,martin3} QCD evolutions at any
$Q^2>Q_0^2$ provided it is valid at the input scale $Q^2=Q_0^2$. 
 
As stated above, the transversity distributions are experimentally entirely
unknown so far.  Being chiral odd, they cannot be directly determined from
the common fully inclusive DIS process, i.e.\ $h_1(x,Q^2)$ is not directly
measurable despite its formal definition in (8.39).  One would need, for
example, to measure a (single) transverse asymmetry in single transversely
polarized {\it {semi}}--inclusive DIS $ep^{\uparrow\downarrow}\to
eh^{\uparrow}X$ with $h=\Lambda$, jet, $\ldots$ 
\cite{artru2,artru,collins1,collins,ji3}. 
This requires, however, a (difficult) measurement of the transverse
polarization of the final state $h^{\uparrow}$.  Old ideas 
\cite{nachtmann,efremov} 
have been revived 
\cite{efremov4,collins1,collins,ji3} for determining the polarization of
an outgoing quark (or gluon), in particular chiral--odd fragmentation
functions, via the hadron distribution in the jet, for example.  Perhaps
a more feasible possibility has been suggested recently 
\cite{jaffe11,jaffe10} to
extract $\delta_Tq$ from DIS two--meson production, e.g.\ $ep^{\uparrow
\downarrow}\to e\pi^+\pi^-X$ via a Collins--angle--like $\phi$ distribution
\cite{collins1} 
by measuring the observable $\vec{\pi}^+ \times \vec{\pi}^-\cdot
\vec{S}_T$, i.e.\ the correlation of the normal to the plane formed by 
the three--momenta $\vec{\pi}^{\pm}$ with the nucleon's transverse spin.
This, however, requires the cross section to be held fully differential
to avoid averaging the meson--meson final state interaction phase to zero.
It is conceivable that HERMES and (in the not too distant future) COMPASS
can perform such measurements.

A more natural way to search for transversity densities is in doubly
transversely polarized hadron--hadron initiated reactions like 
$p^{\uparrow}p^{\uparrow\downarrow}\to\mu^+\mu^-,\, \gamma, \, jj,\, 
c\bar{c},\, \ldots X$ 
\cite{ralston,artru,jaffe8,jaffe4,bunce,cortes1,ji4,vogelsang5,
contogouris3,kamal,kamal1,vogelsang6,jaffe12,martin2,martin3,barone1,robinett}. 
Here the chirality changing densities (cf.\ Fig.\ 54) appear automatically
in the initial states without extra suppressions as illustrated for
Drell-Yan dilepton production in Fig.\ 55.  The expected double transverse
asymmetries $A_{TT}$ turn out, however, to be prohibitively small,
$A_{TT}\ll A_{LL}$ (typically, smaller by about an order of magnitude),
i.e.\ much smaller than the doubly longitudinally polarized asymmetries
(inlcuding the DIS $A_1$ and $A_2$) considered thus far.  Single transverse
asymmetries $A_T$ measured in reactions 
like $pp^{\uparrow\downarrow}\to j^{\uparrow}X$ might be sizeable
(about 10\%) 
\cite{stratmann3}, but require again a delicate measurement of the
polarizations of the outgoing quarks and gluon via the hadron distribution
in the final jet $j^{\uparrow}$.  Such experiments could be performed
at HERA--$\vec{\rm N}$(phase I) and RHIC.  Whereas these purely transverse
asymmetries $A_{TT}$ and $A_T$ measure solely 
$\delta_T\!\!\stackrel{(-)}{q}$, a mixed longitudinal--transverse
asymmetry $A_{LT}$, down by a factor $M/Q$, gives access to $h_L(x,Q^2)
\equiv\frac{1}{2}h_1+h_2$ provided $g_1$ and $g_T=g_1+g_2$ are known
\cite{jaffe8,jaffe4,ji4}. 
It should be noted that most of these processes have
been analyzed in LO--QCD, with the exception of transversely polarized
Drell-Yan dimuon production which has been already extended to the NLO
within different factorization schemes 
\cite{vogelsang5,contogouris3,kamal,vogelsang6,martin3}. 

\begin{figure}
\begin{center}
\epsfig{file=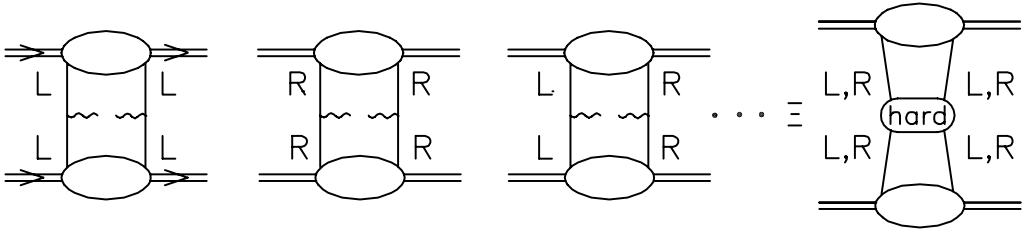,height=3cm}
\bigskip
\caption{Chiral structure of hadronic Drell-Yan 
production
of lepton pairs with invariant mass $Q^2\equiv q^2>0$ produced by the
virtual photons.  Chirality flip ($R\leftrightarrow L$) occurs without
suppression due to the two initial hadrons (non--perturbative bound
states) involved in the process.}
\label{figchdy}
\end{center}
\end{figure}

Due to the lack of any experimental information, all these studies and
expectations for transverse asymmetries are based on theoretical model
estimates for $\delta_T\!\!\!\stackrel{(-)}{q}$ and the other subleading
transversity densities.  In the non--relativistic quark model, 
$\delta_T q(x,Q^2) = \delta q(x,Q^2)$ due to rotational invariance,
whereas in the bag model \cite{jaffe8,jaffe4,stratmann2,scopetta} 
$\delta_T q(x,Q^2)$
\raisebox{-0.1cm}{$\stackrel{>}{\sim}$} $\delta q(x,Q^2)$ 
in the medium to large $x$--region ---
both expectations hold presumably at some non--perturbative 
bound--state--like scale $Q^2={\cal{O}}(\Lambda^2)$.  \mbox{Somewhat} smaller
results, 
$\delta_Tq(x,Q^2)$ \raisebox{-0.1cm}{$\stackrel{<}{\sim}$} $\delta q(x,Q^2)$, 
are obtained by a chiral chromodi--electric confinement model 
\cite{barone2} and
by the chiral soliton approach 
\cite{pobylitsa} at $Q^2\simeq0.3$ GeV$^2$, and
with QCD sum rules 
\cite{ioffe2} at $Q^2$\raisebox{-0.1cm}{$\stackrel{>}{\sim}$} 
1 GeV$^2$.  It seems
that transversity densities are sizeable and not too different from
the (longitudinal) helicity distributions.  Anyway, $\delta_T q\neq
\delta q$ $(h_1\neq g_1)$ directly probes relativistic effects in the
wave function.  In particular, $h_1$ could develop a very different
small--$x$ behavior as compared to $g_1$ 
\cite{barone2,hisai,kirschner,scopetta}. 

There have also been attempts 
\cite{jaffe4,he,soffer,schmidt,barone2} to 
estimate
the nucleon's ``tensor charge'', i.e.\ the total transversity $\Delta_T
q(Q^2)$ carried by quarks,
%Gl. 8.52
\begin{equation}
\int_0^1\left[ \delta_T q(x,Q^2) -\delta_T\bar{q}(x,Q^2) \right]\, dx
   \equiv \Delta_T q(Q^2)\, ,
\end{equation}
which is a flavor non--singlet valence quantity (quarks {\it {minus}}
antiquarks, since the tensor current is C--odd) and measures a simple
local operator analogous to the axial charge \makebox{(cf.\ Eq.\ (5.33))}
%Gl. 8.53
\begin{equation}
\langle PS|\bar{q}i\sigma_{\mu\nu}\gamma_5q|PS\rangle =
  \frac{1}{M}\, (S_{\mu}P_{\nu}-S_{\nu}P_{\mu})\Delta_T q\, .
\end{equation}
Unlike its vector and (NS) axial--vector equivalents, the tensor charge
is not conserved in QCD, so it has an anomalous dimension at one--loop.
To understand the physical meaning of the tensor charge it is helpful
to make a comparison between the chirally odd and even matrix elements
in the rest frame of the nucleon, $P^{\mu}=(M,\vec{0})$ and $S^{\mu}=
(0,\vec{S})$
%Gl. 8.54 u. 8.55
\begin{eqnarray}
\langle PS|\bar{q}i\sigma_{i0}\gamma_5 q|PS\rangle = \langle PS|\bar{q}
         \Sigma_i q|PS \rangle & = & S_i\Delta_T q\\
\langle PS|\bar{q}\, \gamma_i\gamma_5 q|PS\rangle  = \langle PS|\bar{q}
          \gamma_0\Sigma_i q|PS \rangle & = & S_i\Delta(q+\bar{q})\, ,
\end{eqnarray}
i.e.\ the spin operator related to axial current differs by an additional
$\gamma_0$ from the expectation value of the Pauli--Lubanski 
``transversity'' operator.  The above estimates imply 
\cite{he,schmidt,barone2} 
$\Delta_T\Sigma(Q^2)\equiv\Delta_T(u+d+s)=0.6$ to 1 at $Q^2$ of about
$5-10$ GeV$^2$ which compares well with a recent lattice calculation 
\cite{aoki},
$\Delta_T\Sigma\simeq 0.6\pm0.1$.  These results are intriguing since they
seem to imply that the tensor charge behaves more like the `naive' quark
model expectation (5.16), $\Delta_T\Sigma=\Delta\Sigma^{SU(6)}=1$, in
contrast to the experimental result $\Delta\Sigma(Q^2)\simeq 0.2$
(cf.\ Table 3, for example) that quarks carry much less of the nucleon's
spin than naively expected.  Furthermore, one of the outstanding puzzles
is how to obtain an independent measure of $\Delta_Tq(Q^2)$ and thereby
formulate a `transversity sum rule' analogous to those that have been so
helpful in the study of $\Delta(q+\bar{q})$ in connection with the spin
of the proton. 
 
Keeping in mind that we need the transversity densities $\delta_T\!\!
\stackrel{(-)}{q}\!\!(x,Q^2)$, besides the \mbox{more} common $f(x,Q^2)$ and
$\delta f(x,Q^2)$, for a complete understanding of the leading twist--2
\mbox{sector} of the nucleon's parton structure, we face the curious situation
of having reached a remarkably advanced theoretical sophistication 
without having any experimental `transversity' measurements whatsoever!

\vspace*{1cm}

\noindent{\bf{Acknowledgements}}

One of us (E.R.) expresses his warmest thanks to M.\ Gl\"uck for numerous
discussions and for a fruitful collaboration on various topics presented
here during the past decade.  He also thanks M.\ Stratmann and 
W.\ Vogelsang for their collaboration during the past years and for their
help in preparing some of the figures presented.  We are also grateful
to G.\ Altarelli for some helpful and clarifying comments and discussions.
Furthermore, B.L.\ would like to thank J.\ Bartels, H.\ Fritzsch, S.\ Forte,
P.\ Gambino, M. G\"ockeler, Y.\ Koike, J.\ Kodaira, G.\ Piller, G.\ Ridolfi,
P.\ Ratcliffe, M.G.\ Ryskin, A.\ Schaefer, J.\ Soffer 
and E.\ Steffens for comments
and discussions in connection with this work.  Finally we thank 
Mrs.\ H.\ Heininger, Mrs.\ R.\ Jurgeleit and Mrs.\ S.\ Laurent 
for their
help in typing the manuscript and in preparing some of the figures.

This work has been supported in part by the `Bundesministerium f\"ur
Bildung, Wissenschaft, Forschung und Technologie', Bonn, as well as by
the 'Deutsche Forschungsgemeinschaft'.

\vspace*{1cm}

\noindent{\underline{Note added}. Very recently the calculation 
of the NLO contributions to polarized photoproduction of heavy 
quarks, $\vec\gamma \vec p\rightarrow c\bar c X$, as discussed 
in Section 6.2, has been completed [I. Bojak and M. Stratmann, 
Univ. Dortmund/Durham report DO--TH 98/04, DTP/98/22 (hep--ph/9804353), 
Phys. Lett. B, to appear]. As expected, among other things, 
the dependence of $\Delta \sigma^c_{\gamma p}$ in 
(\ref{65n3}) on $\mu_F$ is considerably reduced in NLO.}

\vspace*{1cm}

%% file: katex
\setcounter{equation}{0}
\section{Appendix: Two--loop 
Splitting Functions and Anomalous Dimensions}

In addition to the well known LO (1--loop)polarized splitting 
functions \cite{altarelli} which are given in Eqs. (\ref{4116}) 
and (\ref{4121}), with their n--th moment, as defined in 
(\ref{4122}), given in (\ref{4126}). 
For completeness we list here the results for 
the NLO (2--loop) splitting functions $\delta P_{ij}^{(1)}$ 
as recently calculated in \cite{mertig,vogelsang2}. As discussed 
in the text (Sect. 4, etc.), these calculations were done in the 
$\overline{MS}$ scheme. To deal with $\gamma_5$ and the $\epsilon$--tensor, 
the HVBM scheme [due to its not fully anticommuting $\gamma_5$, 
additional finite renormalizations are required in order to 
guarantee the conservation of the flavor non--singlet axial vector 
current, Eq. (\ref{4322a})] or equivalently the reading point 
method has been used. The flavor non--singlet splitting functions 
in (\ref{4211}) are the same as for the unpolarized case 
\cite{curci} 
\begin{eqnarray}
\delta P_{NS \pm}^{(1)}(x)&=&P_{NS \pm}^{(1)}(x)= 
P_{qq}^{(1)}(x) \pm P_{q\bar q}^{(1)}(x)  \nonumber \\
&=& C_F^2 [-(2\ln x \ln (1-x)+\frac{3}{2}\ln x )\delta p_{qq}(x)
          -(\frac{3}{2} +\frac{7}{2} x)\ln x \nonumber \\ 
& &          -\frac{1}{2} (1+x)\ln^2 x -5(1-x) 
          +(\frac{3}{8} -\frac{\pi^2}{2} +6\zeta (3))\delta (1-x)]
\nonumber \\ 
& & +C_F C_A[(\frac{1}{2}\ln^2 x  +\frac{11}{6}\ln x +\frac{67}{18} 
             -\frac{\pi^2}{6})\delta p_{qq}(x) +(1+x)\ln x 
             +\frac{20}{3} (1-x) 
\nonumber \\
& &              +(\frac{17}{24} +\frac{11}{18}\pi^2-3\zeta (3))\delta (1-x)] 
\nonumber \\ 
& & +C_F T_f[-(\frac{2}{3}\ln x+\frac{10}{9})\delta p_{qq}(x) 
             -\frac{4}{3} (1-x) -(\frac{1}{6}+\frac{2}{9}\pi^2)\delta (1-x)] 
\nonumber \\ 
& & \pm (C_F^2-\frac{1}{2}C_F C_A)[2(1+x)\ln x +4(1-x) 
              +2S_2(x)\delta p_{qq}(-x)]    \,  .  
\label{a001}
\end{eqnarray}
Note that the '$\pm$' meaning of $\delta P_{NS \pm}$ has been 
interchanged in \cite{vogelsang2}. The flavor--singlet splitting 
functions in (\ref{4215}) are given by \cite{mertig,vogelsang2} 
(see Footnote \ref{fn3} and \ref{fn5}) 
\begin{equation}
\delta P_{qq}^{(1)}(x)=\delta P_{NS -}^{(1)}(x)+\delta P_{PS,qq}^{(1)}(x) 
\label{a002}
\end{equation}
with 
\begin{equation}
\delta P_{PS,qq}^{(1)}(x)=2C_F T_f[1-x-(1-3x)\ln x 
                  -(1+x)\ln^2 x]  \/ ,
\label{a003}
\end{equation}                          
and  
%and anomalous dimensions as 
%calculated by \cite{mertig,vogelsang2}. As discussed in the 
%main text, this calculation was done in the $\overline{MS}$--scheme 
%using the reading point method as the $\gamma_5$ prescription . 
%As a shorthand we introduce
%One then has  
\begin{eqnarray}
%\Delta P_{qq}^{\pm,(1)} &=& P_{qq}^{\mp,(1)}  \:\:\: , \label{p1ns} \\
%\Delta P_{qq}^{S,(1)} (x) &=& 2 C_F T_f \Bigg[ \left( 1 - x \right)
%-\left( 1 - 3 x \right) \ln x  - \left( 1 + x \right)
%\ln^2 x \Bigg]  \:\:\: , \label{p1s} \\
2f\delta P_{qg}^{(1)} (x) &=& C_F T_f \Bigg[ -22 + 27 x - 9 \ln x +
     8 \left( 1 - x \right) \ln (1-x) \nonumber \\
&& +
     \delta p_{qg}(x) \left( 2 \ln^2 (1-x) -
      4 \ln (1-x) \ln x + \ln^2 x - \frac{2}{3} \pi^2 \right)
     \Bigg] \nonumber \\
&&+ C_A T_f \Bigg[ 2 \left( 12 - 11 x \right)  -
     8 \left( 1 - x \right) \ln (1-x) + 2 \left( 1 + 8 x \right) \ln x
     \nonumber \\
&& \left. - 2 \left( \ln^2 (1-x) - \frac{\pi^2}{6} \right) \delta p_{qg}(x)
- \left( 2 S_2 (x) - 3 \ln^2 x \right)  \delta p_{qg}(-x) \Bigg] \right.
\:\:\: , \label{p1qg} \\
\delta P_{gq}^{(1)} (x) &=& C_F T_f \left[ -{{4}\over 9} (x+4)
- \frac{4}{3} \delta p_{gq}(x) \ln (1-x) \right] \nonumber
\\ &&+ C_F^2
   \left[ - \frac{1}{2} -
   \frac{1}{2} \left( 4 - x \right) \ln x - \delta p_{gq}(-x) \ln (1-x)
\right. \nonumber \\
&& \left. + \left( - 4 - \ln^2 (1-x) + \frac{1}{2} \ln^2 x \right)
    \delta p_{gq}(x) \right] \nonumber \\
&& + C_F C_A \left[ \left( 4 - 13 x \right) \ln x +
   \frac{1}{3} \left( 10 + x \right) \ln (1-x) +
  \frac{1}{9} \left( 41 + 35 x \right) \right. \nonumber \\
&& + \frac{1}{2} \left( -2 S_2 (x) + 3 \ln^2 x \right)
     \delta p_{gq}(-x) \nonumber \\
&&+ \left. \left( \ln^2 (1-x) - 2 \ln (1-x) \ln x - \frac{\pi^2}{6} \right)
  \delta p_{gq}(x) \right]  \label{p1gq} \\
\delta P_{gg}^{(1)} (x) &=&
  - C_F T_f
  \Bigg[ 10 \left( 1 - x \right)  + 2 \left( 5 - x \right) \ln x  +
     2 \left( 1 + x \right) \ln^2 x
     +\delta (1-x) \Bigg]  \nonumber \\
&& 
  - C_A T_f \left[ 4 \left( 1 - x \right)  + \frac{4}{3}
     \left( 1 + x \right) \ln x + \frac{20}{9} \delta p_{gg}(x)
     +\frac{4}{3} \delta (1-x) \right] \nonumber \\
&& + C_A^2 \Bigg[ \frac{1}{3} \left( 29 - 67 x \right) \ln x -
     \frac{19}{2} \left( 1 - x \right) +
     4 \left( 1 + x \right) \ln^2 x -2 S_2 (x) \delta p_{gg}(-x)
\nonumber \\
&& + \left( {{67}\over 9} - 4 \ln (1-x) \ln x +
        \ln^2 x - \frac{\pi^2}{3} \right) \delta p_{gg}(x)
 +\left( 3 \zeta(3) +\frac{8}{3} \right) \delta (1-x) \Bigg] \; ,
\nonumber \\
\label{p1gg} 
%\\
%\Delta C_q(x) &=& C_F \Bigg[(1+x^2) \left[\frac{\ln (1-x)}{1-x}
%\right]_{\!\!+}
%-\frac{3}{2} \frac{1}{[1-x]_+} -\frac{1+x^2}{1-x} \ln x +
%\nonumber \\
%& & \hspace*{0.75cm}
%+\, 2 + x - \left(\frac{9}{2}+\frac{\pi^2}{3}\right) \delta (1-x)
%\Bigg]  \label{cqend} \:\:\: ,  \\
%\Delta C_g(x) &=& 2 T_f \left[ (2x-1) \left(\ln \frac{1-x}{x}-1\right)+
%2(1-x)\right] \:\:\: \label{cgend} 
.
\end{eqnarray}
where $C_F=4/3$, $C_A=3$, $T_f=fT_R=f/2$, $\zeta (3)\approx 1.202057$ 
and
\begin{eqnarray}
\delta p_{qq} (x) &=& {2\over (1-x)_+} -x-1 \:\:\: , \nonumber \\ 
\delta p_{qg} (x) &=& 2 x-1  \:\:\: ,  \qquad \quad 
\delta p_{gq} (x) = 2 - x  \:\:\: , \nonumber \\
\delta p_{gg} (x) &=& \frac{1}{(1-x)_+} - 2 x + 1  \label{pgg}  \:\:\:
.
\end{eqnarray}
and the '+' description has obviously to be omitted for 
$\delta p_{ii} (-x)$. Furthermore, 
%The unpolarized NS pieces $P_{qq}^{\mp, (1)}$ appearing in 
%Eq. (\ref{p1ns}) can be found
%in \cite{curci}, and
\begin{equation}
S_2(x)= \int_{\frac{x}{1+x}}^{\frac{1}{1+x}} \frac{dz}{z}
\ln \big(\frac{1-z}{z}\big)  \; .
\label{a008}
\end{equation}
For relating the above results to those of \cite{mertig} 
the following 
expression for $S_2(x)$ is needed: 
\begin{equation}
S_2 (x) = -2 {\rm Li}_2 (-x)-2 \ln x \ln (1+x)+\frac{1}{2} \ln^2 x-
\frac{\pi^2}{6}
\label{a009}
\end{equation}
where ${\rm Li}_2 (x)$ is the usual Dilogarithm 
\cite{devoto}. 
The coefficient functions relevant for a consistent NLO($\overline{MS}$) 
analysis of $g_1(x,Q^2)$ in (\ref{429}) are given by Eqs. 
(\ref{424q}) and (\ref{424}). It should be recalled that 
convolutions involving the $(~~)_+$ distributions can be 
conveniently calculated numerically with the help of Eq. (\ref{4118}). 

The n--th moments of these splitting functions, defined in 
(\ref{4122}), which are needed for the evolution equations 
(\ref{4216}) and (\ref{4217}) in Mellin--moment space, are as follows. 
The moments of the LO splitting functions, $\delta P_{ij}^{(0)n}$, 
are given in (\ref{4126}). The moments of the $\delta P_{NS\pm}^{(1)}(x)$ 
in (\ref{a001}), being the same as for the unpolarized case 
\cite{curci,floratos} (see Footnote \ref{fn4}), are 
\begin{eqnarray}
-\delta P_{NS \pm}^{(1)n}&=&
   C_F^2 [ 2\frac{2n+1}{n^2 (n+1)^2}S_1(n) 
         +2(2S_1(n)-\frac{1}{n(n+1)})(S_2(n)-S_2'(\frac{n}{2})) 
\nonumber \\ 
& &          +3S_2(n)+8\tilde{S}(n)-S_3'(\frac{n}{2}) 
         -\frac{3n^3+n^2-1}{n^3 (n+1)^3}-\frac{3}{8} 
         \mp 2\frac{2n^2+2n+1}{n^3 (n+1)^3} ] 
\nonumber \\ 
& &   +C_F C_A [\frac{67}{9}S_1(n)
             -(2S_1(n)-\frac{1}{n (n+1)})(2S_2(n)-S_2'(\frac{n}{2})) 
             -\frac{11}{3}S_2(n)-4\tilde{S}(n)+\frac{1}{2}S_3'(\frac{n}{2}) 
\nonumber \\ 
& &        -\frac{1}{18}\frac{151n^4+236n^3+88n^2+3n+18}{n^3 (n+1)^3} 
       -\frac{17}{24}\pm \frac{2n^2+2n+1}{n^3 (n+1)^3} ] 
\nonumber \\ 
& & +C_F T_f[-\frac{20}{9}S_1(n)+\frac{4}{3}S_2(n)
           +\frac{2}{9}\frac{11n^2+5n-3}{n^2 (n+1)^2}+\frac{1}{6}] 
\/ . 
\label{a010}
\end{eqnarray}
The moments of the flavor singlet splitting functions in 
(\ref{a002})--(\ref{p1gg}) are given by \cite{mertig,glueck2} 
(see Footnote \ref{fn5}) 
\begin{equation}
\delta  P_{qq}^{(1)n}=\delta  P_{NS -}^{(1)n}+\delta  P_{PS,qq}^{(1)n}
\label{a011}
\end{equation}
%The NLO anomalous dimensions can be obtained 
%from the above x--dependent Altarelli--Parisi functions by taking 
%moments.  
%The spin-dependent NLO ($\overline{\rm{MS}}$) two-loop flavor non-singlet
%anomalous dimensions $\delta\gamma_{NS}^{(1)n}(\eta)$, required 
%for the evolution of $\delta q_{NS\,\eta=\pm}^n(Q^2)$,
%are the same as found for the spin-averaged case, $\delta\gamma_{NS}^{(1)n}
%(\eta)=\gamma_{NS}^{(1)n}(\eta)$ with $\gamma_{NS}^{(1)n}(\eta=\pm 1)$
%being given by eq.(B.18) of \cite{floratos}. Note that
%$\delta\gamma_{NS}^{(1)n}(\eta=+1)$ governs the evolution of the NS
%combinations $\delta q - \delta \bar{q}$, while
%$\delta\gamma_{NS}^{(1)n}(\eta=-1)$ refers to the combination NS 
%$\delta q +\delta\bar{q}$.
%The NLO flavor singlet anomalous dimensions $\delta\gamma_{ij}^{(1)n}$ in
%the $\overline{\rm{MS}}$ scheme are as follows
%\cite{mertig,vogelsang2} :
%                 
%\begin{equation}  
%\delta\gamma_{qq}^{(1)n} =\gamma_{NS}^{(1)n}(\eta=-1) + \delta
%\gamma_{PS,qq}^{(1)n}
%\end{equation}    
%                 
%with $\gamma_{NS}^{(1)n}(\eta=-1)$ being again given by Eq. (B.18) of
%\cite{floratos} and 
%        
with
\begin{equation}  
-\delta P_{PS,qq}^{(1)n} = 2 C_F T_f\;
\frac{n^4+2 n^3 + 2 n^2 + 5 n + 2}{n^3 (n + 1)^3} \\
\end{equation}    
and     
\begin{eqnarray}  
\nonumber         
-2f \delta P_{qg}^{(1)n} &=&  C_F T_f \left[ 2\, \frac{n-1}{n (n+1)}
\left( S_2(n)- S_1^2(n) \right) + 4 \,
\frac{n-1}{n^2 (n+1)} S_1(n)  \right. \\
\nonumber         
&& \left. - \frac{5 n^5+5 n^4 -10 n^3 -n^2 +3n -2}{n^3 (n+1)^3} \right] \\
\nonumber         
&+& 2 C_A T_f \left[ \frac{n-1}{n (n+1)}
\left(-S_2(n)+S_2^{'}\left(\frac{n}{2}\right)+S_1^2(n)\right) - \frac{4}
{n(n+1)^2} S_1(n)  \right. \\
&& \left. - \frac{n^5+n^4-4n^3+3n^2-7n -2}{n^3(n+1)^3}\right]\\
\nonumber         
-\delta P_{gq}^{(1)n} &=&
4 C_F T_f \left[- \frac{n+2}{3n(n+1)} S_1(n) +\frac{5 n^2+12 n+ 4}
{9 n (n+1)^2} \right] \\
\nonumber         
&+&  C_F^2 \left[ \frac{n+2}{n (n+1)} \left(S_2(n)+S_1^2(n)\right)-
 \frac{3 n^2+7 n +2}{n (n+1)^2} S_1(n) \right. \\
\nonumber         
&&\left. +\frac{9n^5+ 30 n^4+24 n^3-7 n^2 - 16 n -4}{2n^3 (n+1)^3}
\right] \\        
\nonumber         
&+&  C_F C_A \left[ \frac{n+2}{n (n+1)} \left(-S_2(n)+S_2^{'}\left(
\frac{n}{2}\right)-S_1^2(n)\right)+
\frac{11 n^2+ 22 n+ 12}{3n^2 (n+1)} S_1(n)  \right. \\
&& \left. - \frac{76 n^5 + 271 n^4 + 254 n^3 + 41 n^2 + 72 n +36}
{9 n^3 (n+1)^3} \right]\\
\nonumber         
-\delta P_{gg}^{(1)n} &=&
 C_F T_f\; \frac{n^6+3 n^5+ 5 n^4+ n^3-8 n^2+2 n+ 4}{n^3 (n+1)^3}\\
\nonumber         
&+& 4 C_A T_f \left[-\frac{5}{9} S_1(n) +
\frac{3 n^4+6 n^3 + 16 n^2 + 13 n - 3}{9 n^2 (n+1)^2}\right] \\
\nonumber         
&+&  C_A^2 \left[-\frac{1}{2}S_3^{'}\left(\frac{n}{2}\right)-2 S_1(n) S_2^{'}
\left(\frac{n}{2}\right) + 4 \tilde{S}(n) +\frac{4}{n (n+1)}
S_2^{'}\left(\frac{n}{2}\right) \right. \\
\nonumber         
&& \left. + \frac{67 n^4 + 134 n^3 + 67 n^2 + 144 n + 72}
{9 n^2 (n+1)^2} S_1(n) \right. \\
&& \left. - \frac{48 n^6 + 144 n^5 + 469 n^4 + 698 n^3 + 7 n^2 + 258 n
+144}{18 n^3 (n+1)^3}\right]
\end{eqnarray}    
where             
\begin{eqnarray}  
S_k(n) & \equiv & \sum_{j=1}^n \frac{1}{j^k}\\
S_k'\left(\frac{n}{2}\right) & \equiv & 2^{k-1} \sum_{j=1}^n
\frac{1+(-)^j}{j^k}
= \frac{1}{2} (1+\eta ) S_k\left(\frac{n}{2}\right)+
\frac{1}{2} (1-\eta ) S_k\left(\frac{n-1}{2}\right)\\
\tilde{S}(n) & \equiv & \sum_{j=1}^n \frac{(-)^j}{j^2} S_1(j)
=  -\frac{5}{8} \zeta (3) +\eta \left[ \frac{S_1(n)}{n^2} +
\frac{\pi^2 }{12} G(n) +\int_0^1 dx\; x^{n-1} \frac{{\rm{Li}}_2(x)}
{1+x}\right]      
\end{eqnarray}    
with $G(n)\equiv \psi\left(\frac{n+1}{2}\right) -
\psi\left(\frac{n}{2}\right)$, $\psi (z)=d\ln \Gamma (z) /dz$  
and $\eta =\pm 1$ for $\delta P_{NS \pm}^{(1)n}$ and 
$\eta =- 1$ for the flavor singlet anomalous  
dimensions. 
The analytic continuations in $n$, required for the
Mellin inversion of these sums to Bjorken-$x$ space [cf. 
Eq. (\ref{4136})], are well known \cite{glueck,glueck1} 
(see also \cite{weigl}). The conventional anomalous dimensions 
$\delta\gamma_{ij}^{(1)n}$ are related to the 
$\delta P_{ij}^{(1)n}$ via $\delta\gamma_{ij}^{(1)n}=-8 
\delta P_{ij}^{(1)n}$ (cf. Footnotes \ref{fn1} and \ref{fn5}). 
The moments of the relevant $\overline{MS}$ coefficient functions 
for $g_1^n(Q^2)$ are given in Eqs. (\ref{4225}) and (\ref{4226}). 
Finally we list for completeness the first moments 
$\Delta P_{ij}^{(1)}\equiv 
\delta P_{ij}^{(1)n=1}=\int_0^1 dx\delta P_{ij}^{(1)}(x)$     
 and the ones of the coefficient functions 
$\Delta C_i\equiv \delta C_i^{n=1}$ : 
\begin{eqnarray}  
\Delta P_{NS -}^{(1)}&=&0\;,\;\;\;
\Delta P_{NS +}^{(1)}=(C_F^2-{1\over 2} C_FC_A)
     ({13\over 2}-\pi^2+4\zeta (3))
\nonumber \\
\Delta P_{qq}^{(1)}&=&-3C_FT_f\;,\;\;\;
\Delta P_{qg}^{(1)}=0
\nonumber \\
\Delta P_{gq}^{(1)}&=&-{9\over 4}C_F^2+{71\over 12}C_FC_A 
                   -{1\over 3} C_FT_f 
\nonumber \\
\Delta P_{gg}^{(1)}&=&{17\over 6} C_A^2-C_FT_f-{5\over 3}C_AT_f 
\equiv {\beta_1\over 4}
\end{eqnarray}    
and 
\begin{equation}
\Delta C_q=-{3\over 2} C_F\;,\;\;\;
\Delta C_g=0\;\;\;.             
\end{equation}